%% file: FSQ-14-002_temp.tex
\pdfoutput=1

\documentclass[11pt,twoside,a4paper,cmspaper,final,collab]{cms-tdr}
\def\svnVersion{464816P}\def\cmsCernNoTag{CERN-EP-2017-327}\def\cmsCernDate{\today}\def\cmsMessage{Published in Physical Review C as \href{http://dx.doi.org/10.1103/PhysRevC.97.064912}{\doi{10.1103/PhysRevC.97.064912.}}}

\begin{document}\cmsNoteHeader{FSQ-14-002}

\hyphenation{had-ron-i-za-tion}
\hyphenation{cal-or-i-me-ter}
\hyphenation{de-vices}
\RCS$Revision: 464643 $
\RCS$HeadURL: svn+ssh://svn.cern.ch/reps/tdr2/papers/FSQ-14-002/trunk/FSQ-14-002.tex $
\RCS$Id: FSQ-14-002.tex 464643 2018-06-15 06:48:52Z sikler $
\ifthenelse{\boolean{cms@external}}{\providecommand{\cmsLeft}{upper\xspace}}{\providecommand{\cmsLeft}{left\xspace}}
\ifthenelse{\boolean{cms@external}}{\providecommand{\cmsRight}{lower\xspace}}{\providecommand{\cmsRight}{right\xspace}}
\ifthenelse{\boolean{cms@external}}{\providecommand{\cmsLLeft}{Upper\xspace}}{\providecommand{\cmsLLeft}{Left\xspace}}
\ifthenelse{\boolean{cms@external}}{\providecommand{\cmsRRight}{Lower\xspace}}{\providecommand{\cmsRRight}{Right\xspace}}
\providecommand{\HYDJET} {{\textsc{hydjet}}\xspace}
\newcommand{\ssNN}{\ensuremath{\sqrt{\smash[b]{s_{_\text{NN}}}}}\xspace}
\newcommand{\Noff}{\ensuremath{N_\text{rec1}}\xspace}
\newcommand{\Nr}{\ensuremath{N_\text{rec2}}\xspace}
\newcommand{\Ntpt}{\ensuremath{N_\text{trk0.4}}\xspace}
\newcommand{\Nt}{\ensuremath{N_\text{trk}}\xspace}
\newcommand{\qi}{\ensuremath{q_\text{inv}}\xspace}
\newcommand{\ql}{\ensuremath{q_\text{L}}\xspace}
\newcommand{\qt}{\ensuremath{q_\text{T}}\xspace}
\newcommand{\qo}{\ensuremath{q_\text{O}}\xspace}
\newcommand{\qs}{\ensuremath{q_\text{S}}\xspace}
\newcommand{\Ri}{\ensuremath{R_\text{inv}}\xspace}
\newcommand{\Rl}{\ensuremath{R_\text{L}}\xspace}
\newcommand{\Rt}{\ensuremath{R_\text{T}}\xspace}
\newcommand{\Ro}{\ensuremath{R_\text{O}}\xspace}
\newcommand{\Rs}{\ensuremath{R_\text{S}}\xspace}
\newcommand{\Rp}{\ensuremath{R_\text{param}}}
\newcommand{\ldg}{Lines are drawn to guide the eye.}
\newlength\cmsTabSkip\setlength{\cmsTabSkip}{2ex}

\cmsNoteHeader{FSQ-14-002}

\title{Bose--Einstein correlations in \texorpdfstring{$\Pp\Pp$, $\Pp$Pb, and
PbPb}{pp, pPb, and PbPb} collisions at \texorpdfstring{$\ssNN =$ 0.9--7\TeV}{sqrt(sNN) = 0.9 - 7 TeV}}

\date{\today}

\abstract{
Quantum statistical (Bose--Einstein) two-particle correlations are measured in
pp collisions at $\sqrt{s}= 0.9$, 2.76, and 7\TeV, as well as in $\Pp$Pb and
peripheral PbPb collisions at nucleon-nucleon center-of-mass energies of 5.02
and 2.76\TeV, respectively, using the CMS detector at the LHC.
Separate analyses are performed for same-sign unidentified charged particles as
well as for same-sign pions and kaons identified via their energy loss in the
silicon tracker.
The characteristics of the one-, two-, and three-dimensional correlation
functions are studied as functions of the pair average transverse momentum
($\kt$) and the charged-particle multiplicity in the event.
For all systems, the extracted correlation radii steadily increase with the
event multiplicity, and decrease with increasing $\kt$. The radii are in the
range 1--5\unit{fm}, the largest values corresponding to very high multiplicity $\Pp$Pb
interactions and to peripheral PbPb collisions with multiplicities similar to
those seen in $\Pp$Pb data.
It is also observed that the dependencies of the radii on multiplicity and
$\kt$ largely factorize. At the same multiplicity, the radii are relatively
independent of the colliding system and center-of-mass energy.  }

\hypersetup{%
pdfauthor={CMS Collaboration},%
pdftitle={Bose-Einstein correlations in pp, pPb, and PbPb collisions at
sqrt\(sNN\) = 0.9-7 TeV },%
pdfsubject={CMS},%
pdfkeywords={CMS, physics, femtoscopy, hadrons, HBT, particle identification}}

\maketitle

\section{Introduction}
\label{sec:intro}

Studies of quantum-statistical correlations of pairs of identical particles produced in
high-energy collisions provide valuable information about the size and shape of
the underlying emitting system.
The general technique is similar to the intensity interference mechanism used
by Hanbury-Brown and Twiss (HBT effect)~\cite{hbt1,hbt2,hbt3} for estimating
angular dimensions of stars. In high-energy collisions, the equivalent of this
astronomical effect in the femtoscopic realm was discovered in
antiproton-proton collisions at $\sqrt{s} = 1.05\GeV$~\cite{gglp}.
The effect is known as Bose--Einstein correlations (BEC) when dealing with
bosonic pairs, or femtoscopy since the characteristic probed lengths are in the
femtometer range. It relates the joint probability of observing two identical
particles to the product of the isolated probabilities of observing each
one independently. The result is a correlation function in terms of the
relative momentum of the particles in the pair, reflecting the so-called length
of homogeneity of the particle-emitting source.

A broad investigation of correlations of like-sign charged particles as a
function of the invariant four-momentum difference of the particles was
performed by CMS using $\Pp\Pp$ collisions at $\sqrt{s} =
0.9\TeV$ \cite{cms-hbt-1st,cms-hbt-2nd}, 2.36\TeV~\cite{cms-hbt-1st}, and
7\TeV~\cite{cms-hbt-2nd}.
The present paper extends the investigation of such femtoscopic correlations
using two different analysis methods. First, correlations in $\Pp\Pp$ collisions at
$\sqrt{s} =2.76$ and 7\TeV are extracted using the ``double ratio''
technique~\cite{cms-hbt-1st,cms-hbt-2nd} for unidentified pairs of same-sign
charged particles with respect to different components of the relative
three-momentum of the pair, thereby allowing the exploration of the source
extent in various spatial directions. This procedure has the advantage of
suppressing non-BEC effects coming from multi-body resonance decays, mini-jets,
and energy-momentum conservation with the help of collision events simulated
without Bose--Einstein correlations. Secondly, BEC effects are also studied
using pairs of identical charged pions and kaons, identified via their energy
loss in the CMS silicon tracker, in $\Pp\Pp$ collisions at $\sqrt{s} = 0.9$, 2.76,
and 7\TeV, $\Pp$Pb collisions at a nucleon-nucleon center-of-mass energy of $\ssNN
= 5.02$\TeV, and peripheral PbPb collisions at $\ssNN = 2.76$\TeV.
The suppression of non-BEC contributions is less model-dependent with this
``particle identification and cluster subtraction'' approach, which also has
different systematic uncertainties than the double ratio method.
In both cases, the characteristics of the one-, two-, and three-dimensional
correlation functions are investigated as functions of pair average transverse
momentum, $\kt$, and charged-particle multiplicity in the pseudorapidity range
$\abs{\eta} < 2.4$.

The paper is organized as follows. The CMS detector is introduced in
Section~\ref{sec:detector}, while track selections and particle identification
are detailed in Section~\ref{sec:data_analysis}. In
Section~\ref{sec:techniques}, the two analysis methods are described. A
compilation of the results is presented in
Sections~\ref{sec:legacy-results-1d2d3d} and
\ref{sec:clusteremov_results_pp_ppb_pbpb}. Finally,
Section~\ref{sec:summ_conclusions} summarizes and discusses the conclusions of
this study.

\section{The CMS detector}
\label{sec:detector}

A detailed description of the CMS detector can be found in Ref.~\cite{cmstdr}.
The CMS experiment uses a right-handed coordinate system, with the origin at
the nominal interaction point (IP) and the $z$ axis along the
counterclockwise-beam direction.
The central feature of the CMS apparatus is a superconducting solenoid of 6 m
internal diameter. Within the 3.8\unit{T} field volume are the silicon pixel and
strip tracker, the crystal electromagnetic calorimeter, and the brass and
scintillator hadronic calorimeter. The tracker measures charged particles
within the $\abs{\eta} < 2.4$ range. It has 1440 silicon pixel and 15\,148 silicon
strip detector modules, ordered in 13 tracking layers. In addition to the
electromagnetic and hadron calorimeters composed of a barrel and two endcap
sections, CMS has extensive forward calorimetry. Steel and quartz fiber hadron
forward (HF) calorimeters cover $3 < \abs{\eta} < 5$.
Beam Pick-up Timing for the eXperiments (BPTX) devices are used to trigger the
detector readout. They are located around the beam pipe at a distance of 175 m
from the IP on either side, and are designed to provide precise information on
the LHC bunch structure and timing of the incoming beams.

\section{Selections and data analysis}
\label{sec:data_analysis}

\subsection{Event and track selections}
\label{subsec:selections}

At the trigger level, the coincidence of signals from both BPTX devices is
required, indicating the presence of two bunches of colliding protons and/or ions
crossing the IP. In addition, at least one track with $\pt > 0.4\GeV$ within
$\abs{\eta} < 2.4$ is required in the pixel detector. In the offline selection, the
presence of at least one tower with energy above 3\GeV in each of the HF
calorimeters, and at least one interaction vertex reconstructed in the tracking
detectors are required. Beam halo and beam-induced background events, which
usually produce an anomalously large number of pixel
hits~\cite{Khachatryan:2010xs}, are suppressed.
The two analysis methods employ slightly differing additional event and
particle selection criteria, which are detailed in
Sections~\ref{subsubsec:legacy_selections} and
\ref{subsubsec:data_driven_selections}.

\subsubsection{Double ratio method}
\label{subsubsec:legacy_selections}

In the case of the double ratio method, minimum bias events are selected in a
similar manner to that described in Refs.~\cite{cms-hbt-1st,cms-hbt-2nd}.
Events are required to have at least one reconstructed primary vertex within
15~cm of the nominal IP along the beam axis and within 0.15~cm transverse to
the beam trajectory. At least two reconstructed tracks are required to be
associated with the primary vertex.
Beam-induced background is suppressed by rejecting events containing more
than 10 tracks for which it is also found that less than 25\% of all the
reconstructed tracks in the event pass the \textit{highPurity} track selection
defined in Ref.~\cite{TRK-11-001}.

A combined event sample produced in $\Pp\Pp$ collisions at $\sqrt{s} = 7\TeV$ is
considered, which uses data from different periods of CMS data taking, i.e.
from the commissioning run (23 million events), as well as from later runs
(totalling 20 million events). The first data set contains almost exclusively
events with a single interaction per bunch crossing, whereas the later ones
have a non-negligible fraction of events with multiple $\Pp\Pp$ collision events
(pileup). In such cases, the reconstructed vertex with the largest number of
associated tracks is selected. An alternative event selection for reducing
pileup contamination is also investigated by considering only events with a
single reconstructed vertex. This study is then used to assess the related
systematic uncertainty.
Minimum bias events in $\Pp\Pp$ collisions at 7\TeV (22 million) and at 2.76\TeV (2
million) simulated with the Monte Carlo (MC) generator
{\PYTHIA}6.426~\cite{Sjostrand:2006za} with the tune Z2~\cite{Field:2011iq} are
also used to construct the single ratios employed in the
double ratio technique. Two other tunes ({\PYTHIA}6 D6T~\cite{Field:2010bc} and
{\PYTHIA}6 Z2*~\cite{Chatrchyan:2013gfi,Khachatryan:2015pea}) are used for
estimating systematic uncertainties related to the choice of the MC tune. The
samples for each of these tunes contain 2 million events. The selected tunes
describe reasonably well the particle spectra and their multiplicity
dependence~\cite{Khachatryan:2010xs,Khachatryan:2010us,Chatrchyan:2012qb}.

The selected events are then categorized by the multiplicity of reconstructed
tracks, $\Noff$, which is obtained in the region $\abs{\eta}<2.4$, after imposing
additional conditions:  $\pt > 0.4\GeV$, $\abs{d_z/\sigma (d_z)}< 3.0$ (impact
parameter significance of the track with respect to the primary vertex along
the beam axis), $\abs{d_{xy}/\sigma (d_{xy})}< 3.0$ (impact parameter significance
in the transverse plane), and $\sigma (\pt)/\pt < 0.1$ (relative $\pt$
uncertainty). The resulting multiplicity range is then divided in three (for $\Pp\Pp$ collisions at 2.76\TeV) or four (for $\Pp\Pp$ collisions at 7\TeV) bins. The
corrected average charged-particle multiplicity, $\langle\Ntpt\rangle$, in the
same acceptance range $\abs{\eta}<2.4$ and $\pt > 0.4\GeV$, is determined using
the efficiency estimated with \PYTHIA, as shown in Table~\ref{tab:ranges-dr}.
While $\Noff$ is a measured quantity used to bin the data, for a given bin in
this variable, the calculated $\langle\Ntpt\rangle$ value (which is only known
on average) facilitates comparisons of the data with models.

\begin{table}[b]
 \topcaption{Ranges in charged-particle multiplicity over $\abs{\eta}<2.4$,
$\Noff$, and corresponding average corrected number of tracks with
$\pt>0.4\GeV$, $\langle\Ntpt\rangle$, in $\Pp\Pp$ collisions at 2.76 and 7\TeV
considered in the double ratio method. The $\langle\Ntpt\rangle$ values are
rounded off to the nearest integer. A dash indicates that
there is not enough data to allow for a good-quality measurement.}

 \label{tab:ranges-dr}

 \centering
 \begin{scotch}{ccc}
   \multirow{2}{*}{$\Noff$} & \multicolumn{2}{c}{$\langle\Ntpt\rangle$} \\ \cline{2-3}
           & 2.76\TeV & 7\TeV \\
  \hline
      0--9 &      7  &    6 \\
    10--14 &     14  &   14 \\
    15--19 &     20  &   20 \\
    20--29 &     28  &   28 \\
    30--79 &     47  &   52 \\[\cmsTabSkip]
     0--9  &      7  &    6 \\
    10--24 &     18  &   19 \\
    25--79 &     42  &   47 \\
   80--110 &     \NA  &  105 \\
 \end{scotch}
\end{table}

In Table~\ref{tab:ranges-dr}, two different sets of $\Noff$ ranges are shown.
The upper five ($\Noff =$ 0--9, 10--14, 15--19, 20--29, 30--79) are used for
comparisons with previous CMS results~\cite{cms-hbt-2nd}.
The larger sample recorded in $\Pp\Pp$ collisions at 7\TeV allowed the analysis to be
extended to include the largest bin in multiplicity shown in
Table~\ref{tab:ranges-dr}.

For the BEC analysis the standard CMS {\it highPurity}~\cite{TRK-11-001} tracks
were used. The additional track selection requirements applied to all the
samples mentioned above follow the same criteria as in
Refs.~\cite{cms-hbt-1st,cms-hbt-2nd}, i.e.  primary tracks  falling in the
kinematic range of $\abs{\eta}<2.4$ with full azimuthal coverage and $\pt >$
200\MeV are used. To remove spurious tracks, primary tracks with a loose
selection on the $\chi^2$ of the track fit ($\chi^2/\mathrm{ndf} < 5$, where
ndf is the number of degrees of freedom) are used.  After fulfilling  these
requirements, tracks are further constrained to have an impact parameter with
respect to the primary vertex of $\abs{d_{xy}} < 0.15\cm$. Furthermore, the
innermost measured point of the track must be within 20\cm of the primary
vertex in the plane transverse to the beam direction. This requirement reduces
the number of electrons and positrons from photon conversions in the detector
material, as well as secondary particles from the decays of long-lived hadrons
($\PKzS$, $\PgL$, etc.). After applying these requirements, a small residual
amount of duplicated tracks remain in the sample. In order to eliminate them,
track pairs with an opening angle smaller than 9\unit{mrad} are rejected, if the
difference of their $\pt$ is smaller than 0.04\GeV, a requirement that is found
not to bias the BEC results~\cite{cms-hbt-1st,cms-hbt-2nd}.

\subsubsection{Particle identification and cluster subtraction method}
\label{subsubsec:data_driven_selections}

The analysis methods (event selection, reconstruction of charged particles in
the silicon tracker, finding interaction vertices, treatment of pileup) used
for this method are identical
to the ones used in the previous CMS papers on the
spectra of identified charged hadrons produced in $\sqrt{s} = 0.9$, 2.76, and
7\TeV pp~\cite{Chatrchyan:2012qb}, and in $\ssNN = 5.02\TeV$
$\Pp$Pb~\cite{Chatrchyan:2013eya} collisions.
In the case of $\Pp$Pb collisions, due to the asymmetric beam energies, the
nucleon-nucleon center-of-mass is not at rest with respect to the laboratory
frame, but moves with a velocity $\abs{\beta}$ = 0.434. Data were taken with both
directions of the colliding proton and Pb beams, and are combined in this
analysis by reversing the $z$ axis for one of the beam directions.

For this method, 9.0, 9.6, and 6.2 million minimum bias events from $\Pp\Pp$ collisions at $\sqrt{s} = 0.9$, 2.76, and 7\TeV, respectively, as well
as 9.0 million minimum bias events from $\Pp$Pb collisions at $\ssNN = 5.02$\TeV
are used.
The data sets have small fractions of events with pileup.
The data samples are complemented with 3.1 million peripheral (60--100\%
centrality) PbPb events at $\ssNN = 2.76$\TeV, where 100\% corresponds to fully
peripheral and 0\% to fully central (head-on) collisions. The centrality
percentages of the total inelastic hadronic cross section for PbPb collisions
are determined by measuring the sum of the energies in the HF calorimeters.

This analysis extends charged-particle reconstruction down to $\pt \approx
0.1\GeV$ by exploiting special tracking algorithms used in previous CMS
studies~\cite{Khachatryan:2010xs,Khachatryan:2010us,Chatrchyan:2012qb,Chatrchyan:2013eya}
that provide high reconstruction efficiency and low background rates.
The acceptance of the tracker is flat within 96--98\% in the range $-2 < \eta <
2$ and $\pt > 0.4\GeV$. The loss of acceptance for $\pt < 0.4\GeV$ is caused by
energy loss and multiple scattering of particles, both depending on the
particle mass. Likewise, the reconstruction efficiency is about 75--90\% for
$\pt > 0.4\GeV$, degrading at low \pt, also in a mass-dependent way. The
misreconstructed-track rate, defined as the fraction of reconstructed primary
tracks without a corresponding genuine primary charged particle, is very small,
reaching 1\% only for $\pt < 0.2\GeV$. The probability of reconstructing
multiple tracks from a single true track is about 0.1\%, mostly due to
particles spiralling in the strong magnetic field of the CMS solenoid.  For the
range of event multiplicities included in the current study, the efficiencies
and background rates do not depend on the multiplicity.

\begin{table}[tbh]
\topcaption{Relation between the number of reconstructed tracks ($\Nr$, $\pt >
0.1\GeV$) and the average number of corrected tracks ($\langle\Nt\rangle$, $\pt
> 0$) in the region $\abs{\eta} < 2.4$ for the 24 multiplicity classes
considered in the particle identification and cluster subtraction method. The
values are rounded off to the nearest integer. The corrected
$\langle\Nt\rangle$ values listed for $\Pp\Pp$ collisions are common to all three
measured energies. A dash  indicates that there is not
enough data to allow for a good-quality measurement.}
 \label{tab:multiClass}
 \centering
 \begin{scotch}{cccc}
  \multirow{2}{*}{$\Nr$} & \multicolumn{3}{c}{$\langle\Nt\rangle$} \\ \cline{2-4}
          & $\Pp\Pp$  & $\Pp$Pb & PbPb \\
  \hline
   0--9   &   7 &   8 &   7 \\
  10--19  &  16 &  19 &  19 \\
  20--29  &  28 &  32 &  32 \\
  30--39  &  40 &  45 &  45 \\
  40--49  &  52 &  58 &  58 \\
  50--59  &  63 &  71 &  71 \\
  60--69  &  75 &  84 &  84 \\
  70--79  &  86 &  96 &  97 \\
  80--89  &  98 & 109 & 110 \\
  90--99  & 109 & 122 & 123 \\
 100--109 & 120 & 135 & 136 \\
 110--119 & 131 & 147 & 150 \\
 120--129 & 142 & 160 & 163 \\
 130--139 & \NA  & 173 & 177 \\
 140--149 & \NA  & 185 & 191 \\
 150--159 & \NA  & 198 & 205 \\
 160--169 & \NA  & 210 & 219 \\
 170--179 & \NA  & 222 & 233 \\
 180--189 & \NA  & 235 & 247 \\
 190--199 & \NA  & 247 & 261 \\
 200--209 & \NA  & 260 & 276 \\
 210--219 & \NA  & 272 & 289 \\
 220--229 & \NA  & 284 & 302 \\
 230--239 & \NA  & 296 & 316 \\
 \end{scotch}
\end{table}

An agglomerative vertex reconstruction algorithm~\cite{Sikler:2009nx} is used,
having as input the $z$ coordinates (and their uncertainties) of the tracks at
the point of closest approach to the beam axis. The vertex reconstruction
resolution in the $z$ direction depends strongly on the number of reconstructed
tracks, but is always smaller than 0.1\unit{cm}. Only tracks associated with a
primary vertex are used in the analysis. If multiple vertices are present, the
tracks from the highest multiplicity vertex are used.

The multiplicity of reconstructed tracks, $\Nr$, is obtained in the region
$\abs{\eta}<2.4$. It should be noted that $\Nr$ differs from $\Noff$. As defined in
Section~\ref{subsubsec:legacy_selections}, $\Noff$ has a threshold of
$\pt>0.4\GeV$ applied to the reconstructed tracks, while $\Nr$ has no such
threshold and also includes a correction for the extrapolation down to $\pt =
0$. Over the range $0 < \Nr < 240$, the events are divided into 24 classes,
defined in Table~\ref{tab:multiClass}. This range is a good match to the multiplicity
in the 60--100\% centrality in PbPb collisions. To facilitate comparisons with
models, the corresponding corrected average charged-particle multiplicity
$\langle\Nt\rangle$ in the same region defined by $\abs{\eta}<2.4$ and down to
$\pt = 0$ is also determined.
For each multiplicity class, the correction from $\Nr$ to $\left< \Nt \right>$
uses the efficiency estimated with MC event generators, followed by the CMS
detector response simulation based on
\textsc{geant4}~\cite{Agostinelli2003250}. The employed event generators are
{\PYTHIA}6.426 (with the tunes D6T and Z2) for $\Pp\Pp$ collisions, and \HIJING
2.1~\cite{Deng:2010mv,Xu:2012au} for $\Pp$Pb collisions.
The corrected data are then integrated over \pt, down to zero yield at $\pt =0$
with a linear extrapolation below $\pt = 0.1\GeV$. The yield in the
extrapolated region is about 6\% of the total yield. The systematic uncertainty
due to the extrapolation is small, well below 1\%. Finally, the integrals for
each $\eta$ slice are summed up. In the case of PbPb collisions, events
generated with the \HYDJET 1.8~\cite{Lokhtin:2008xi} MC event generator are
simulated and reconstructed. The $\Nr-\Nt$ relationship is parametrized using a
fourth order polynomial.

\begin{figure*}[tbh]
 \centering
  \includegraphics[width=0.49\textwidth]{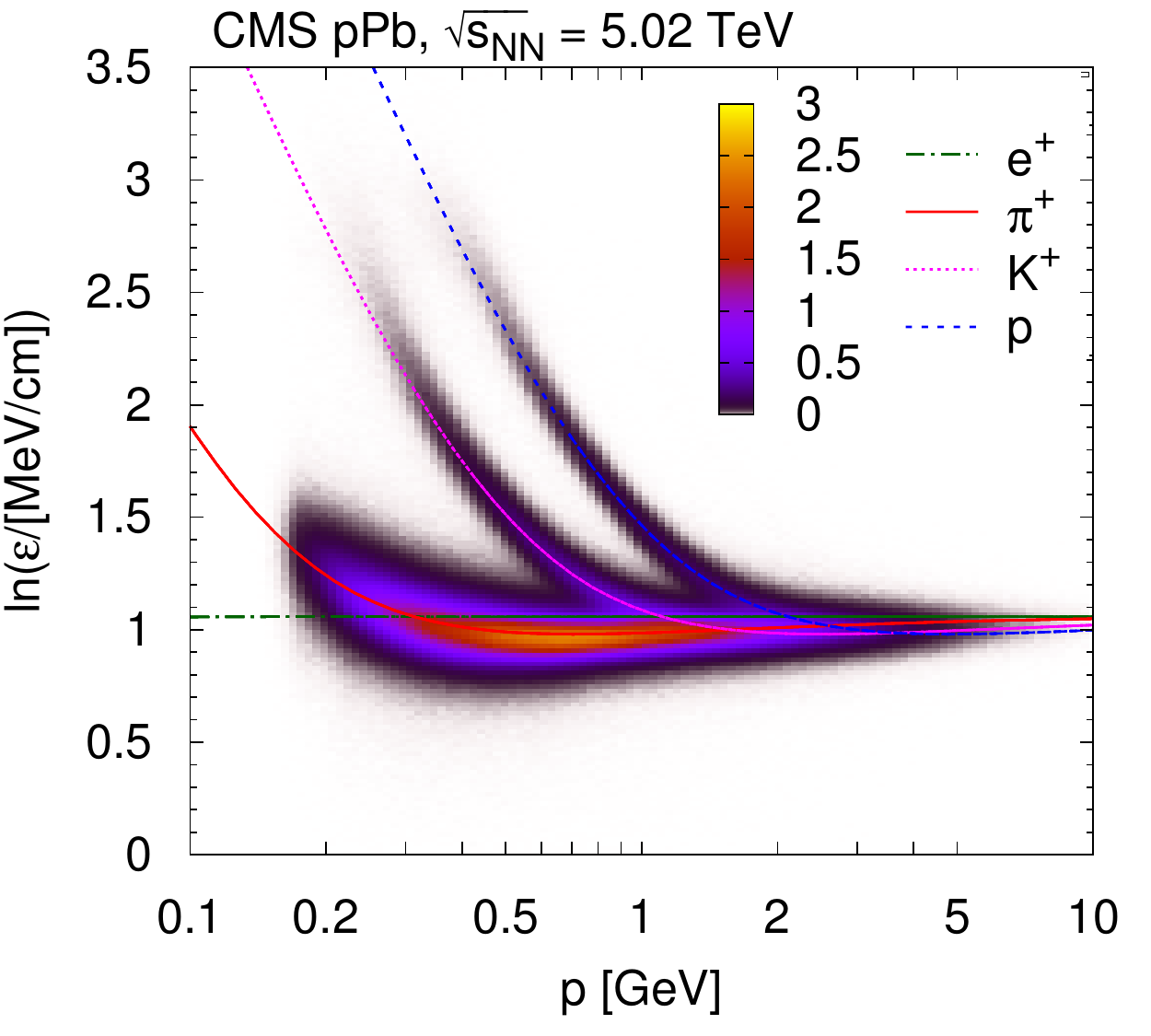}
  \includegraphics[width=0.49\textwidth]{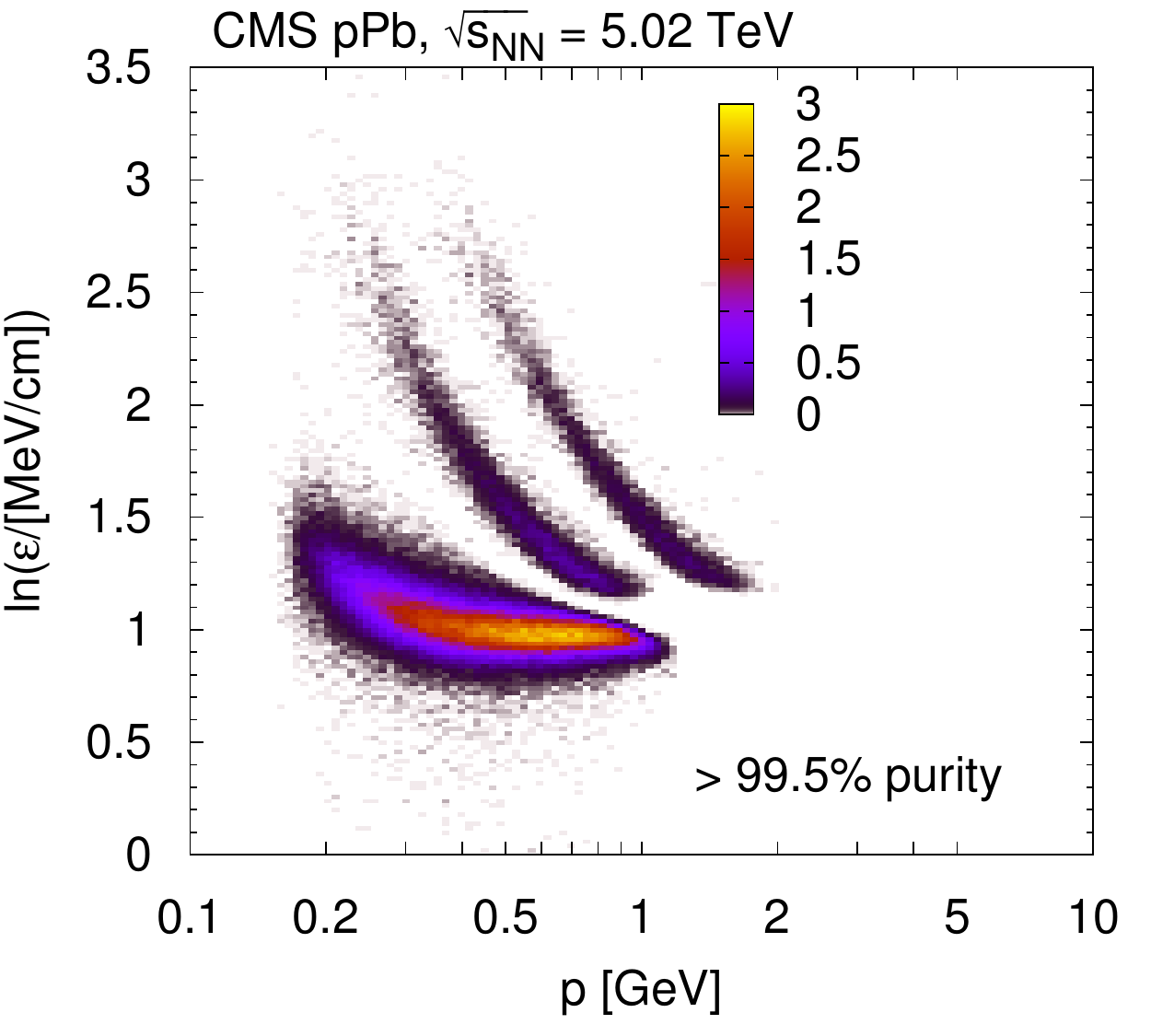}
 \caption{Logarithm of the most probable energy loss rate $\varepsilon$
normalized at a reference path length $l_0 = 450\mum$ in $\Pp$Pb collisions at
$\ssNN = 5.02$\TeV. The distribution of $\ln\varepsilon$ is shown as a function
of total momentum $p$ for positively charged particles (left), as well as for
identified charged particles (both charges) with high purity selection ($>$
99.5\%, right). The linear $z$ axis scale is shown in arbitrary units. The
curves show the expected $\ln\varepsilon$ for electrons, pions, kaons, and
protons (full theoretical calculation, Eq.~(33.11) in
Ref.~\cite{Olive:2016xmw}).}
 \label{fig:elossHistos}
\end{figure*}

\subsection{Particle identification}
\label{subsubsec:pid}

The reconstruction of charged particles in CMS is limited by the acceptance of
the tracker and by the decreasing tracking efficiency at low momentum.
Particle-by-particle identification using specific ionization losses in the
tracker is possible in
the momentum range $p < 0.15\GeV$ for electrons, $p < 1.15\GeV$ for pions and
kaons, and $p < 2.00\GeV$ for protons (note that protons were not used for
correlation studies in this analysis).
In view of the $(\eta,\pt)$ regions
where pions, kaons, and protons can all be identified, only particles in the
range $-1 < \eta < 1$ (in the laboratory frame) are used for this measurement.

For the identification of charged particles, the estimated most probable energy
loss rate $\varepsilon$ at a reference path length ($l_0 = 450\mum$) is
used~\cite{Sikler:2011yy}. For an accurate determination of $\varepsilon$ and
its variance $\sigma^2$ for each individual track, the responses of all readout
chips in the tracker are calibrated with multiplicative gain correction
factors. The procedures for gain calibration and track-by-track determination
of $\varepsilon$ values are the same as in previous CMS
analyses~\cite{Chatrchyan:2012qb,Chatrchyan:2013eya}.

Charged particles in the region $-1 < \eta < 1$ and $0 < \pt < 2\GeV$ are
sorted into bins with a width of 0.1 units in $\eta$, and $0.05\GeV$ in $\pt$.
Since the ratios of particle yields do not change significantly with the
charged-particle multiplicity of the
event~\cite{Chatrchyan:2012qb,Chatrchyan:2013eya}, the data are not subdivided
into bins of $\Nr$. The relative abundances of different particle species in a
given $(\eta,\pt)$ bin are extracted by minimizing the joint log-likelihood
\ifthenelse{\boolean{cms@external}}{
\begin{multline}
 -2\ln{\mathcal{L}}\\
    = - 2 \sum_{i = \text{tracks}} \ln \left(\sum_{k = \Pgp,\PK,\Pp}
       P_k \exp\left[- \frac{(\ln\varepsilon_i - \mu_k)^2}
                            {2 (\xi \sigma_i)^2}\right]\right),
 \label{eq:logl}
\end{multline}
}{
\begin{equation}
 -2\ln{\mathcal{L}}
    = - 2 \sum_{i = \text{tracks}} \ln \left(\sum_{k = \Pgp,\PK,\Pp}
       P_k \exp\left[- \frac{(\ln\varepsilon_i - \mu_k)^2}
                            {2 (\xi \sigma_i)^2}\right]\right),
 \label{eq:logl}
\end{equation}
}
where $\mu_k$ are the expected means of $\ln\varepsilon$ for the
different particle species, and $P_k$ are their relative probabilities. The
value of $-2\ln{\mathcal{L}}$ is minimized by varying $P_k$ and $\mu_k$ starting
from reasonable initial values. In the above formula $\varepsilon_i$ and
$\sigma_i^2$ are the estimated most probable value and its estimated variance,
respectively, for each individual track.
Since the estimated variance can slightly differ from its true value, a scale
factor $\xi$ is introduced. The $\mu_k$ values are used to determine a unique
$\ln\varepsilon(p/m)$ function. A third-order polynomial closely approximates
the collected $(p/m, \ln\varepsilon)$ pairs, within the corresponding
uncertainties. This information is reused by fixing $\mu_k$ values according to
the polynomial, and then re-determining $P_k$.

{\sloppy
In this analysis a very high purity particle identification is achieved by
requiring $P_k/(\sum_{k = \Pgp,\PK,\Pp} P_k) > 0.995$. If none of the particle
type assumptions yields a $P_k$ value in this range, the particle is regarded
as unidentified. This requirement ensures that less than 1\% of the examined
particle pairs are misidentified.
The degree of purity achieved in particle identification is indicated by
distributions of $\ln\varepsilon$ as a function of total momentum $p$
for $\Pp$Pb collisions at $\ssNN = 5.02$\TeV in Fig.~\ref{fig:elossHistos}. The
left plot shows the initial distribution for positive particles. The plot for
negatively charged particles is very similar. This distribution is compared to
the theoretically expected energy loss for electrons, pions, kaons, and
protons. In Fig.~\ref{fig:elossHistos}, the distribution of $\ln\varepsilon$ as
a function of total momentum for identified charged particles with high purity
is shown in the right plot. The plots for $\Pp\Pp$ and PbPb interactions are very
similar.
\par}

\section{The Bose--Einstein correlations}
\label{sec:techniques}

The correlation functions are investigated in one (1D), two (2D), and three
(3D) dimensions in terms of $\qi = \sqrt{(\vec{p}_1 - \vec{p}_2)^2 - (E_1 -
E_2)^2}$, as well as in projections of the relative
three-momentum of the pair, $\vec{q} = \vec{p}_1 - \vec{p}_2$, in two (in terms
of  longitudinal and transverse momenta, $\ql$, $\qt$), and three (in terms of
out-short-long momenta, $\ql$, $\qs$, $\qo$) directions.
In the center-of-mass (CM) frame the variables are defined as:
\begin{align*}
 \vec{k}_\mathrm{T} &= \frac{\vec{p}_\text{T,1} + \vec{p}_\text{T,2}}{2}, & \\
 \vec{q}_\mathrm{T} &= \vec{p}_\mathrm{T,1} - \vec{p}_\mathrm{T,2},  & \ql &=
|p_\mathrm{L,1} -p_\mathrm{L,2}|\\
 \vec{q}_\mathrm{O} &= \left(\vec{\qt} \cdot \vec{e}_{\kt}\right)
                                    \vec{e}_{\kt},
 &\vec{q}_\mathrm{S} & = \vec{q}_\mathrm{T} - \vec{q}_\mathrm{O} = \vec{e}_{\kt} \times \vec{q}_\mathrm{T},
\end{align*}
where $\vec{e}_{\kt}\equiv \vec{k}_\mathrm{T}/\kt$ is a unit vector
along the direction of the pair average transverse momentum,
$\vec{k}_\mathrm{T}$.
In the case of $\Pp\Pp$ collisions, and when dealing with the double ratio technique,
the multidimensional investigations are carried out both in the CM frame and in
the local co-moving system (LCMS). The latter is the frame in which the
longitudinal component of the pair average momentum, $k_\text{L} =
(k_\text{L,1} + k_\text{L,2})/2$, is zero.
In the remainder of this paper, a common notation is used to
refer to the magnitudes of the average and relative momentum vectors,
$\kt = |\vec{k}_\mathrm{T}|$, $\qt = |\vec{q}_\mathrm{T}|$, $\qo =
|\vec{q}_\mathrm{O}|$, and $\qs = |\vec{q_\mathrm{S}}|$.

\subsection{Double ratio technique}
\label{subsec:doubleratios}

The analysis procedure using the double ratio technique is the same as
that described in Refs.~\cite{cms-hbt-1st,cms-hbt-2nd}, where no particle
identification is considered. However, the contamination by non-pions is
expected to be small, since pions are the dominant type of hadrons in the
sample (the ratio of kaons to pions is about 12\%, and of protons to pions is
roughly 6\%~\cite{Chatrchyan:2012qb}). The unidentified kaons and protons would
contaminate mainly the low relative momentum region of the correlation
functions. The impact of this level of contamination is covered by the systematic
uncertainties.

For each event, the signal containing the BEC is formed by pairing same-sign
tracks in the same event originating from the primary vertex, with $\abs{\eta} <
2.4$ and $\pt > 0.2\GeV$. The distributions in terms of the relative momentum
of the pair (the invariant $\qi$, or the components $\qt$, $\ql$ or $\qo$,
$\qs$, $\ql$) are divided into bins of the reconstructed charged-particle
multiplicity $\Ntpt$, and of the pair average transverse momentum $\kt$.
The distributions are determined in the CM and LCMS reference frames.

The background distribution or the reference sample can be constructed in
several ways, most commonly formed by mixing tracks from different events
(mixed-event technique), which can also be selected in several ways. In the first
method employed in this analysis, the reference sample is constructed by
pairing particles (all charge combinations) from mixed events and within the
same $\eta$ range, as in Refs.~\cite{cms-hbt-1st,cms-hbt-2nd}. In this case,
the full $\abs{\eta} < 2.4$ interval is divided in three subranges: $-2.4 < \eta <
-0.8$, $-0.8 \le \eta < 0.8$, and $0.8 \le \eta < 2.4$. Alternative techniques
considered for estimating the systematic uncertainties associated with this
choice of reference sample are discussed in Sec.~\ref{syst-fsq-13002}.

The correlation function is initially defined as a single ratio (SR), having
the signal pair distribution as numerator and the reference sample as
denominator, with the appropriate normalization,
\begin{equation}
  \text{SR} (q) = C_2 (q) =
  \left(\frac{\mathcal{N}_{\text{ref}}}{\mathcal{N}_{\text{sig}}}\right)
  \;
  \left(\frac{\text{d}N_\text{sig}/\text{d}q}
             {\text{d}N_\text{ref}/\text{d}q}\right),
 \label{eq:1D-singleratios-gen}
\end{equation}
where $C_2(q)$ is two-particle correlation function,
$\mathcal{N}_{\text{sig}}$ is the integral of the signal content, whereas
$\mathcal{N}_{\text{ref}}$ is the equivalent for the reference sample, both
obtained by summing up the pair distributions for all the events in the sample.
For obtaining the parameters of the BEC effect in this method, a double ratio
(DR) is constructed~\cite{cms-hbt-1st,cms-hbt-2nd} as
\ifthenelse{\boolean{cms@external}}{
\begin{multline}
 \text{DR} (q) = \frac{\text{SR} (q)}{\text{SR} (q)_\mathrm{MC}}\\
  =
  \left[\left(\frac{\mathcal{N}_{\text{ref}}}
                   {\mathcal{N}_{\text{sig}}}\right)
  \left(\frac{\rd{}N_\text{sig}/\rd{}q}
             {\rd{}N_\text{ref}/\rd{}q}\right)\right]\\
  \Big/
  \left[\left(\frac{\mathcal{N}_{\text{ref}}}
                   {\mathcal{N}_{\text{sig}}}\right)_\mathrm{MC}
        \left(\frac{\rd{}N_\mathrm{MC}/\rd{}q}
                   {\rd{}N_{\text{MC, ref}}/\rd{}q}\right)\right],
\label{eq:doubleratio}
\end{multline}
}
{
\begin{equation}
 \text{DR} (q) = \frac{\text{SR} (q)}{\text{SR} (q)_\mathrm{MC}} =
  \left[\left(\frac{\mathcal{N}_{\text{ref}}}
                   {\mathcal{N}_{\text{sig}}}\right)
  \left(\frac{\rd{}N_\text{sig}/\rd{}q}
             {\rd{}N_\text{ref}/\rd{}q}\right)\right]
  \Big/
  \left[\left(\frac{\mathcal{N}_{\text{ref}}}
                   {\mathcal{N}_{\text{sig}}}\right)_\mathrm{MC}
        \left(\frac{\rd{}N_\mathrm{MC}/\rd{}q}
                   {\rd{}N_{\text{MC, ref}}/\rd{}q}\right)\right],
\label{eq:doubleratio}
\end{equation}
}
where $\text{SR} (q)_\mathrm{MC}$ is the single ratio as in
Eq.~\eqref{eq:1D-singleratios-gen}, but computed with simulated events
generated without BEC effects.

Since the reference samples in single ratios for data and simulation are
constructed with exactly the same procedure, the bias due to such
background construction could be, in principle, reduced when the DR is taken.
However, even in this case, the correlation functions are still sensitive to
the different choices of reference sample, leading to a spread in the
parameters fitted to the double ratios, that is considered as a systematic
uncertainty~\cite{cms-hbt-1st,cms-hbt-2nd}. On the other hand, by selecting MC
tunes that closely describe the behavior seen in the data, this DR technique
should remove the contamination due to non-Bose--Einstein correlations.

Table~\ref{tab:widths-inid} shows the ranges and bin widths of relative
momentum components used in fits to the double ratios. The bins in $\kt$ are
also listed, which coincide with those in Refs.~\cite{cms-hbt-1st,cms-hbt-2nd}.

\begin{table}[th]
 \topcaption{Bin widths chosen for the various variables studied with the
double ratio method.}
 \label{tab:widths-inid}
 \centering
 \begin{scotch}{lcc}
   Variable       & \multicolumn{2}{c}{Bin ranges}  \\
   $\kt$  [\GeVns{}]   & \multicolumn{2}{c}{0.2--0.3, 0.3--0.5, 0.5--1.0}  \\[\cmsTabSkip]
   & Range  & Bin width \\ \cline{2-3}
   \qi\   [\GeVns{}]           & 0.02--2.0 & $0.01{^{^{(*)}}}$/0.02 \\
   \qt, \ql\   [\GeVns{}]      & 0.02--2.0 & 0.05 \\
   \qo, \qs, \ql\   [\GeVns{}] & 0.02--2.0 & 0.05 \\
  \hline
  \multicolumn{3}{r}{$^{^{(*)}}$ for results integrated over $\Ntpt$ and
$\kt$ bins} \\
 \end{scotch}
\end{table}

\subsection{Particle identification and cluster subtraction technique}
\label{sec:partpairs}

Similarly to the case of unidentified particles discussed in
Section~\ref{subsec:doubleratios}, the identified hadron pair distributions are
binned in the number of reconstructed charged particles $\Nr$ of the event, in
the pair average transverse momentum $\kt$, and also in the relative momentum variables
in the LCMS of the pair in terms of $\qi$, $(\ql,\qt)$ or $(\ql,\qo,\qs)$. The
chosen bin widths for pions and kaons are shown in Table~\ref{tab:widths}.

\begin{table}[b]
\topcaption{Chosen bin widths for the various variables studied for pions and
kaons in the particle identification and cluster subtraction method.}
 \label{tab:widths}
 \centering
 \begin{scotch}{lccc}
  \multirow{2}{*}{Variable} & \multirow{2}{*}{Range}
                            & \multicolumn{2}{c}{Bin width} \\ \cline{3-4}
                            & & \Pgp & \PK \\
  \hline
  $\Nr$                &   0--240 & 10   & 60   \\
  \kt\ [\GeVns{}]           & 0.2--0.7 & 0.1  & 0.1  \\
  \qi\ [\GeVns{}]           & 0.0--2.0 & 0.02 & 0.04 \\
  \ql, \qt\ [\GeVns{}]      & 0.0--2.0 & 0.04 & 0.04 \\
  \ql, \qo, \qs\ [\GeVns{}] & 0.0--2.0 & 0.04 & 0.04 \\
 \end{scotch}
\end{table}

The construction of the signal distribution is analogous to that described
above: all valid particle pairs from the same event are taken and the
corresponding distributions are obtained.
Several types of pair combinations are used to study the BEC: \Pgpp\Pgpp,
\Pgpm\Pgpm, \PKp\PKp, and \PKm\PKm; while \Pgpp\Pgpm\ and \PKp\PKm\ are
employed to correct for non-BEC contributions, as described in
Section~\ref{subsubsec:cluster-removal}.
For the reference sample an event mixing procedure is adopted, in which
particles from the same event are paired with particles from 25 preceding
events.
Only events belonging to the same multiplicity ($\Nr$) class are mixed. Two
additional cases are also investigated. In one of them particles from the same
event are paired, but the laboratory momentum vector of the second particle is
rotated around the beam axis by 90 degrees. Another case considers pairs of
particles from the same event, but with an opposite laboratory momentum vector
of the second particle (mirrored). Based on the goodness-of-fit distributions
of correlation function fits, the first event mixing prescription is used,
while the rotated and mirrored versions, which give considerably worse
$\chi^2/\mathrm{ndf}$ values, are employed in the estimation of the systematic
uncertainty.

The measured two-particle correlation function
$C_2 (q)$ is itself the single ratio of signal and background distributions, as
written in Eq.~\eqref{eq:1D-singleratios-gen}. This single ratio contains the
BEC from quantum statistics, while non-BEC effects also contribute. Such
undesired contributions are taken into account as explained in
Section~\ref{subsec:non-be-effects}. A joint functional form combining all
effects is fitted in order to obtain the radius parameter and the correlation
function intercept, as discussed in
Section~\ref{sub:fitting_functions}.

\subsection{Corrections for non-BEC effects}
\label{subsec:non-be-effects}

\subsubsection{Effect of Coulomb interaction and correction}
\label{subsubsec:coulomb-effects}

The BEC method employed for the study of the two-particle correlations reflects
not only the quantum statistics of the pair of identical particles, but is also
sensitive to final-state interactions. Indeed, the charged-hadron correlation
function may be distorted by strong, as well as by Coulomb effects. For pions,
the strong interactions can usually be neglected in femtoscopic
measurements~\cite{Sinyukov:1998fc}.

The Coulomb interaction affects the two-particle relative momentum distribution
in a different way in the case of pairs with same-sign (SS) and opposite-sign
(OS) electric charge. This leads to a depletion (enhancement) in the
correlation function caused by repulsion (attraction) of SS (OS) charges. The
effect of the mutual Coulomb interaction is taken into account by the factor
$K$, the squared average of the relative wave function $\Psi$, as $K(\qi) =
\int \rd^3\vec{r} \; f(\vec{r}) \; |\Psi(\vec{k},\vec{r})|^2$, where
$f(\vec{r})$ is the source intensity of emitted particles and $\vec{k}$ is the
relative momentum of the pair. Pointlike sources, $f(\vec{r}) =
\delta(\vec{r})$, result in an expression for $K(\qi)$ coincident with the
Gamow factor~\cite{Gamow1928} which, in case of same sign and opposite sign
charges, is given by
\ifthenelse{\boolean{cms@external}}{
\begin{equation}
\begin{aligned}
 G^\text{SS}_w(\zeta) &= |\Psi^\text{SS}(0)|^2
                        = \frac{2\pi\zeta}{\re^{2\pi\zeta}-1},\\
 G^\text{OS}_w(\zeta) &= |\Psi^\text{OS}(0)|^2
                        = \frac{2\pi\zeta}{1-\re^{-2\pi\zeta}},
\label{eq:gamow}
\end{aligned}
\end{equation}
}{
\begin{equation}
 G^\text{SS}_w(\zeta) = |\Psi^\text{SS}(0)|^2
                        = \frac{2\pi\zeta}{\re^{2\pi\zeta}-1},
 \qquad
 G^\text{OS}_w(\zeta) = |\Psi^\text{OS}(0)|^2
                        = \frac{2\pi\zeta}{1-\re^{-2\pi\zeta}},
\label{eq:gamow}
\end{equation}
}
where $\zeta=\alpha m/\qi$, is the Landau parameter, $\alpha$ is the
fine-structure constant, $m$ is the particle mass, and $\qi$ is the invariant
relative momentum of the pair~\cite{GyuKaufWil:1979}.

For extended sources, a more elaborate treatment is
needed~\cite{PhysRevD.33.72,Biyajima:1995ig}, and the Bowler--Sinyukov
formula~\cite{Bowler:1991vx,Sinyukov:1998fc} is most commonly used. Although
this does not differ significantly from the Gamow factor correction in the case
of pions, this full estimate is required for kaons. The possible screening of
Coulomb interaction, sometimes seen in data of heavy ion
collisions~\cite{Anchishkin:1996vp,Wong:1996if}, is not considered. The value
of zeta is typically $\zeta \ll 1$ in the $q$ range studied in this analysis.
The absolute square of the confluent hypergeometric function of the first kind
$F$, present in $\Psi$, can be approximated as $|F|^2 \approx 1 + 2\zeta
\operatorname{Si}(x)$, where $\operatorname{Si}$ is the sine integral
function~\cite{AbramowitzStegun64}.
Furthermore, for Cauchy-type source functions ($f(r) = R/\{2 \pi^2 \left[r^2
+(R/2)^2\right]^2\}$, where $f$ is normalized, such that $\int f(r)\,
\rd^3 r = 1$), the Coulomb-correction factor $K$ is well described by the formula
\begin{equation}
 K(\qi) = G_w(\zeta) \left[1 + \pi \; \zeta \;\qi R / (1.26 + \qi R)\right],
 \label{eq:full_coulomb}
\end{equation}
for $\qi$ in $\GeV$, and $R$ in fm.
The factor $\pi$ in the
approximation comes from the fact that, for large $kr$ arguments,
$\operatorname{Si}(kr) \to \pi/2$. Otherwise it is a simple but
accurate approximation of the result of a numerical calculation, with
deviations below 0.5\%.

\subsubsection{Clusters: mini-jets, multi-body decays of resonances}
\label{subsubsec:cluster-removal}

\begin{figure}[t]

 \centering
  \includegraphics[width=0.49\textwidth]{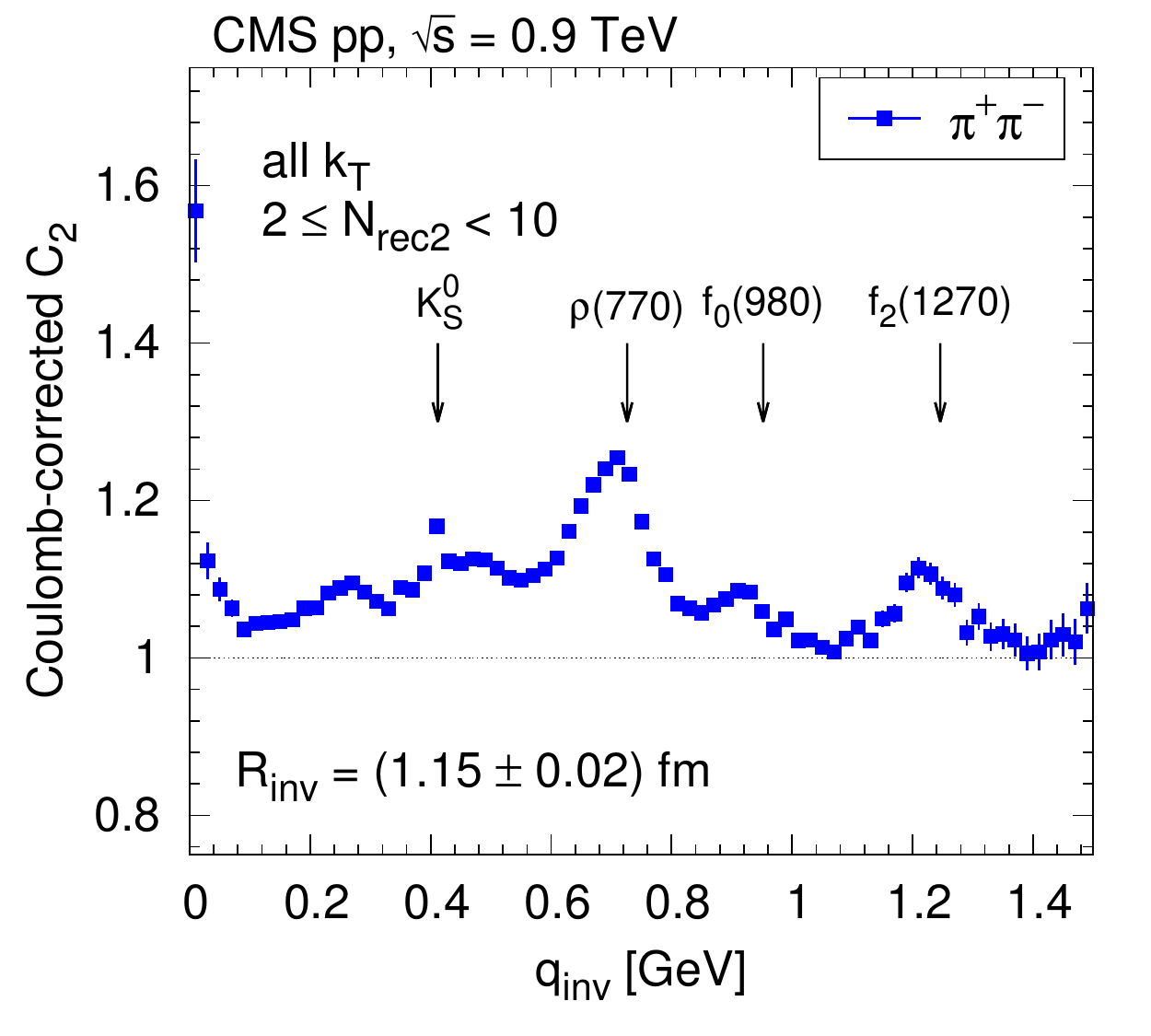}
  \includegraphics[width=0.49\textwidth]{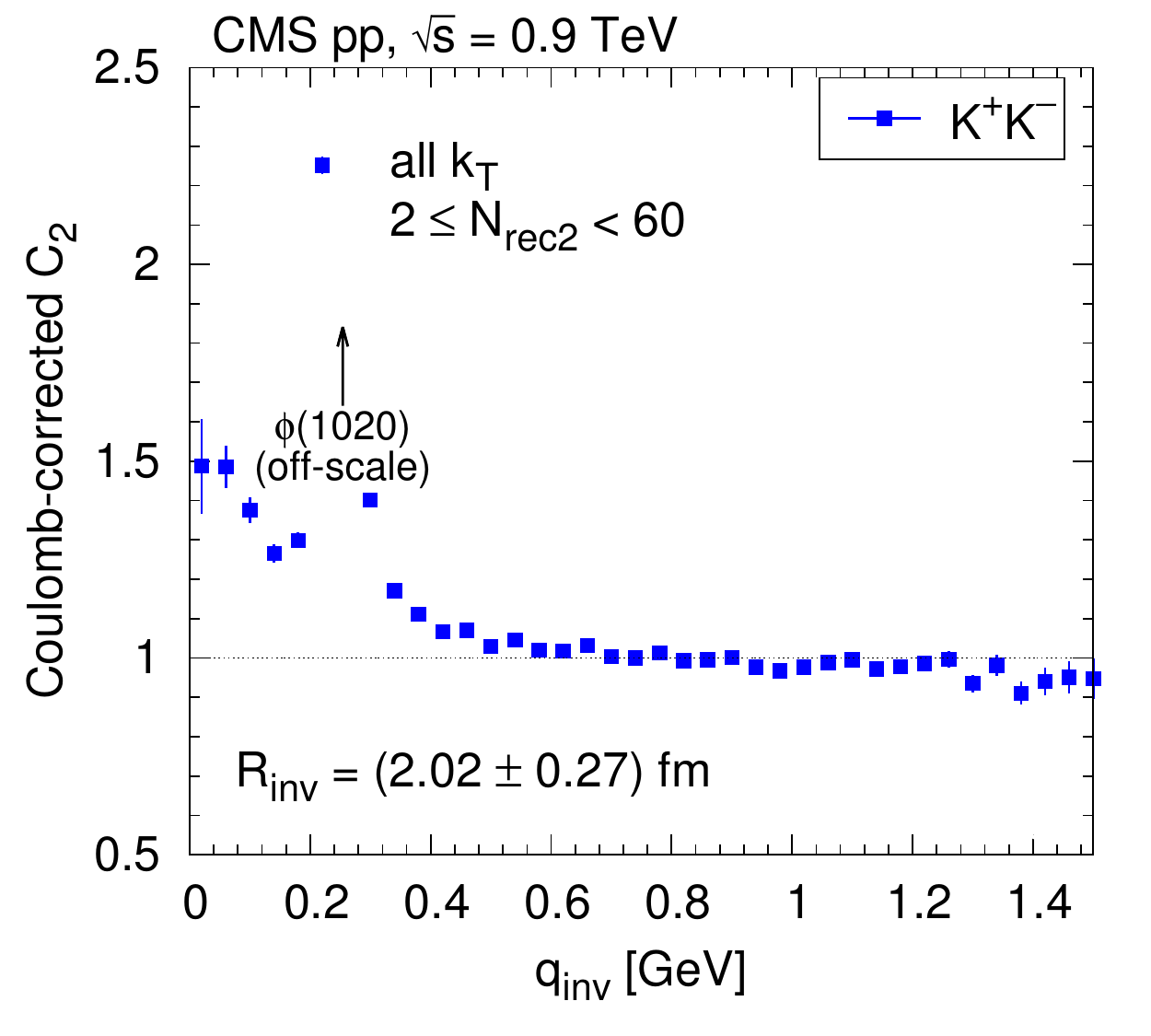}

 \caption{Contribution of resonance decays to the measured Coulomb-corrected
correlation function of \Pgpp\Pgpm\ (for $2 \le \Nr < 10$, \cmsLeft, blue squares)
and \PKp\PKm\ (for $2 \le \Nr < 60$, \cmsRight, blue squares). The $\Pgf$ peak
(\cmsRight panel) is off-scale.}

 \label{fig:resonances}

\end{figure}

{\tolerance=1200
The measured opposite-sign correlation functions show contributions from
various hadronic resonances~\cite{alice}. Selected Coulomb-corrected correlation
functions for low $\Nr$ bins are shown in Fig.~\ref{fig:resonances}. The
observed resonances include \PKzS, \Pgr, $\mathrm{f_0(980)}$, and
$\mathrm{f_2(1270)}$ decaying to \Pgpp\Pgpm, and \Pgf\ decaying to \PKp\PKm.
Photon conversions into \Pep\Pem\ pairs, when misidentified as pion
pairs, can also appear as a peak at very low \qi\ in the \Pgpp\Pgpm\ spectrum.
With increasing $\Nr$ values the effect of resonances diminishes, since their
contribution is quickly exceeded by the combinatorics of unrelated particles.
\par}

\begin{figure}[!t]
 \centering
  \includegraphics[width=0.49\textwidth]{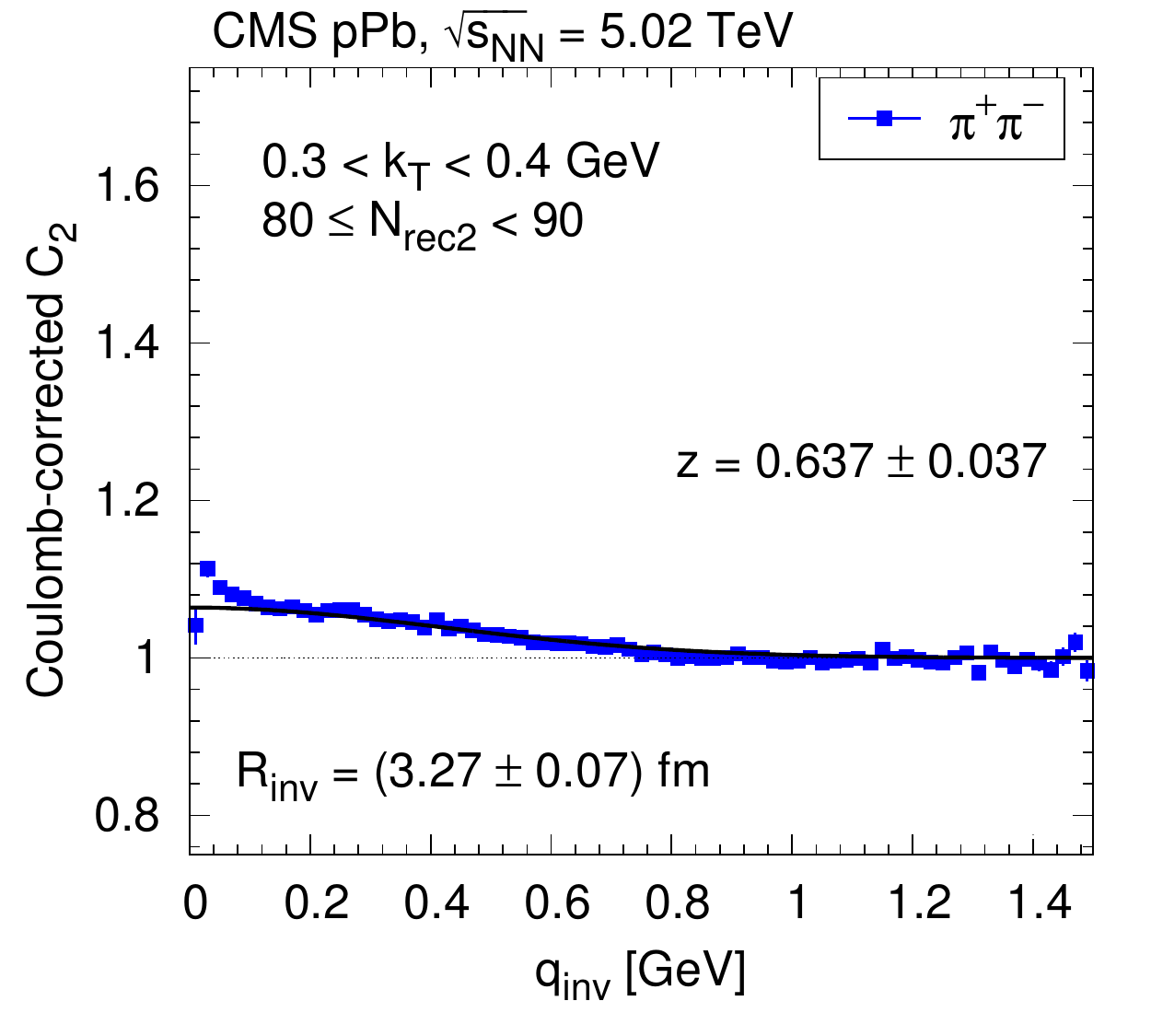}
  \includegraphics[width=0.49\textwidth]{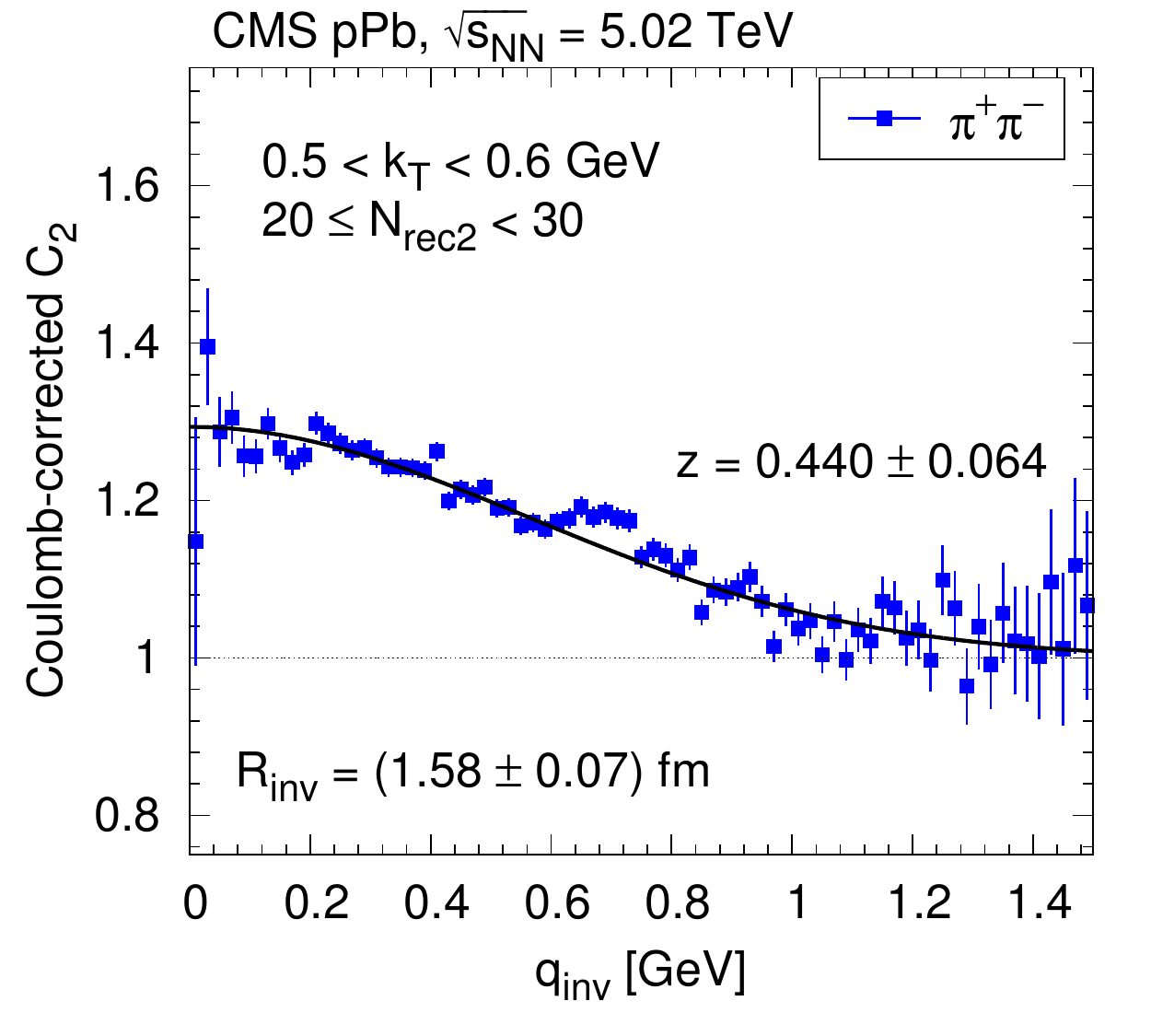}
 \caption{Contribution of clusters (mini-jets and multi-body decays of
resonances) to the measured Coulomb-corrected correlation function of
\Pgpp\Pgpm for $\Pp$Pb interactions at $\ssNN = 5.02$\TeV in two bins of event
multiplicity and $\kt$. The solid curves show fits according to
Eq.~\eqref{eq:minijet}.}
 \label{fig:minijet_vs_Nrec_and_kt}
\end{figure}

The Coulomb-corrected OS correlation functions are not always close to
unity even at low $\qi$, but show a Gaussian-like hump
(Fig.~\ref{fig:minijet_vs_Nrec_and_kt}). That structure has a varying amplitude
with a stable scale ($\sigma$ of the corresponding Gaussian) of about 0.4\GeV.
This feature is often related to particles emitted inside low-momentum
mini-jets, but can be also attributed to the effect of multi-body decays of
resonances. In the following we will refer to such an effect as due to
fragmentation of clusters, or cluster
contribution~\cite{Alver:2008aa,Khachatryan:2010gv,CMS:2012qk}.
For the purpose of evaluating and later eliminating that contribution, the
one-dimensional OS correlation functions are fitted with the following form:
\begin{equation}
 \label{eq:minijet}
 C^{+-}_2(\qi) = c_\text{OS} \, K^{+-}(\qi) \;
  \left[1 + \frac{b}{\sigma_b \sqrt{2\pi}}
             \exp\left(- \frac{\qi^2}{2\sigma_b^2}\right)\right]
\end{equation}
where $b$ and $\sigma_b$ are $\Nr$- and \kt-dependent parameters, and
$c_\text{OS}$ is a normalization constant. The values of $b$ and $\sigma_b$ are
parametrized using:
\ifthenelse{\boolean{cms@external}}{
\begin{equation}
 \label{eq:b_sigma_b}
 \begin{aligned}
 b(\Nr,\kt &= \frac{b_0}{\Nr^{n_b}} \exp\left(\frac{\kt}{k_0}\right), \\
 \sigma_b(\Nr,\kt) &=
  \left[\sigma_0 + \sigma_1 \exp\left(-\frac{\Nr}{N_0}\right)\right] \kt^{n_\text{T}}.
\end{aligned}
\end{equation}
}{
\begin{equation}
 \label{eq:b_sigma_b}
 b(\Nr,\kt = \frac{b_0}{\Nr^{n_b}} \exp\left(\frac{\kt}{k_0}\right),
 \qquad
 \sigma_b(\Nr,\kt) =
  \left[\sigma_0 + \sigma_1 \exp\left(-\frac{\Nr}{N_0}\right)\right] \kt^{n_\text{T}}.
\end{equation}
}
The amplitude $b$ is inversely proportional to a power of
$\Nr$ (with the exponent $n_b$ in the range 0.80--0.96). The parameter $b_0$
is much bigger for 5\TeV $\Pp$Pb than for 2.76\TeV PbPb data; for $\Pp\Pp$ collisions the
dependence is described by $0.28 \ln(\sqrt{s}/0.48\TeV)$. For $\Pp$Pb collisions
$b_0$ is about 1.6--1.7 times higher than would be predicted from the $\Pp\Pp$ curve
at the same $\ssNN$. At the same time $b$ increases exponentially with \kt, the
parameter $k_0$ being in the range of 1.5--2.5\GeV.
The Gaussian width $\sigma_b$ of the hump at low \qi first decreases with
increasing $\Nr$, but for $\Nr > 70$ remains constant at about 0.35--0.55\GeV.
The width increases with $\kt$ as a power law, with the exponent $n_\text{T}$
in the range 0.19--0.33. The expression $b \sigma_b^2 \Nr$ is proportional to
the fraction of particles that have a cluster-related partner. Our data show
that this fraction does not change substantially with $\Nr$, but increases with
\kt and $\sqrt{s}$ for $\Pp\Pp$ collisions.
In summary, the object that generates this structure is assumed to always emit
particles in a similar way, with a relative abundance that increases with the
center-of-mass energy.

\begin{figure}[!t]
 \centering
  \includegraphics[width=0.49\textwidth]{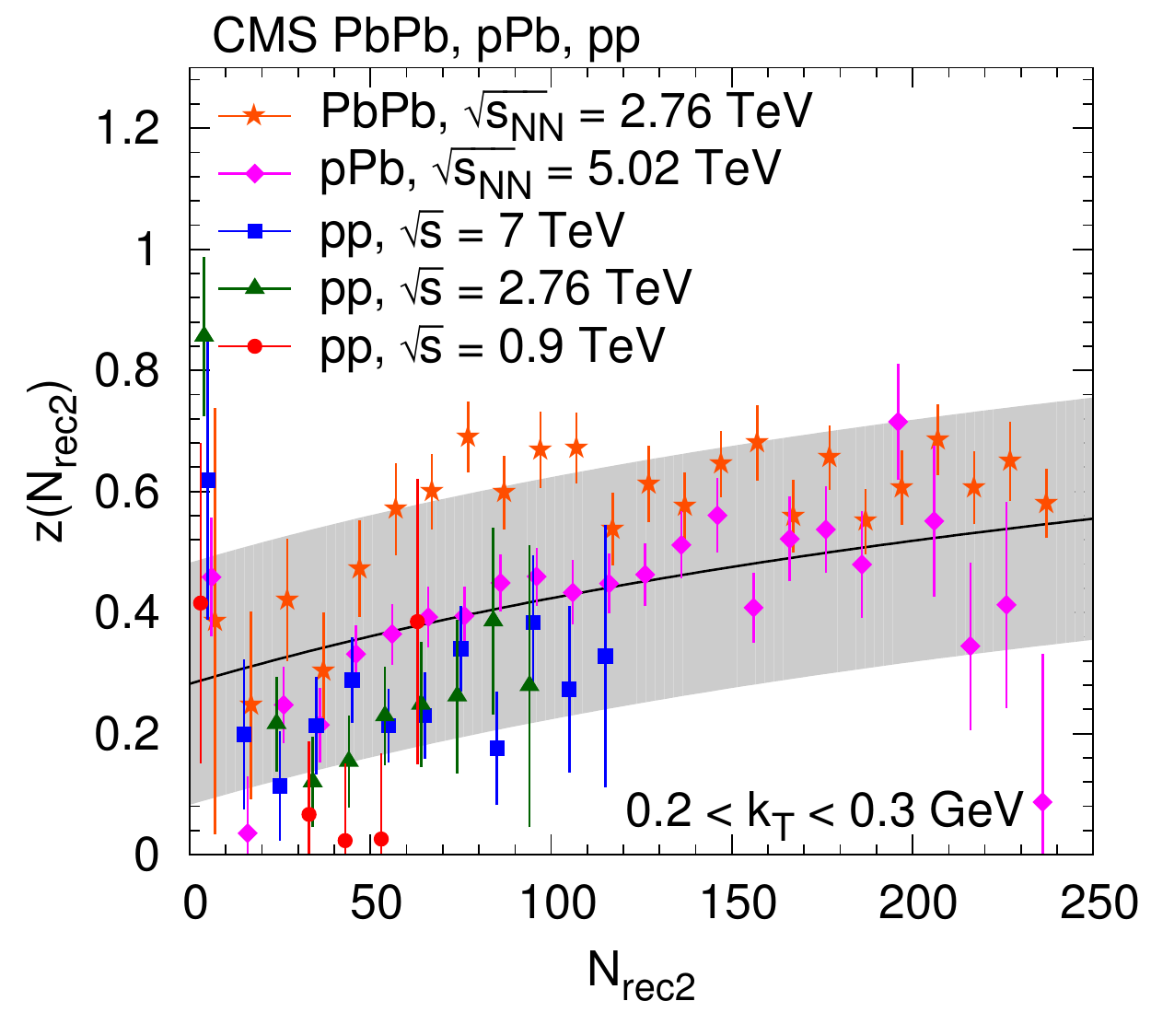}
  \includegraphics[width=0.49\textwidth]{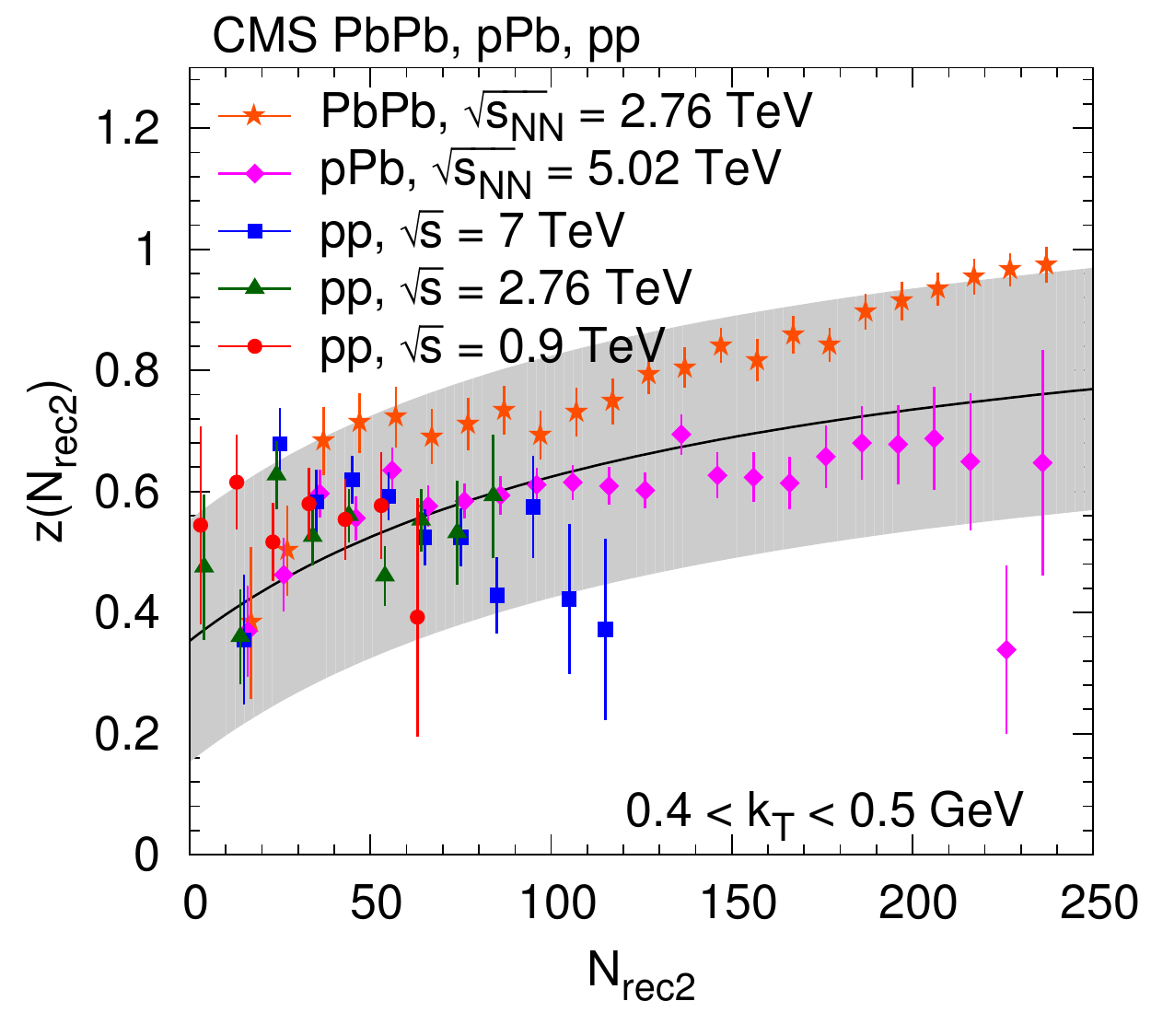}
 \caption{Dependence of the relative amplitude $z(\Nr)$ of the cluster
contribution (between SS and OS pairs), as a function of $\Nr$ for
two $\kt$ bins. Fit results (points with error bars)
from all collision types are plotted together with combinatorics-motivated fits
(solid curves) and their estimated systematic uncertainties ($\pm$0.2 bands).}
 \label{fig:zz}
\end{figure}

The cluster contribution can also be extracted in the case of the SS
correlation function, if the momentum scale of the BEC effects and that of the
cluster contribution ($\approx$0.4\GeV) are different enough. An important
constraint in both mini-jet and multi-body resonance decays is the conservation
of electric charge that results in a stronger correlation for OS pairs than for
SS pairs.  Consequently, the cluster contribution is expected to be present as well
for SS pairs, with similar shape but a somewhat smaller amplitude. The
form of the cluster-related contribution obtained from OS pairs, multiplied by
a relative amplitude $z(\Nr, \kt)$ (ratio of measured OS to SS contributions),
is used to fit the SS correlations.
The dependence of this relative amplitude on $\Nr$ in $\kt$ bins is shown in
Fig.~\ref{fig:zz}. Results (points with error bars) from all collision systems
are plotted together with a parametrization ($(a \Nr + b)/(1 + a\Nr +b)$) based
on the combinatorics of SS and OS particle pairs (solid curves). Instead of the
points, these fitted curves are used to describe the $\Nr$ dependence of $z$ in
the following. The $\pm$0.2 grey bands are chosen to cover most of the
measurements and are used in determining a systematic uncertainty due to the
subtraction of the cluster contribution.

Thus, the fit function for SS pairs is
\ifthenelse{\boolean{cms@external}}{
\begin{multline}
  C^{\pm\pm}_2(\qi) = c_\text{SS} \, K^{\pm\pm}(\qi)\\
   \times\left[1 + z(\Nr) \frac{b}{\sigma_b \sqrt{2\pi}}
    \exp\left(- \frac{\qi^2}{2\sigma_b^2}\right)\right] \;
  C_\text{2,BE}(\qi),
 \label{eq:pid_c2_1d}
\end{multline}
}{
\begin{equation}
  C^{\pm\pm}_2(\qi) = c_\text{SS} \, K^{\pm\pm}(\qi) \;
   \left[1 + z(\Nr) \frac{b}{\sigma_b \sqrt{2\pi}}
    \exp\left(- \frac{\qi^2}{2\sigma_b^2}\right)\right] \;
  C_\text{2,BE}(\qi),
 \label{eq:pid_c2_1d}
\end{equation}
}
where $K^{\pm\pm}$ describes the Coulomb effect, the squared brackets
encompass the cluster contribution, and $C_\text{2,BE}(\qi)$ contains the
quantum correlation to be extracted.
Only the normalization $c_\text{SS}$ and the parameters of $C_\text{2,BE}(\qi)$
are fitted, all the other variables (those in $K^{\pm\pm}$, $z(\Nr)$, $b$, and
$\sigma_b$) were fixed using Eq.~\eqref{eq:b_sigma_b}.

\begin{figure}[ht]
 \centering
  \includegraphics[width=0.49\textwidth]{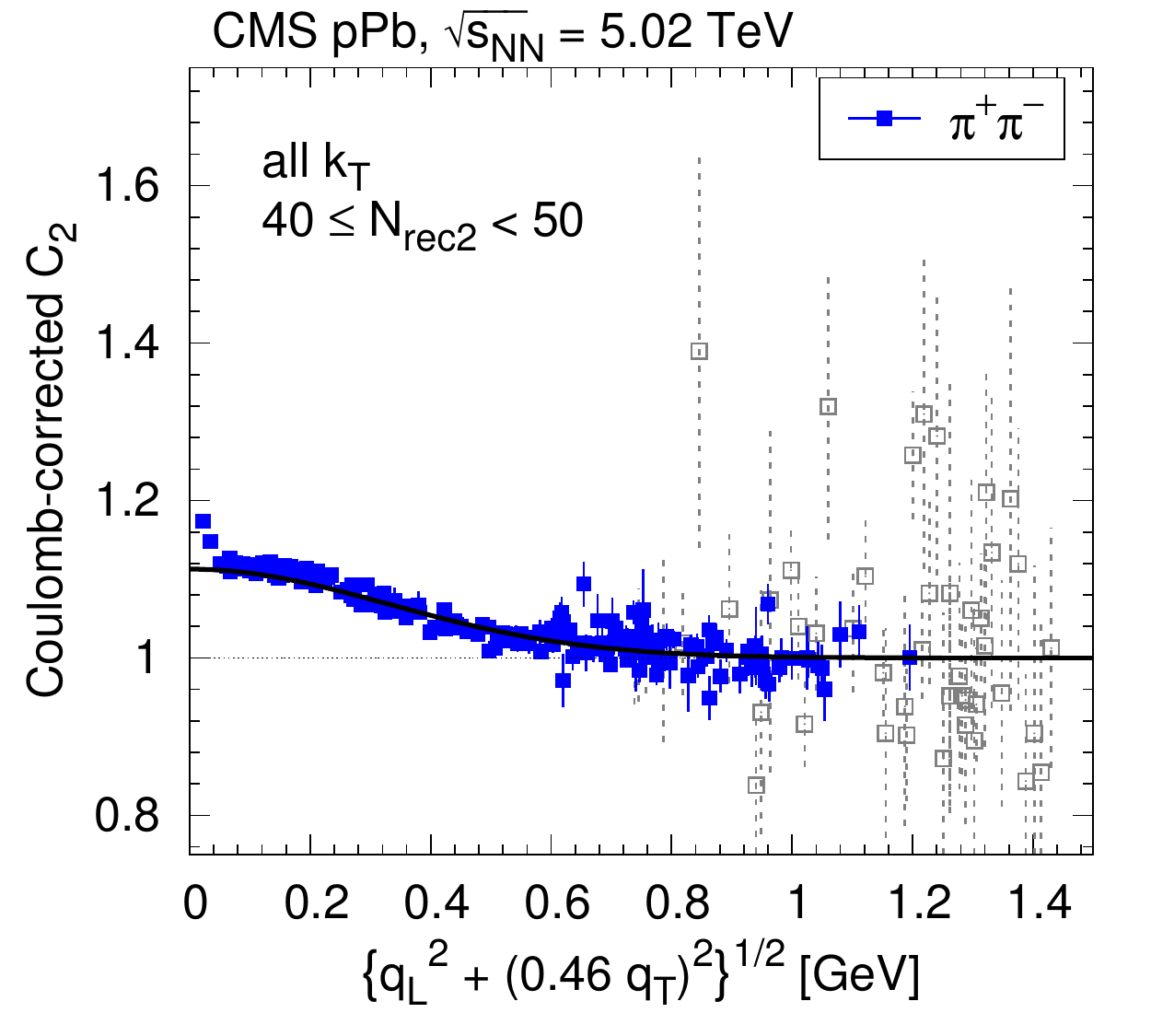}
  \includegraphics[width=0.49\textwidth]{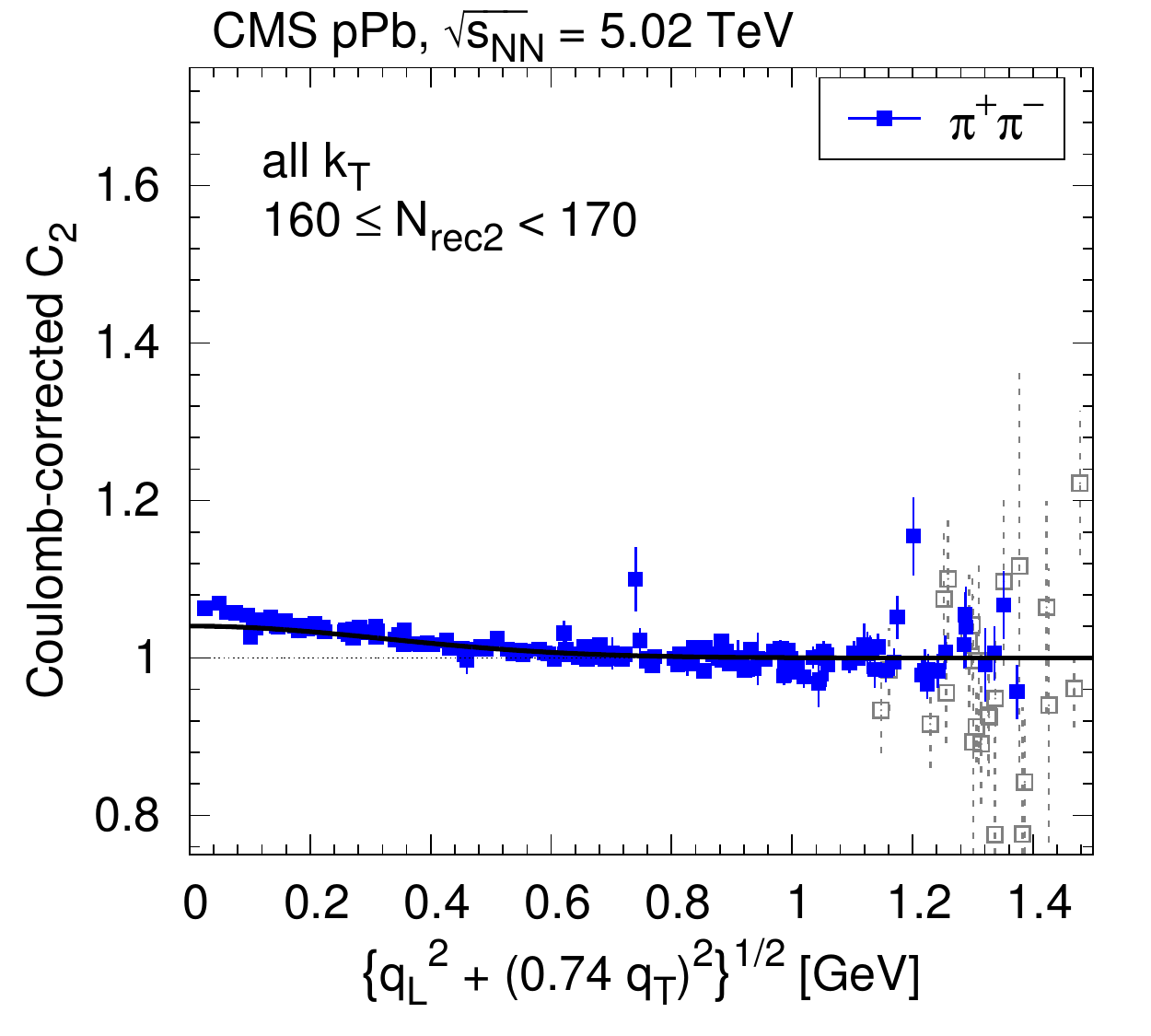}
\caption{Coulomb-corrected OS correlation function of pions (blue squares)
as a function of a combination of $\ql$ and $\qt$, in two selected $\Nr$ bins
for all $\kt$ values (\cmsLeft: $40 \le \Nr < 50$, \cmsRight: $160 \le \Nr < 170$). For
better visibility, only a randomly selected small fraction of the points is
plotted. In addition, those with statistical uncertainty higher than 10\% are
plotted with light grey color. The solid curves indicate the prediction of the
parametrized cluster contribution.}
 \label{fig:pi_pPb_5TeVc/q3ix__op}
\end{figure}

In the case of two and three dimensions, the measured OS correlation functions
are constructed as a function of a length given by a weighted
sum of $\vec{q}$ components, instead of $\qi$.
The Coulomb-corrected OS correlation functions of pions as a function of a
combination of $\ql$ and $\qt$, $q_h = \sqrt{\ql^2 + (a \qt)^2}$, in selected
$\Nr$ bins for all \kt, are shown in Fig.~\ref{fig:pi_pPb_5TeVc/q3ix__op}.
The dependence of the $a$ factor on $\Nr$ and \kt is described by:
\begin{equation}
 a(\Nr,\kt) = a_0 - b_0 \exp\left(- \frac{\Nr}{N_a} - \frac{\kt}{k_a} \right).
 \label{eq:a}
\end{equation}
This particular form is motivated by the physics results pointing to
an elongated source, shown later in
Section~\ref{sec:clusteremov_results_pp_ppb_pbpb}. The small peak at $q_h
\approx 0$ is from photon conversions where both the electron and the positron
are misidentified as a pion. The very low $q_h$ region is not included in the
fits.
Instead of the measurements, the fitted curves shown in
Fig.~\ref{fig:pi_pPb_5TeVc/q3ix__op} are used to describe the $\Nr$
and $\kt$ dependence of these parameters in the following. Hence the fit
function for SS pairs in the multidimensional case is:
\ifthenelse{\boolean{cms@external}}{
\begin{multline}
 C^{\pm\pm}_2(\vec{q}) = c \, K^{\pm\pm}(\qi) \\
   \times \left[1 + z(\Nr) \frac{b}{\sigma_b \sqrt{2\pi}}
             \exp\left(- \frac{q_h^2}{2\sigma_b^2}\right)\right] \;
   C_\text{2,BE}(\vec{q}).
 \label{eq:pid_c2_23d}
\end{multline}
}{
\begin{equation}
 C^{\pm\pm}_2(\vec{q}) = c \, K^{\pm\pm}(\qi) \;
   \left[1 + z(\Nr) \frac{b}{\sigma_b \sqrt{2\pi}}
             \exp\left(- \frac{q_h^2}{2\sigma_b^2}\right)\right] \;
   C_\text{2,BE}(\vec{q}).
 \label{eq:pid_c2_23d}
\end{equation}
}
In the formula above, only the normalization $c$ and the parameters of
$C_\text{2,BE}(\vec{q})$ are fitted, all the other variables (those in
$K^{\pm\pm}$, $z(\Nr)$, $b$, $\sigma_b$, and $a$) are fixed using the
parametrization described above.

There are about two orders of magnitude less kaon pairs than pion pairs, and
therefore a detailed study of the cluster contribution is not possible. We
assume that their cluster contribution is similar to that of pions, and that
the parametrization deduced for pions with identical parameters can be used
(for $b$ and $\sigma_b$ in Eq.~\eqref{eq:b_sigma_b}, and for $a$ in
Eq.~\eqref{eq:a}). This choice is partly justified by OS kaon correlation
functions shown later, but based on the observed difference, the systematic
uncertainty on $z(\Nr)$ for kaons is increased to 0.3.

\subsection{Fitting the correlation function}
\label{sub:fitting_functions}

The parametrizations used to fit the correlation functions in 1D, 2D, and
3D in terms of $\qi$, $(\ql, \qt)$, and $(\qs, \ql, \qo)$, respectively, are
the following~\cite{cms-hbt-1st,cms-hbt-2nd,Csorgo:2003uv}:
\begin{equation}
 C_\text{2,BE}(\qi) = C [1 + \lambda \re^{- ( \qi\Ri )^a}] \, (1 + \delta_\text{inv} \; \qi),
 \label{eq:1d-levy}
\end{equation}
is the fit function employed in the 1D case,  while in the 2D case it has the
form:
\ifthenelse{\boolean{cms@external}}{
\begin{multline}
 C_\text{2,BE}(\qt,\ql) =
 C \Bigl\{1 + \lambda\exp\bigl[
  -|\qt^2 \Rt^2 + \ql^2 \Rl^2 \\+ 2 \qt \ql R_{LT}^2|^{a/2}
  \bigr]\Bigr\} \, (1 + \delta_T \qt + \delta_L \ql),
\label{eq:2d-levy}
\end{multline}
}{
\begin{equation}
 C_\text{2,BE}(\qt,\ql) = C \left\{1 + \lambda \exp{\left[
  -\left|\qt^2 \Rt^2 + \ql^2 \Rl^2 + 2 \qt \ql R_{LT}^2
  \right|^{a/2}
  \right]}\right\} \, (1 + \delta_T \qt + \delta_L \ql),
\label{eq:2d-levy}
\end{equation}
}
and, finally, the form used for 3D is:
\ifthenelse{\boolean{cms@external}}{
\begin{multline}
 C_\text{2,BE}(\qs,\ql,\qo) =
 C \Bigl\{1 + \lambda \exp\bigl[
  - |\qs^2 \Rs^2 + \ql^2 \Rl^2 \\ + \qo^2 \Ro^2 + 2 \qo \ql R_\mathrm{LO}^2|^{a/2}\bigr]\Bigr\}  \\
  \times (1+ \delta_S \qs + \delta_L \ql + \delta_O \qo).
\label{eq:3d-levy}
\end{multline}
}{
\begin{equation}
\begin{aligned}
 C_\text{2,BE}(\qs,\ql,\qo) = &C \left\{1 + \lambda \exp{\left[
  - \left|\qs^2 \Rs^2 + \ql^2 \Rl^2 + \qo^2 \Ro^2 + 2 \qo \ql R_\mathrm{LO}^2
  \right|^{a/2}
  \right]}\right\}  \\
  & \times (1+ \delta_S \qs + \delta_L \ql + \delta_O \qo).
\label{eq:3d-levy}
\end{aligned}
\end{equation}
}
The radius parameters $\Ri$, $(\Rt,\Rl)$, $(\Rs,\Rl,\Ro)$, correspond to the
lengths of homogeneity in 1D, 2D and 3D, respectively. In the above
expressions, $\lambda$ is the intercept parameter (intensity of the correlation
at $q = 0$) and $C$ is the overall normalization. The polynomial factors on
the right side, proportional to the fit variables $ \delta_\text{inv},
\delta_T, \delta_L, \delta_S, \delta_O$, are introduced for accommodating
possible deviations of the baseline from unity at large values of these
variables, corresponding to long-range correlations. They are only used in the
double ratio method, their values are uniformly set to zero in the case of the
particle identification and cluster subtraction method that contains no such
factors.

In Eq.~\eqref{eq:2d-levy}, $\Rt = \Rt^{^G} + \tau \beta_T \cos\phi$ and $\Rl =
\Rl^{^G} + \tau \beta_L$, where $R_{T,L}^{^G}$ are the geometrical transverse
and longitudinal radii, respectively, whereas $\beta_T=\kt/k^0$ and
$\beta_L=k_\text{L}/k^0$ are the transverse and longitudinal
velocities, respectively; $\phi$ is the angle between the directions of
$\vec{q}_\mathrm{T}$ and $\vec{k}_\mathrm{T}$, and $\tau$ is the source lifetime. Similarly, in
Eq.~\eqref{eq:3d-levy} $\Rs^2 = (R_S^{^G})^2$, $ \Rl^2 = (R_L^{^G})^2 + \tau^2
\beta_L^2$, $\Ro^2 = (R_O^{^G})^2 + \tau^2 \beta_T^2$, and $R_{LO}^2 = \tau^2
\beta_L^2 \beta_T^2$ are the corresponding radius parameters in the 3D case.
When the analysis is performed in the LCMS, the cross-terms $\qt \ql$ and $\qo
\ql$ do not contribute for sources symmetric along the longitudinal direction.

Theoretical studies show that for the class of L\'evy stable distributions
(with index of stability $0 < a \le 2$)~\cite{Csorgo:2003uv} the BEC function
has a stretched exponential shape, corresponding to $a=1$ in
Eqs.~\eqref{eq:2d-levy} and \eqref{eq:3d-levy}. This is to be distinguished
from the simple exponential parametrization, in which the coefficients of the
exponential would be given by $\sum_i R_i q_i$. In Section
\ref{subsec:doubratio_results-3d}, the parameters resulting from the fits
obtained with the simple and the stretched exponentials will be discussed in
the
3D case. If $a$ is treated as a free parameter, it is usually between $a=1$ and
$a = 2$ (which corresponds to the Gaussian distribution~\cite{Csorgo:2003uv}).
In particular, for $a=1$, the exponential term in expressions given by
Eqs.~\eqref{eq:2d-levy} and \eqref{eq:3d-levy} coincides with the Fourier
transform of the source function, characterized by a Cauchy distribution.

The measurements follow Poisson statistics, hence the correlation functions are
fitted by minimizing the corresponding binned negative log-likelihood.

\subsection{Systematic uncertainties}
\label{subsec:systematics}

\subsubsection{Double ratio method}
\label{syst-fsq-13002}

Various sources of systematic uncertainties are investigated with respect to
the double ratio results, similar to those discussed in Refs.~\cite{cms-hbt-1st,cms-hbt-2nd}:
the MC tune used, the reference sample employed to estimate the single ratios,
the Coulomb correction function, and the selection requirements applied to the
tracks used in the analysis. From these, the major sources of systematic
uncertainty come from the choice of the MC tune, followed by the type of
reference sample adopted. In the latter case,
alternative techniques to those considered in this analysis for constructing the reference sample
are studied:
mixing tracks from different events selected at random, taking tracks from different events
with similar invariant mass to that in the original event (within 20\%,
calculated using all reconstructed tracks), as well as combining oppositely charged
particles from the same event. Each of these procedures produces different shapes for the single
ratios, and also leads to significant differences in the fit parameters.
The exponential function in Eq.~\eqref{eq:1d-levy}
(L\'evy type with $a=1$) is adopted as the fit function in the systematic studies.

The results of the analysis do not show any significant dependence on the
hadron charges, as obtained in a study separating positive and negative charges
in the single ratios. Similarly, the effect of pileup, investigated by
comparing the default results to those where only single-vertex events are
considered, does not introduce an extra source of systematic uncertainty.

The sources of systematic uncertainties discussed above are considered to be
common to $\Pp\Pp$ collisions at both 2.76 and 7\TeV. In the 1-D case the
overall systematic uncertainty associated with $\Ri$ and that associated with
$\lambda$ are estimated using the change in the fitted values found when
performing the variations in procedure mentioned above for the four major
sources of uncertainties at both energies.  This leads to 15.5\% and 10\% for
the systematic uncertainties in the fit radius and intercept parameters,
respectively, which are considered to be common to all $\Nt$ and $\kt$ bins.
The estimated values in the multidimensional cases were of similar order or
smaller than those found in the 1-D case. Therefore, the systematic
uncertainties found in the 1-D case were also applied to the 2-D and 3-D
results.

\subsubsection{Particle identification and cluster subtraction method}
\label{syst-hin-14013}

The systematic uncertainties are dominated by two sources: the construction of
the background distribution, and the amplitude $z$ of the cluster contribution
for SS pairs with respect to that for OS ones.
Several choices for the construction of the background distribution are
studied. The goodness-of-fit distributions clearly favor the event mixing
prescription, while the rotated version gives poorer results, and the mirrored
background yields the worst performance. The associated systematic
uncertainty is calculated by performing the full analysis using the mixed event
and the rotated variant, and calculating the root-mean-square of the final
results (radii and $\lambda$).

Although the dependence of the ratio of the mini-jet contribution $z$ (between
SS and OS pairs), as a function of $\Nr$ could be well described by
theory-motivated fits in all $\kt$ regions, the extracted ratios show some
deviations within a given colliding system, and also between systems. In order
to cover those variations, the analysis is performed by moving the fitted ratio
up and down by 0.2 for pions, and by 0.3 for kaons. The associated
systematic uncertainties are calculated by taking the average deviation of the
final results (radii and $\lambda$) from their central values.

In order to suppress the effect of multiply-reconstructed particles and
misidentified photon conversions, the low-$q$ region ($q < 0.02\GeV$) is
excluded from the fits. To study the sensitivity of the results to the size of
the excluded range, its upper limit is doubled and tripled. Both the radii and
$\lambda$ decreased slightly with increasing exclusion region. As a result, the
contribution of this effect to the systematic uncertainty is estimated to be
0.1\unit{fm} for the radii and 0.05 for $\lambda$.

The systematic uncertainties from all the sources related to the MC
tune, Coulomb corrections, track selection, and the differences between
positively and negatively charged particles are similar to those found in case
of the double ratio method above. The systematic uncertainties from various
sources are added in quadrature.

\subsection{Comparisons of the techniques}
\label{sec:consistency_checks}

\begin{figure}[t]
 \centering
  \includegraphics[width=0.49\textwidth]{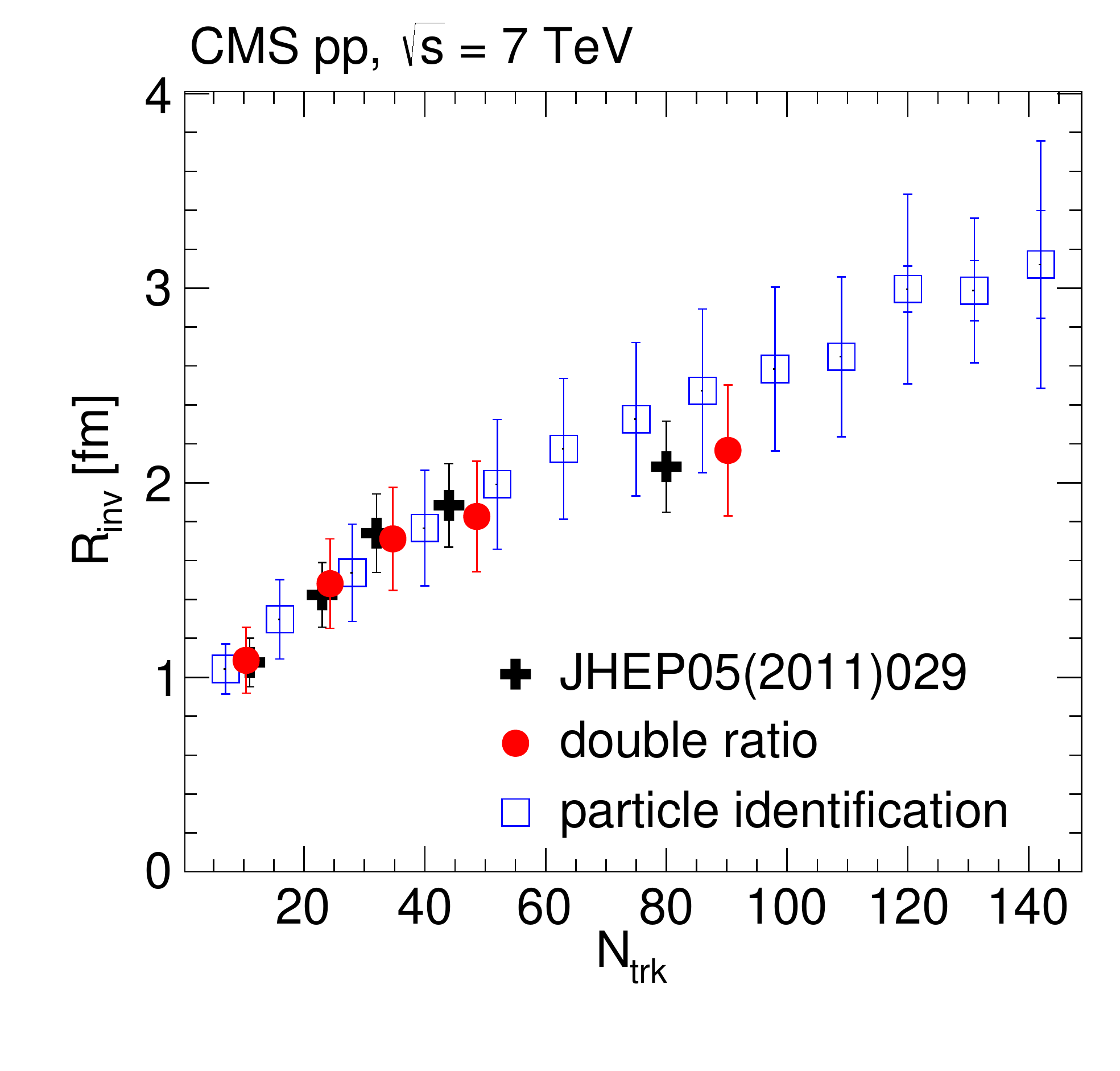}
  \includegraphics[width=0.49\textwidth]{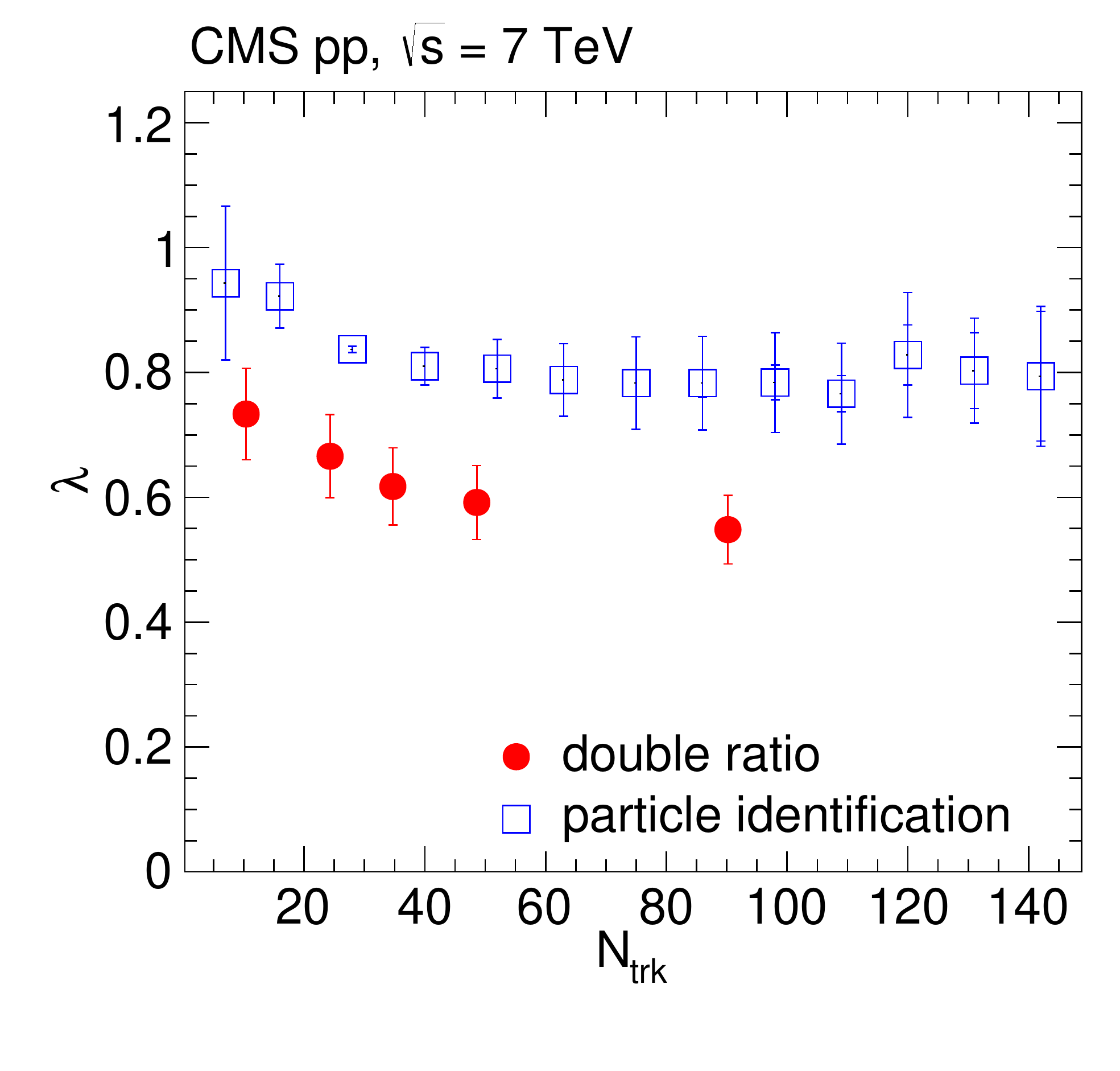}
 \caption{Dependence on track multiplicity of the radius (\cmsLeft) and intercept
parameter (\cmsRight), from the exponential fits to the 1D correlation functions
via Eq.~\eqref{eq:1d-levy}, obtained in the double ratio (filled markers) and
the particle identification (for pions, empty squares) methods in $\Pp\Pp$ collisions
at 7\TeV, together with the results from Ref.~\cite{cms-hbt-2nd}. The inner
(outer) error bars correspond to statistical (systematic) uncertainties.}
 \label{fig:comp-dobratio-clusterrm}
\end{figure}

In order to illustrate the level of consistency between both BEC analyses, the
fit results for 1D $\Ri$ and $\lambda$ obtained employing the double ratio and
the particle identification and cluster subtraction methods are compared. In
the double ratio method, mixed events having similar multiplicities within the
same $\eta$ ranges are used as the reference sample. The non-BEC effects are
corrected with the double ratio technique, as discussed in
Section~\ref{subsec:doubleratios}, which also reduces the bias due to the
choice of reference sample.
In the particle identification and cluster subtraction method, single ratios
are measured considering mixed pairs from 25 events chosen at random, whose
contamination from resonance decays and mini-jet contributions are corrected as
discussed in Sections~\ref{sec:partpairs} and \ref{subsubsec:cluster-removal}.
Regarding the final-state Coulomb corrections, no significant difference is
observed if applying the Gamow factor, Eq.~\eqref{eq:gamow}, or the full
Coulomb correction, Eq.~\eqref{eq:full_coulomb}, to the correlation
functions in $\Pp\Pp$ collisions. In both cases, the 1D correlations are fitted with
an exponential function ($a=1$ in Eq.~\eqref{eq:1d-levy}), from which the
radius and intercept parameters are obtained.
Since in the particle identification and cluster subtraction method
$\langle\Nt\rangle$ is fully corrected, as discussed in
Section~\ref{subsubsec:data_driven_selections} and summarized in
Table~\ref{tab:multiClass}, an additional correction is applied to the values
of $\langle\Ntpt\rangle$ in Table~\ref{tab:ranges-dr} in the double ratio
method, thus allowing the comparison of both sets of results plotted at the
corresponding value of $\langle\Nt\rangle$.
Such a correction is applied with a multiplicative factor obtained from the
ratio of the total charged hadron multiplicities in both analyses,
${N_\text{ch} ({\pt} \to 0)}/ {N_\text{ch} ({\pt} > 0.4\GeV)}
\sim 1.7$. The measurements are plotted in
Fig.~\ref{fig:comp-dobratio-clusterrm} for charged hadrons in the double ratio
method and from identified charged pions from the particle identification and
cluster subtraction method, as well as those for charged hadrons from
Ref.~\cite{cms-hbt-2nd}. In the latter, the correction factor is obtained in a
similar fashion as ${N_\text{ch} ({\pt} \to 0)}/ {N_\text{ch}
({\pt} > 0.2\GeV)} \sim 1.1$, reflecting the $\pt > 0.2\GeV$ requirement
applied in that case.

The radius parameter $\Ri$ (Fig.~\ref{fig:comp-dobratio-clusterrm}, \cmsLeft)
steadily increases with the track multiplicity. The results using the two
different correlation techniques agree within the uncertainties, and they also
agree well with our previous results~\cite{cms-hbt-2nd}.
In Fig.~\ref{fig:comp-dobratio-clusterrm} (\cmsRight), the corresponding results
for the intercept parameter, $\lambda$, are shown integrated over all \kt. The
overall behavior of the correlation strength is again similar in both cases,
showing first a decrease with $\Nt$ and then a flattening, approaching a
constant for large values of the track multiplicity. The differences in the
absolute values of the fit parameter $\lambda$ in the two approaches are
related to the differences in $\pt$ (and $\kt$) acceptances for the tracking
methods used in the two analyses, and the availability of particle
identification, with both factors lowering the fit value of $\lambda$ for the
double ratio method.

\section{Results with the double ratio technique in \texorpdfstring{\Pp\Pp}{pp} collisions}
\label{sec:legacy-results-1d2d3d}

\subsection{One-dimensional results}
\label{subsec:legacy-results-1d}

\begin{figure}[btp]
 \centering
  \includegraphics[width=0.49\textwidth]{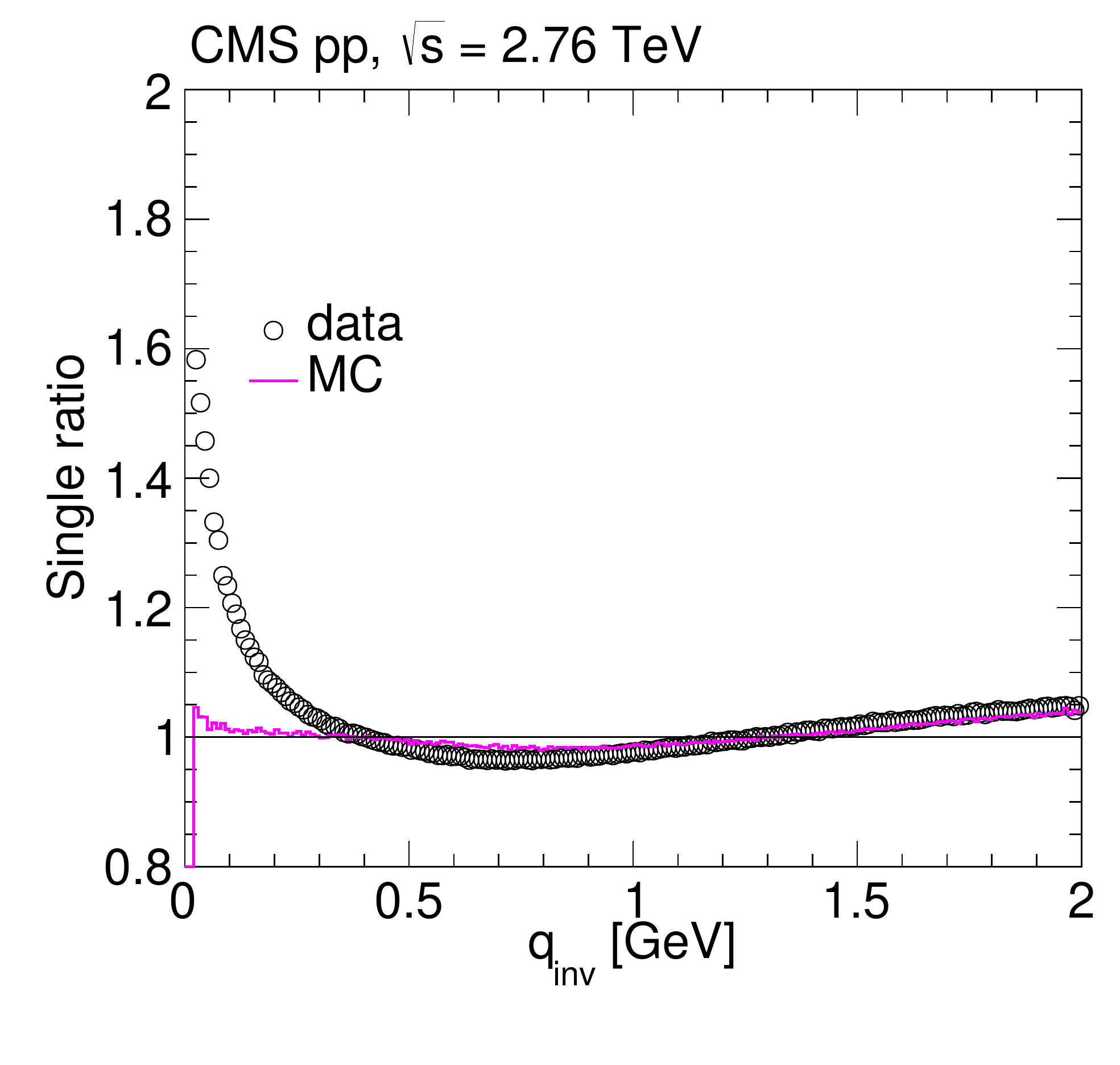}
  \includegraphics[width=0.49\textwidth]{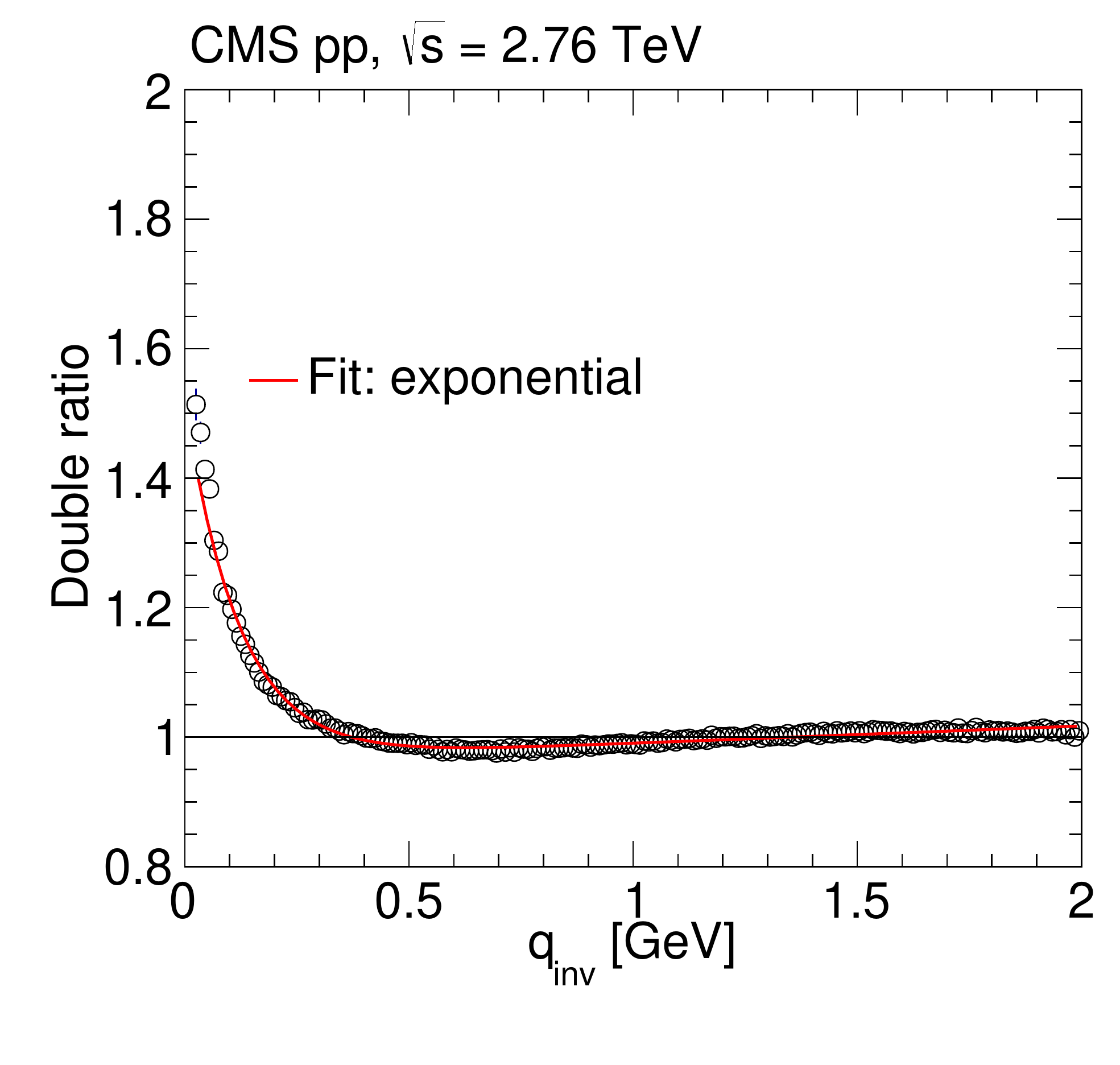}
 \caption{\cmsLLeft: 1D single ratios as a function of $\qi$ for data and simulation
({\PYTHIA}6 Z2 tune) in $\Pp\Pp$ collisions at $\sqrt{s} = 2.76\TeV$. \cmsRRight:
Corresponding 1D double ratio versus $\qi$ fitted to the superimposed
exponential fit function. Only statistical uncertainties are included, commonly
of the order or smaller than the marker size.}
 \label{fig:1D-singleratios}
\end{figure}

Single and double ratios for $\Pp\Pp$ collisions at $\sqrt{s} = 2.76$\TeV
are shown in Fig.~\ref{fig:1D-singleratios}, integrated over the
charged-particle multiplicity, $\Ntpt$, and the pair momentum, $\kt$. The fit
to the double ratio is performed using the exponential function in
Eq.~\eqref{eq:1d-levy}, with $a=1$.
The corresponding fit values of the
invariant radii, $\Ri$, and the intercept parameter, $\lambda$, are shown in
Table~\ref{tab:1d-lambdas-radii-276-7tev}, together with those for $\Pp\Pp$
collisions at $\sqrt{s} = 7$\TeV; these being compatible within the
uncertainties at these two energies.

\begin{figure*}[htb]
 \centering
  \includegraphics[width=0.49\textwidth]{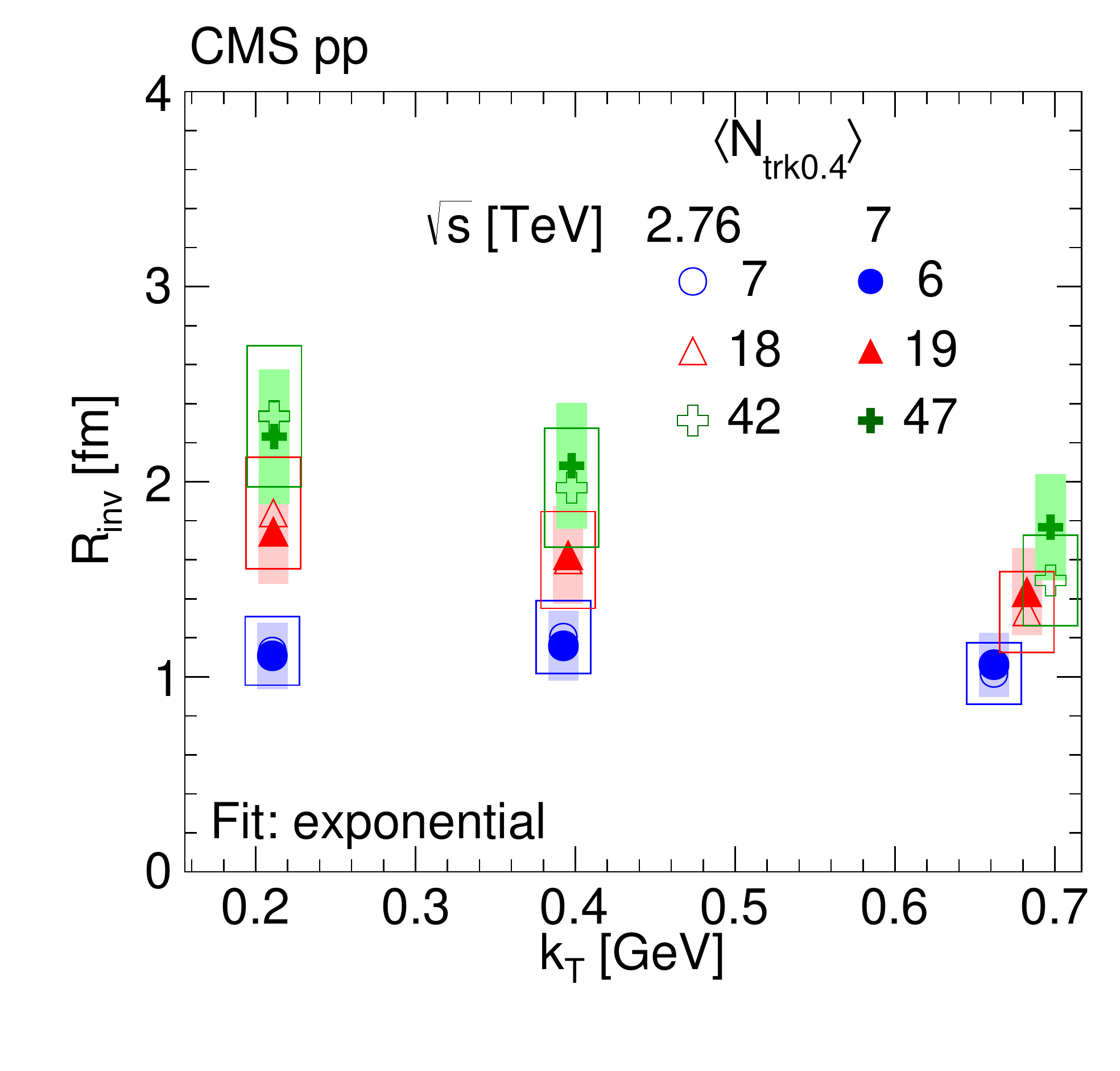}
  \includegraphics[width=0.49\textwidth]{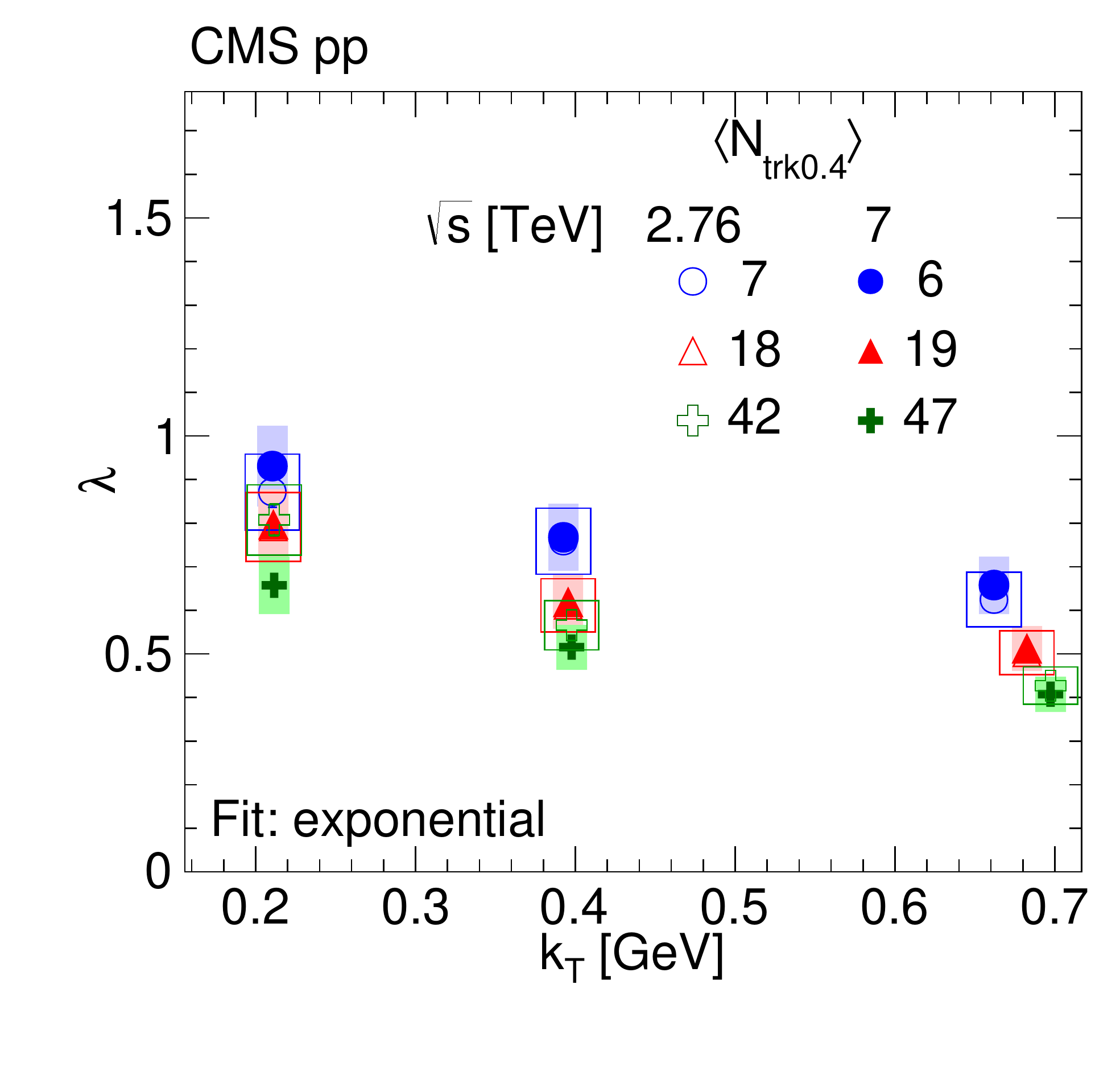}
  \includegraphics[width=0.49\textwidth]{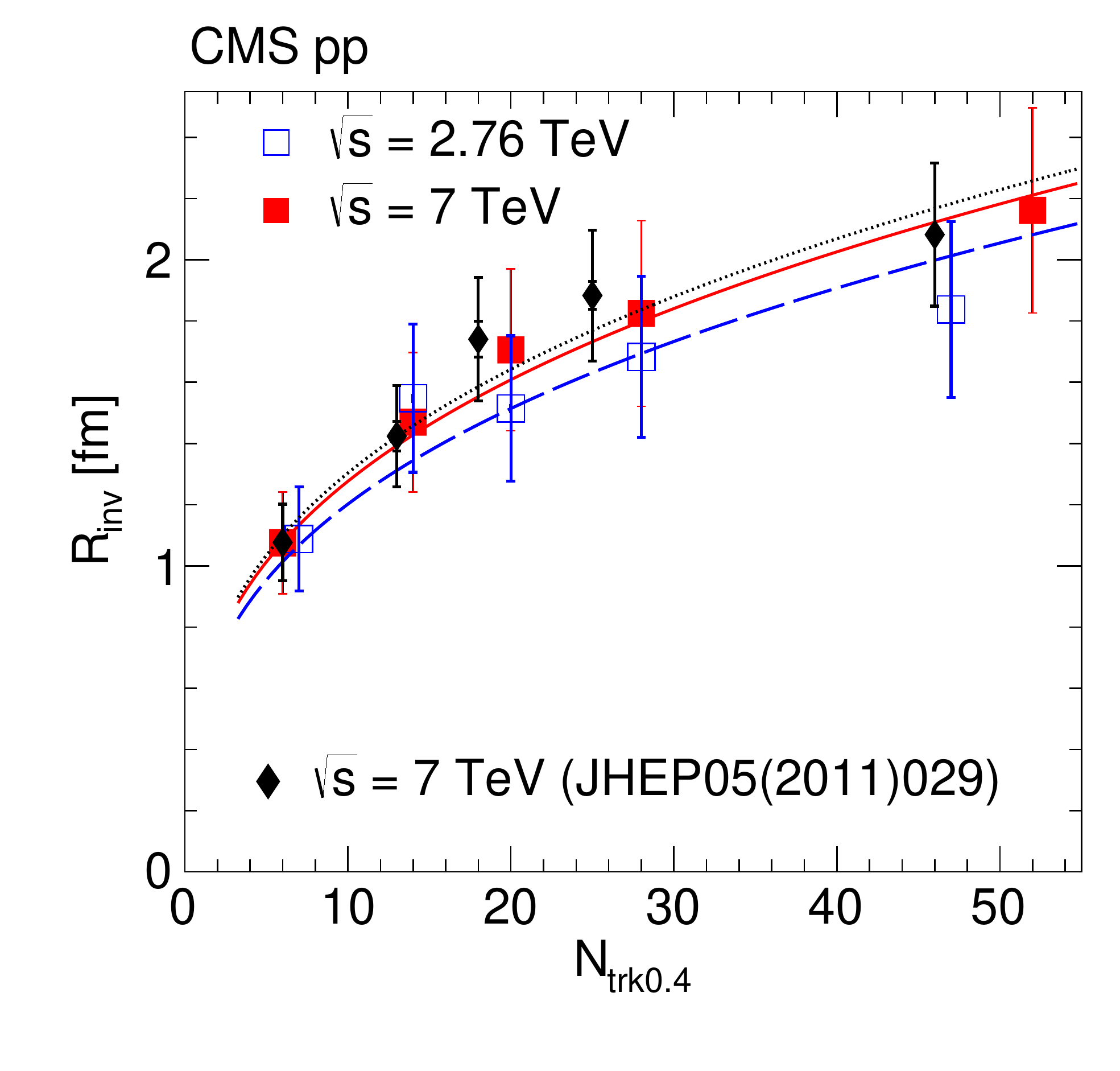}
 \caption{Top: $\Ri$(left) and $\lambda$ (right) as a function of $\kt$
obtained from an exponential fit to BEC data in three multiplicity bins in $\Pp\Pp$ collisions at 2.76\TeV (open symbols) and 7\TeV (filled symbols). The boxes indicate the systematic uncertainties.
Bottom: Radius parameter $\Ri$ as a function of $\Ntpt$ for $\Pp\Pp$ collisions at 2.76\TeV (open squares) and 7\TeV (filled squares), where the systematic uncertainties are shown as error bars and the statistical uncertainties are smaller than the
marker size. Previous results at 7\TeV (filled diamonds)~\cite{cms-hbt-2nd} are
also shown, with their corresponding
statistical (inner error bars) and statistical plus
systematic uncertainties added in quadrature (outer error bars).
The curves indicate the respective fits to a $\Ntpt^{1/3}$ functional form.}
 \label{fig:fits-to-dr-nch-kt-276-7tev}
\end{figure*}

\begin{table*}[bht]
\topcaption{Fit parameters of the 1D BEC function in $\Pp\Pp$ collisions at $\sqrt{s}
= 2.76$ and 7\TeV.}
 \label{tab:1d-lambdas-radii-276-7tev}
 \centering
 \begin{scotch}{lcc}
             & $\Pp\Pp$ $\sqrt{s} = 2.76\TeV$ &
               $\Pp\Pp$ $\sqrt{s} = 7\TeV$ \\
  \hline
  $\lambda$  &  $0.577 \pm 0.007\stat\pm 0.058\syst$ &
                $0.545 \pm 0.001\stat\pm 0.054\syst$ \\
  $\Ri$ [fm] &  $1.624 \pm 0.013\stat\pm 0.252\syst$ &
                $2.022 \pm 0.004\stat\pm 0.313\syst$ \\
 \end{scotch}
\end{table*}

The upper panel of Fig.~\ref{fig:fits-to-dr-nch-kt-276-7tev} shows the results
for the invariant radius $\Ri$ (left) and the intercept parameter $\lambda$
(right), obtained with the exponential fit function in three bins of both
$\Ntpt$ and $\kt$, for $\Pp\Pp$ collisions at $\sqrt{s} = 2.76$ and 7\TeV. The
results corresponding to the two energies again show good agreement in the
various ($\Ntpt$, $\kt$) bins considered, extending the observation made with
respect to the integrated values in Table~\ref{tab:1d-lambdas-radii-276-7tev}
to different bin combinations of these variables. The parameter $\lambda$
decreases with increasing $\kt$, and its dependence on the charged
multiplicity $\Ntpt$ follows a similar trend. The fit invariant radius, $\Ri$,
decreases with $\kt$, a behavior previously observed in several different collision systems and energy
ranges~\cite{cms-hbt-1st,cms-hbt-2nd,e735,star,phobos,phenix,alice}, and
expected for expanding emitting sources. In the case of heavy ion collisions,
as well as for deuteron-nucleus and proton-nucleus collisions, both at the BNL
RHIC and at the CERN LHC, this observed trend of the fit radii is compatible
with a collective behavior of the source.
The radius fit parameter is also investigated in
Fig.~\ref{fig:fits-to-dr-nch-kt-276-7tev} (bottom) for results integrated over
$\kt$ bins and in five intervals of the charged-particle multiplicity, as in
Table~\ref{tab:ranges-dr}. The value of $\Ri$ steadily increases with
multiplicity for $\Pp\Pp$ collisions at 2.76 and 7\TeV, as previously observed at
7\TeV in Ref.~\cite{cms-hbt-2nd}.
The various curves are fits to the data, and are proportional to
$(\Ntpt)^{1/3}$, showing also that the results are approximately independent on
the center-of-mass energy.

\begin{figure*}[tbh]
 \centering
  \includegraphics[width=\textwidth]{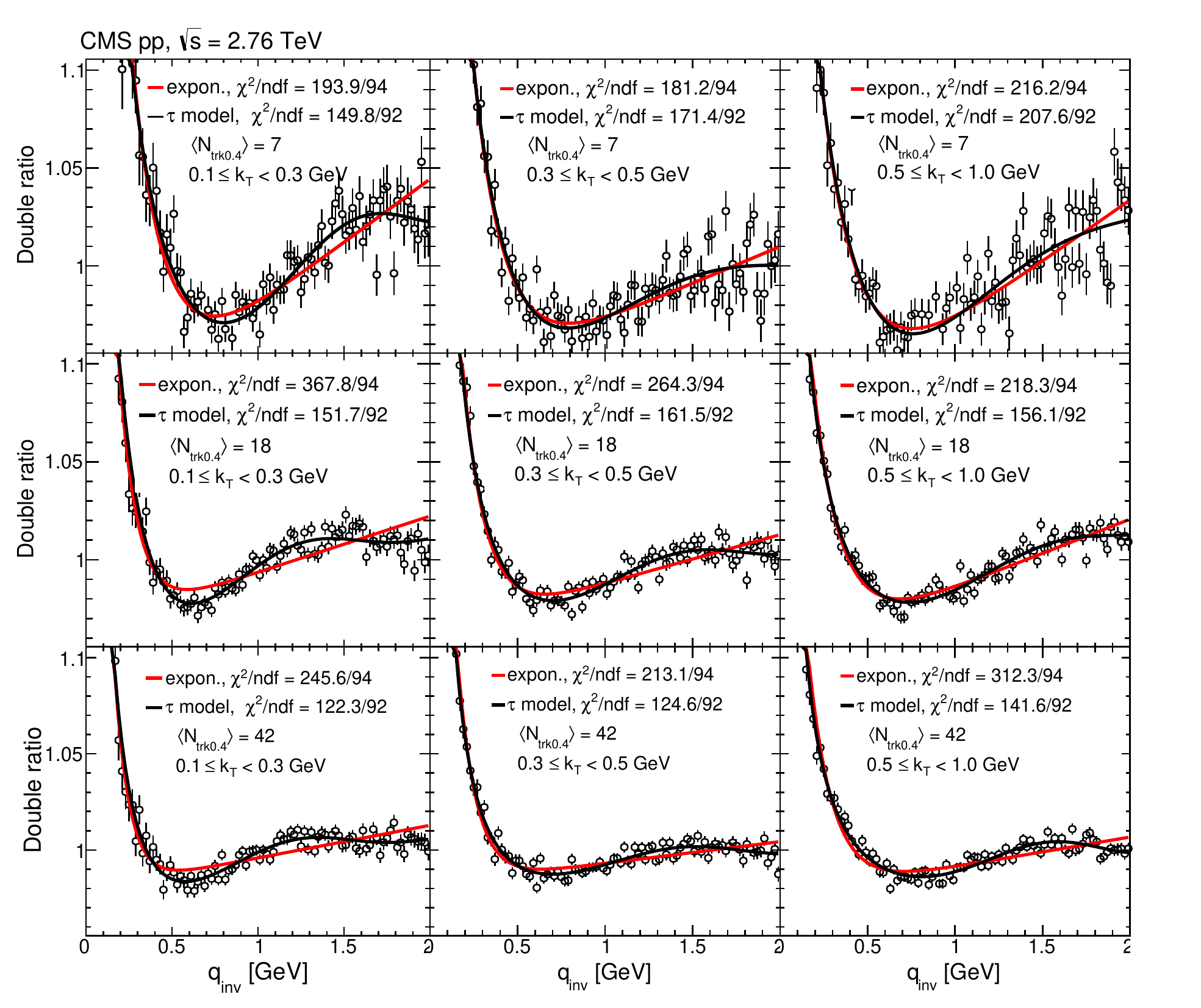}
 \caption{1D correlation functions (with the $y$-axis cut off at $\sim$1.1)
versus $\qi$ obtained with the double ratio method in $\Pp\Pp$ at 2.76\TeV for
increasing ranges of $\kt$ (left to right) and of $\Ntpt$ (top to bottom). Data
are fitted to the exponential (red curve) and the $\tau$ model (black curve)
functions.  The error bars show the statistical uncertainties.}
 \label{fig:dr-5bins-dip-7tev-nch-kt}
\end{figure*}

As discussed in Ref.~\cite{cms-hbt-2nd}, the exponential fit alone in
Eq.~\eqref{eq:1d-levy} is not able to describe the overall behavior of the
correlation function in the range $0.2 < \qi < 2\GeV$, for which an additional
long-range term proportional to $\qi$ is required.  Although the general trend
can be described by this parametrization, it misses the turnover of the
baseline at large values of $\qi$. The resulting
poor fit quality
is partially due to very small statistical uncertainties which restrict the variation of the
parameters, but mainly due to the presence of an
anticorrelation, i.e. correlation function values below the expected asymptotic
minimum at unity, which is observed in the intermediate $\qi$ range ($0.4 < \qi < 1.1\GeV$).
It is unlikely that this anticorrelation is due to
the contamination of non-pion pairs in the sample, which would mainly give
some contribution in the lower $\qi$ region.

The anticorrelation depth depends on the range of $\Ntpt$ considered, as
reported in Ref.~\cite{cms-hbt-2nd}. This is investigated further in this analysis by
considering different $\kt$ bins for $\Pp\Pp$ collisions at 2.76 and 7\TeV. The
results for the lower energy are shown in Fig.~\ref{fig:dr-5bins-dip-7tev-nch-kt}, where the data
and the exponential fit are shown together with those based on Eq. (\ref{eq:alevy}).
As previously observed in $\Pep\Pem$ collisions~\cite{L3}
and also in Ref.~\cite{cms-hbt-2nd}, this anticorrelation feature is compatible
with effects as suggested by the $\tau$ model~\cite{Htau}, where particle
production has a broad distribution in proper time and the phase space
distribution of the emitted particles is dominated by strong correlations of
the space-time coordinate and momentum components. In this model the time
evolution of the source is parameterized by means of a one-sided asymmetric
L\'evy distribution, leading to the fit function:
\ifthenelse{\boolean{cms@external}}{
\begin{multline}
 C_\text{2,BE}(q) = C \Bigl\{1 + \lambda \bigl[\cos[
  (q r_0)^2 + \\
  \tan(\nu \pi/4) (q r_\nu)^{\nu}
  ] \exp\{-(qr_\nu)^\nu\}\bigr]
\Big\}\, (1+ \delta\; q),
 \label{eq:alevy}
\end{multline}
}{
\begin{equation}
 C_\text{2,BE}(q) = C \bigl\{1 + \lambda \left[\cos\left[
  (q r_0)^2 + \tan(\nu \pi/4) (q r_\nu)^{\nu}
  \right] \exp\{-(qr_\nu)^\nu\}\right]
\bigr\} \, (1+ \delta\; q),
 \label{eq:alevy}
\end{equation}
}
where the parameter $r_0$ is related to the proper time of the onset
of particle emission, $r_\nu$ is a scale parameter entering in both the
exponential and the oscillating factors, and $\nu$ corresponds to the L\'evy
index of stability.

The appearance of such anticorrelations may be due to the fact that hadrons are
composite, extended objects. The authors of Ref.~\cite{bialas} propose a simple
model in which the final-state hadrons are not allowed to interpenetrate each
other, giving rise to correlated hadronic positions. This assumption results in
a distortion in the correlation function, leading to values below unity.

Fitting the double ratios with
Eq.~\eqref{eq:alevy} in the 1D case considerably improves the values of the
$\chi^2/\mathrm{ndf}$ as compared to those fitted with the exponential function
in Eq.~\eqref{eq:1d-levy}, for $\Pp\Pp$ collisions at both 2.76 and
7\TeV~\cite{cms-hbt-2nd}. From Fig.~\ref{fig:dr-5bins-dip-7tev-nch-kt}, it can be seen that the fit
based on Eq. (\ref{eq:alevy})
describes the overall behavior of the measurements more
closely in all the bins of $\Ntpt$ and $\kt$ investigated. In all but one of
those bins, the fit based on the $\tau$ model results in $1 < \chi^2/\text{ndf}
< 2$.
In addition,
a significant improvement is also observed when 2D and 3D global fits to the
correlation functions are employed, mainly when the L\'evy fit with $a=1$ is adopted
(Sections~\ref{subsec:doubratio_results-2d} and
\ref{subsec:doubratio_results-3d}).

\begin{figure}[thb]
 \centering
  \includegraphics[width=0.49\textwidth]{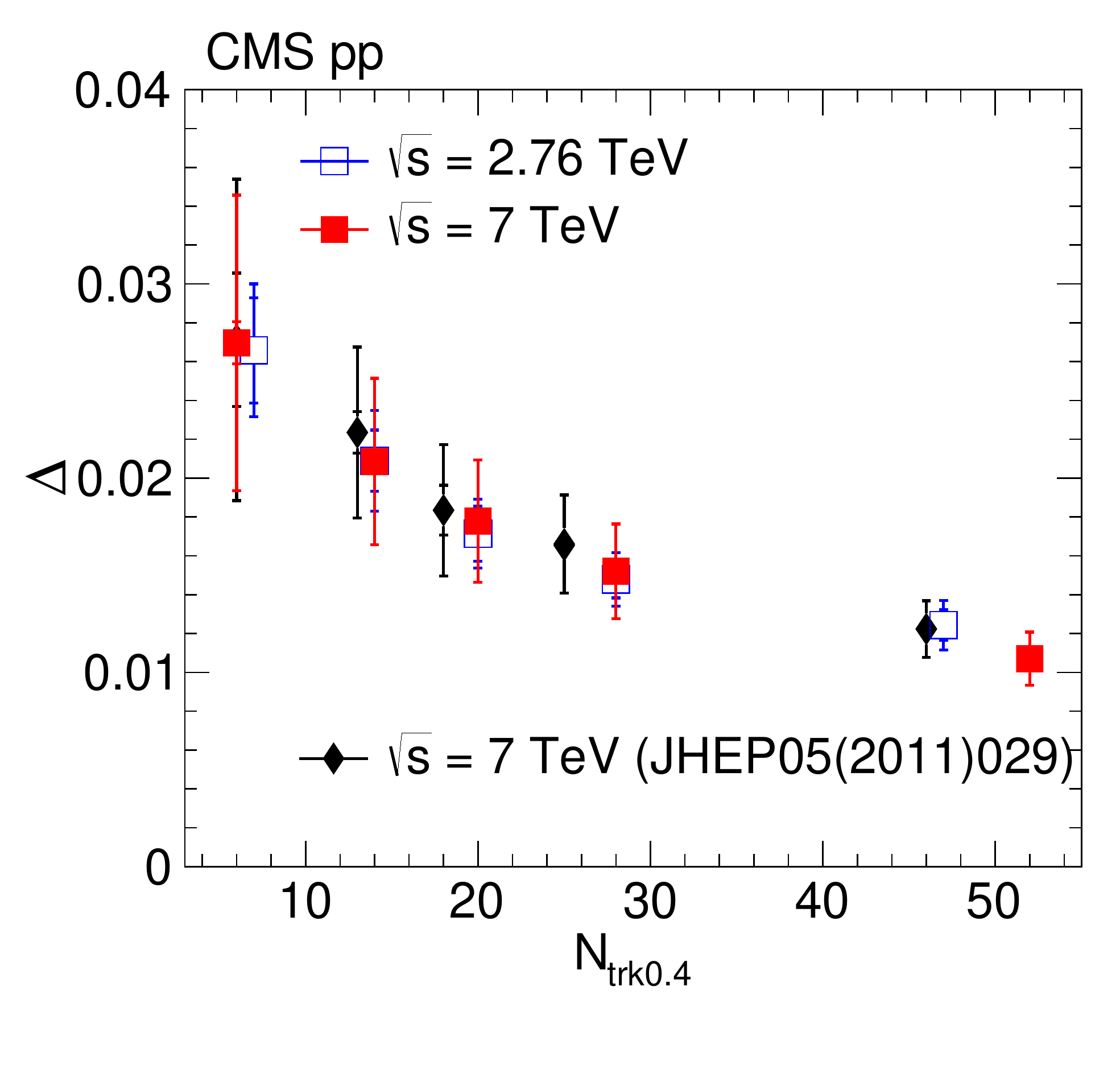}
  \includegraphics[width=0.49\textwidth]{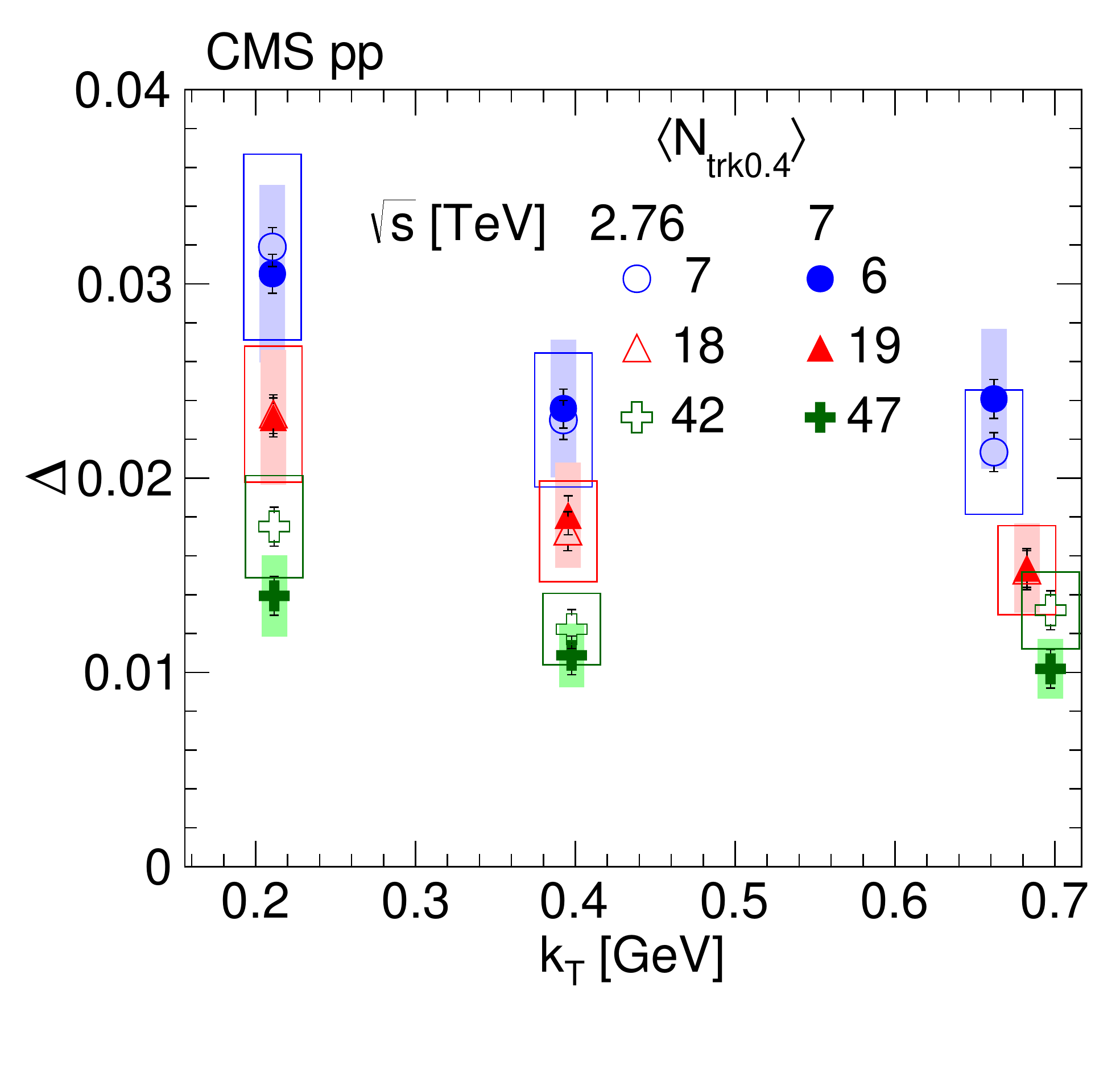}
 \caption{\cmsLLeft: Anticorrelation depth $\Delta$ as a function of $\Ntpt$
integrated over all $\kt$, in $\Pp\Pp$ collisions at 2.76\TeV (open squares) and
7\TeV (filled squares), together with previous results at 7\TeV
(filled diamonds)~\cite{cms-hbt-2nd}. Inner (outer) error bars represent the
statistical (plus systematic, added in quadrature) uncertainties. \cmsRRight:
Corresponding anticorrelation depth as a function of $\kt$ for various $\Ntpt$
ranges, in $\Pp\Pp$ at 2.76\TeV (open symbols) and 7\TeV (filled symbols). The boxes
indicate the systematic uncertainties.}
 \label{fig:dip-depth-nch-kt}
\end{figure}

The polynomial form $C(1 + \delta\; q)$ in Eq.~\eqref{eq:alevy} is used as a
baseline to quantify the depth of the dip~\cite{cms-hbt-2nd}. This is estimated
by the difference, $\Delta$, of the baseline function and the fit based on
Eq.~\eqref{eq:alevy} at the minimum, leading to the results shown in
Fig.~\ref{fig:dip-depth-nch-kt}. The \cmsLeft plot shows the present results for $\Pp\Pp$ collisions at 2.76 and at 7\TeV as a function of the multiplicity $\Ntpt$
integrated over $\kt$ bins, exhibiting their consistency at different energies,
as well as close similarity with the previous results~\cite{cms-hbt-2nd}. The
\cmsRight plot in Fig.~\ref{fig:dip-depth-nch-kt} shows the current results
corresponding to the 2.76 and 7\TeV data further scrutinized in different
$\Ntpt$ and $\kt$ bins.  From this plot, it can be seen that the depth of the
dip decreases with increasing $\Ntpt$ for both energies; however, the
dependence on $\kt$ seems to decrease and then flatten out. In any case, the dip
structure observed in the correlation function is a small effect, ranging from
$\approx$3\% in the lower $\Ntpt$ and $\kt$ range investigated, to $\approx$1\% in the largest one.

In summary, the BEC results obtained in $\Pp\Pp$ collisions and discussed in this
section are shown to have a similar behavior to those observed in such systems
in previous measurements at the LHC, by CMS \cite{cms-hbt-1st,cms-hbt-2nd}, as
well as by the ALICE~\cite{Aamodt:2011kd} and ATLAS~\cite{atlas2015}
experiments. This behavior is reflected in the decreasing radii for increasing
pair average transverse momentum $\kt$, and is also observed in $\Pep\Pem$
collisions at LEP~\cite{opal1987}. Such behavior is compatible with nonstatic,
expanding emitting sources. Correlations between the particle production points
and their momenta, $(x,p)$ correlations, appear as a consequence of this
expansion, generating a dependence of the correlation function on the average
pair momentum, and no longer only on their relative momentum; $(x,p)$
correlations were modeled in Refs.~\cite{Htau,bialas} as well. Similar behavior
is also observed in various collision systems and energy
ranges~\cite{cms-hbt-1st,cms-hbt-2nd,e735,star,phobos,phenix,alice}. In the
case of heavy ion, deuteron-nucleus, and proton-nucleus collisions, both at
RHIC and at the LHC, these observations are compatible with a collective
behavior of the source.

\subsection{Two-dimensional results}
\label{subsec:doubratio_results-2d}

The 1D analysis is extended to the 2D case in terms of the components ($\qt,
\ql$) of the pair relative momentum, for data samples at $\sqrt{s} = 2.76$ and
7\TeV. In the 2D and 3D cases the measurement is conducted both in the CM frame
and in the LCMS. For clarity, the asterisk symbol (*) is added to the variables
defined in the CM frame when showing the results in this section.  In the case
of longitudinally symmetric systems, the LCMS has the advantage
(Section~\ref{sub:fitting_functions}) of omitting the cross-term proportional
to $\qt \ql$ in the 2D, or $\ql \qo$ in the 3D fit functions.

\begin{figure*}[bp]
 \centering
 \includegraphics[width=0.49\textwidth]{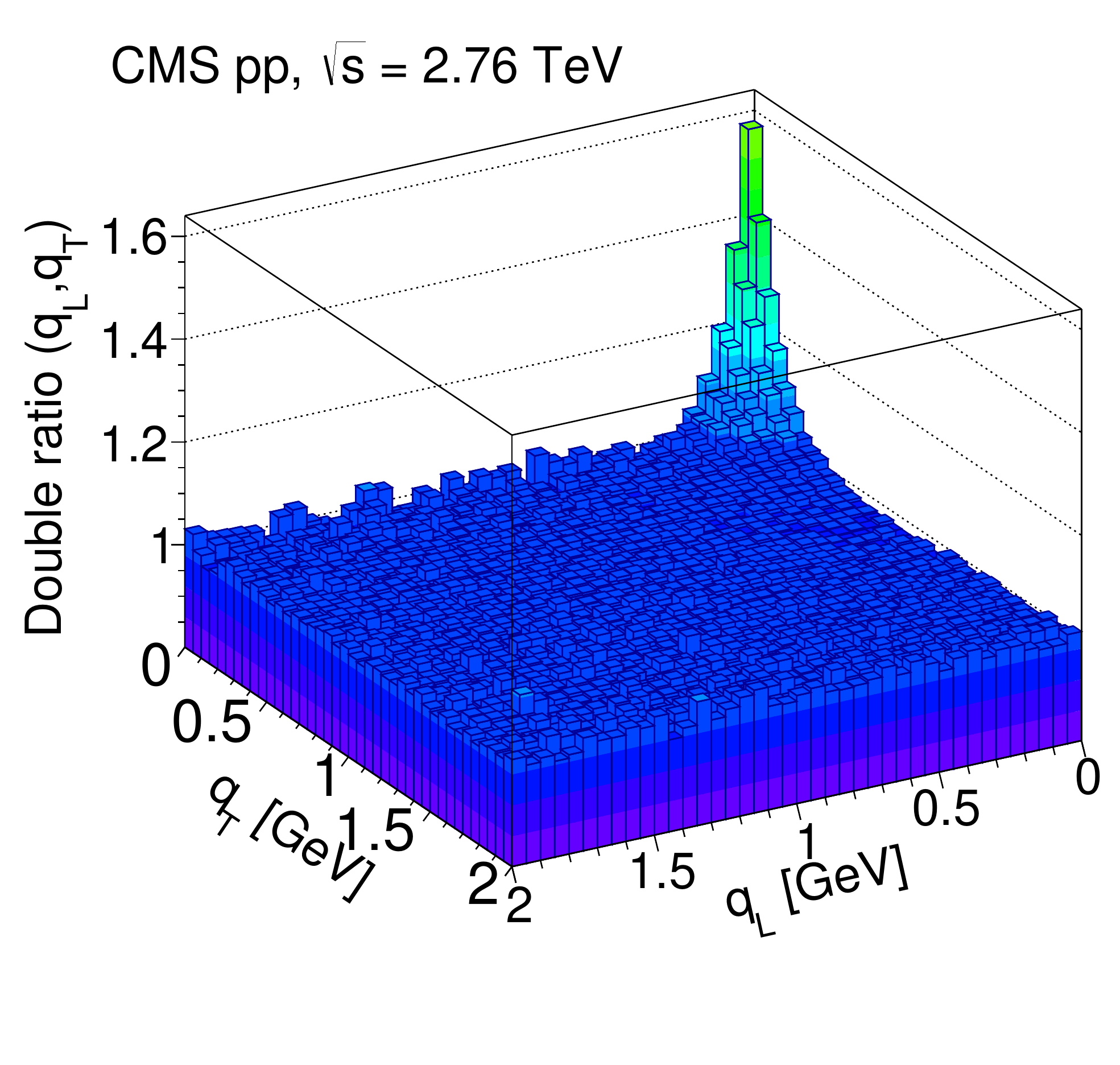}
 \includegraphics[width=0.49\textwidth]{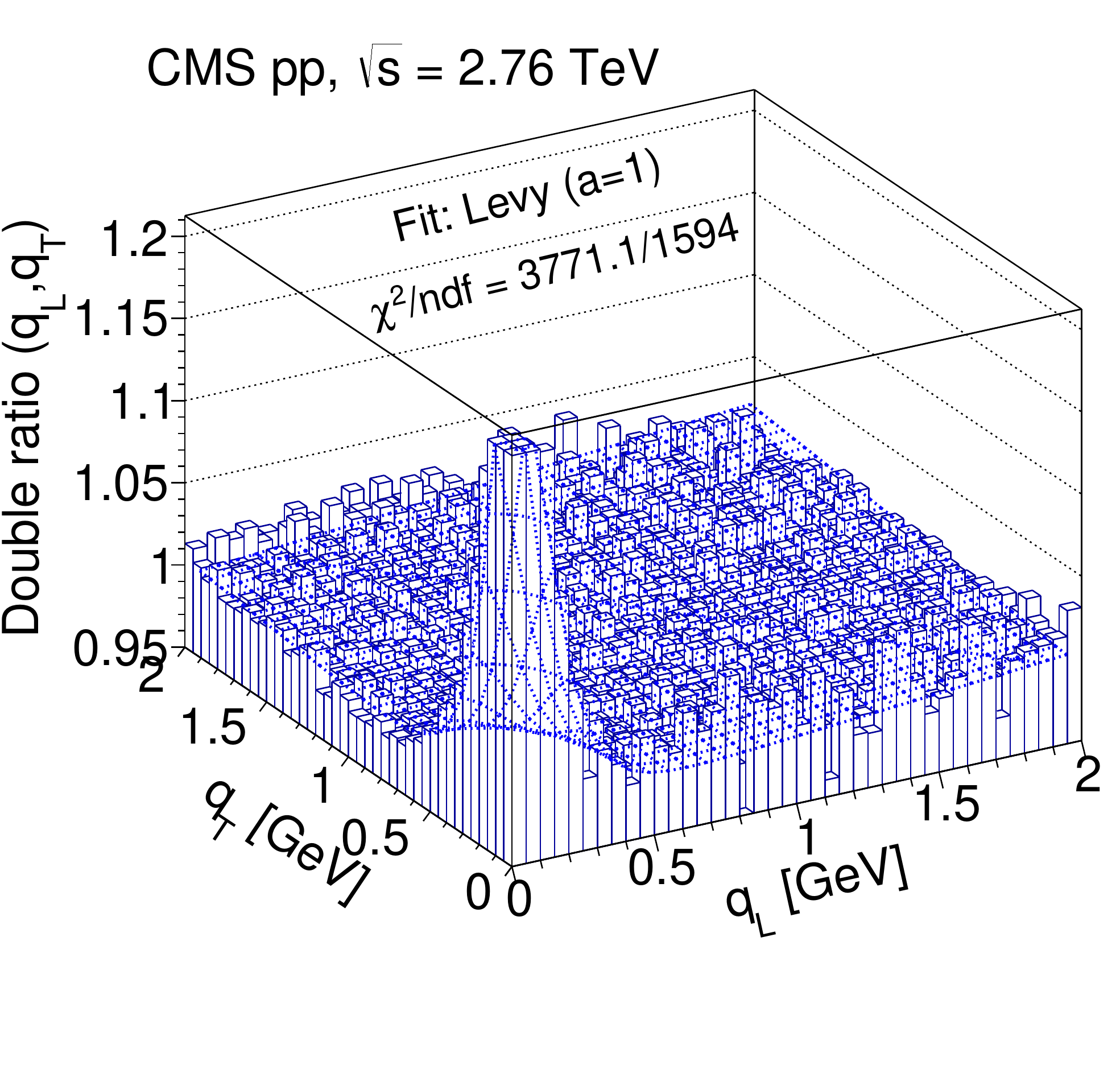}
 \includegraphics[width=0.49\textwidth]{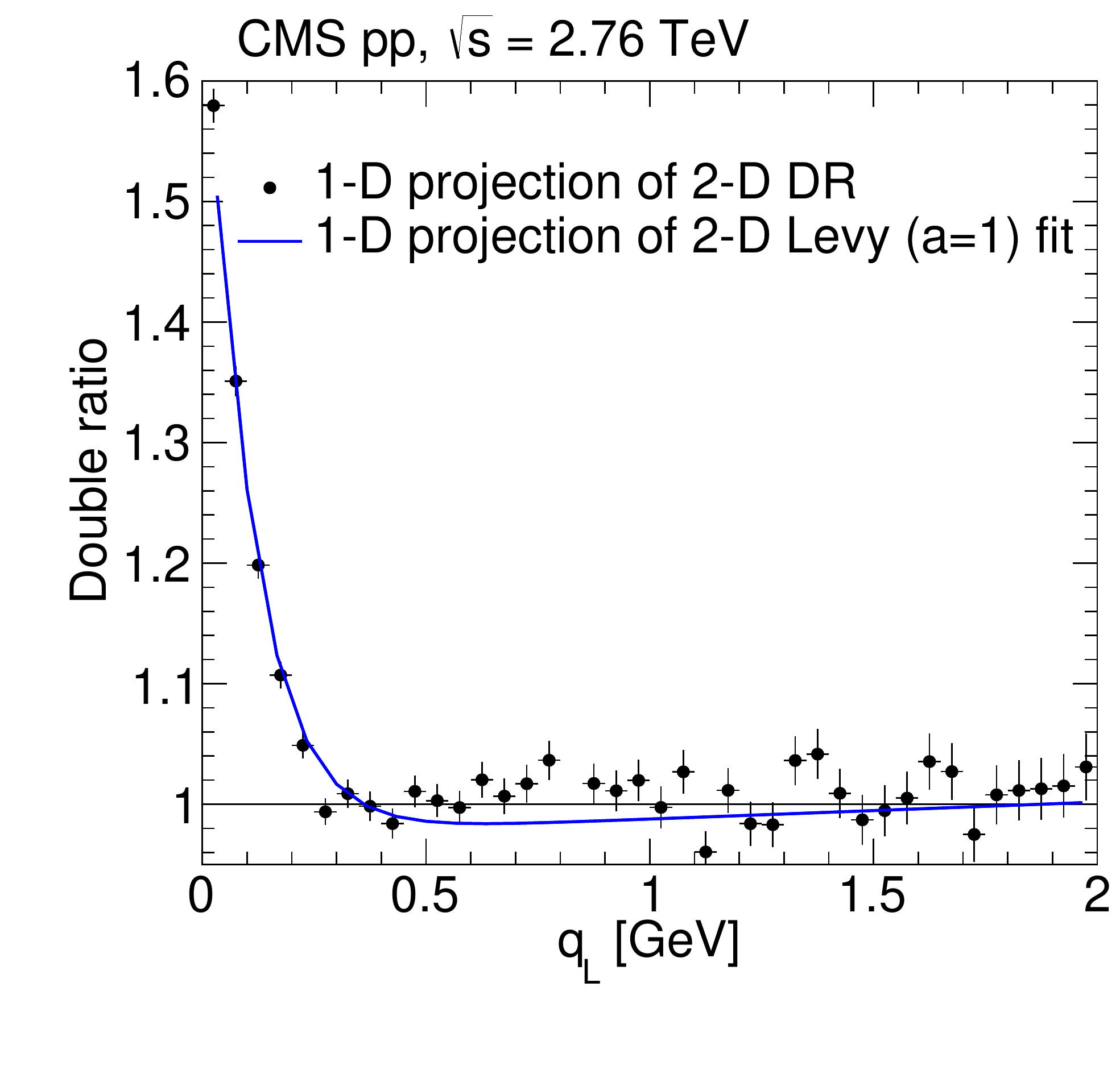}
 \includegraphics[width=0.49\textwidth]{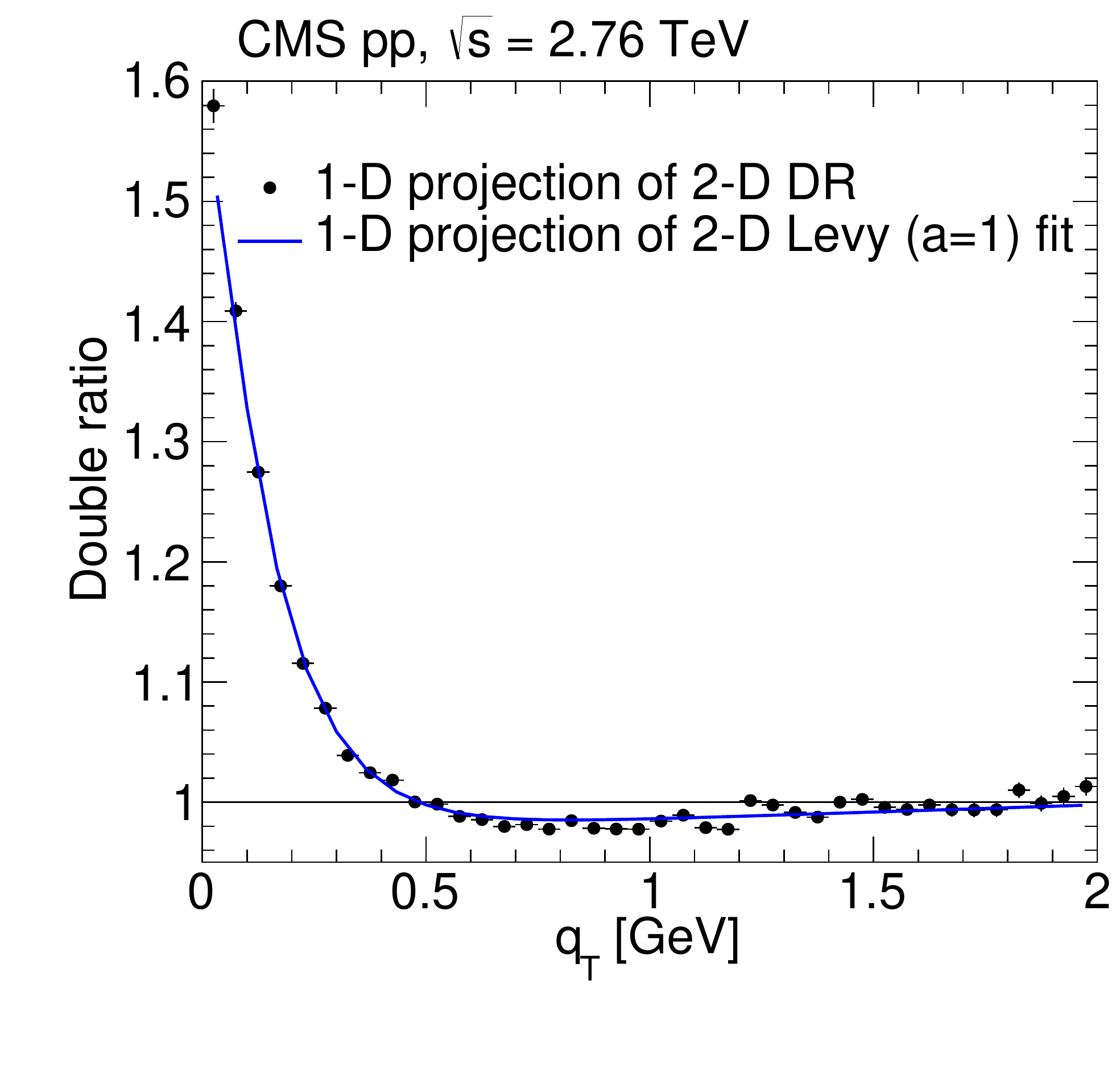}
 \caption{Double ratio results measured in the LCMS, integrated over all
$\Ntpt$ and $\kt$ bins, in $\Pp\Pp$ collisions at $\sqrt{s} = 2.76\TeV$. Top: 2D
correlation functions as a function of $(\qt, \ql)$ without (left) and with
(right) the 2D L\'evy (with $a = 1$) fit function superimposed. Bottom:
Corresponding 1D projections in terms of $\ql$ (for $\qt < 0.05\GeV$, left)
and $\qt$ (for $\ql < 0.05\GeV$, right). The error bars indicate statistical
uncertainties.}
 \label{fig:2d_doubleratios-project-276-7tev-lcms}
\end{figure*}

The 2D correlation function provides information on the
behavior of the emitting source along and transverse to the beam direction.
Typical examples are illustrated in
Fig.~\ref{fig:2d_doubleratios-project-276-7tev-lcms} for $\Pp\Pp$ collisions at
$\sqrt{s} = 2.76\TeV$, in the LCMS. The upper panel shows the 2D plot of the
double ratio in terms of $\qt$ and $\ql$, with the right plot showing the 2D
L\'evy fit (with $a=1$) defined in Eq. (\ref{eq:2d-levy}), superimposed on the
2D correlation function. The lower plots show 1D projections in terms of $\ql$
(left) and $\qt$ (right) of the 2D double ratios, and the corresponding 1D
projections of the 2D L\'evy (with $a=1$) fit function. When plotting in terms
of $\ql$, only the first bin in $\qt$ (for $\qt < 0.05\GeV$) is considered, and
vice-versa. This choice is made to exhibit the largest values of the 2D
correlation function in each direction, since the BEC decreases with increasing
$\qt, \ql$.
The functions shown in the 1D figures are not fits to that data, but are projections
of the global 2D-stretched exponential fit to the 2D correlation function. This
explains the poor description of the data by the L\'evy fit with $a=1$,
in Fig.~\ref{fig:2d_doubleratios-project-276-7tev-lcms}, left.
The statistical uncertainties are also larger in the $\ql$ case, as well as the
fluctuations in the results along this direction.

\begin{figure*}[tbhp]
 \centering
  \includegraphics[width=0.49\textwidth]{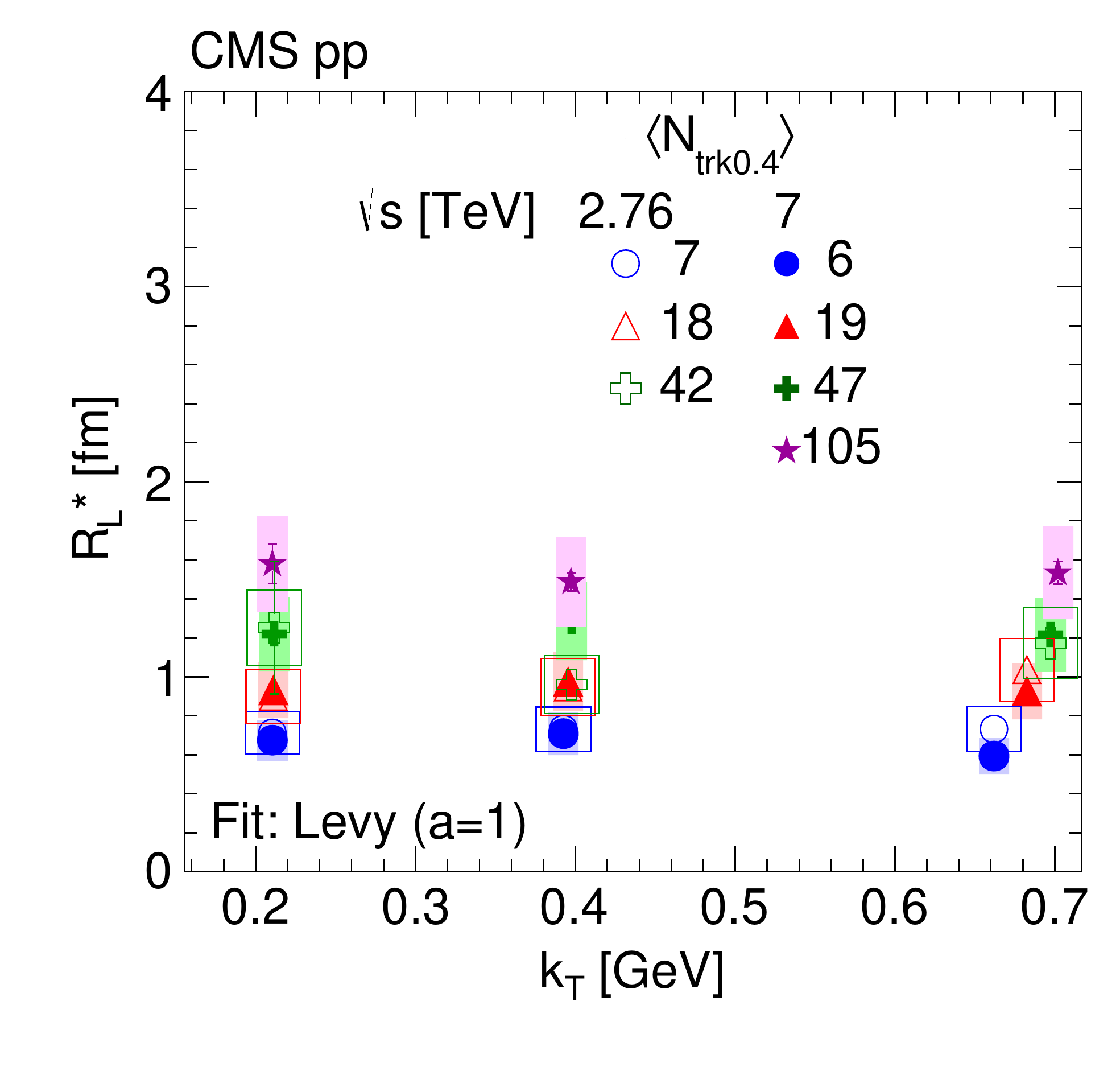}
  \includegraphics[width=0.49\textwidth]{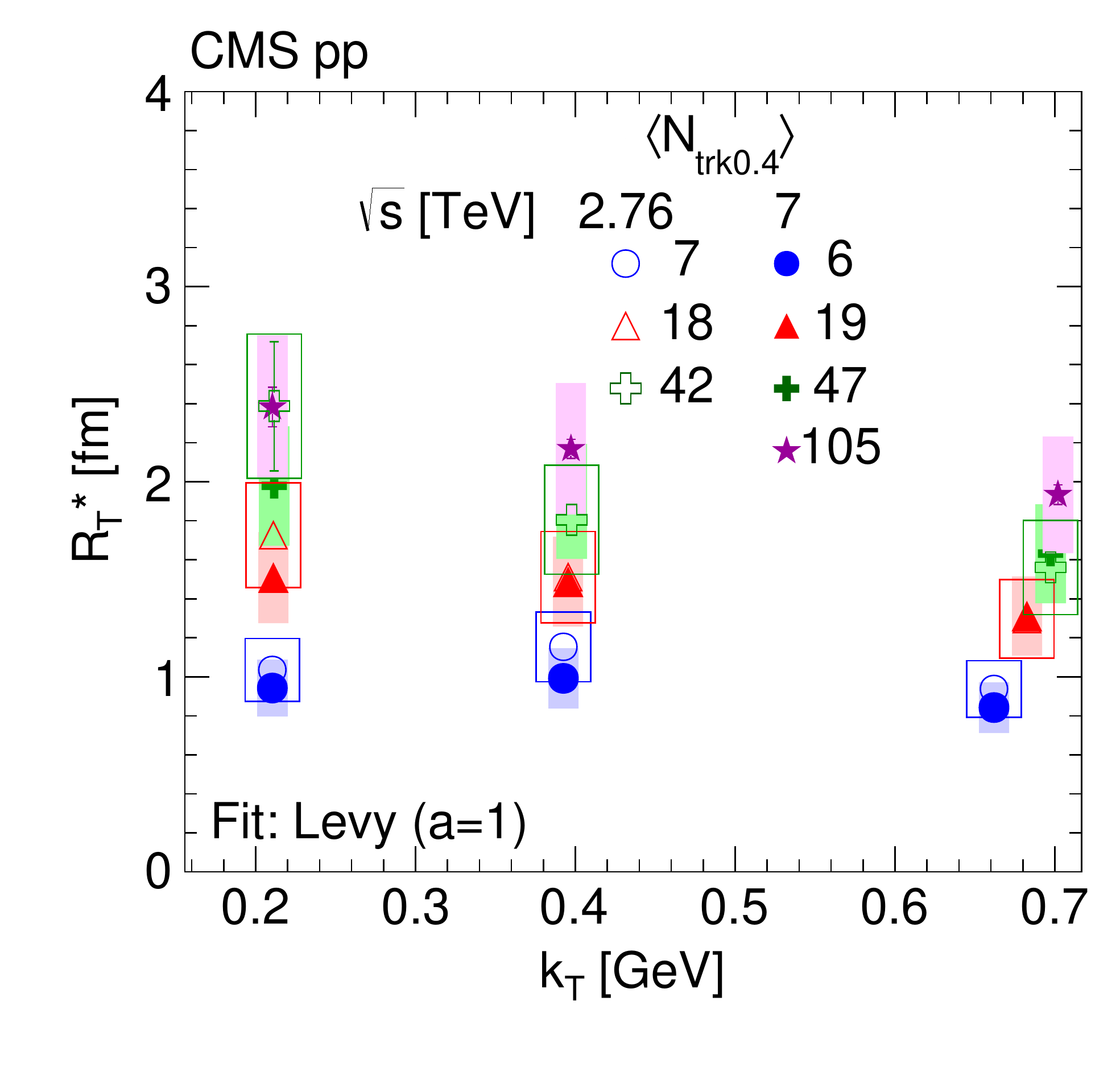}

  \includegraphics[width=0.49\textwidth]{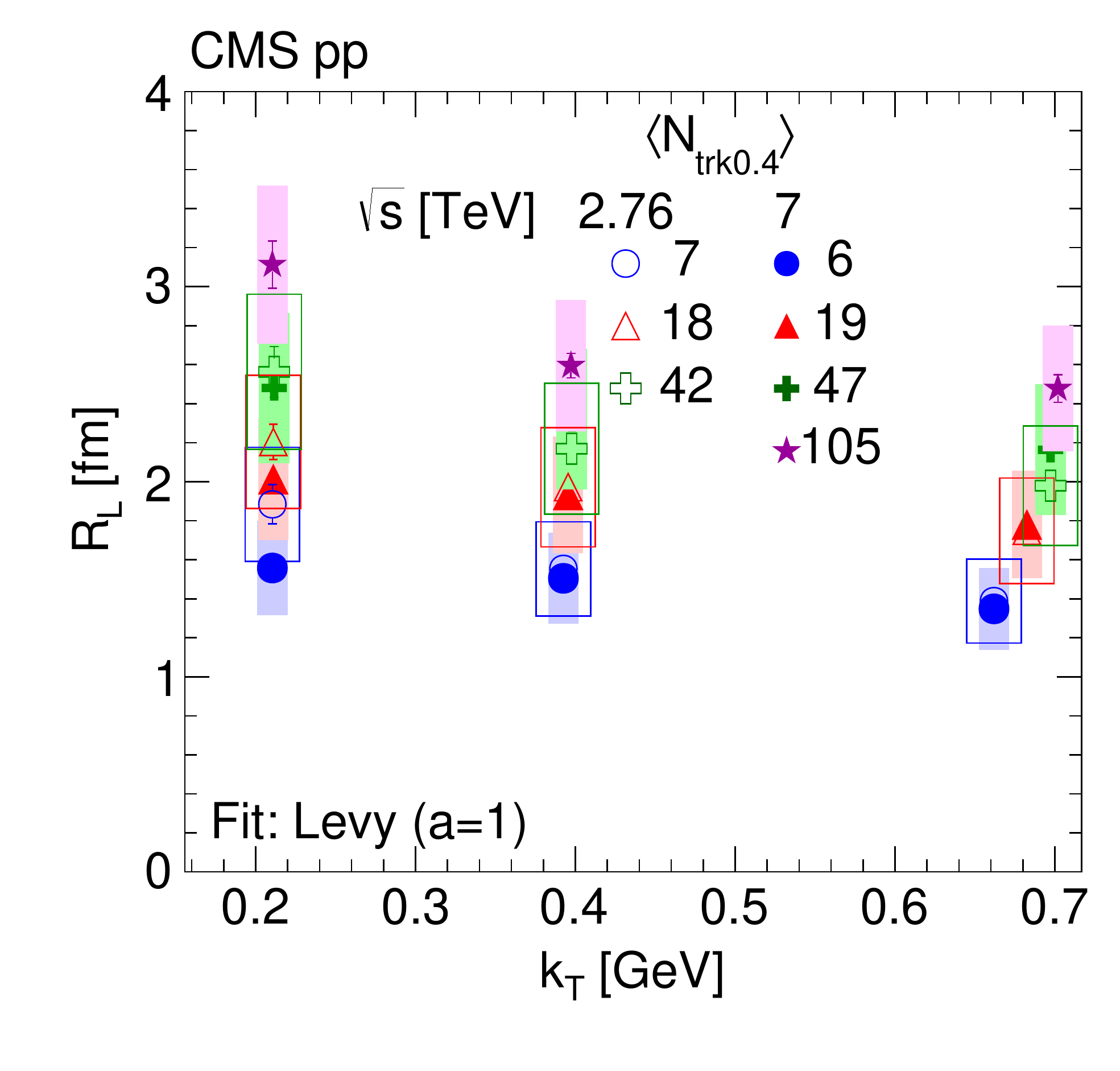}
  \includegraphics[width=0.49\textwidth]{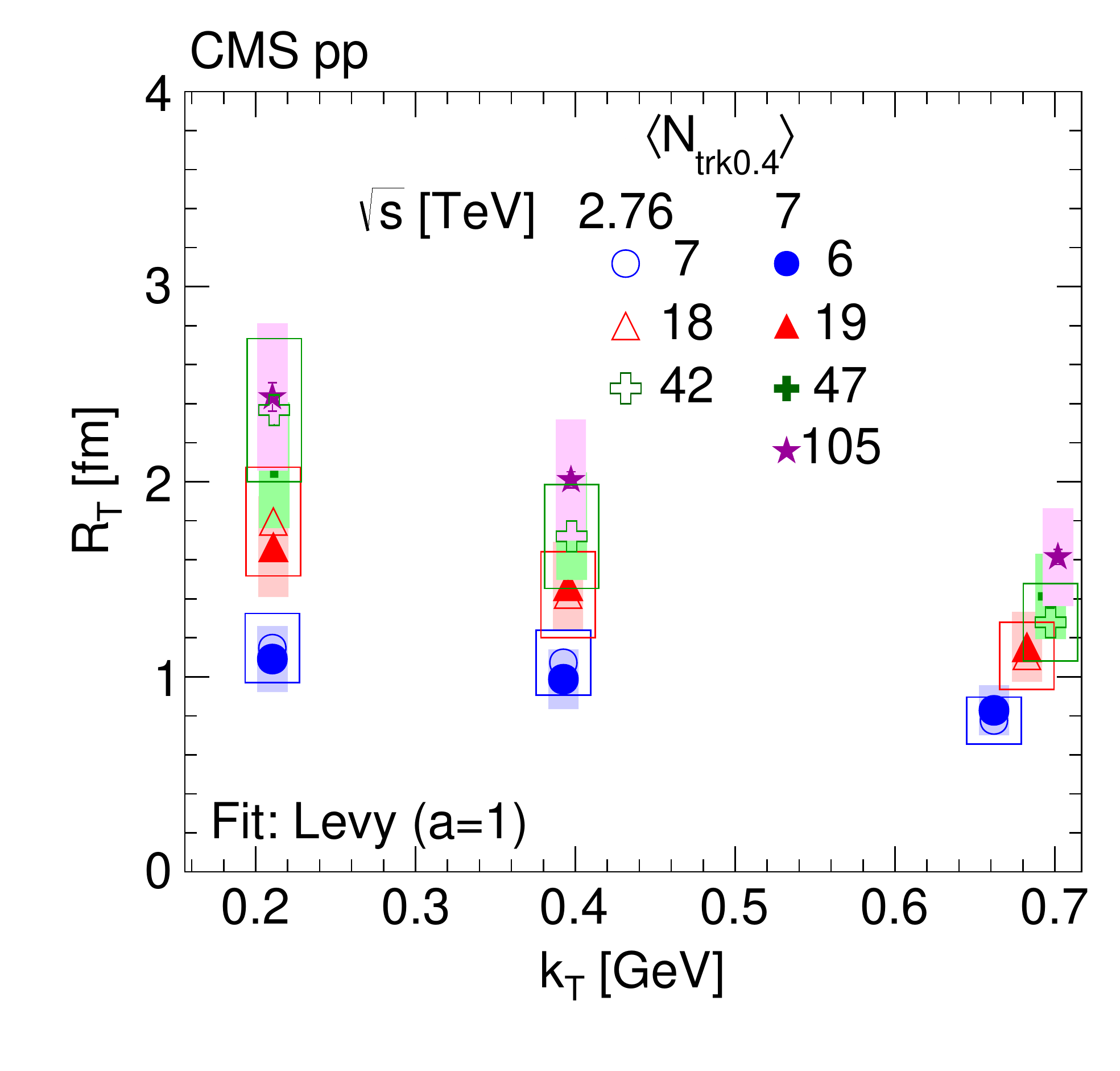}

 \caption{Radius parameters as a function of $\kt$, for different $\Ntpt$ bins,
obtained from data fitted to the stretched exponential function in $\Pp\Pp$ collisions at $\sqrt{s} = 2.76\TeV$ (open symbols) and 7\TeV (filled symbols) in
the CM frame (top) and in the LCMS (bottom). The boxes indicate the systematic
uncertainties.}
 \label{fig:2d_radii_rl_rt-276_7tev-cm-lcms}
\end{figure*}

Analogously to the studies performed in 1D, the correlation functions are also
investigated in 2D in terms of the components $(\qt^{(*)}, \ql^{(*)})$, in
three intervals of $\kt$, and in different $\Ntpt$ bins. The results from the
stretched exponential fit to the double ratios are compiled in
Fig.~\ref{fig:2d_radii_rl_rt-276_7tev-cm-lcms}, performed both in the CM frame
(upper) and in the LCMS (lower). The behavior is very similar in both frames,
clearly showing that all fit parameters $\Rl^{(*)}$ and $\Rt^{(*)}$ increase
with increasing multiplicity $\Ntpt$. On the other hand, the behavior with
respect to $\kt$ varies, showing low sensitivity to this parameter for the
lower $\Ntpt$ bins. The radius parameters $\Rl^{(*)}$ and $\Rt^{(*)}$ decrease
with $\kt$ for larger multiplicities, a behavior similar to that observed in
the 1D case, and expected for expanding sources.
The observation that $\Rl \text{(LCMS)} > \Rl^* \text{(CM)}$ for the same bins
of $\Ntpt$ and $\kt$ can be related to the Lorentz-contraction of the
longitudinal radius in the CM frame.

\begin{table*}[bth]
\topcaption{Fit parameters of the 2D BEC function in $\Pp\Pp$ collisions at $\sqrt{s}
= 2.76$ and 7\TeV in the LCMS.}
 \label{tab:radii-276-7tev}
 \centering
 \begin{scotch}{lcc}
                & $\Pp\Pp$ $\sqrt{s} = 2.76\TeV$ &
                  $\Pp\Pp$ $\sqrt{s} = 7\TeV$ \\
   \hline
   $\lambda$    &  $0.830 \pm 0.010\stat\pm 0.083\syst$ &
                   $0.700 \pm 0.002\stat\pm 0.070\syst$ \\
   $\Rt$ [fm]   &  $1.498 \pm 0.013\stat\pm 0.232\syst$ &
                   $1.640 \pm 0.003\stat\pm 0.254\syst$ \\
   $\Rl$ [fm]   &  $1.993 \pm 0.022\stat\pm 0.309\syst$ &
                   $2.173 \pm 0.006\stat\pm 0.337\syst$ \\
 \end{scotch}
\end{table*}

\begin{figure*}[thb]
 \centering
  \includegraphics[width=0.49\textwidth]{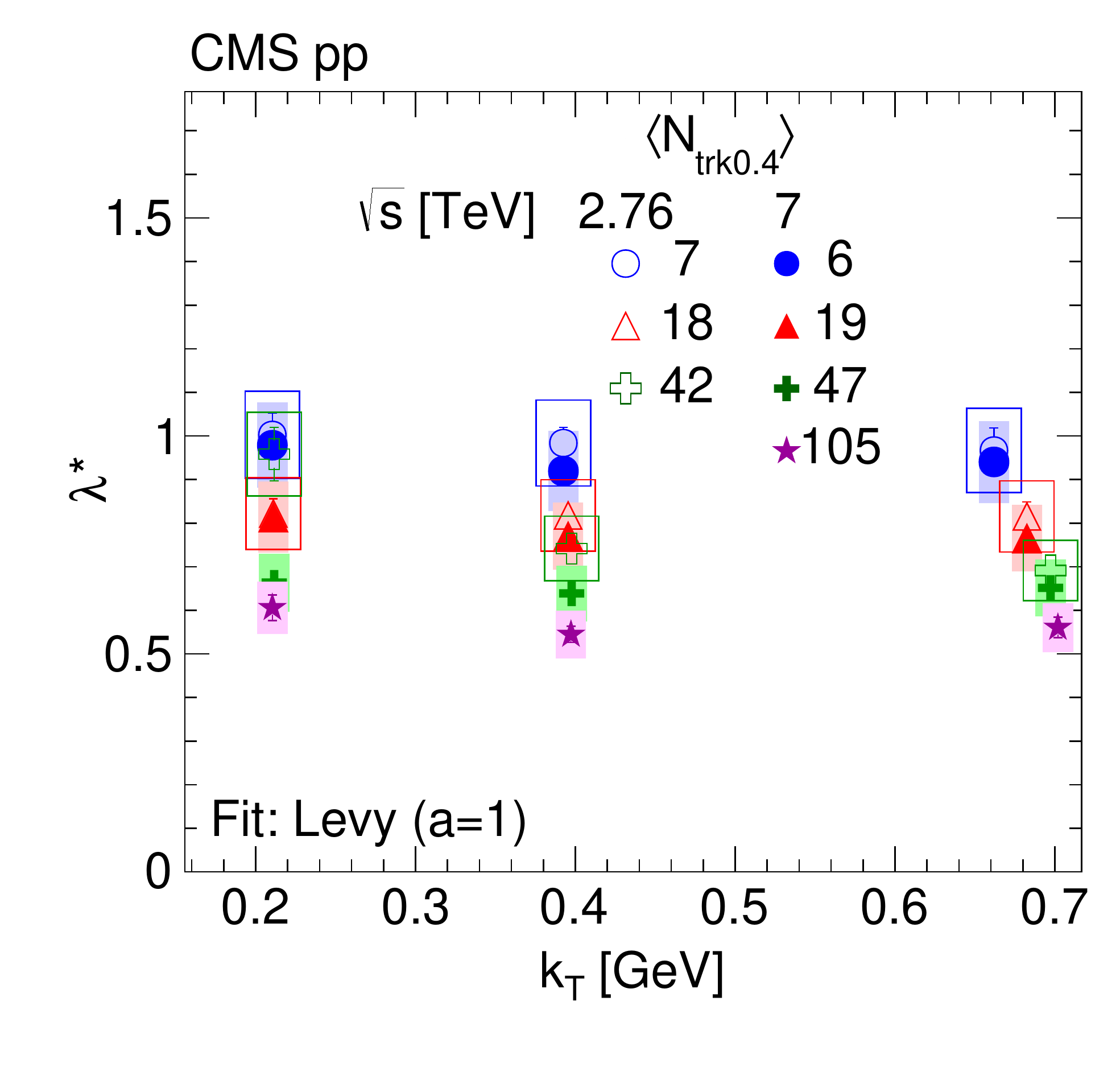}
  \includegraphics[width=0.49\textwidth]{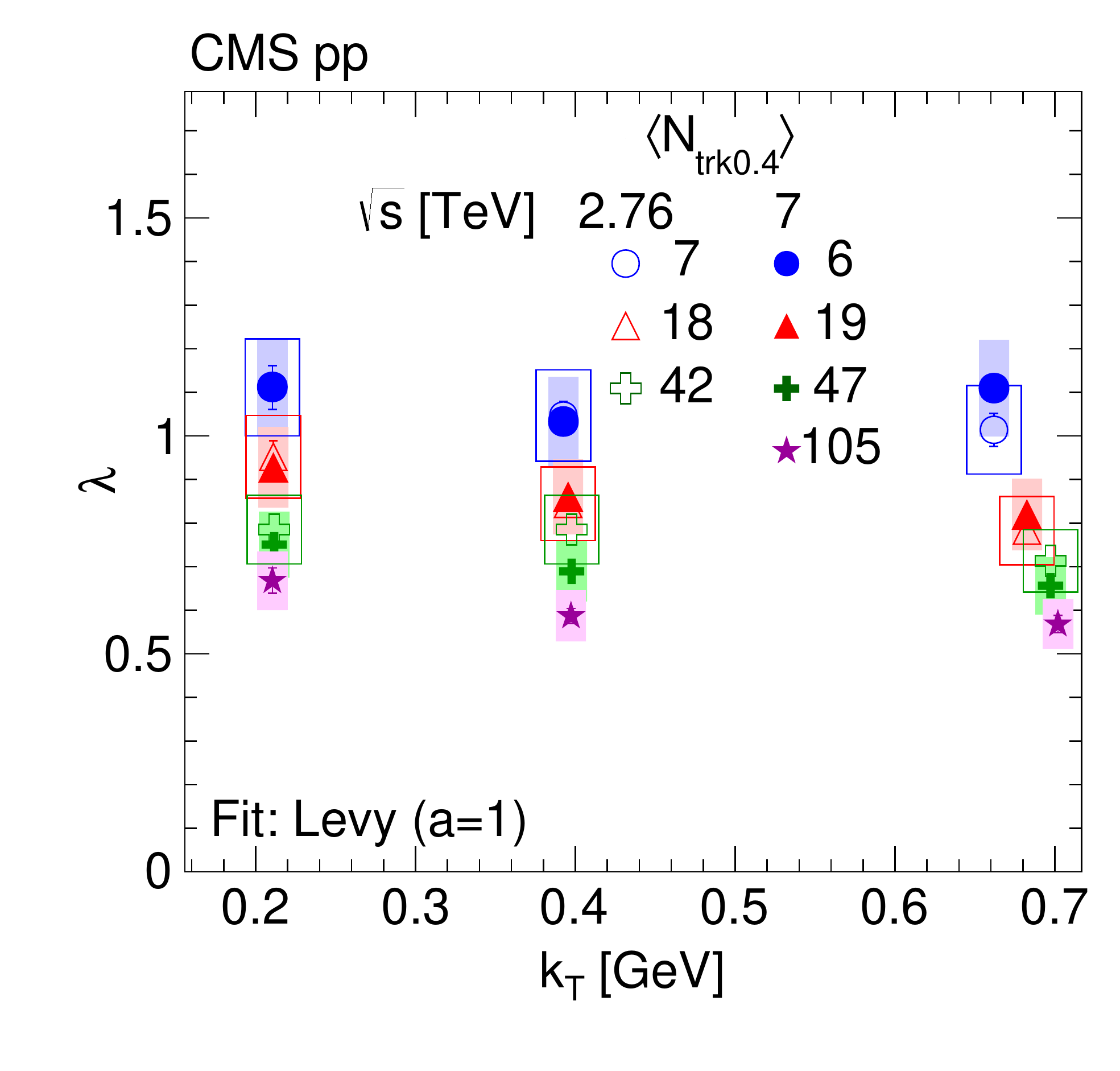}
 \caption{Intercept parameter $\lambda$ as a function of $\kt$, for different
$\Ntpt$ bins, obtained from data fitted to the stretched exponential
function in $\Pp\Pp$ collisions at $\sqrt{s} = 2.76\TeV$ (open symbols) and 7\TeV
(filled symbols) in the CM frame (left) and in the LCMS (right). The boxes
indicate the systematic uncertainties.}
 \label{fig:lambda-cm-lcms}
\end{figure*}

Figure~\ref{fig:lambda-cm-lcms} shows the results for the correlation strength
parameter, $\lambda^{(*)}$, obtained with the same fit procedure, for both the
CM and LCMS frames. No significant sensitivity of the intercept is seen as a
function of $\kt$. However, within each $\kt$ range, $\lambda^{(*)}$ slowly
decreases with increasing track multiplicity, in a similar way in both frames.

\begin{figure*}[thb]
 \centering
  \includegraphics[width=0.49\textwidth]{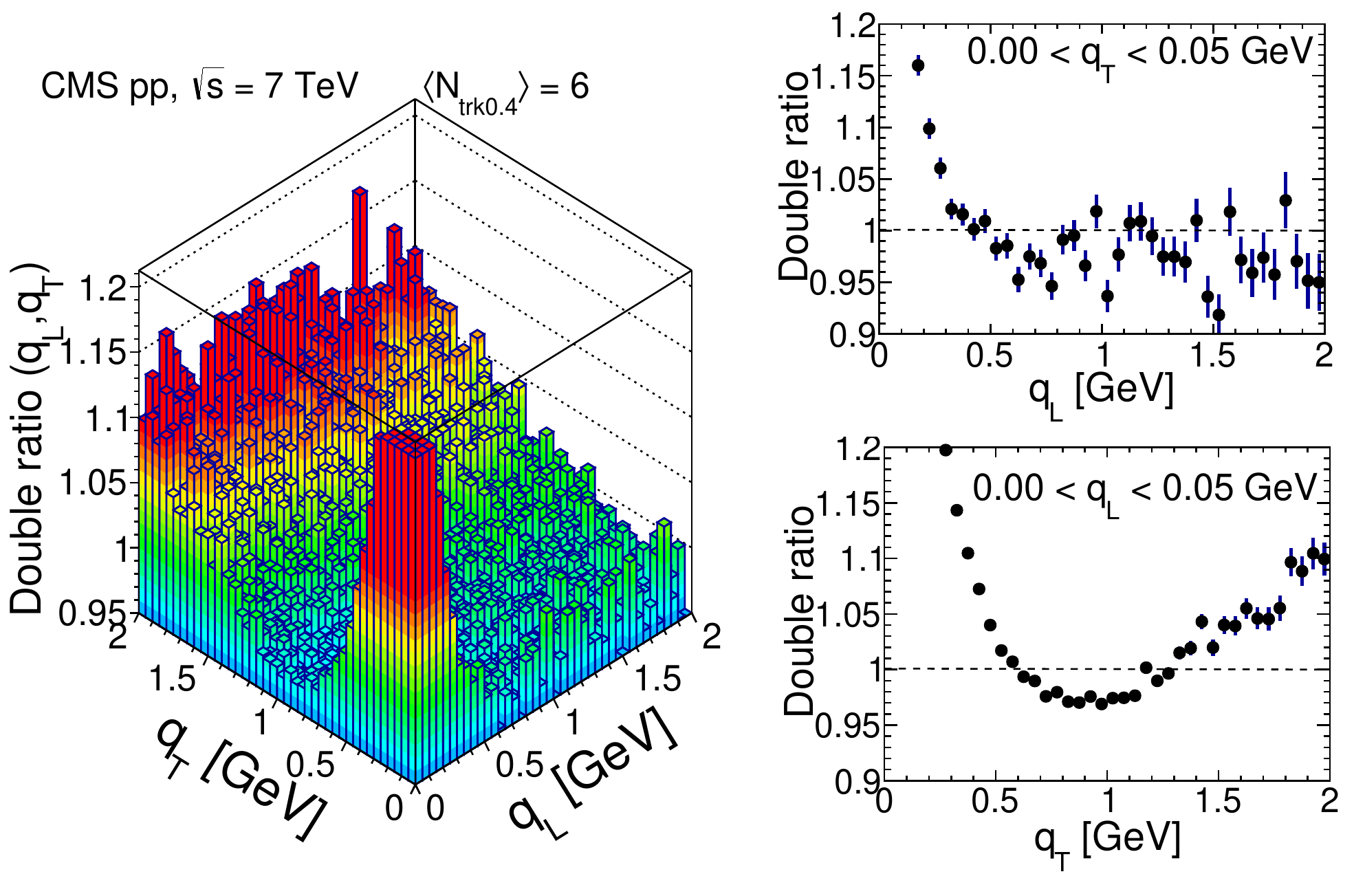}
  \includegraphics[width=0.49\textwidth]{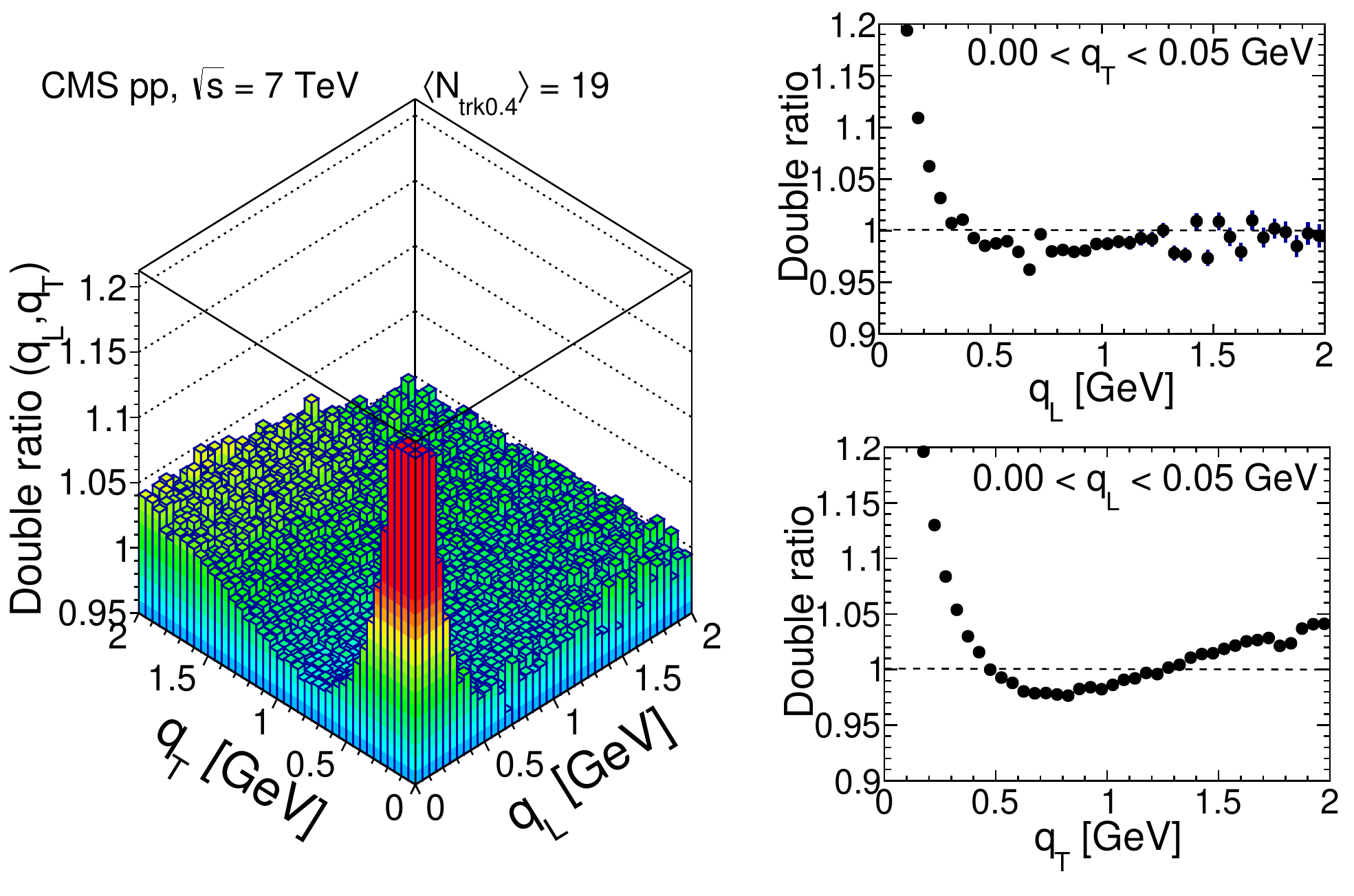}
\vspace*{1cm}
  \includegraphics[width=0.49\textwidth]{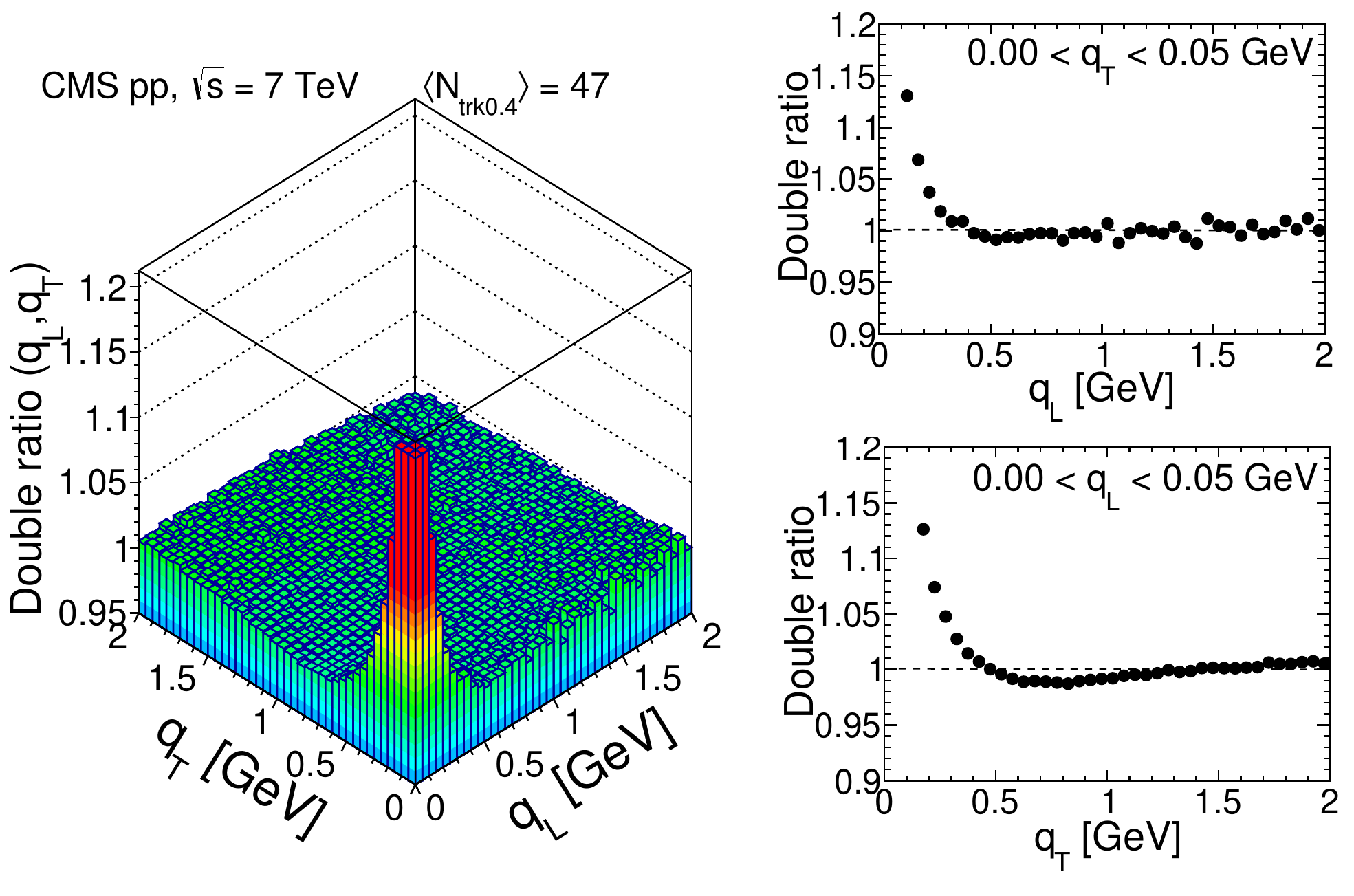}
  \includegraphics[width=0.49\textwidth]{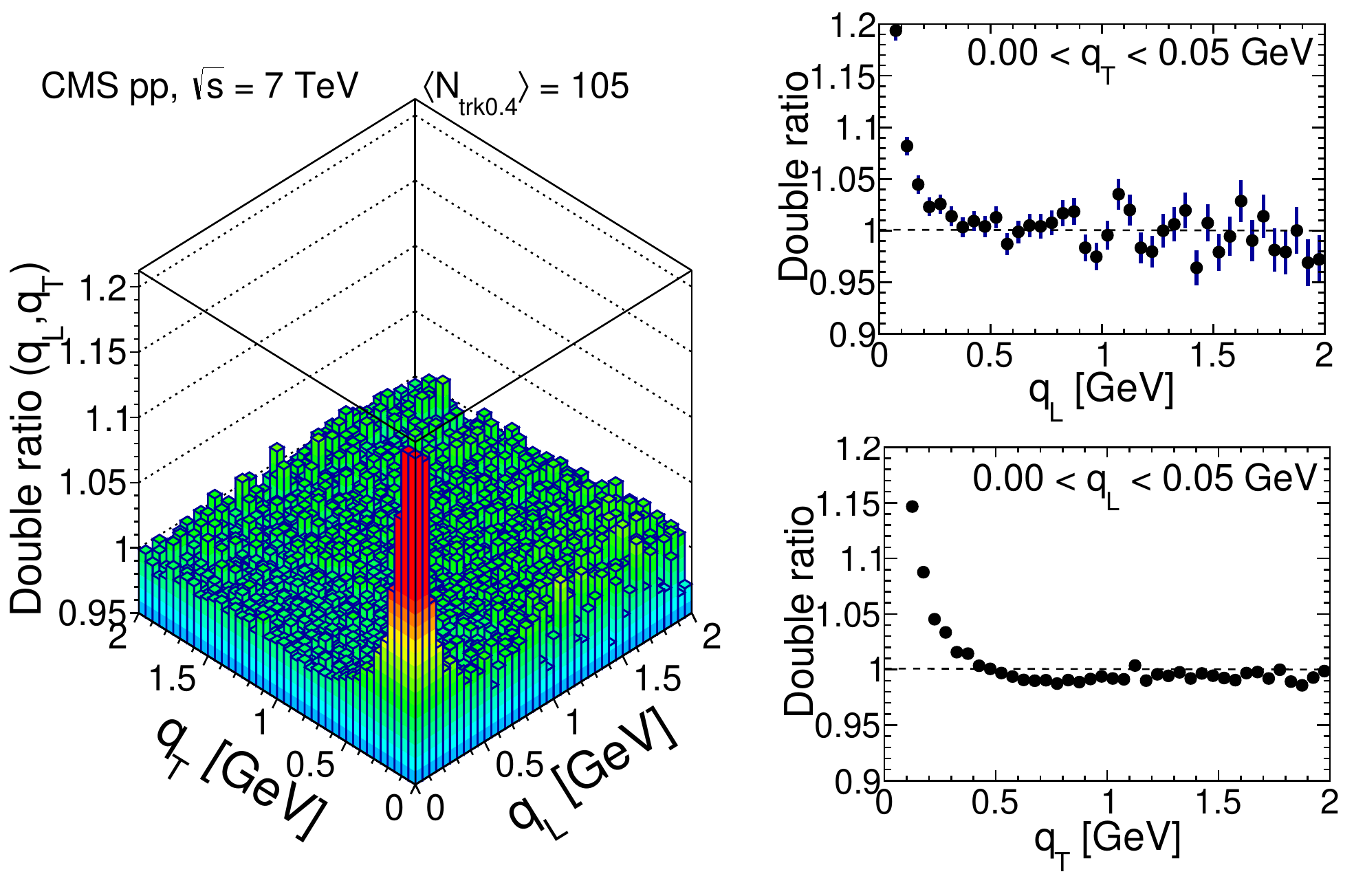}
 \caption{Results obtained in the LCMS for the 2D double ratios in the $(\ql,
\qt)$ plane, zoomed along the correlation function axis, for four charged
multiplicity bins, $\Ntpt$ (integrated over all $\kt$), increasing from upper
left to lower right. The BEC peak is cut off at $\sim$1.2 for a better
visualization of the directional behavior. The 1D projections in $\ql$ (for
$\qt < 0.05\GeV$) and $\qt$ (for $\ql < 0.05\GeV$) are shown to the right of
the bidimensional plots, with error bars corresponding to statistical
uncertainties.}
 \label{fig:2d-dip-structure-lcms}
\end{figure*}

The values obtained with the stretched exponential fit in the LCMS for the
radius parameters $\Rt, \Rl$, together with the intercept $\lambda$,
corresponding to data integrated both in $\Ntpt$ and $\kt$, are collected in
Table~\ref{tab:radii-276-7tev} for $\Pp\Pp$ collisions at 2.76 and 7\TeV. The values
fitted to the cross-term $R_\text{LT}$ in the CM frame are of order 0.5\unit{fm} or
less, and in the LCMS such terms are negligible ($\mathcal{O}[10^{-7}]$), as
expected. From Table~\ref{tab:radii-276-7tev} it can be seen that, in the LCMS
and at both energies, $\Rl \approx 4 \Rt/3$, suggesting that the source is
longitudinally elongated.

In Fig.~\ref{fig:2d-dip-structure-lcms} the 2D results for the double ratios
versus ($\ql, \qt$) in the LCMS are shown zoomed along the correlation function
axis, with the BEC peak cut off around 1.2, for four charged-multiplicity bins,
$\Ntpt$. These plots illustrate the directional dependence of the
$\kt$-integrated correlation functions along the beam and transverse to its
direction, as well as the variation of their shapes when increasing the
multiplicity range (from the upper left to the lower right plot) of the events
considered. The edge effects seen for $\qt > 1.5\GeV$ in the upper left plot
(for $\langle\Ntpt\rangle = 6$) are an artifact of the cutoff at 1.2, being
much smaller than the true value of the peak in the low ($\ql, \qt$) region,
and correspond to fluctuations in the number of pairs in those bins, due to
a lower event count than in the other $\Ntpt$ intervals. The $\Ntpt$ ranges are
the same as in
Figs.~\ref{fig:2d_radii_rl_rt-276_7tev-cm-lcms}--\ref{fig:lambda-cm-lcms},
although the fit parameters shown in those figures are obtained considering
differential bins in both $\Ntpt$ and $\kt$.
Figure~\ref{fig:2d-dip-structure-lcms} also shows the 1D projections of the 2D
double ratios in terms of $\qt$ and $\ql$ (for values $q_\text{L,T} <
0.05\GeV$). The shallow anticorrelation seen in the 1D double
ratios in terms of $\qi$ seems to be also present in the 2D plots, suggesting
slightly different depths and shapes along the $\ql$ and $\qt$ directions.

\clearpage

\subsection{Three-dimensional results}
\label{subsec:doubratio_results-3d}

\begin{figure*}[tbhp]
 \centering
\includegraphics[width=0.32\textwidth]{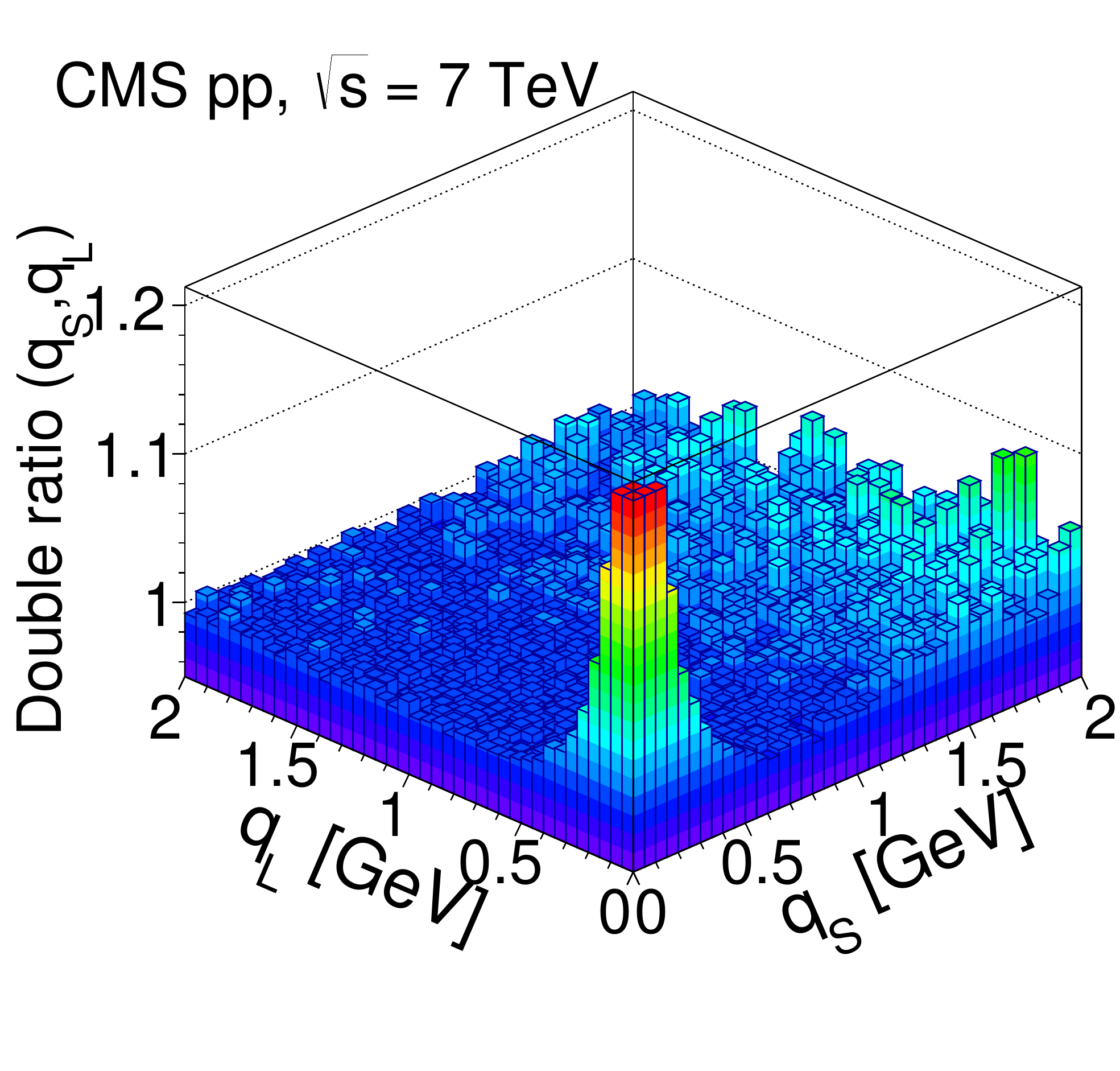}
\includegraphics[width=0.32\textwidth]{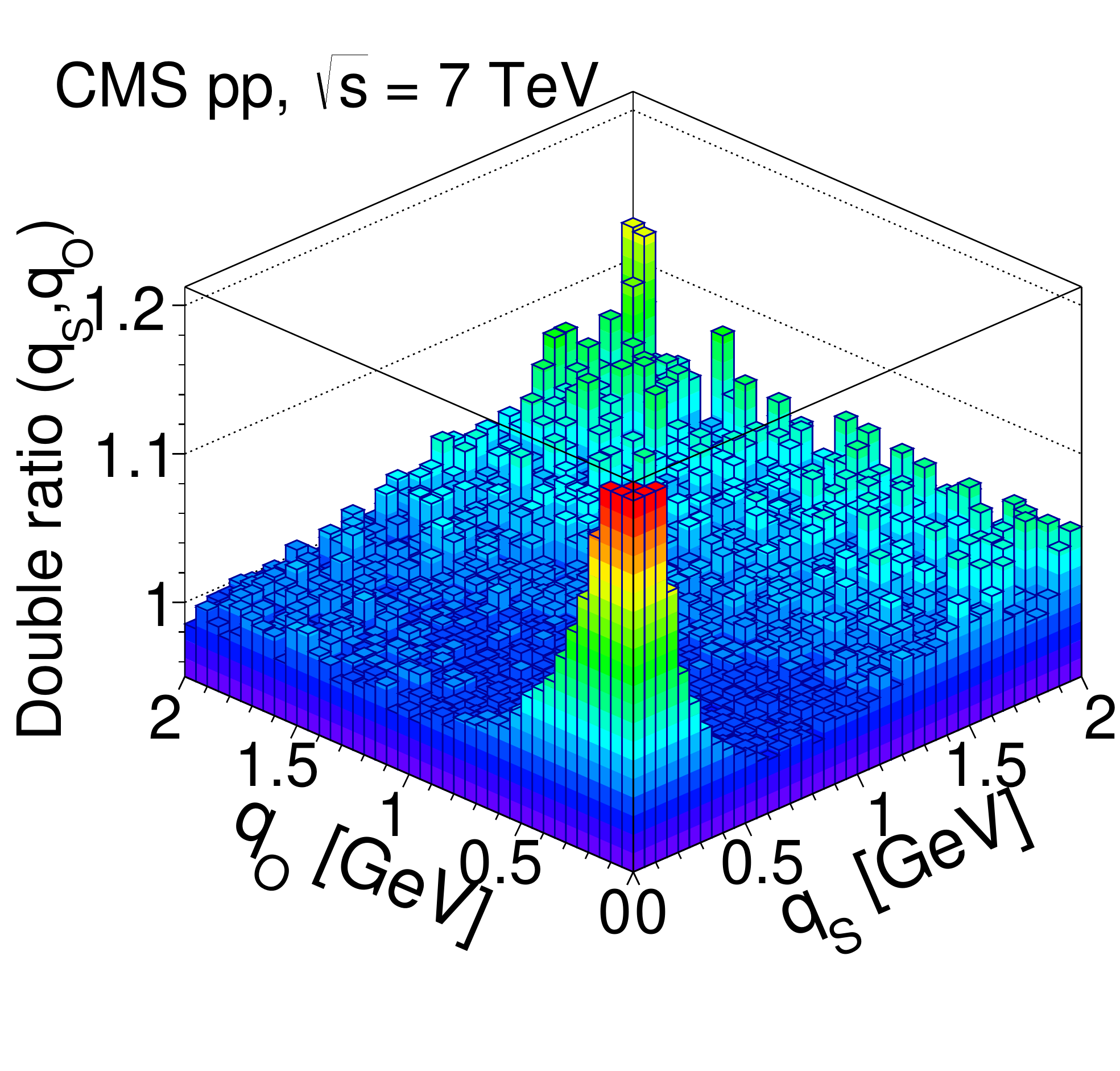}
\includegraphics[width=0.32\textwidth]{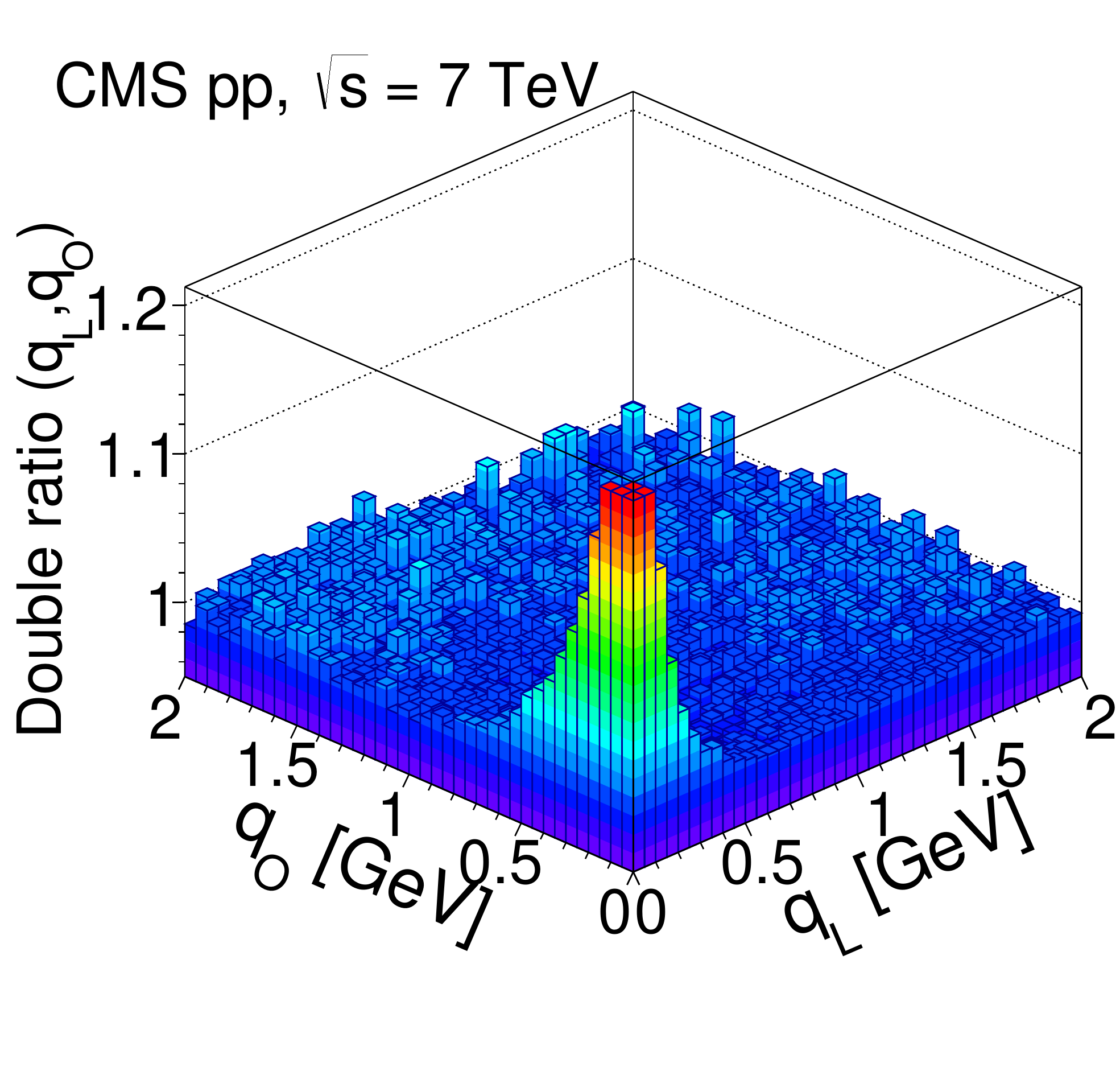}
\includegraphics[width=0.32\textwidth]{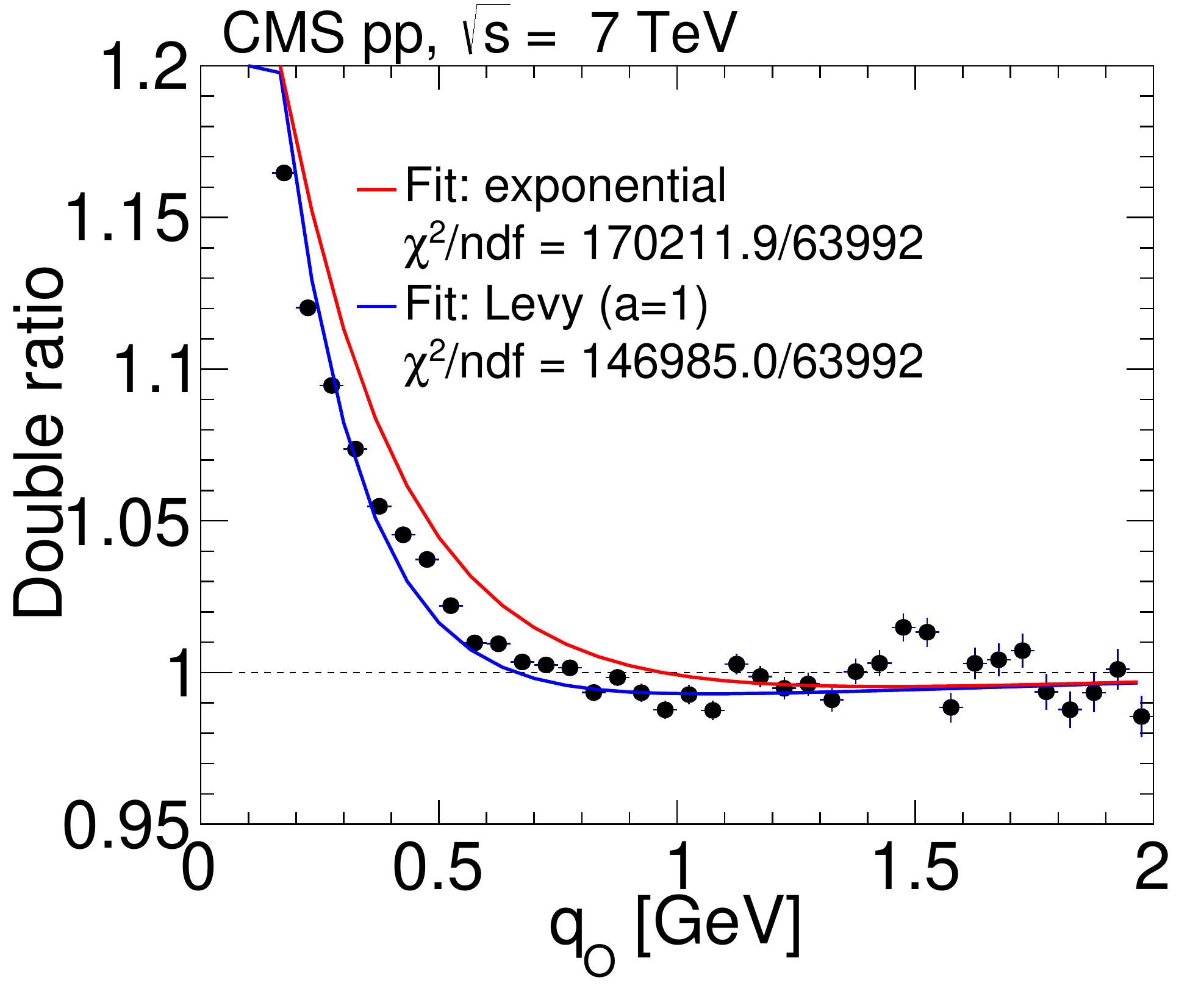}
\includegraphics[width=0.32\textwidth]{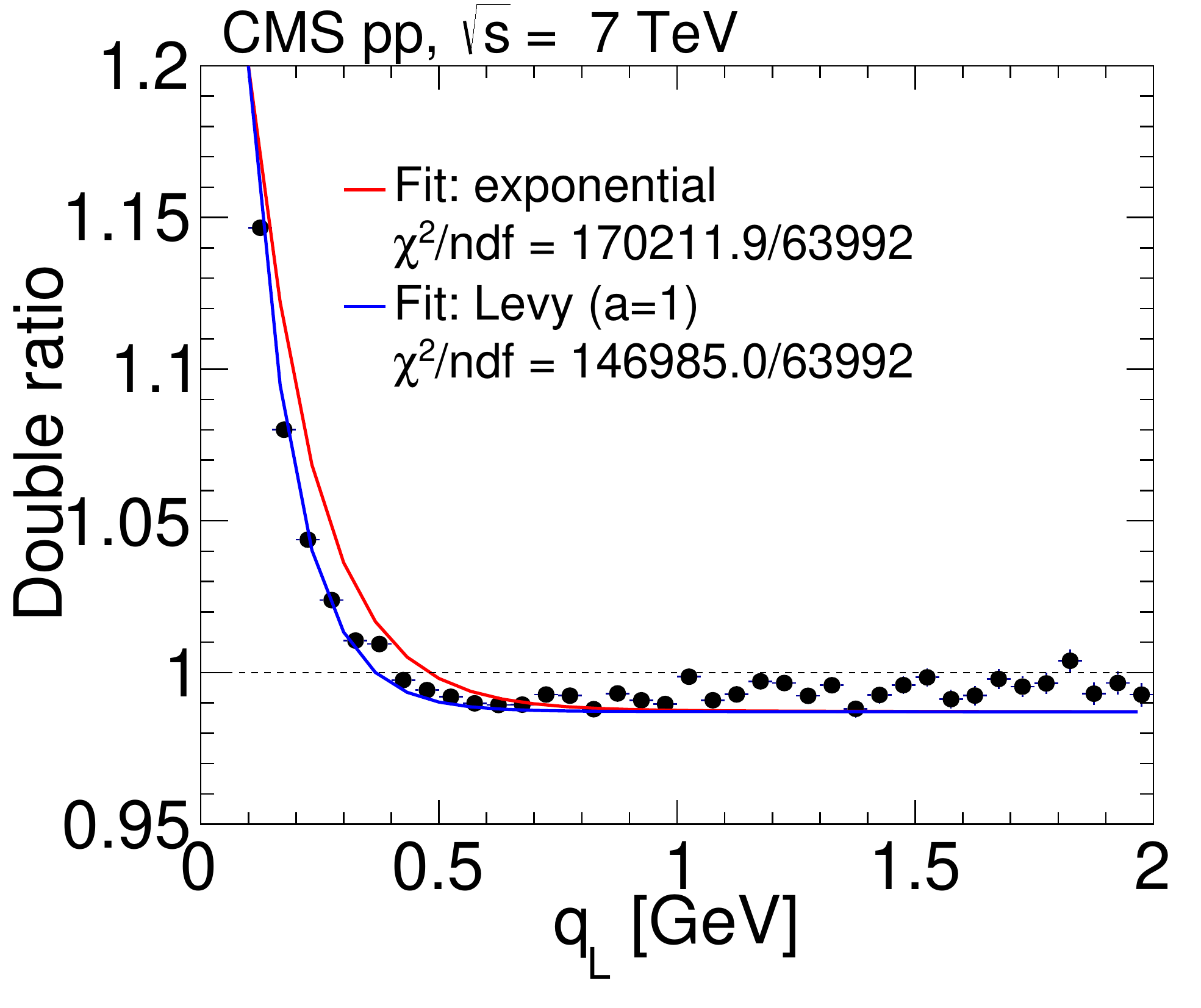}
\includegraphics[width=0.32\textwidth]{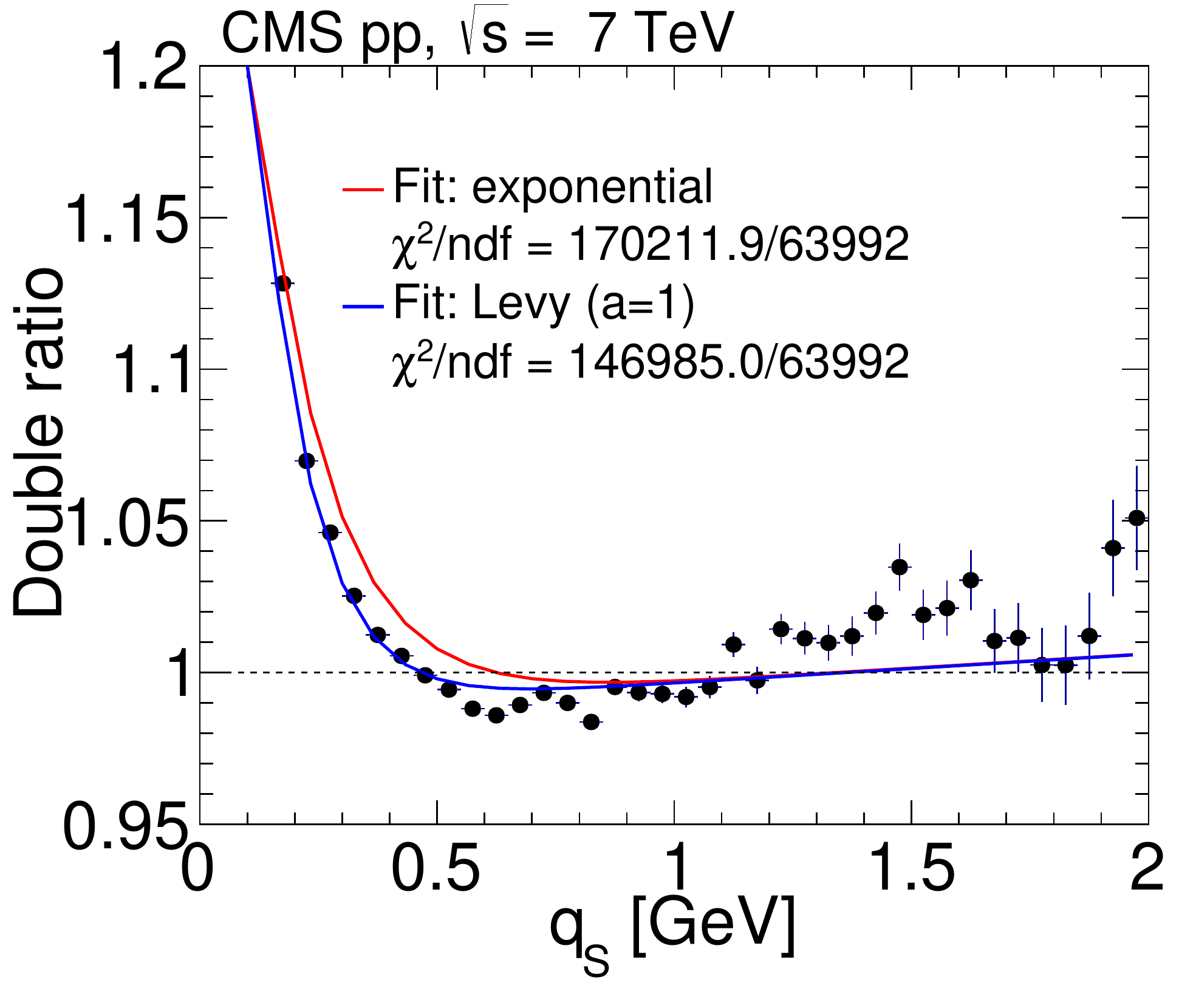}
 \caption{Results obtained in the LCMS for the 3D double ratios in $\Pp\Pp$ collisions at 7\TeV, integrated over $\Ntpt$ and $\kt$. Top: 2D projections over
the $(\ql,\qs)$, $(\qo,\qs)$, and $(\qo,\ql)$ planes, for $\qo < 0.05\GeV$
(left), $\ql < 0.05\GeV$ (middle), and $\qs < 0.05\GeV$ (right).
Bottom: 1D projections of the same data, with the complementary
two variables constrained to be within the first bin ($q_{i,j} < 0.05\GeV$).
The error bars show the corresponding statistical uncertainties. All plots are
zoomed along the correlation function axis with the BEC peak cut off above
1.2.}
 \label{fig:3dcorr-2dproj-lcms}
\end{figure*}

{\tolerance=1200
The double ratios can be further investigated in 3D as a function of the
relative momentum components $(\qo, \qs, \ql)$. In order to obtain the
corresponding fit parameters, $(\Ro, \Rs, \Rl)$, as well as the correlation
function strength, $\lambda$, the fits are performed to the full 3D double
ratios and, if desired for illustration purposes, projected onto the directions
of the variables $\qo, \qs, \ql$, similarly to the projections done for the
measurements. The 3D correlation functions can be visualized through 2D
projections in terms of the combinations $(\qs, \ql)$, $(\ql, \qo)$, and $(\qo,
\qs)$, with the complementary components constrained to $\qo < 0.05\GeV$,
$\qs < 0.05\GeV$, and $\ql < 0.05\GeV$, respectively, corresponding to
the first bin in each of these variables. For clarity, the asterisk symbol (*) is added
to the variables defined in the CM frame when showing the results in this section.
An illustration of the 2D projections is shown in the upper panel of
Fig.~\ref{fig:3dcorr-2dproj-lcms}, with data from $\Pp\Pp$ collisions at 7\TeV,
integrated in all $\Ntpt$ and $\kt$ ranges. Similarly to
Fig.~\ref{fig:2d-dip-structure-lcms}, the results are zoomed along the 3D
double ratio axis, with the BEC peak cut off around 1.2 for better
visualization of the directional behavior of the correlation function. The
higher values seen around $(\qo, \qs) \sim (2,2)\GeV$ in the upper middle plot
are again artifacts of cutting the double ratio at 1.2, this value being much
smaller than the true value of the peak in the low ($\qo, \qs$) region, and
correspond to fluctuations in the number of pairs in those bins.
\par}

The lower panel in Fig.~\ref{fig:3dcorr-2dproj-lcms} shows the corresponding 1D
projections of the full 3D correlation function and fits. Similarly to what was
stressed regarding Fig.~\ref{fig:2d_doubleratios-project-276-7tev-lcms}, the
functions shown in Fig.~\ref{fig:3dcorr-2dproj-lcms} are projections of the
global 3D-stretched exponential fit to the 3D correlation function. This
explains the poor description of the 1D data by the L\'evy fit with $a=1$.

The values of the radii in the Bertsch--Pratt
parametrization~\cite{Bertsch:1989vn,Pratt:1986cc} for $\Pp\Pp$ collisions at both
2.76 and 7\TeV, obtained with the stretched exponential fit in the LCMS and
integrating over all $\Ntpt$ and $\kt$ ranges, are summarized in
Table~\ref{tab:radii-3d-7tev}, together with the corresponding intercept fit
parameter.

\begin{table*}[bht]
\topcaption{Parameters of the 3D BEC function, obtained from a stretched
exponential fit, in $\Pp\Pp$ collisions at $\sqrt{s} = 2.76$ and 7\TeV in the LCMS.}
 \label{tab:radii-3d-7tev}
 \centering
 \begin{scotch}{lcc}
             & $\Pp\Pp$ $\sqrt{s} = 2.76\TeV$ &
               $\Pp\Pp$ $\sqrt{s} = 7\TeV$ \\
  \hline
   $\lambda$    & $0.799 \pm 0.001\stat\pm 0.080\syst$ &
                  $0.679 \pm 0.002\stat\pm 0.068\syst$ \\
   $\Ro$ [fm]   & $1.129 \pm 0.016\stat\pm 0.175\syst$ &
                  $1.256 \pm 0.003\stat\pm 0.195\syst$ \\
   $\Rs$ [fm]   & $1.773 \pm 0.017\stat\pm 0.275\syst$ &
                  $1.842 \pm 0.004\stat\pm 0.286\syst$ \\
   $\Rl$ [fm]   & $2.063 \pm 0.022\stat\pm 0.320\syst$ &
                  $2.120 \pm 0.004\stat\pm 0.329\syst$ \\
  \end{scotch}
\end{table*}

Comparing the LCMS radii in the 3D case for $\Pp\Pp$ collisions at 7\TeV, the
hierarchy $\Rl > \Ro$ and $\Rl \gtrsim \Rs$ can be seen (although the values of
$\Rl$ and $\Rs$ are also compatible with each other within their statistical
and systematic uncertainties). Therefore, as in the 2D case, the 3D source also
seems to be more elongated along the longitudinal direction in the LCMS, with
the relationship between the radii being given by $\Rl \gtrsim \Rs > \Ro$. A
similar inequality holds for $\Pp\Pp$ collisions at 2.76\TeV.

The data sample obtained in $\Pp\Pp$ collisions at 2.76\TeV in 2013 is considerably
smaller than at 7\TeV and would lead to large statistical uncertainties if the
double ratios in the 3D case are measured differentially in $\Ntpt$ and $\kt$,
so this study is not considered at 2.76\TeV.

\begin{figure*}[htb]
 \centering
  \includegraphics[width=0.49\textwidth]{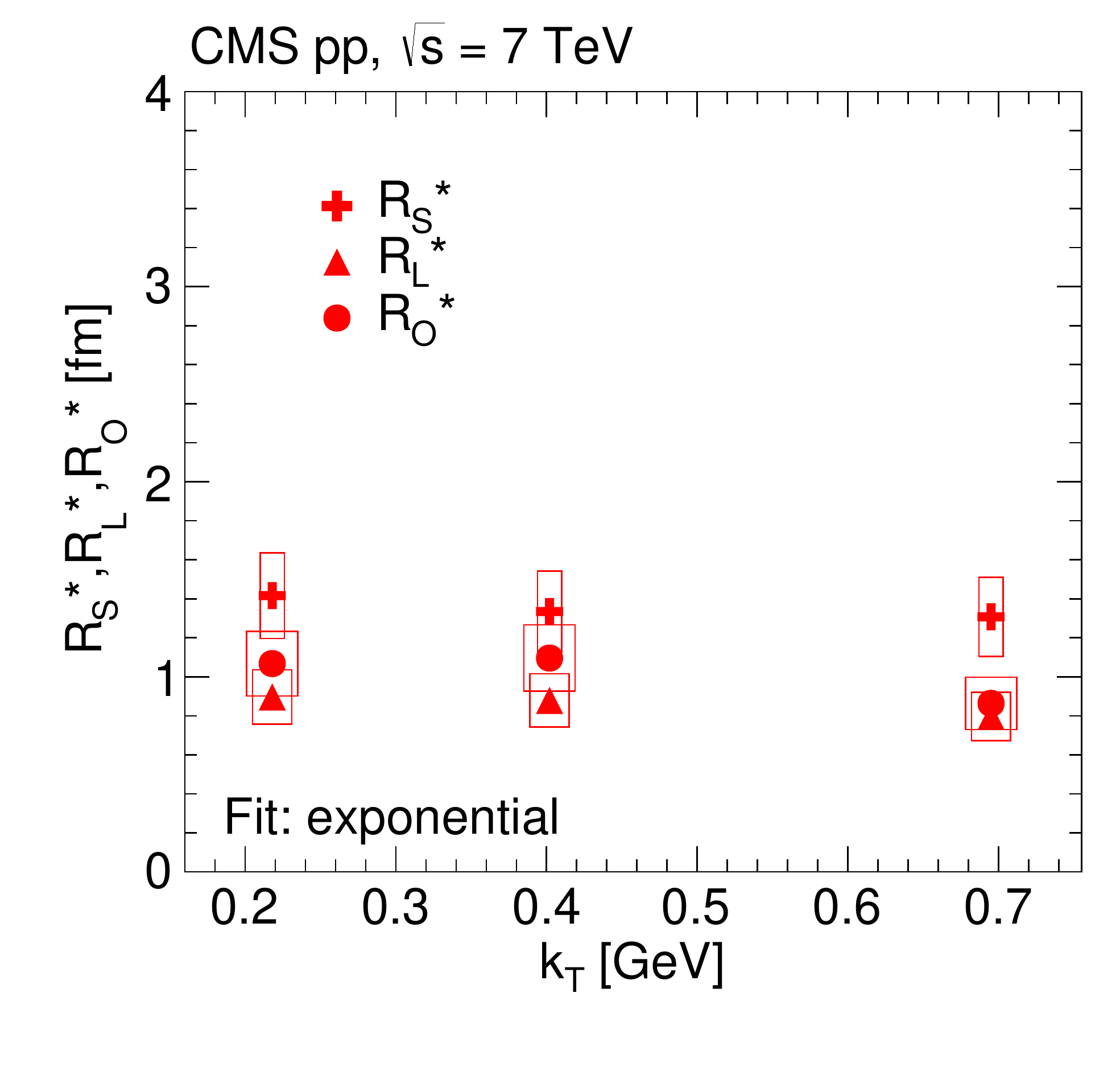}
  \includegraphics[width=0.49\textwidth]{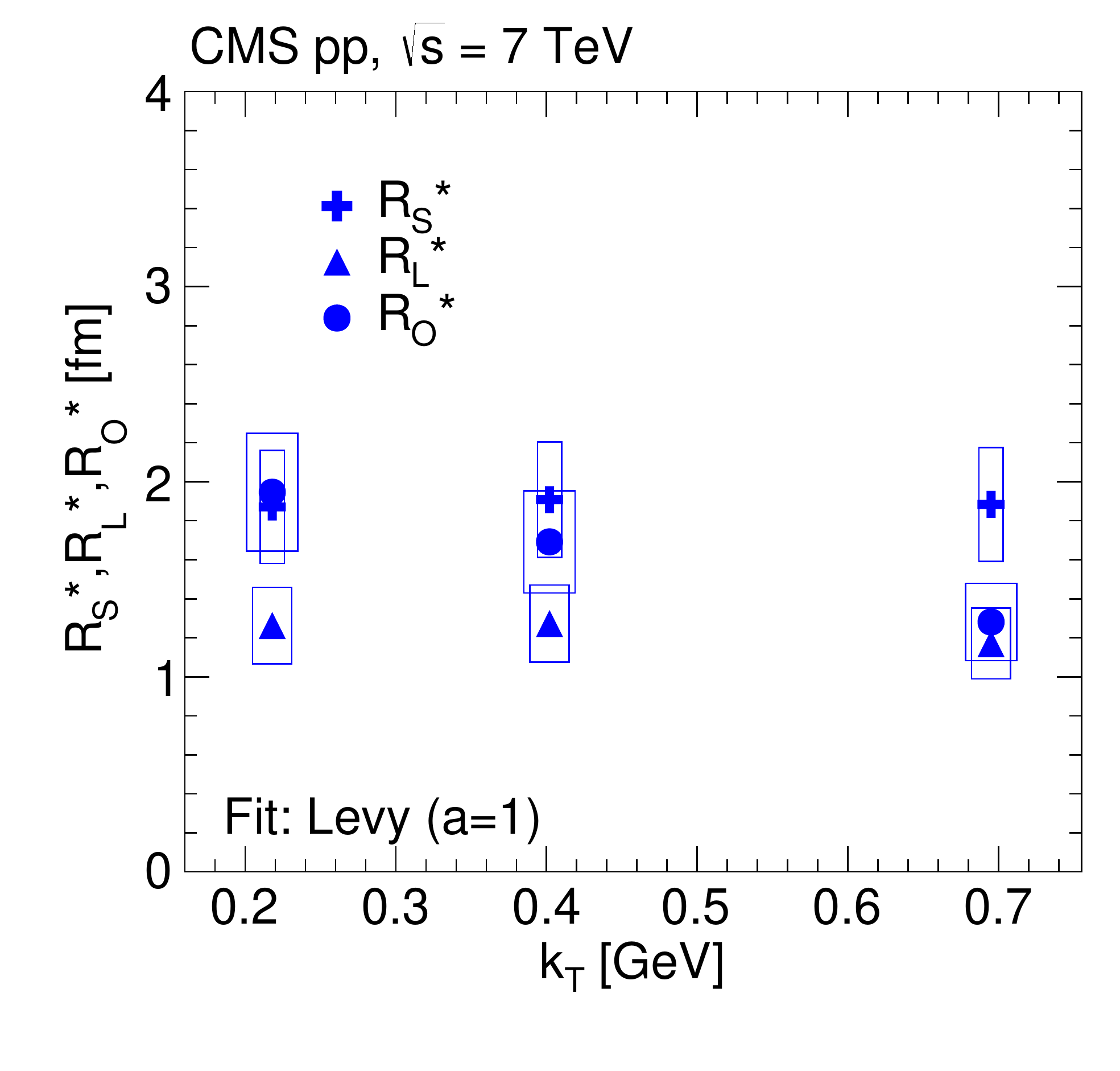}
  \includegraphics[width=0.49\textwidth]{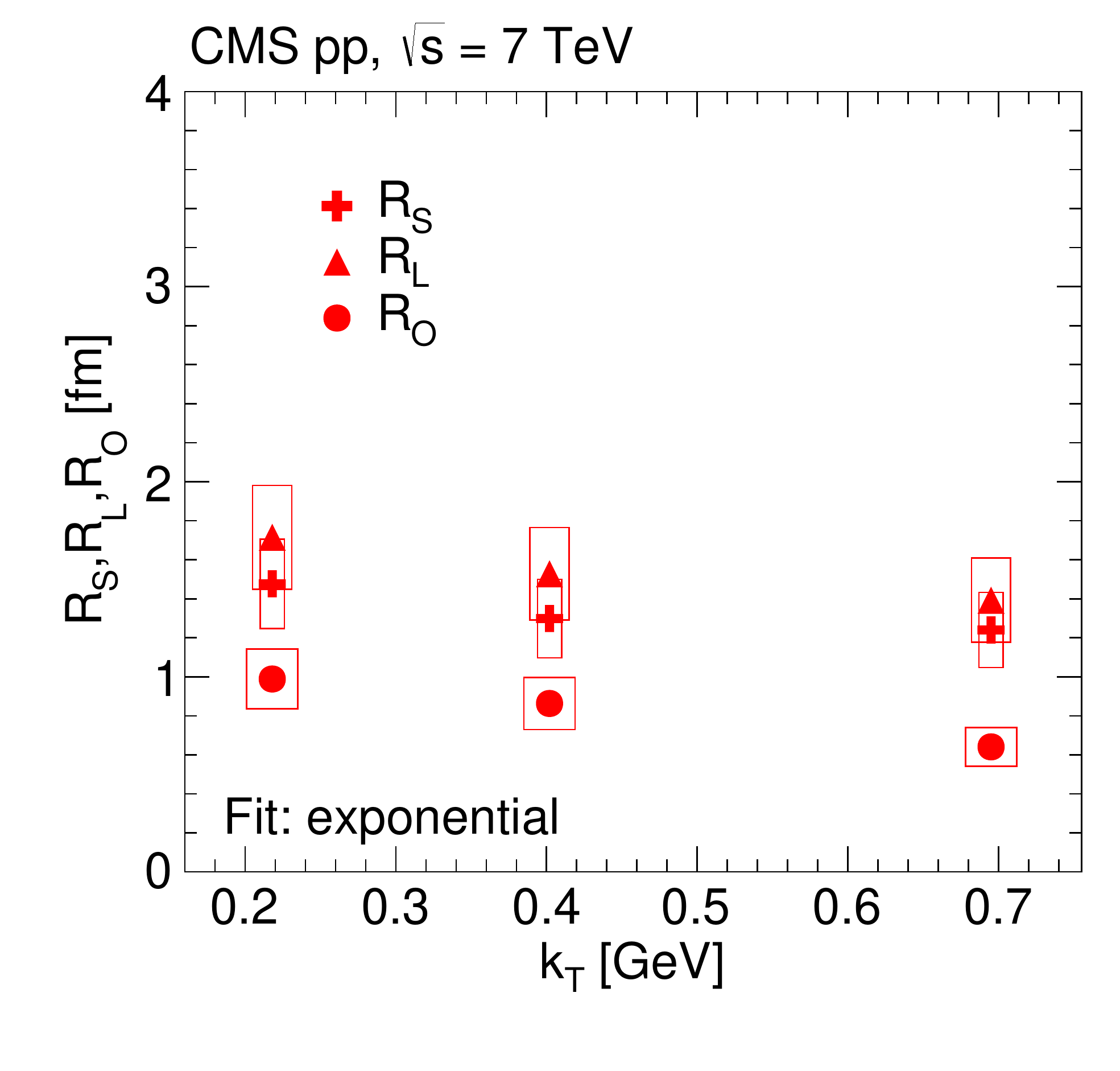}
  \includegraphics[width=0.49\textwidth]{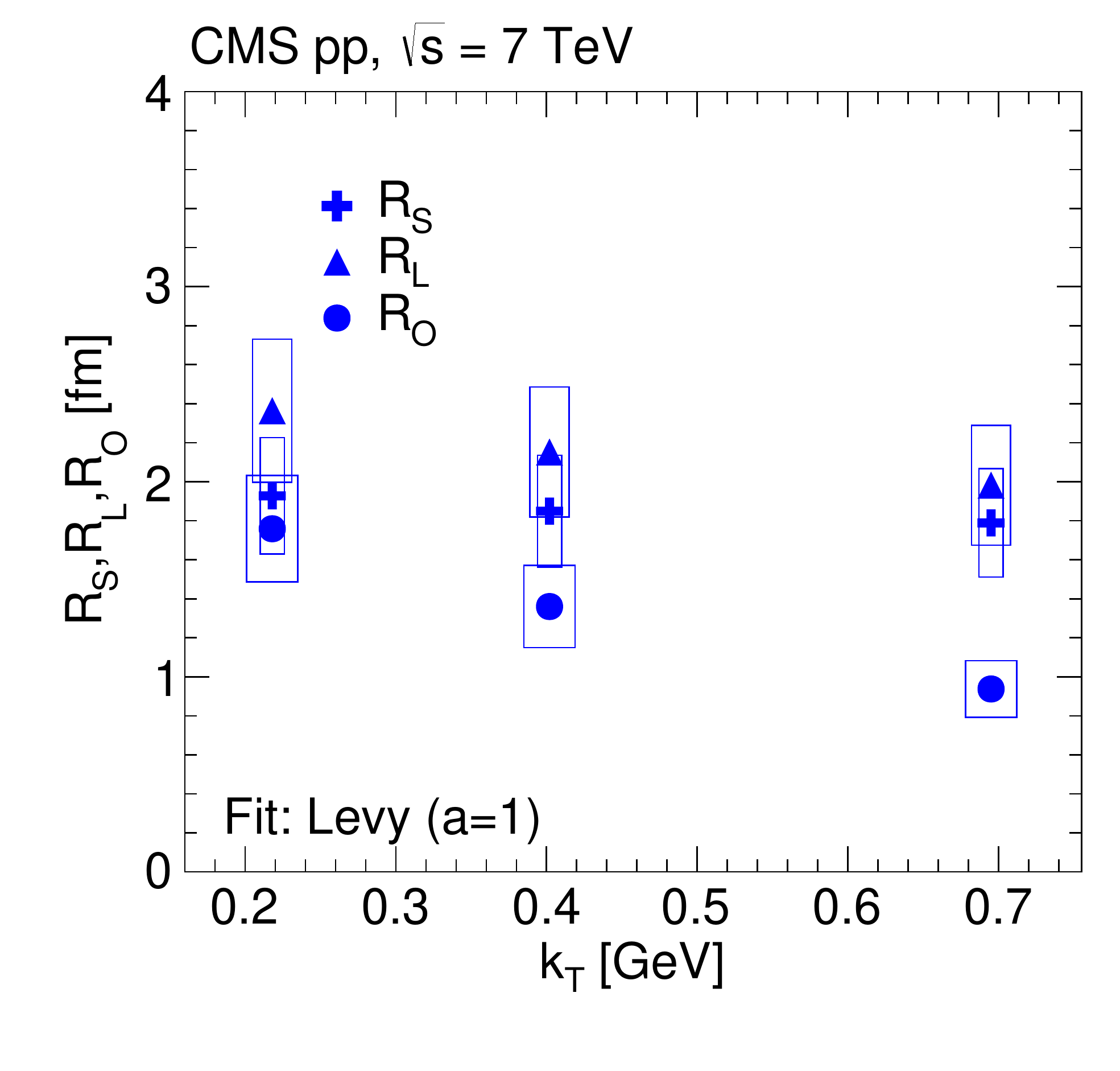}
 \caption{3D radii parameters as a function of $\kt$ (integrated over all
$\Ntpt$), obtained from double ratios fitted to the exponential (left) and
L\'evy (with $a = 1$, right) functions for $\Pp\Pp$ collisions at $\sqrt{s} = 7$\TeV
in the CM frame (top) and in the LCMS (bottom). The boxes indicate the
systematic uncertainties.}
 \label{fig:fits_exp-gauss-levy-qsqlqo-kt_cm-lcms-7tev}
\end{figure*}

For the data from $\Pp\Pp$ collisions at 7\TeV, the fits to the double ratios are
investigated in three bins of the pair average transverse momentum,
$\kt$ (integrating over charged multiplicity, $\Ntpt$). The fit parameters are
obtained with exponential and L\'evy (with $a=1$) type functions. The
corresponding results are compiled in
Fig.~\ref{fig:fits_exp-gauss-levy-qsqlqo-kt_cm-lcms-7tev}, both in the CM frame
and in the LCMS, showing that the values are very sensitive to the
expression adopted for the fit function, as also suggested in
Fig.~\ref{fig:3dcorr-2dproj-lcms} by the 1D projections of the global 3D fit
performed to the 3D double ratios.

In the LCMS, the behavior of $\Rs$ with increasing pair average transverse
momentum is very similar to that of $\Rs^*$ in the CM frame, and the fit
values are consistent within statistical and systematic uncertainties, as
would be expected, since the boost to the LCMS is in the longitudinal
direction. It can also be seen in both frames that the {\sl sideways} radius is
practically independent of $\kt$, within the uncertainties. The {\sl outwards}
fit parameter $\Ro^{(*)}$ shows a similar decrease with respect to $\kt$ in
both frames. However, in the CM frame the $\Ro^{*}$ fit values are slightly
larger than those in the LCMS ($\Ro$) for the same ($\Ntpt,\kt$) bins. This is
compatible with the dependence of $\Ro^{(*)}$ on the source lifetime (as seen
in Section~\ref{sub:fitting_functions}), which is expected to be higher in the
CM frame than in the LCMS, because of the Lorentz time dilation. In the LCMS,
$\Rl$ decreases with increasing $\kt$. It has larger values than in the CM
frame for both energies, suggesting an effect related to the Lorentz boost,
similar to the 2D case. Also in this frame, its decrease with increasing $\kt$
is more pronounced for the same reason.

\begin{figure*}[htb]
 \centering
  \includegraphics[width=0.49\textwidth]{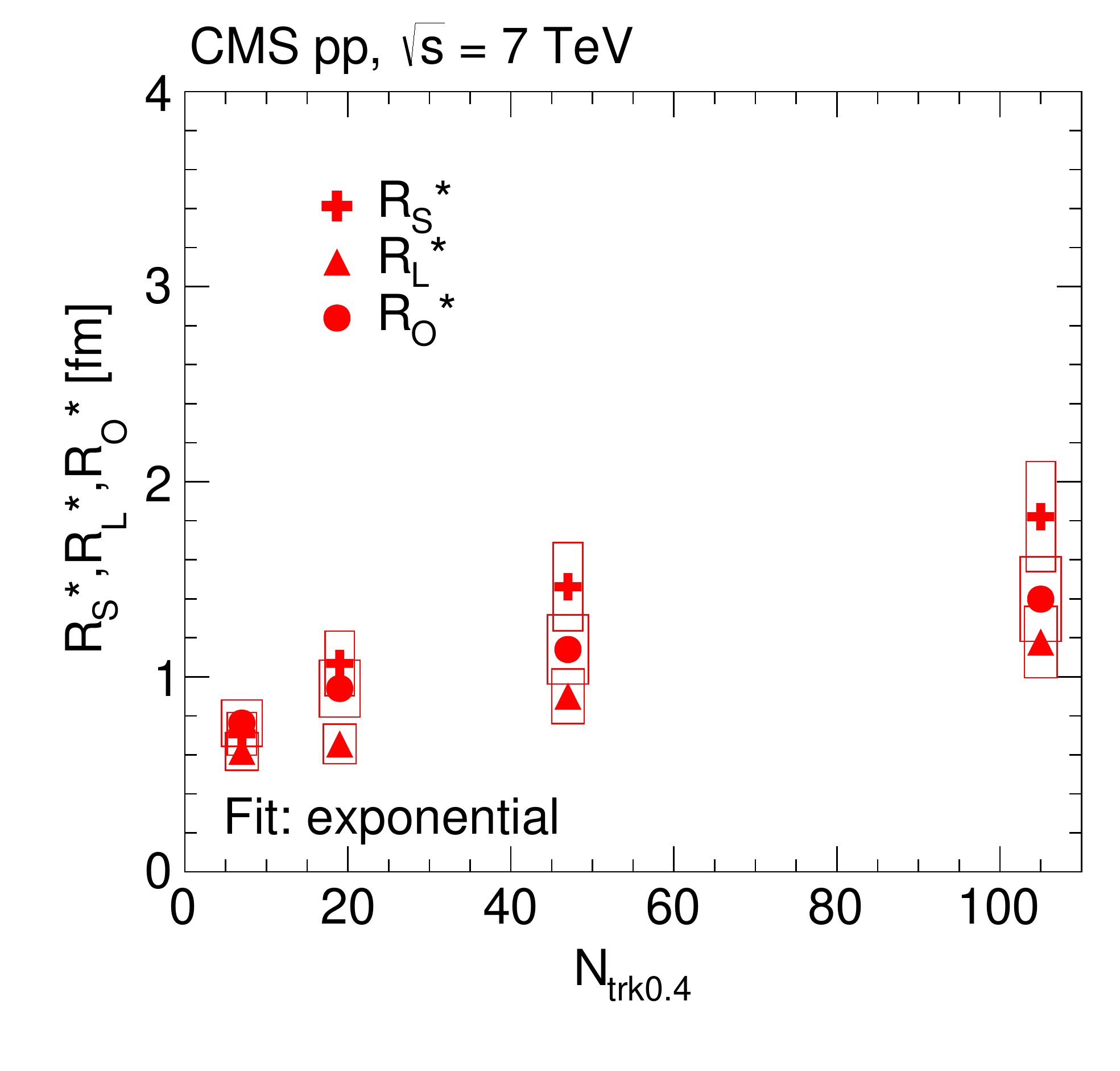}
  \includegraphics[width=0.49\textwidth]{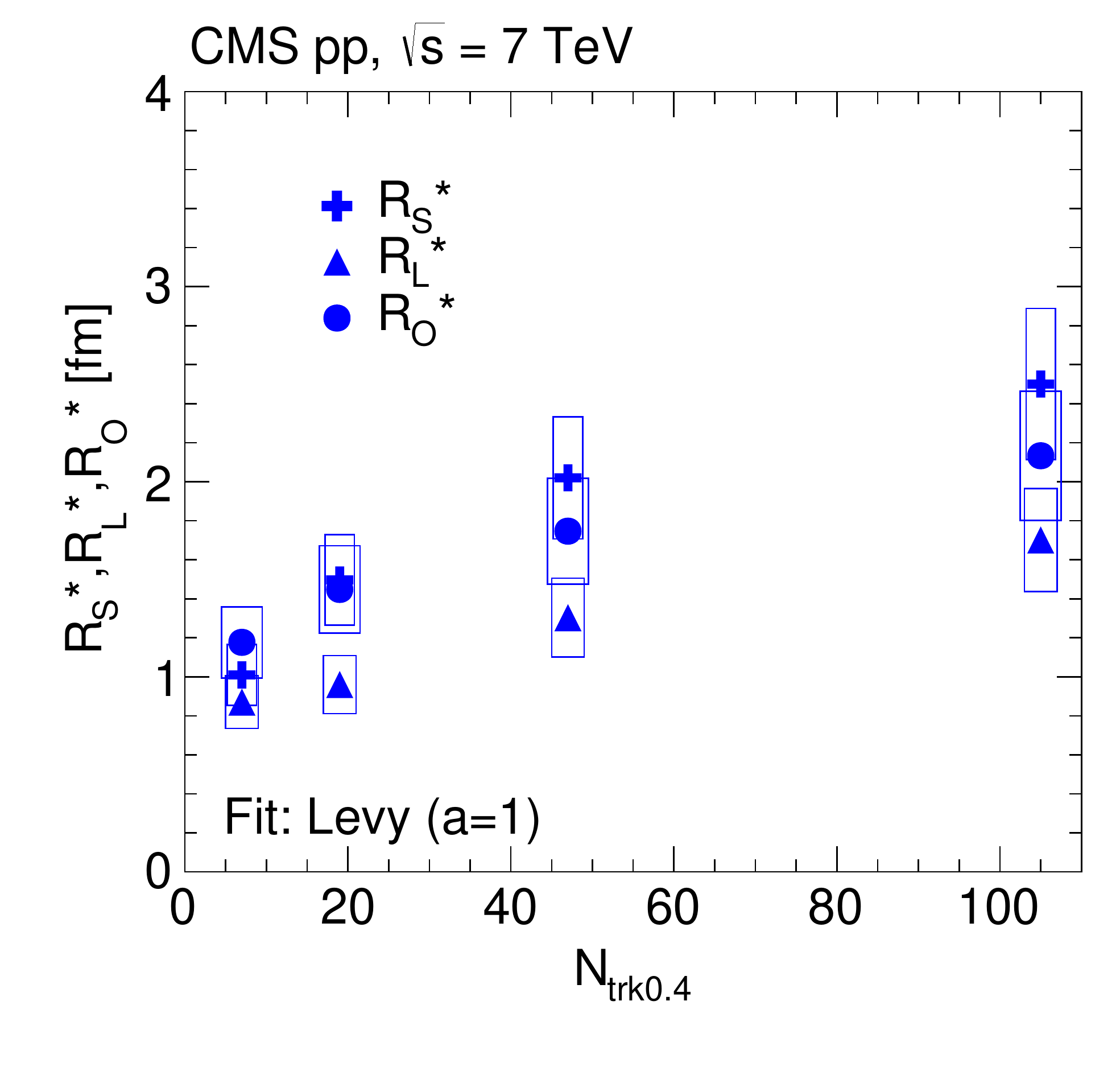}
  \includegraphics[width=0.49\textwidth]{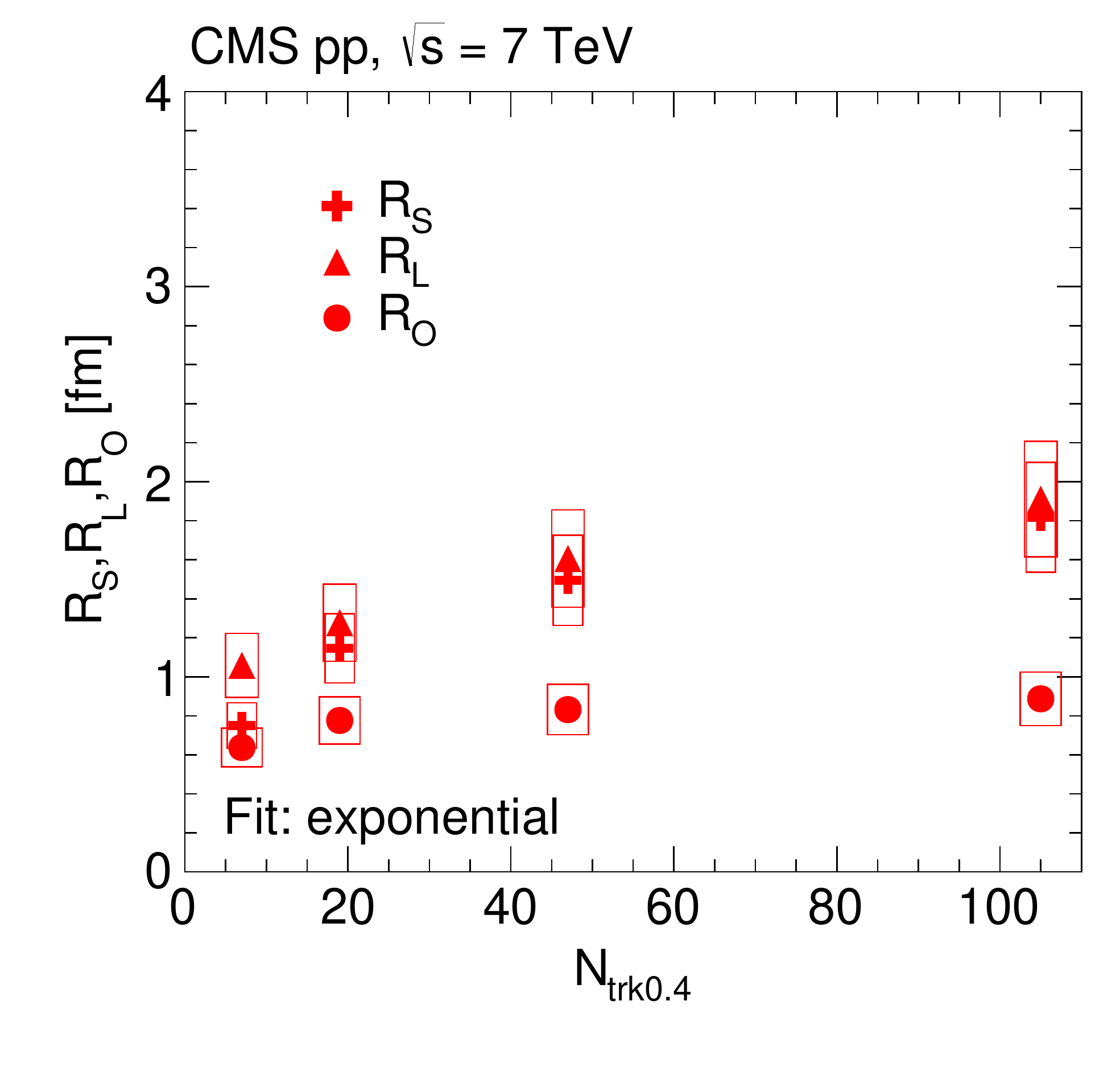}
  \includegraphics[width=0.49\textwidth]{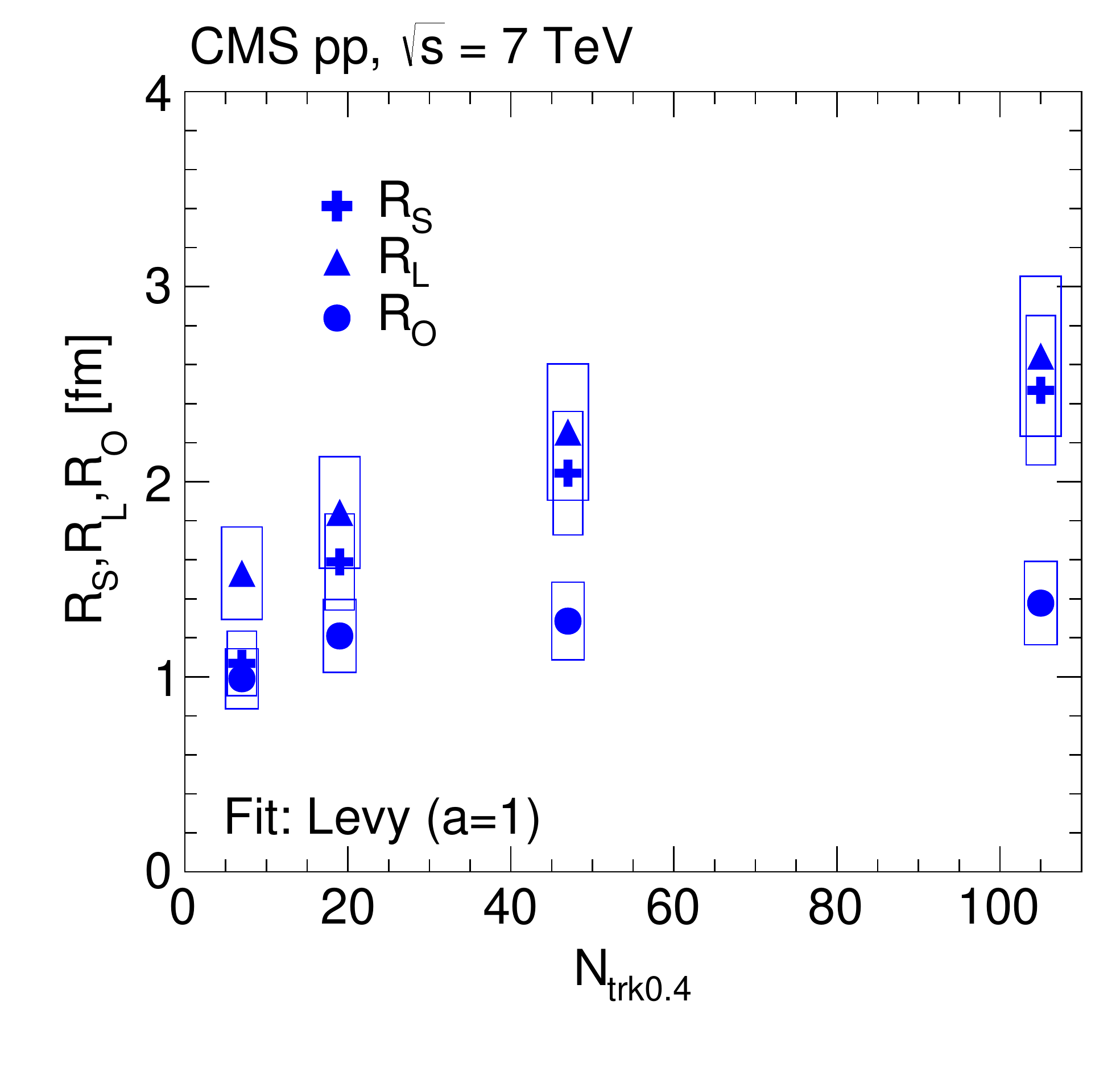}
 \caption{3D radii parameters as a function of $\Ntpt$ (integrated over all
$\kt$), obtained from double ratios fitted to the exponential (left) and
L\'evy (with $a = 1$, right) functions for $\Pp\Pp$ collisions at $\sqrt{s} = 7$\TeV
in the CM frame (top) and in the LCMS (bottom). The boxes indicate the
systematic uncertainties.}
 \label{fig:fits_exp-gauss-levy-qsqlqo-nch_cm-lcms-7tev}
\end{figure*}

The fits to the double ratios are also studied in four bins of $\Ntpt$
integrating over all $\kt$, and the corresponding results are shown in
Fig.~\ref{fig:fits_exp-gauss-levy-qsqlqo-nch_cm-lcms-7tev}, both in the CM
frame (upper) and in the LCMS (lower). A clear trend can be seen, common to
both fit functions and in all directions of the relative momentum components:
the radii $\Rs^{(*)}$, $\Rl^{(*)}$, and $\Ro^{(*)}$ increase with increasing
average multiplicity, $\Ntpt$, similar to what is seen in the 1D and 2D
cases.

\begin{figure*}[htb]
 \centering
  \includegraphics[width=0.49\textwidth]{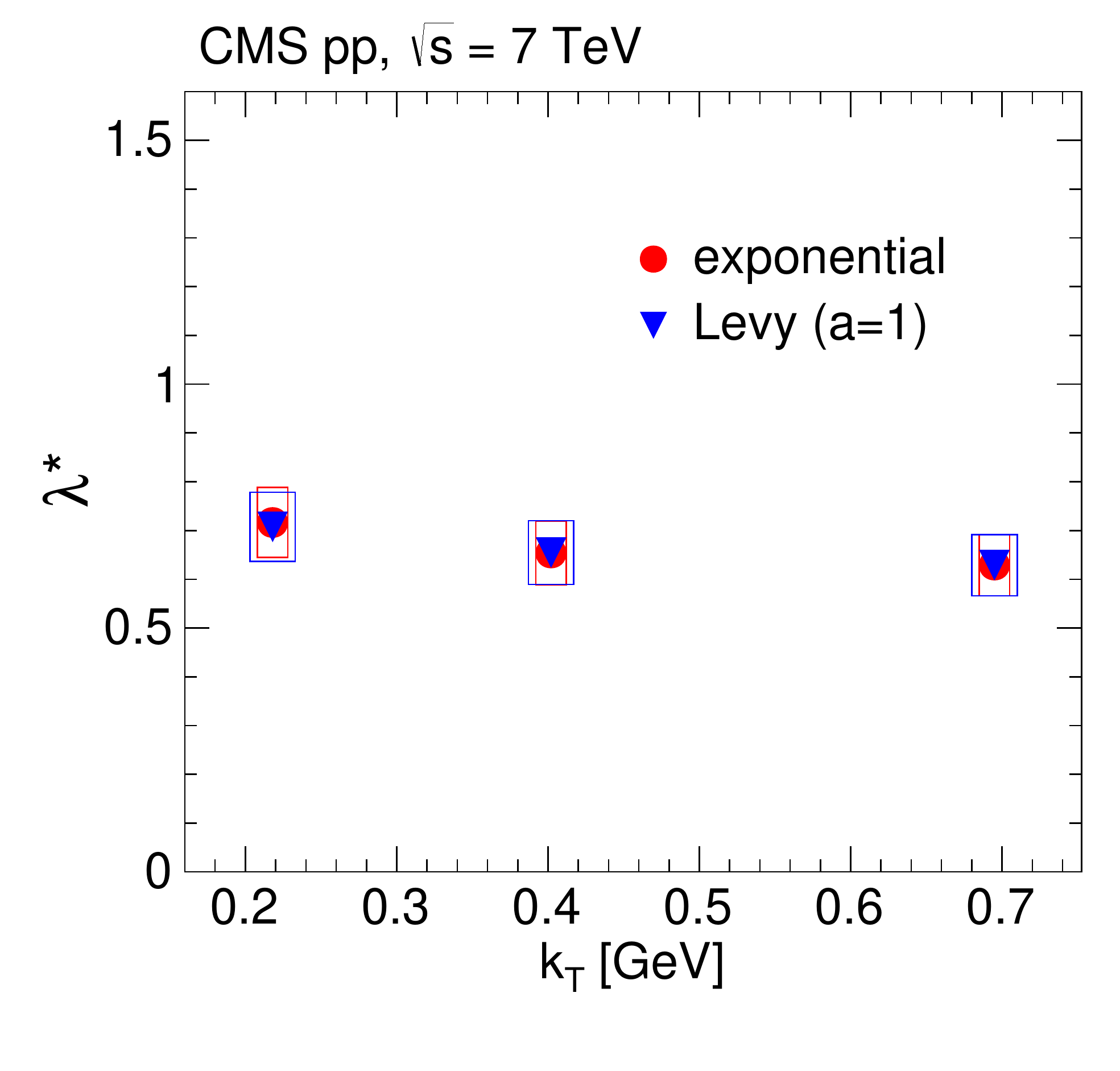}
  \includegraphics[width=0.49\textwidth]{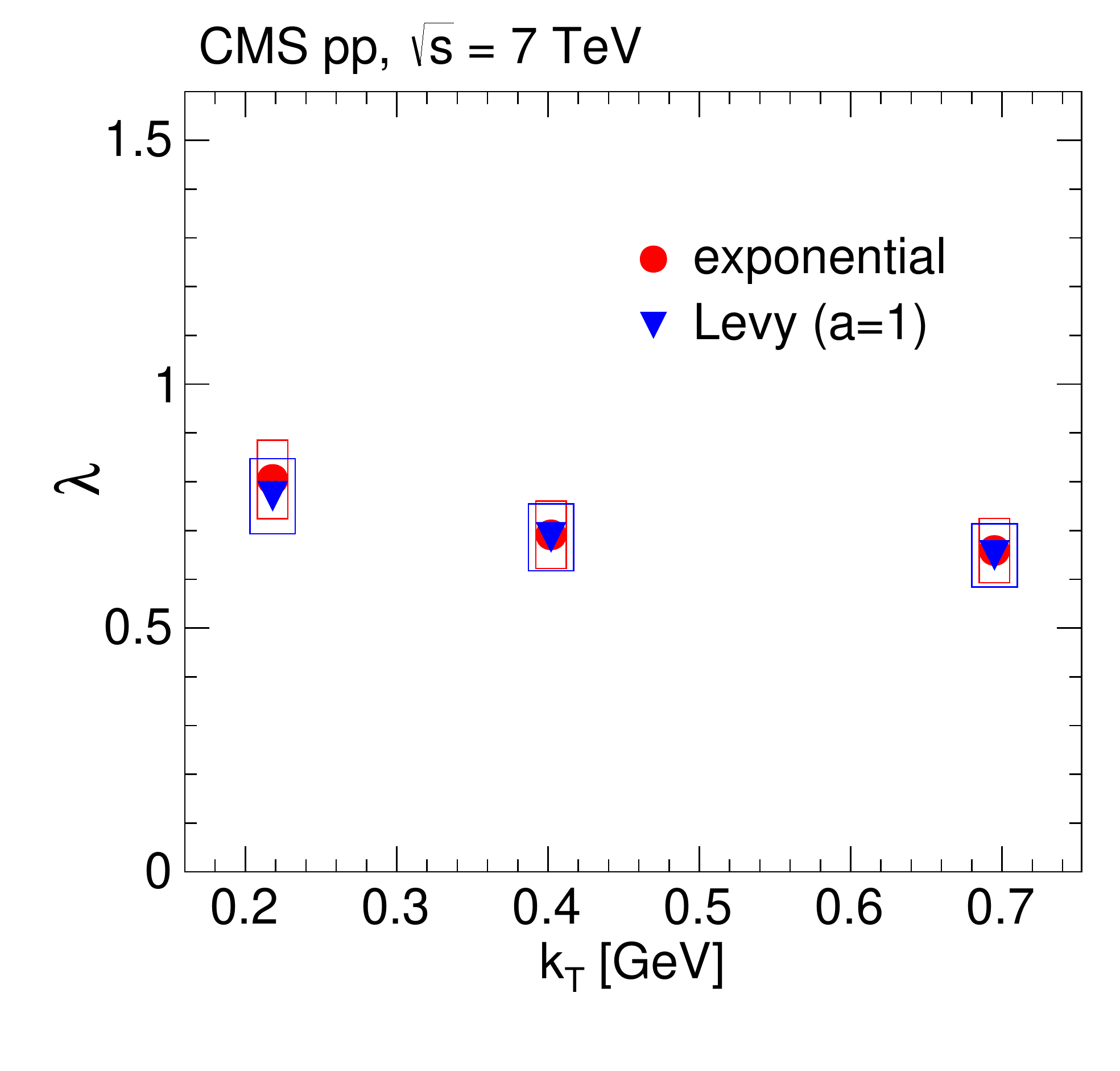}

  \includegraphics[width=0.49\textwidth]{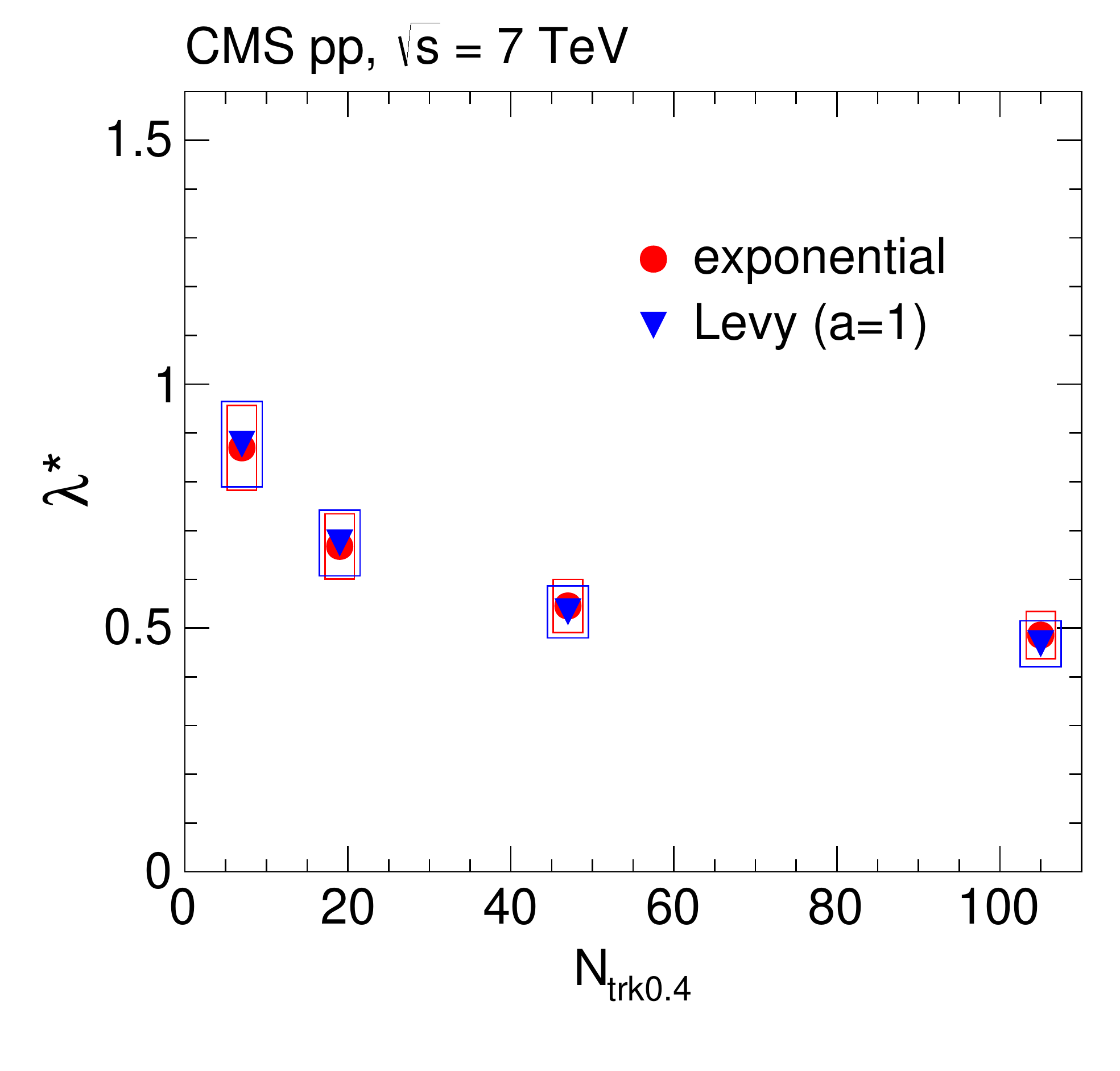}
  \includegraphics[width=0.49\textwidth]{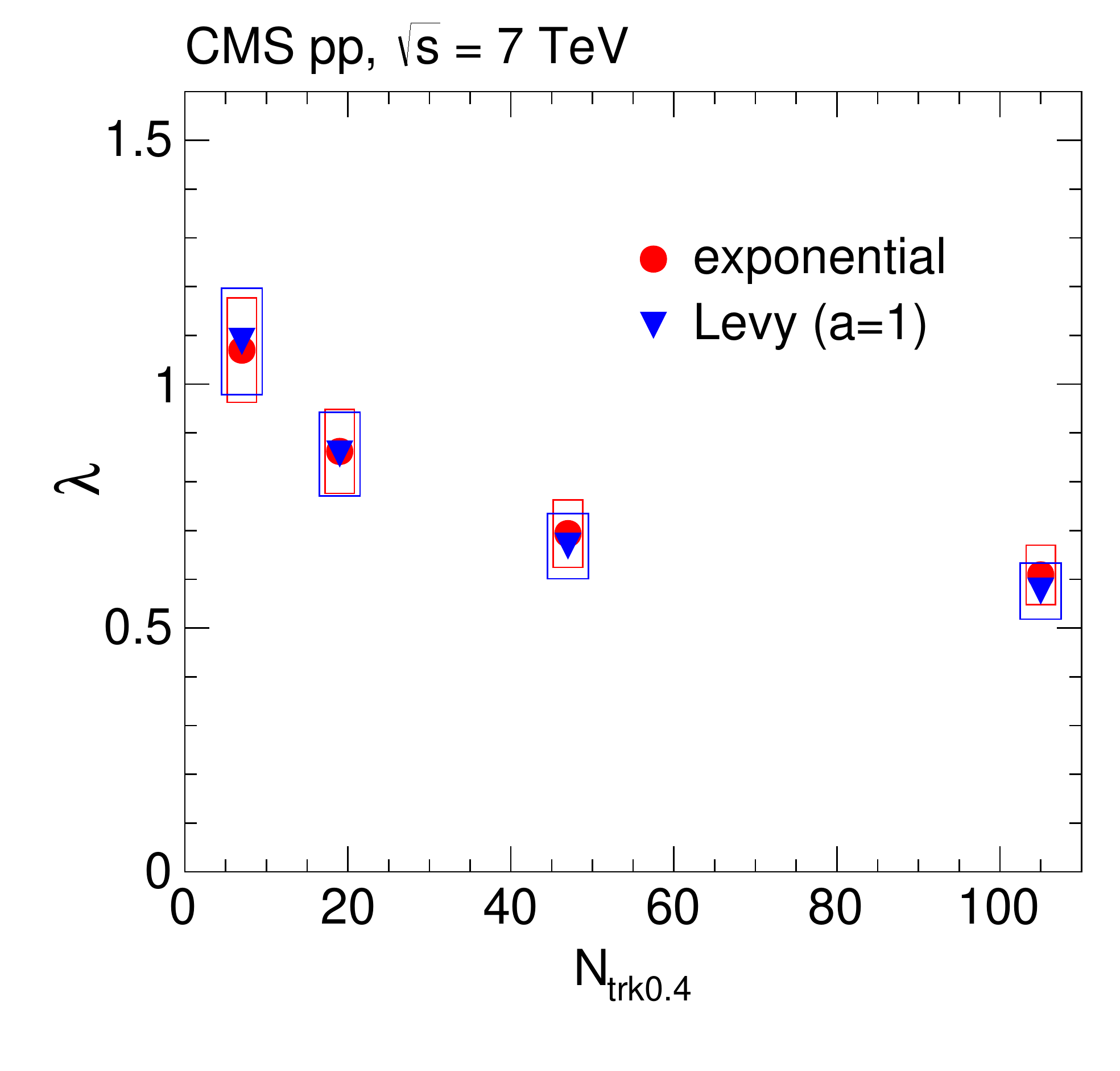}
 \caption{Intercept parameter $\lambda^{(*)}$ in the CM frame (left) and in the
LCMS (right) as functions of $\kt$ (integrated over $\Ntpt$, top) and of $\Ntpt$
(integrated over $\kt$, bottom), obtained from fits to the double ratios with
the exponential and the L\'evy (with $a = 1$) functions in $\Pp\Pp$ collisions at
7\TeV. The boxes indicate the systematic uncertainties.}
 \label{fig:lambda-3d-ktnch-cm-lcms}
\end{figure*}

The intercept parameter $\lambda^{(*)}$ is also studied as a function of $\kt$
and $\Nt$, both in the CM frame and in the LCMS. The corresponding results are
shown in Fig.~\ref{fig:lambda-3d-ktnch-cm-lcms}. A moderate decrease with
increasing $\kt$ is observed. As a function of $\Ntpt$, $\lambda^{(*)}$ first
decreases with increasing charged-particle multiplicity and then it seems to
saturate.

\begin{figure*}[tbh]
 \centering
  \includegraphics[width=0.49\textwidth]{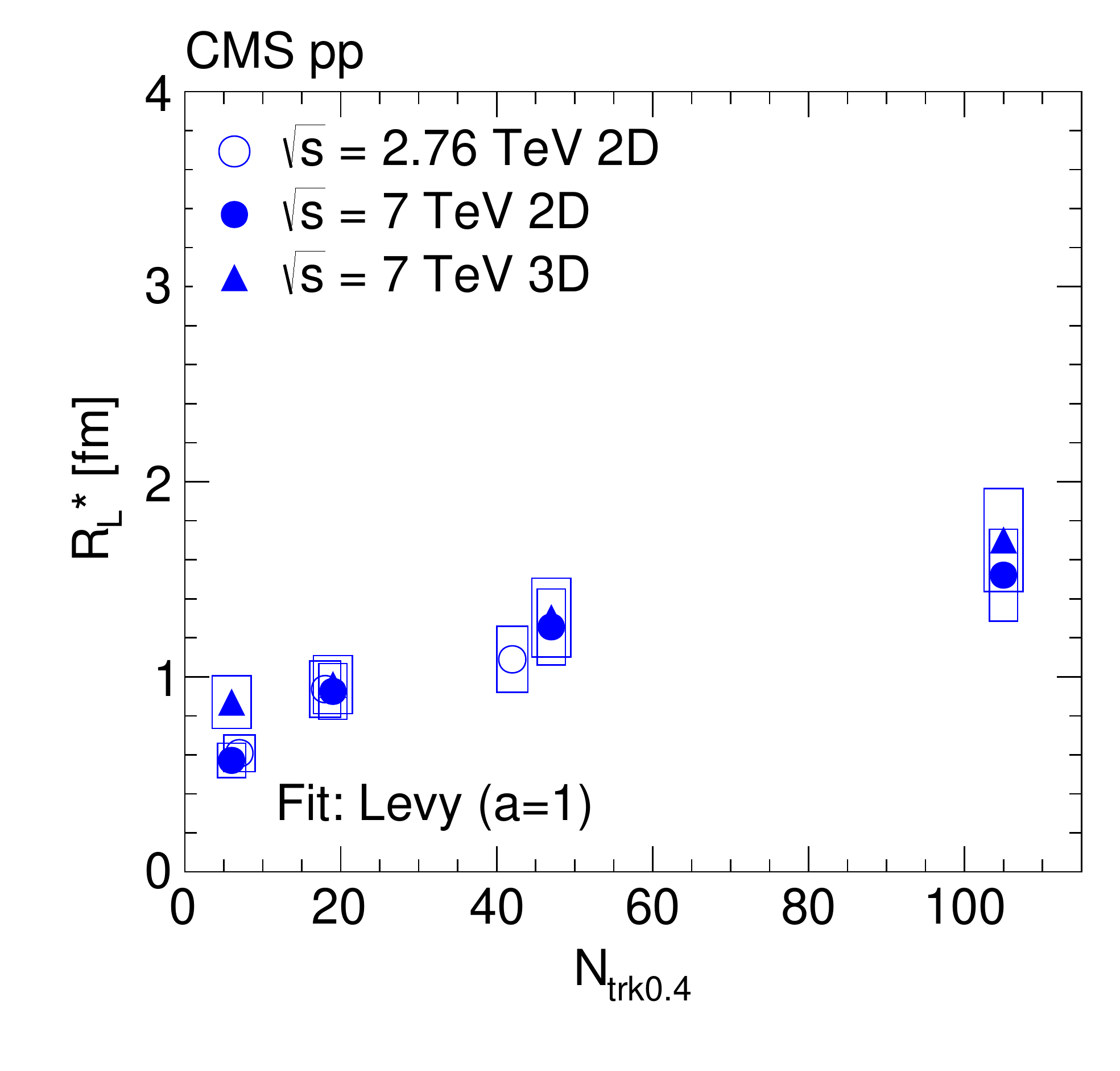}
  \includegraphics[width=0.49\textwidth]{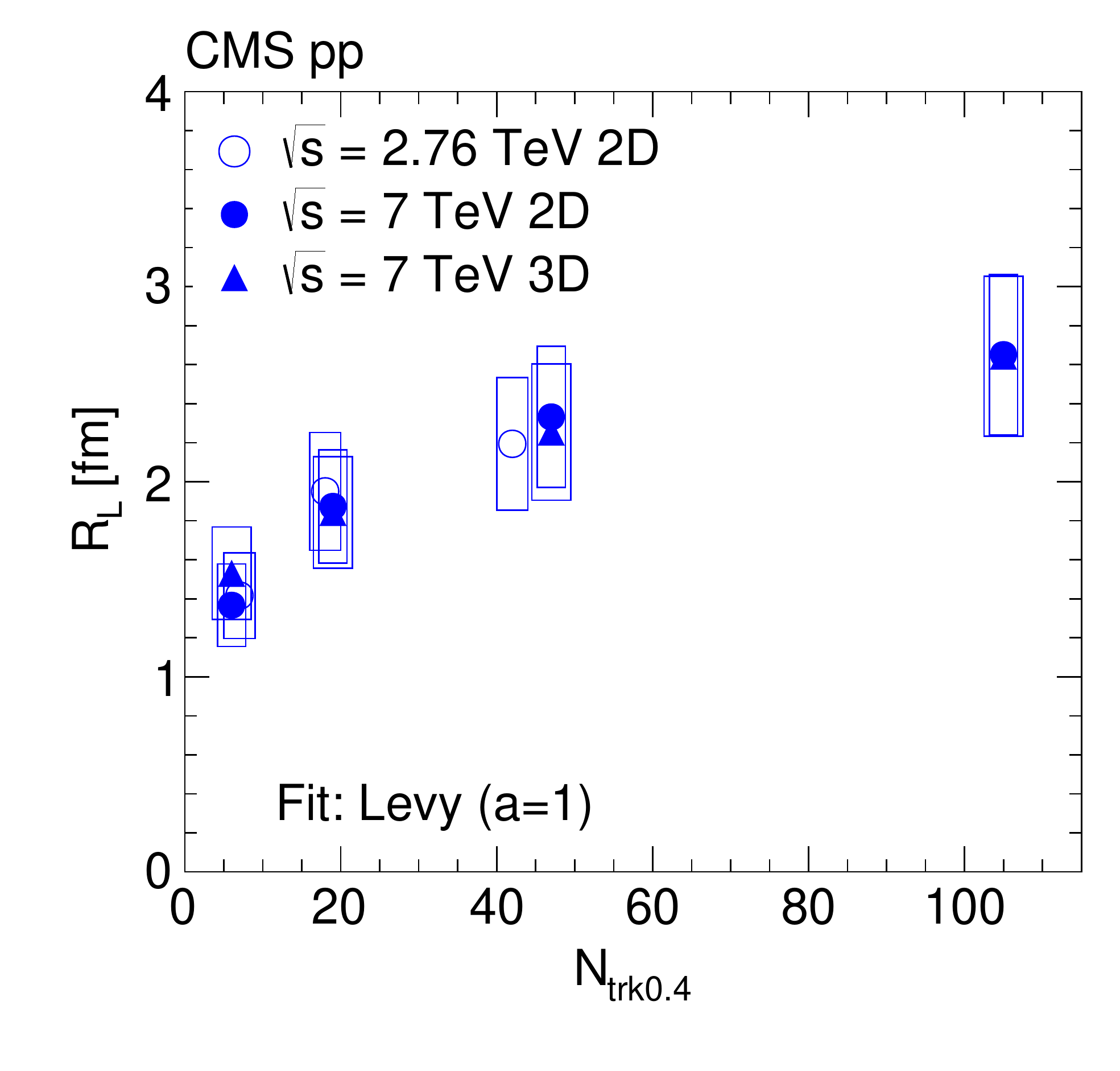}
 \caption{Longitudinal radius parameter as a function of $\Ntpt$ (integrated
over all $\kt$) in the CM frame ($\Rl^*$, left) and in the LCMS ($\Rl$, right),
obtained from 2D correlation functions fitted as functions of $\qt$ and $\ql$,
compared to the values obtained in the 3D case (fits as a function of $\qs$,
$\ql$, and $\qs$), in $\Pp\Pp$ collisions at $\sqrt{s} = 2.76$\TeV (open symbols) and
7\TeV (closed symbols). The boxes indicate the systematic uncertainties. }
 \label{fig:rL-2d-3d-comparision-cm-lcms}
\end{figure*}

Since the longitudinal radius parameter represents the same length of
homogeneity along the beam direction, it would be expected that the
corresponding fit values in the 2D and 3D cases are comparable. In fact,
they are consistent within the experimental uncertainties, as can be seen in
Fig.~\ref{fig:rL-2d-3d-comparision-cm-lcms}, except in the first bin of $\Ntpt$
investigated in the CM frame, where the central $\Rl$ value in 3D is roughly
0.3\unit{fm} larger than the corresponding $\Rl$ value in 2D, for $\Pp\Pp$ collisions at
both 2.76 and 7\TeV. This difference, however, may reflect a better description
of the source in the 3D case, where the transverse component is decomposed into
two independent orthogonal directions, whereas it may not be fully accommodated
by considering just one transverse parameter as in the 2D case. In addition,
it can be seen in Fig.~\ref{fig:rL-2d-3d-comparision-cm-lcms} that there is an
approximate energy independence of $\Rl^*$ ($\Rl$) as a function of
$\Ntpt$, for results at $\sqrt{s} = 2.76$ and 7\TeV, similar to what is
observed in the 1D case.

\section{Results with particle identification and cluster subtraction in \texorpdfstring{\Pp\Pp,
$\Pp$Pb, and PbPb}{pp, pPb, and PbPb} collisions}
\label{sec:clusteremov_results_pp_ppb_pbpb}

\begin{figure*}[!!hp]

 \centering
  \includegraphics[width=0.45\textwidth]{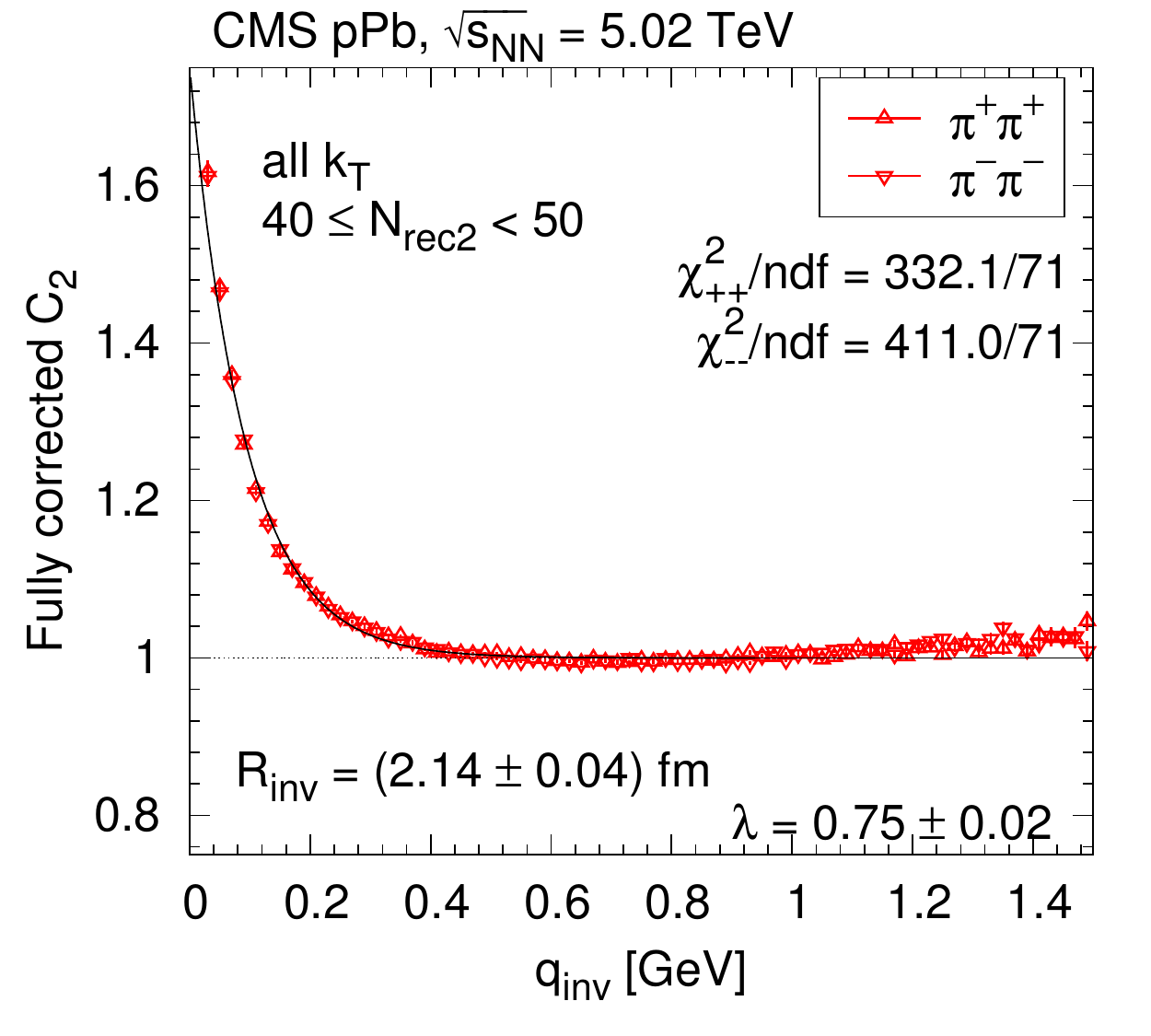}
  \includegraphics[width=0.45\textwidth]{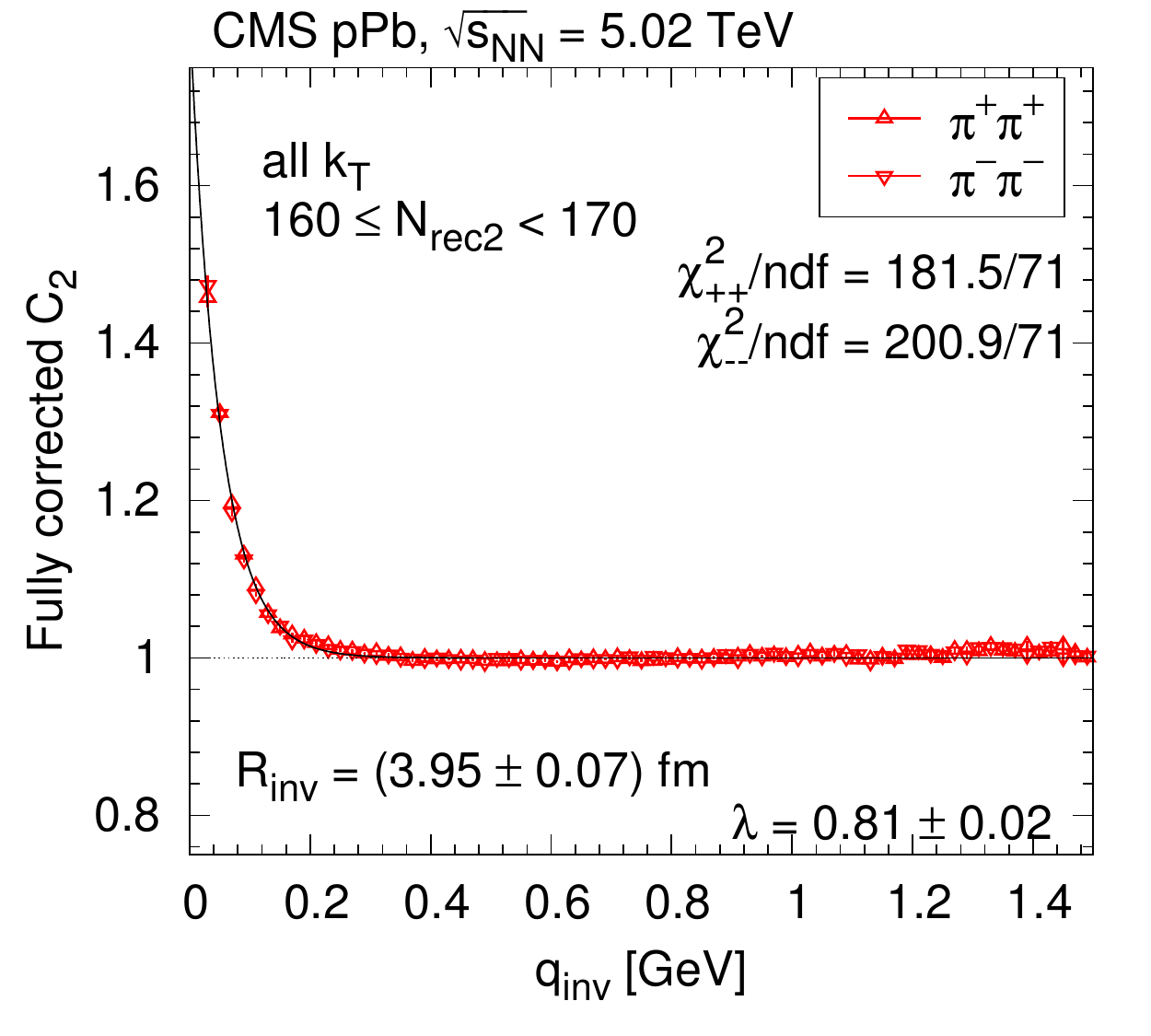}
  \includegraphics[width=0.45\textwidth]{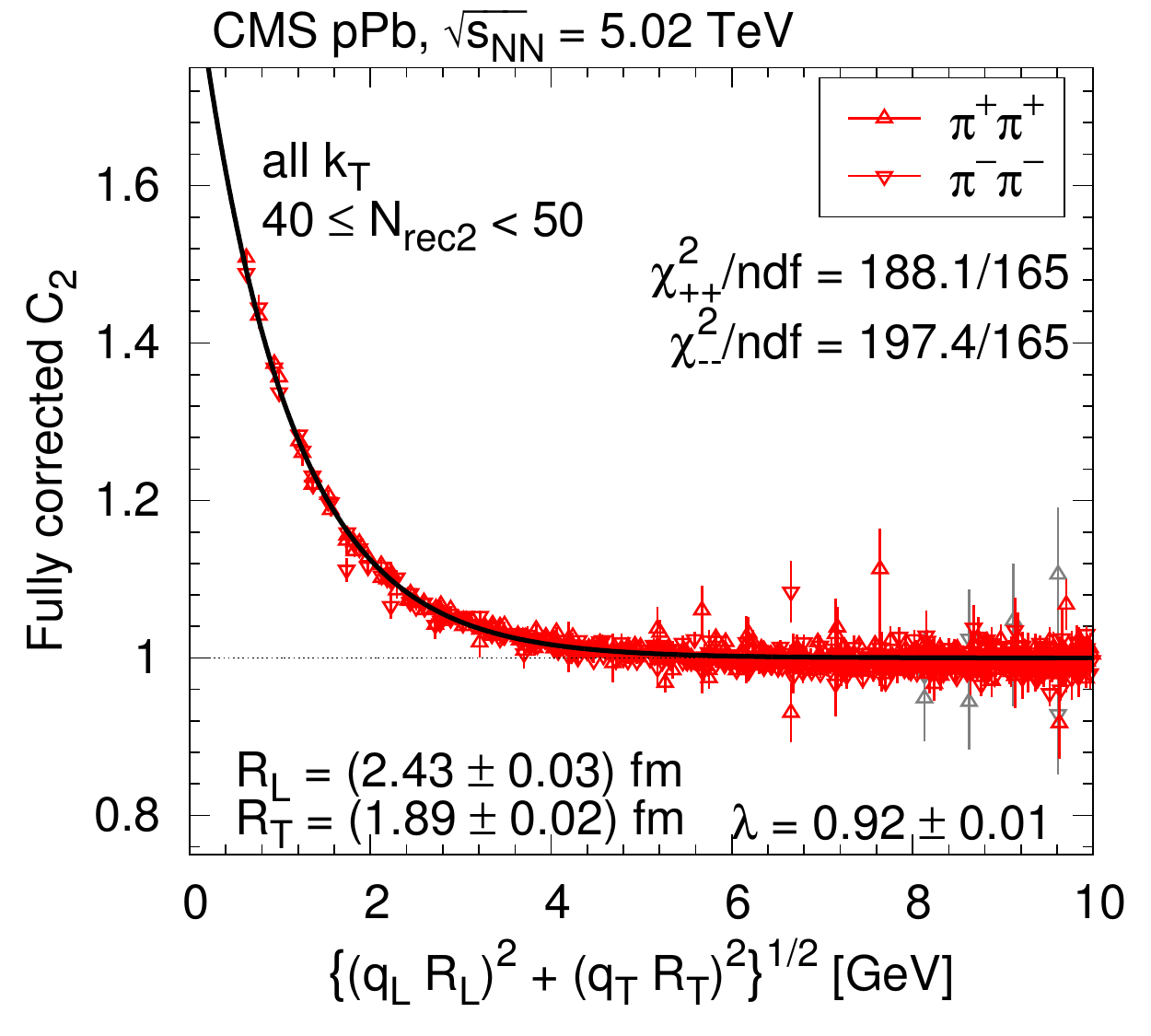}
  \includegraphics[width=0.45\textwidth]{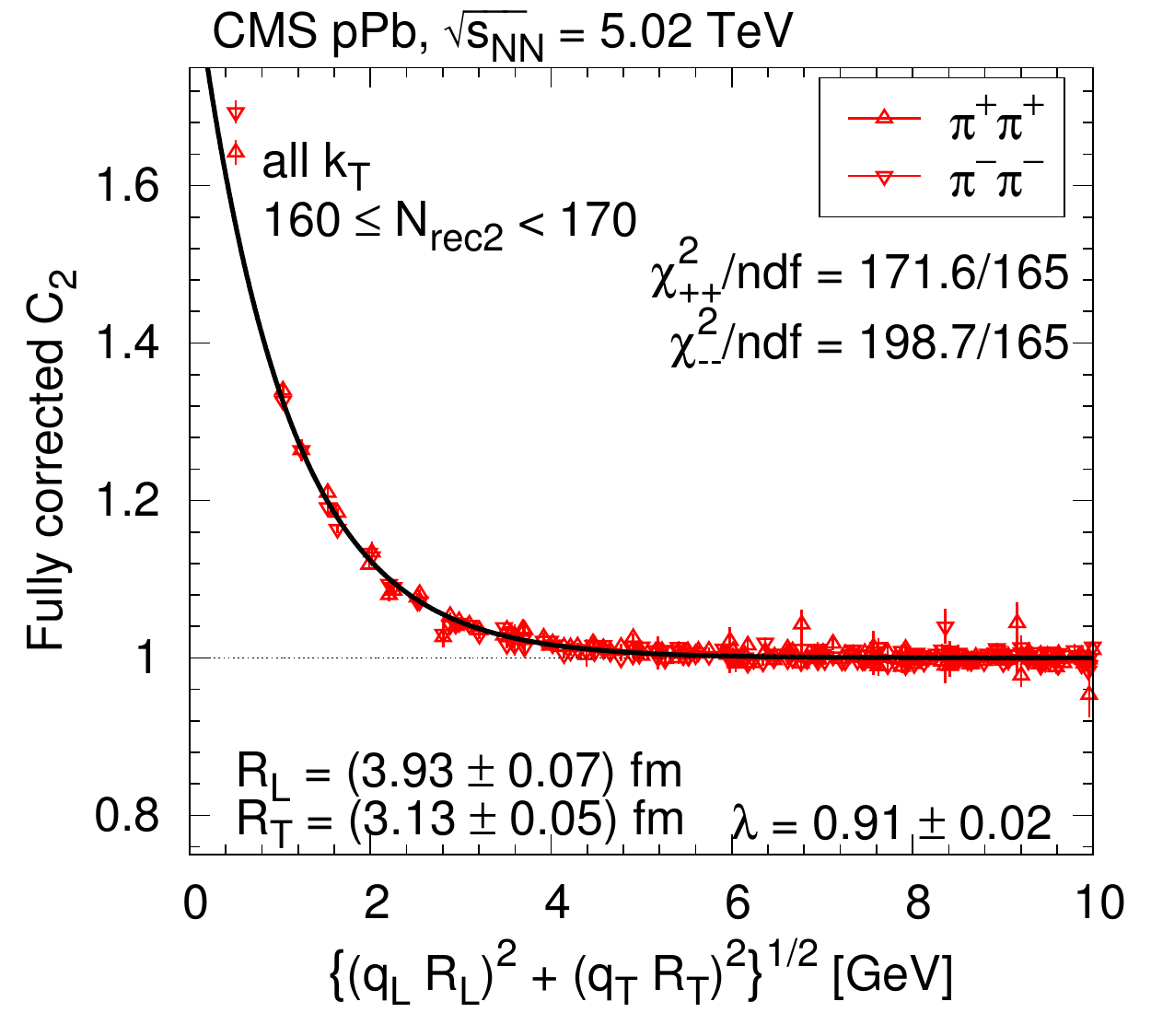}
  \includegraphics[width=0.45\textwidth]{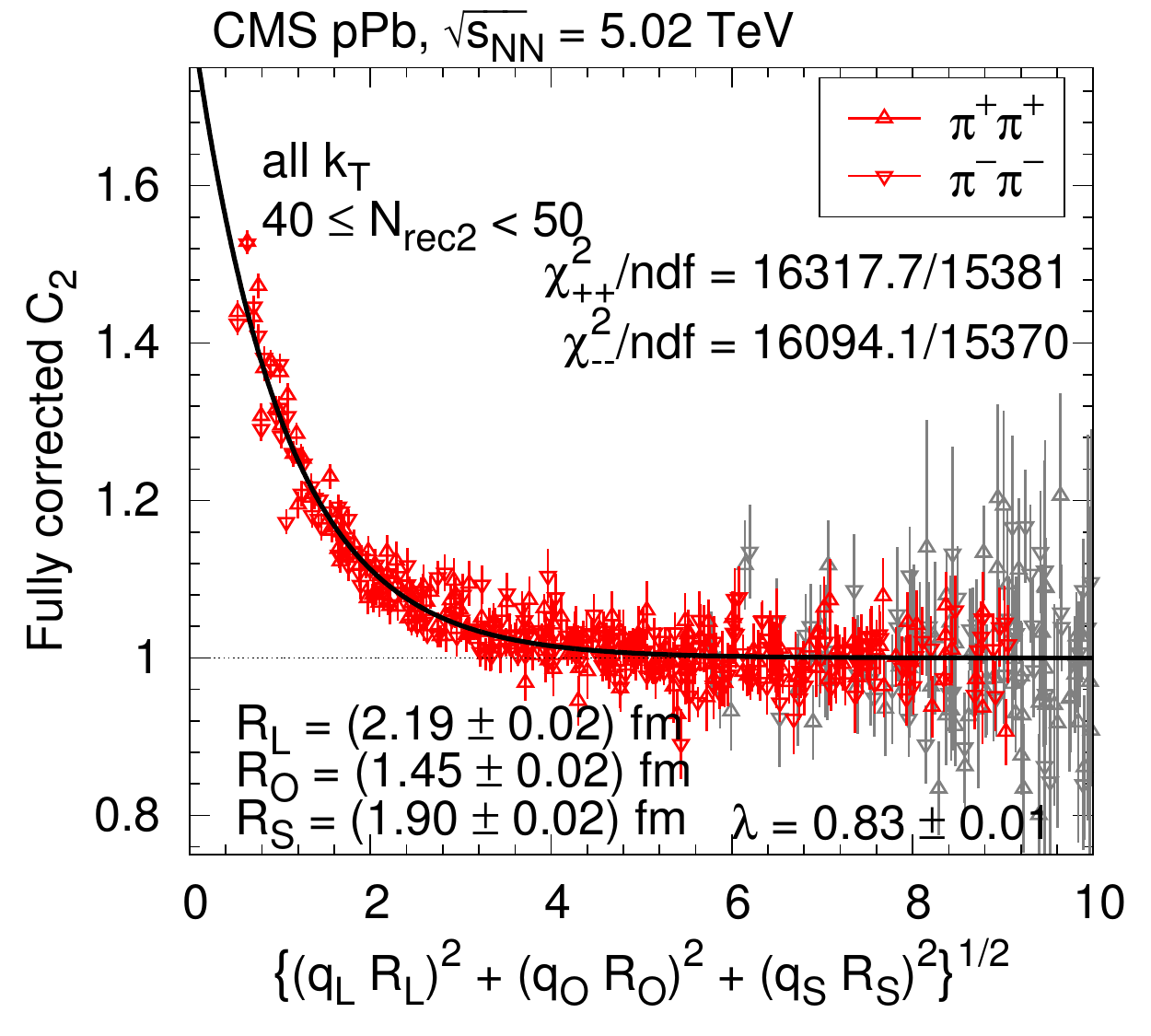}
  \includegraphics[width=0.45\textwidth]{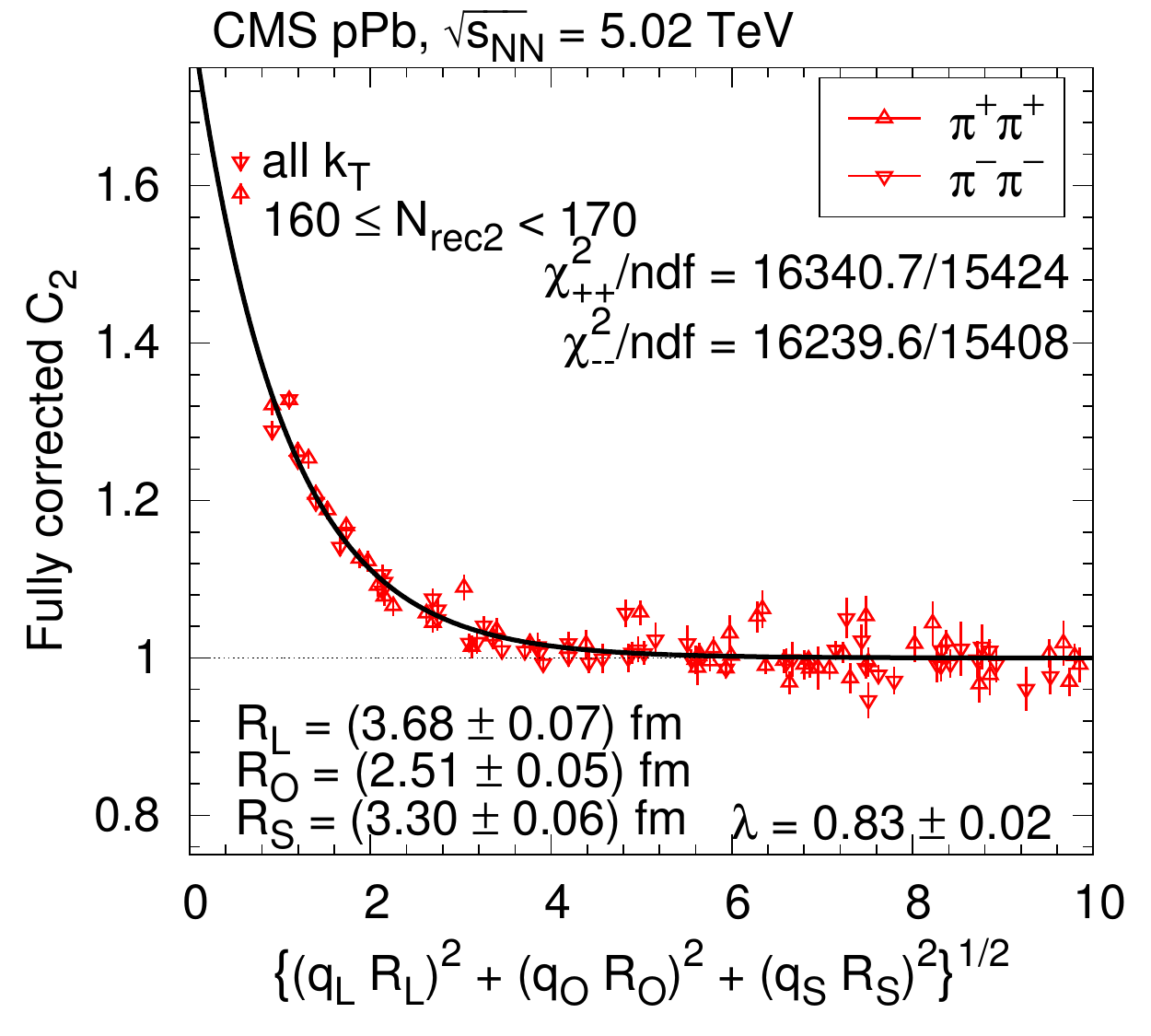}
 \caption{SS correlation function (integrated over all $\kt$) measured in $\Pp$Pb
collisions at 5.02\TeV as a function of $\qi$ or of the combined momentum, in
two selected $\Nr$ bins (left and right), for pion pairs (red triangles) in one
(top), two (middle), and three (bottom) dimensions, corrected for Coulomb
interaction and cluster contributions (mini-jets and multi-body resonance
decays). For better visibility, only a fraction of the points is plotted, and
those with statistical uncertainty higher than 10\% are in light-grey color.
The solid curves indicate fits with the stretched exponential parametrization.}
 \label{fig:pi_pPb_5TeVc/qmix_}
\end{figure*}

\begin{figure*}[t]
 \centering
  \includegraphics[width=0.49\textwidth]{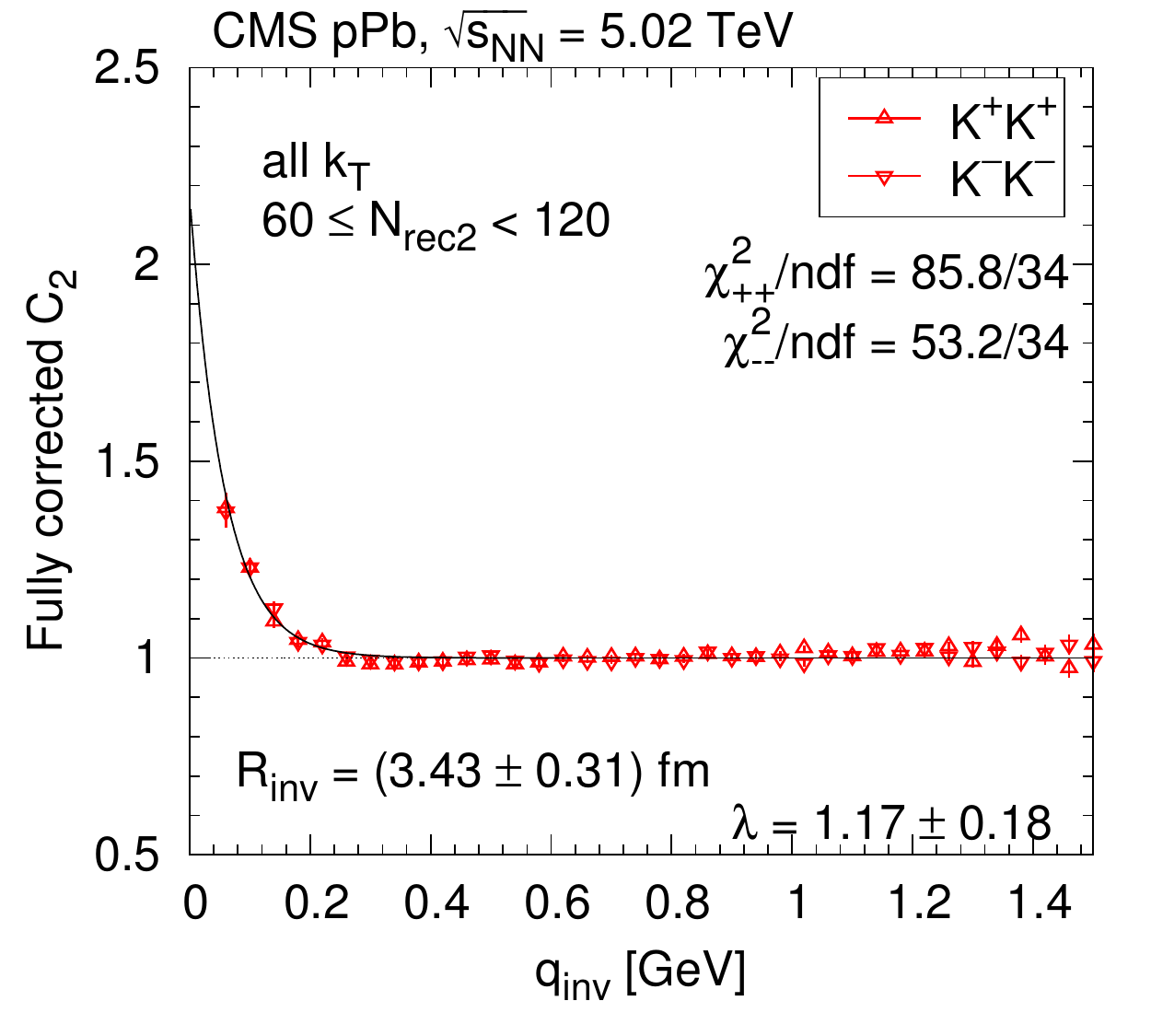}
  \includegraphics[width=0.49\textwidth]{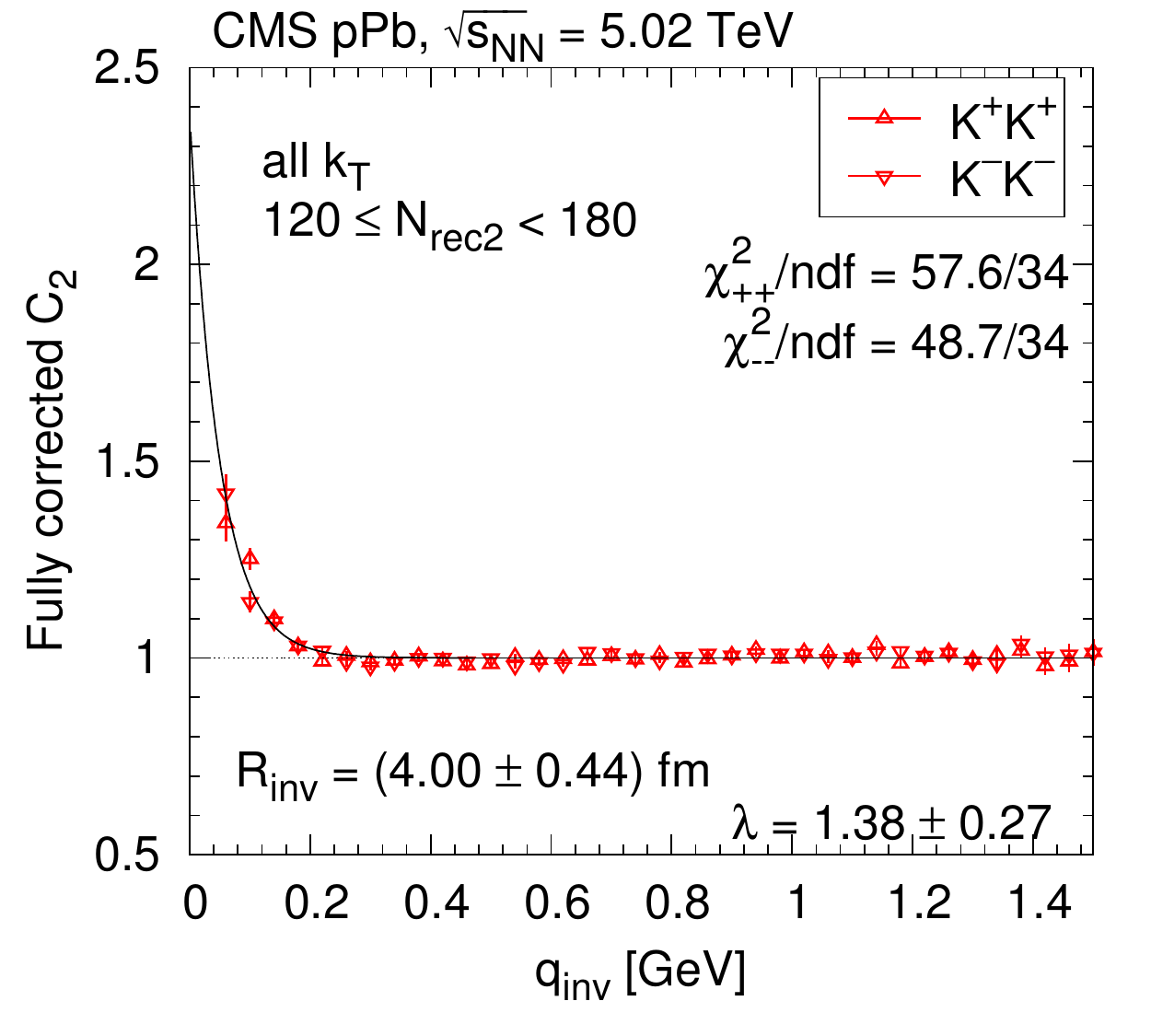}
  \includegraphics[width=0.49\textwidth]{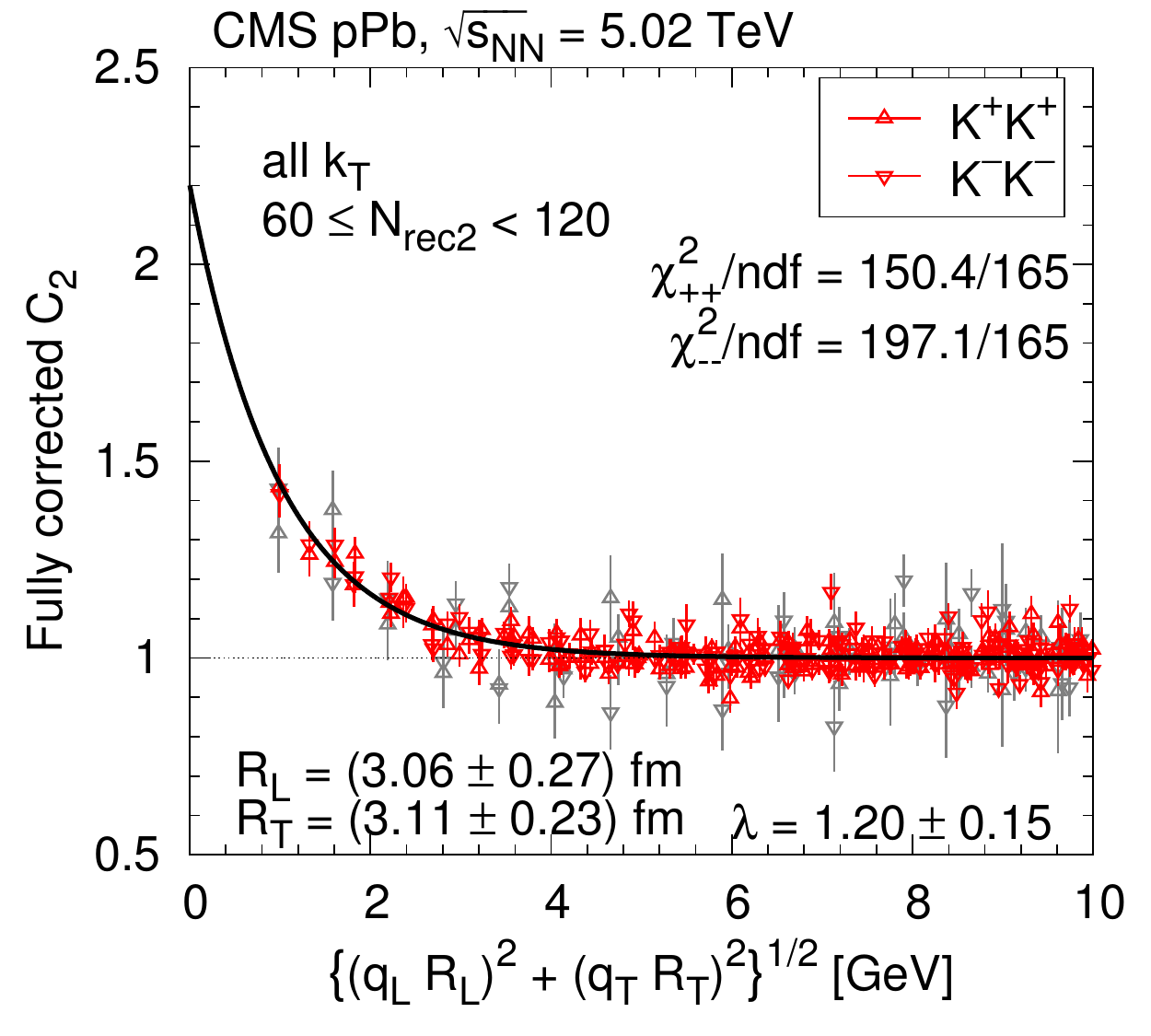}
  \includegraphics[width=0.49\textwidth]{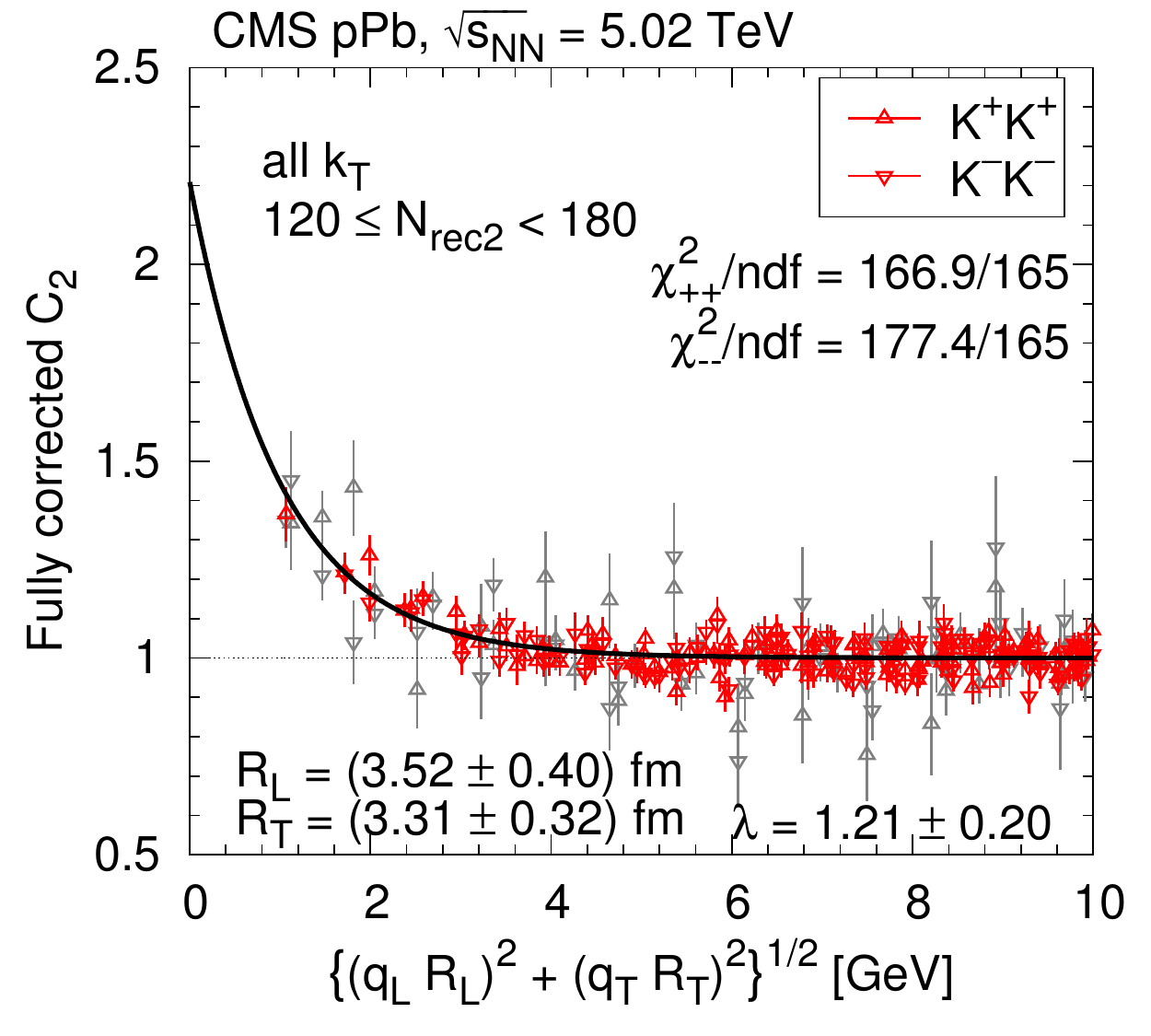}
 \caption{SS correlation function (integrated over all $\kt$) measured in $\Pp$Pb
collisions at 5.02\TeV as a function of $\qi$ or of the combined momentum, in
two selected $\Nr$ bins (left and right), for kaon pairs (red triangles) in one
(top) and two (bottom) dimensions, corrected for Coulomb interaction and
cluster contributions (mini-jets and multi-body resonance decays). For better
visibility, only a fraction of the points is plotted, and those with
statistical uncertainty higher than 10\% are in light-grey color. The solid
curves indicate fits with the stretched exponential parametrization.}
 \label{fig:ka_pPb_5TeVc/qmix_}
\end{figure*}

A selection of correlation functions obtained using the mixed-event
prescription, as described in Section~\ref{sec:partpairs}, and the
corresponding fits are shown in Figs.~\ref{fig:pi_pPb_5TeVc/qmix_} (pions) and
\ref{fig:ka_pPb_5TeVc/qmix_} (kaons).
The SS correlation functions, corrected for Coulomb interaction and cluster
contribution (mini-jets and multi-body resonance decays), as a function of \qi,
$(\ql,\qt)$, and $(\ql, \qo, \qs)$ in selected $\Nr$ bins are plotted for
all \kt. The solid curves indicate fits with the stretched exponential
parametrization.
Although the probability of reconstructing multiple tracks from a single true
track is very small (about 0.1\%), the bin with the smallest $|\vec{q}|$ is not
used in the fits in order to avoid regions potentially containing pairs of
multiply-reconstructed tracks.

The fits are mostly of good quality, the corresponding $\chi^2/\mathrm{ndf}$
values being close to one in the 2D and 3D cases, for pions and kaons alike.
The stretched exponential parametrization provides an excellent match to the
data. The 1D correlation function shows a wide minimum around 0.5\GeV
(Fig.~\ref{fig:pi_pPb_5TeVc/qmix_}, upper panel). The amplitude of that very
shallow dip is of the order of a few percent only, and is compatible with the
previously discussed results.

\begin{figure*}
 \centering
  \includegraphics[width=0.45\textwidth]{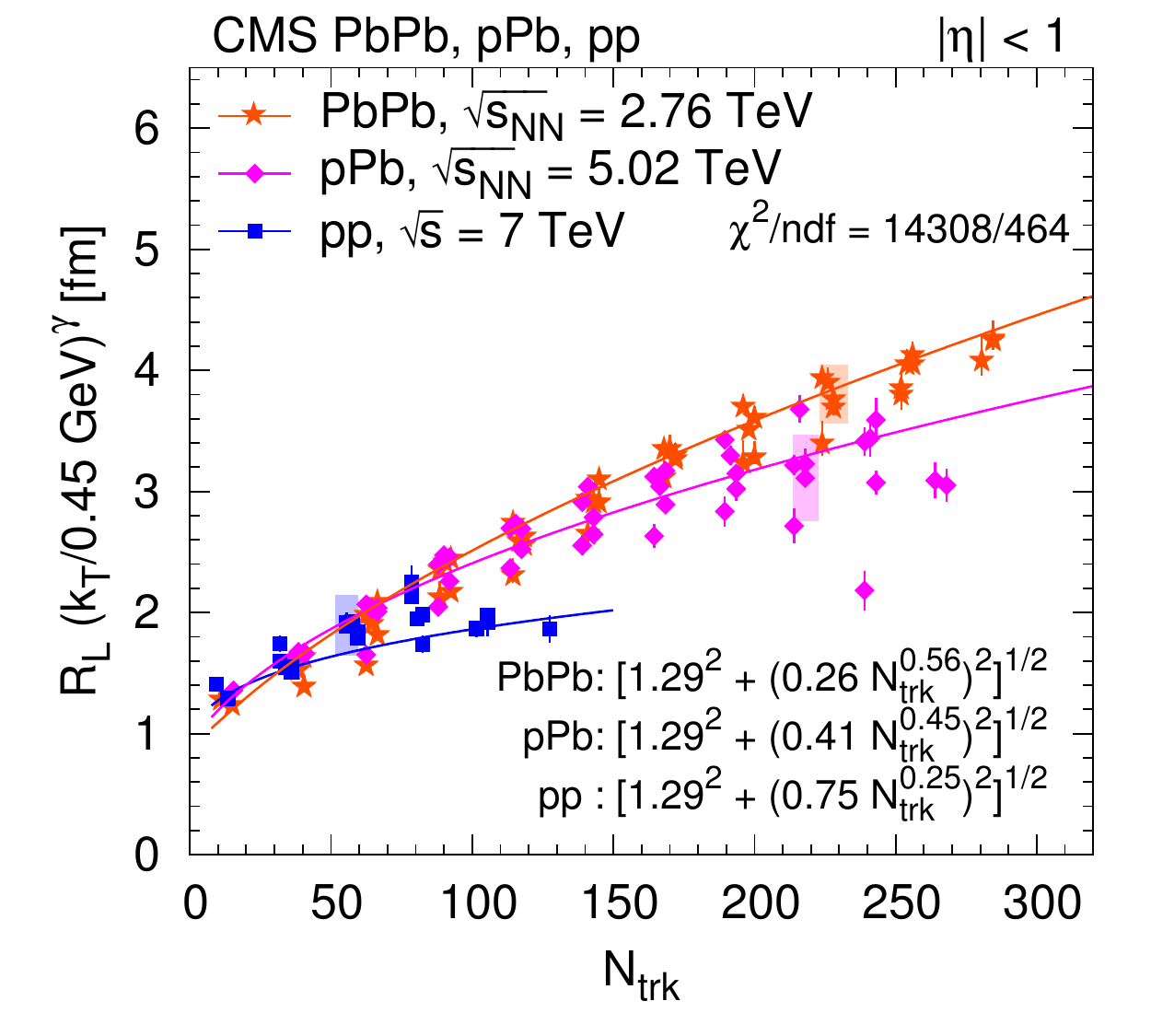}
  \includegraphics[width=0.45\textwidth]{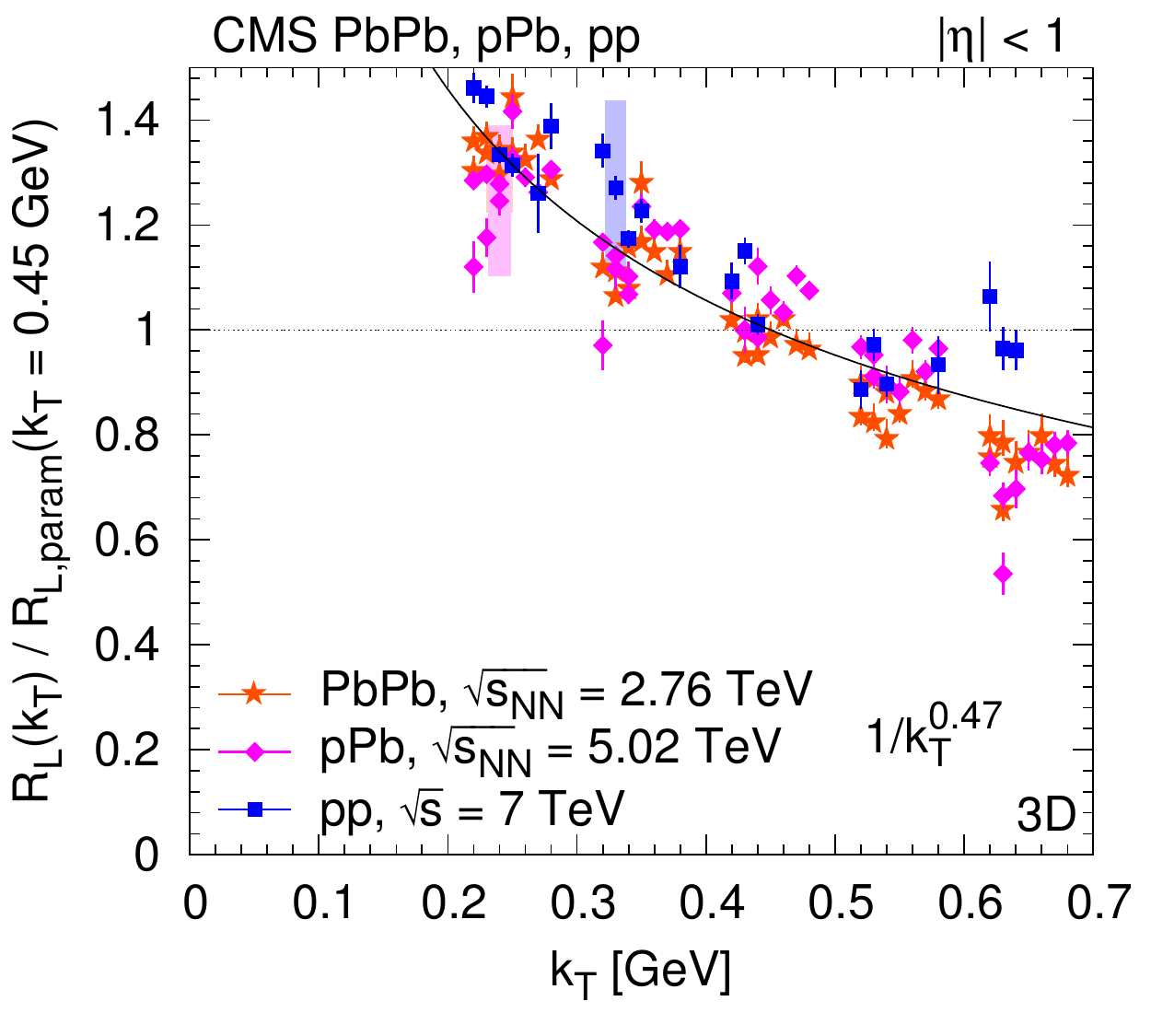}
  \includegraphics[width=0.45\textwidth]{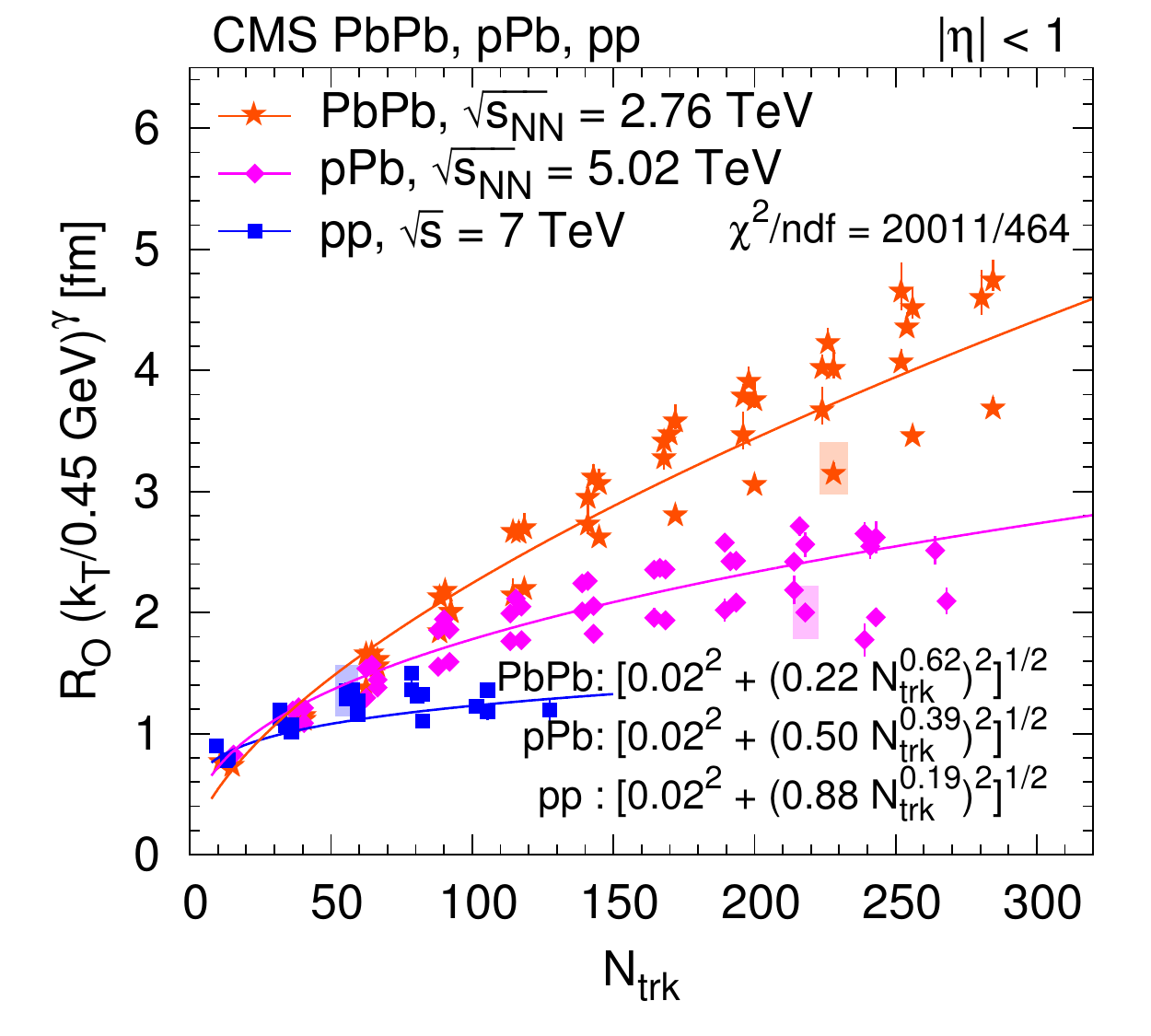}
  \includegraphics[width=0.45\textwidth]{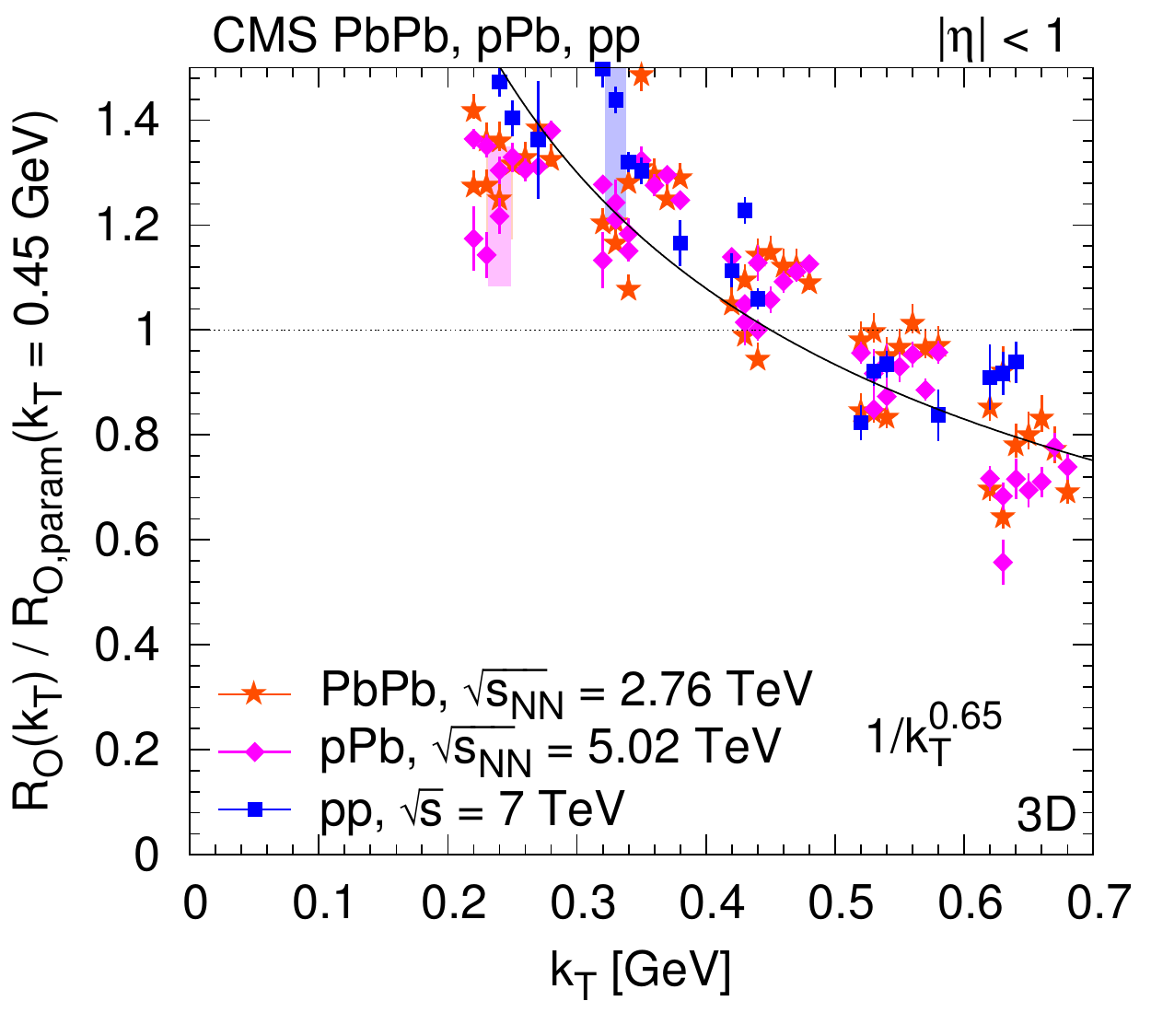}
  \includegraphics[width=0.45\textwidth]{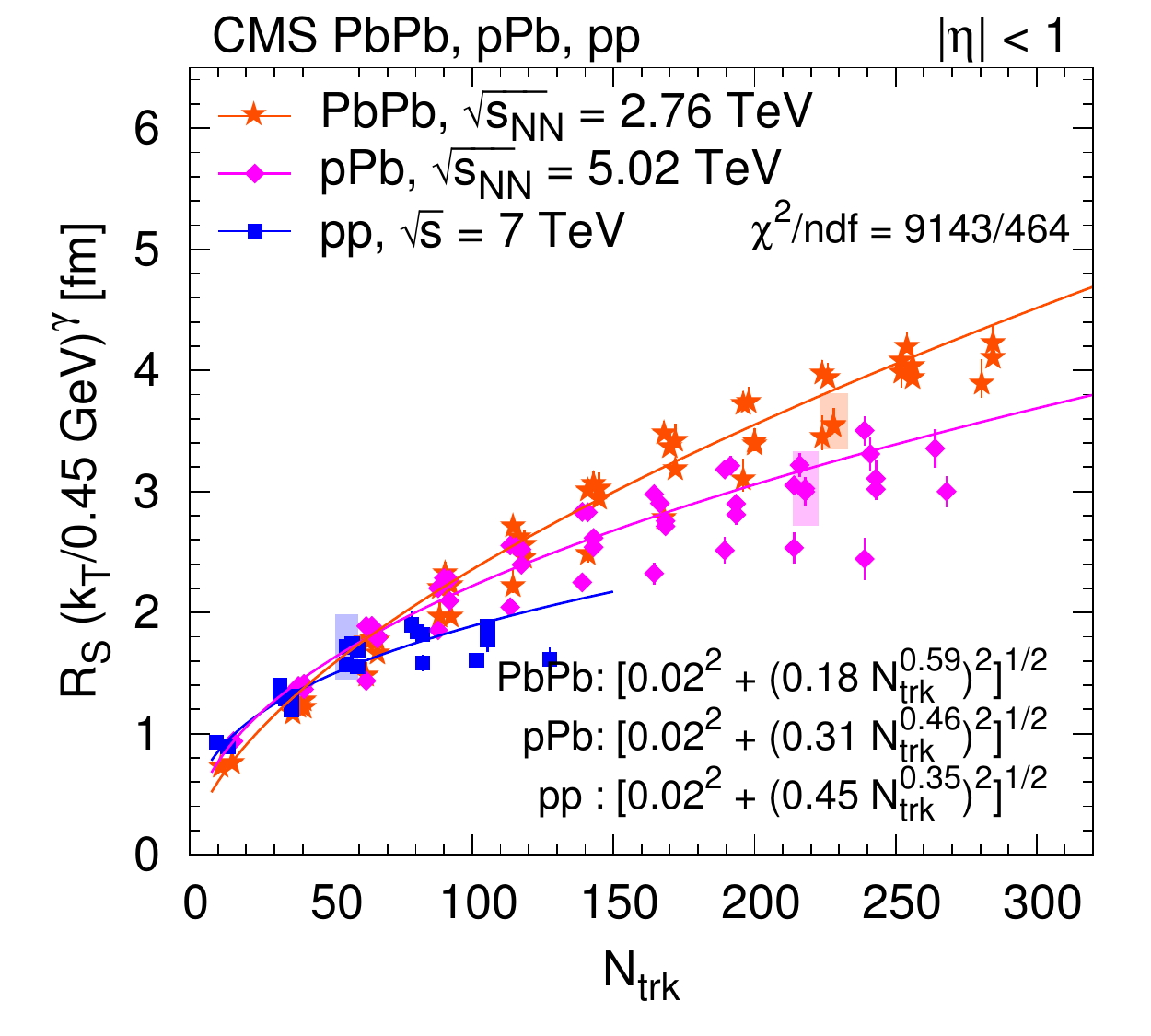}
  \includegraphics[width=0.45\textwidth]{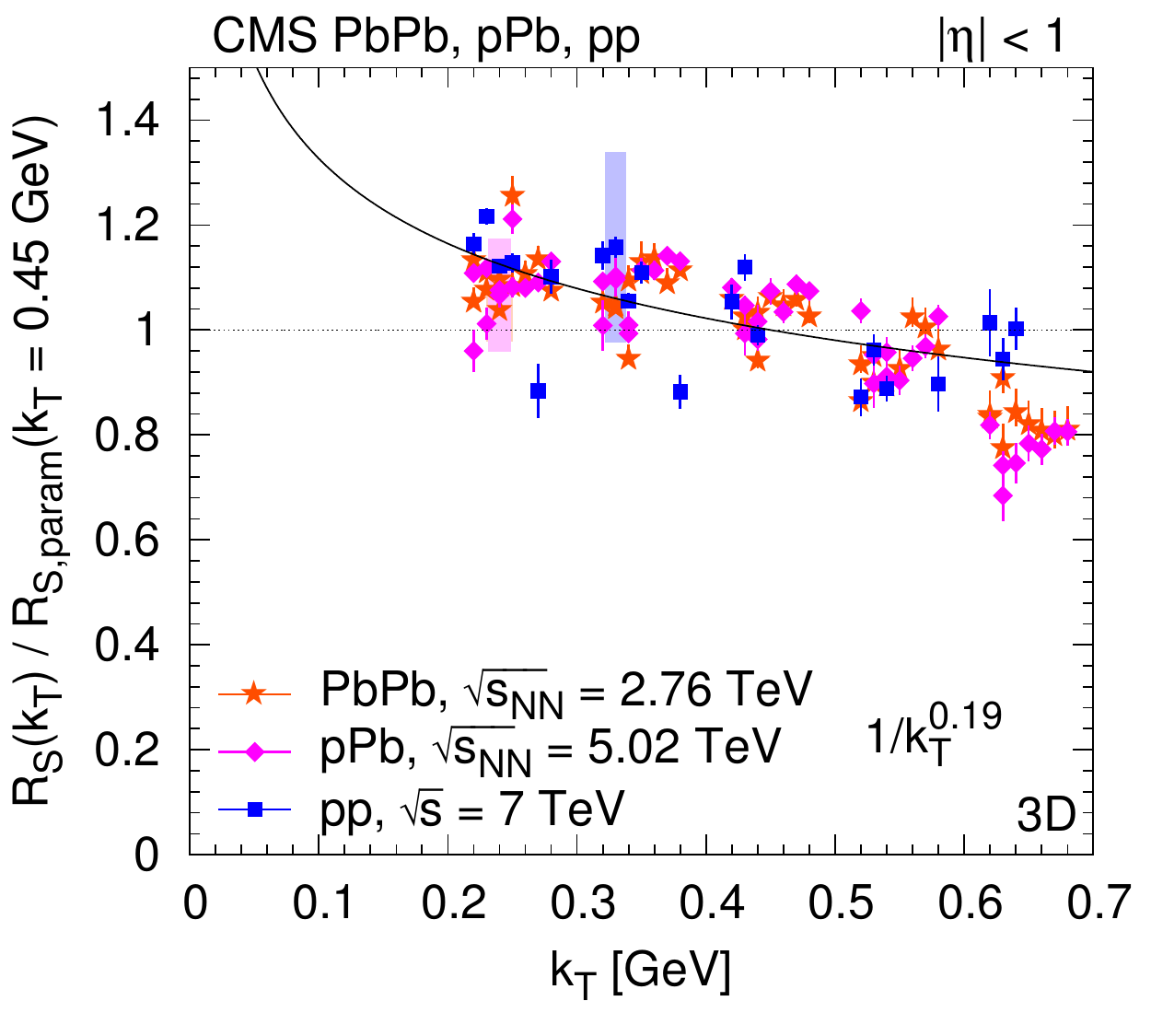}
 \caption{Left: Radius parameters for pions as a function of $\Nt$ scaled to
$\kt = 0.45\GeV$ with help of the parametrization as $R (\kt/0.45\GeV)^\gamma$
(Eq.~\eqref{eq:frakR}). Right: Ratio of the radius parameter and the value of
the parametrization $\Rp$ (Eq.~\eqref{eq:frakR}) at $\kt = 0.45\GeV$ as a
function $\kt$. (For better visibility, points are shifted to left and to right
with respect to the center of the $\kt$ bin. Systematic uncertainties are
indicated with shaded boxes, only for a fraction of the points.) Radii from
the 3D $(\ql,\qo,\qs)$ analysis are shown, from upper to lower: $\Rl$, $\Ro$,
$\Rs$. Fit results are indicated in the figures, see text for details.}
 \label{fig:radius3_scale}
\end{figure*}

The characteristics of the extracted one-, two-, and three-dimensional
correlation functions of identified pions and kaons in $\Pp\Pp$, $\Pp$Pb, and peripheral
PbPb collisions are presented in
Figs.~\ref{fig:pi_radius_qinv}--\ref{fig:radius3_scale} as functions of the
pair average transverse momentum $\kt$ and of the corrected charged-particle
multiplicity $\Nt$ of the event (in the range $\abs{\eta} < 2.4$ in the laboratory
frame, with integrated results extrapolated to $\pt = 0$).
In all plots, the results of
positively and negatively charged pions and kaons are averaged.
The central values for the radii and intercept parameter $\lambda$ are given by
markers. The statistical uncertainties are indicated by vertical error bars,
while the combined systematic uncertainties (choice of background method;
uncertainty in the relative amplitude $z$ of the cluster contribution; low $q$
exclusion) are given by the filled boxes. Unless indicated, the lines are drawn
to guide the eye (cubic splines whose coefficients are found by weighting the
measurements with the inverse of their squared statistical uncertainty).

\subsection{Dependencies on \texorpdfstring{$\Nt$ and $\kt$}{Ntrk and kT}}

\begin{figure}
 \centering
  \includegraphics[width=0.49\textwidth]{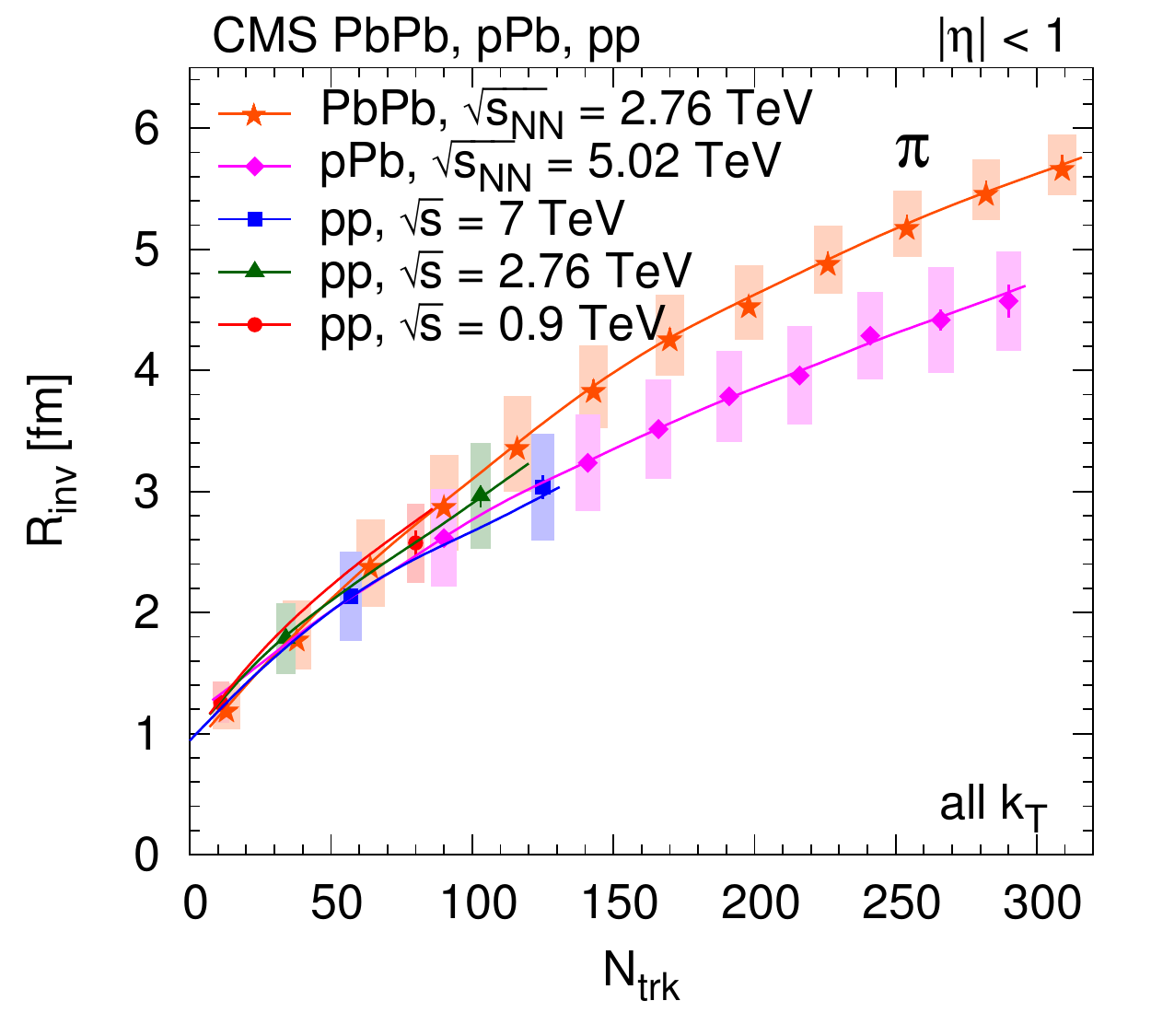}
  \includegraphics[width=0.49\textwidth]{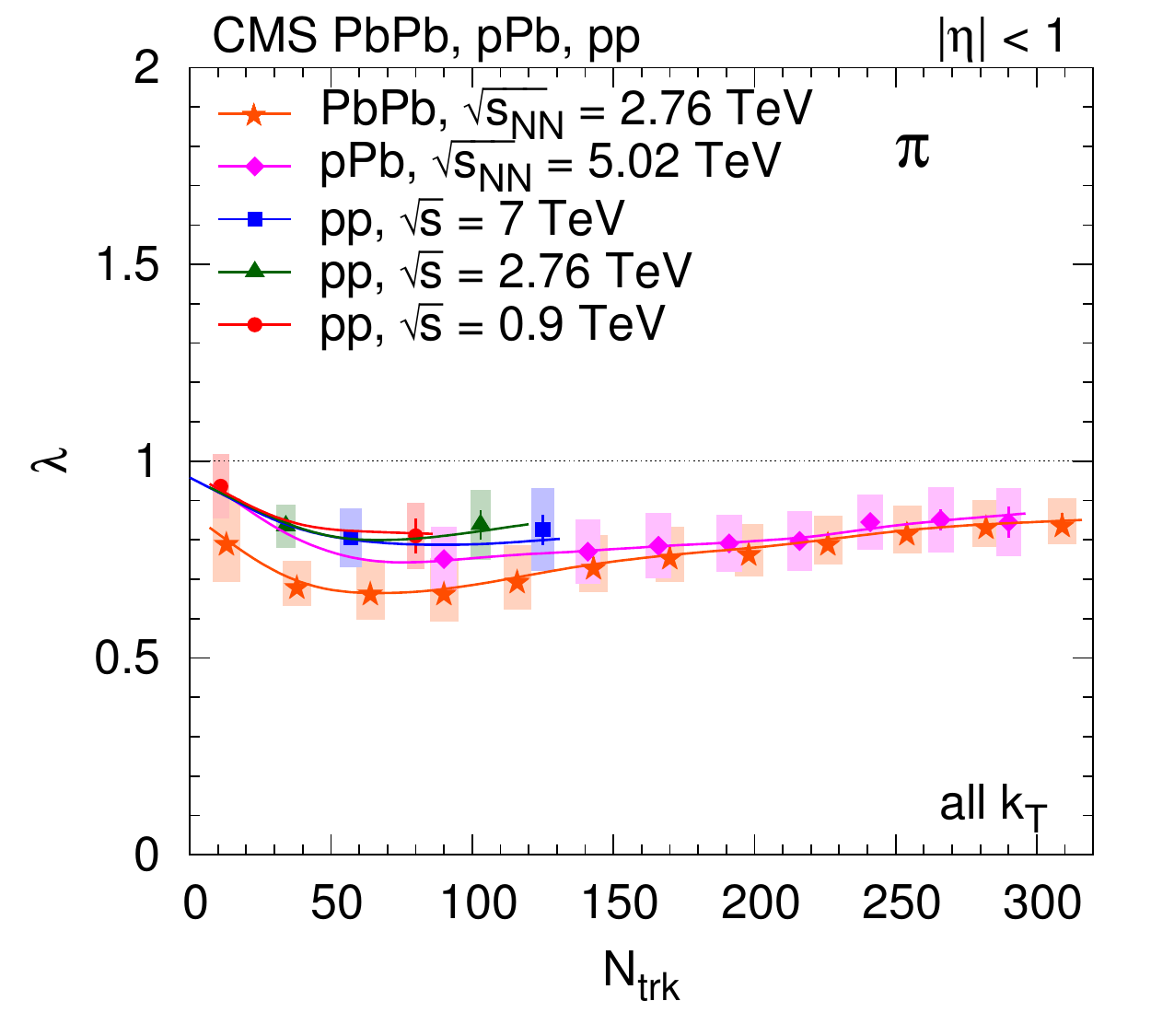}
 \caption{Track-multiplicity ($\Nt$) dependence of the 1D pion radius parameter
$\Ri$ (\cmsLeft) and intercept parameter $\lambda$ (\cmsRight), obtained from 1D fits
(integrated over all $\kt$) for all collision systems studied.
For better visibility, only every second measurement is plotted.  Lines are
drawn to guide the eye.}
 \label{fig:pi_radius_qinv}
\end{figure}

\begin{figure}[!ht]
 \centering
  \includegraphics[width=0.49\textwidth]{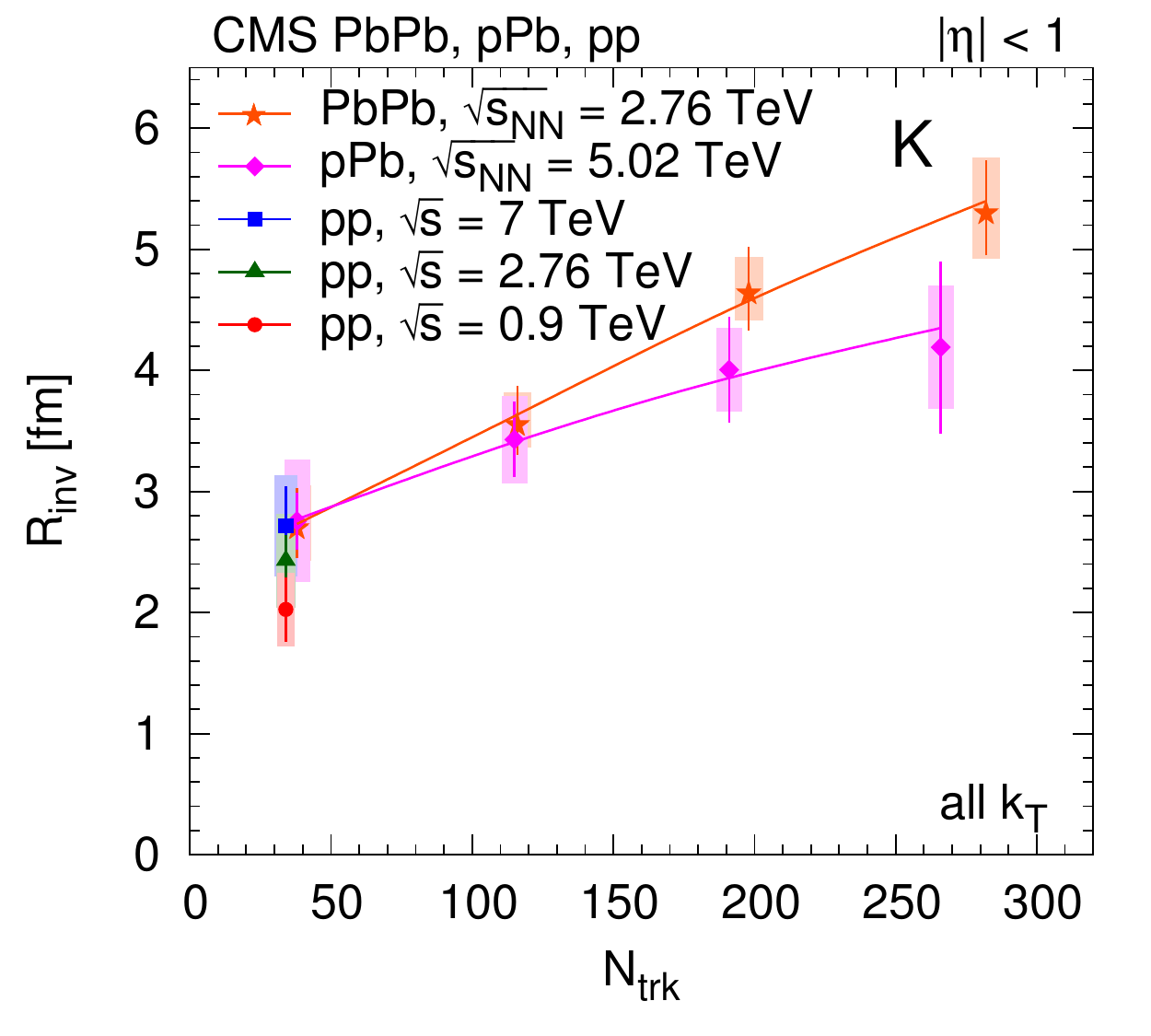}
  \includegraphics[width=0.49\textwidth]{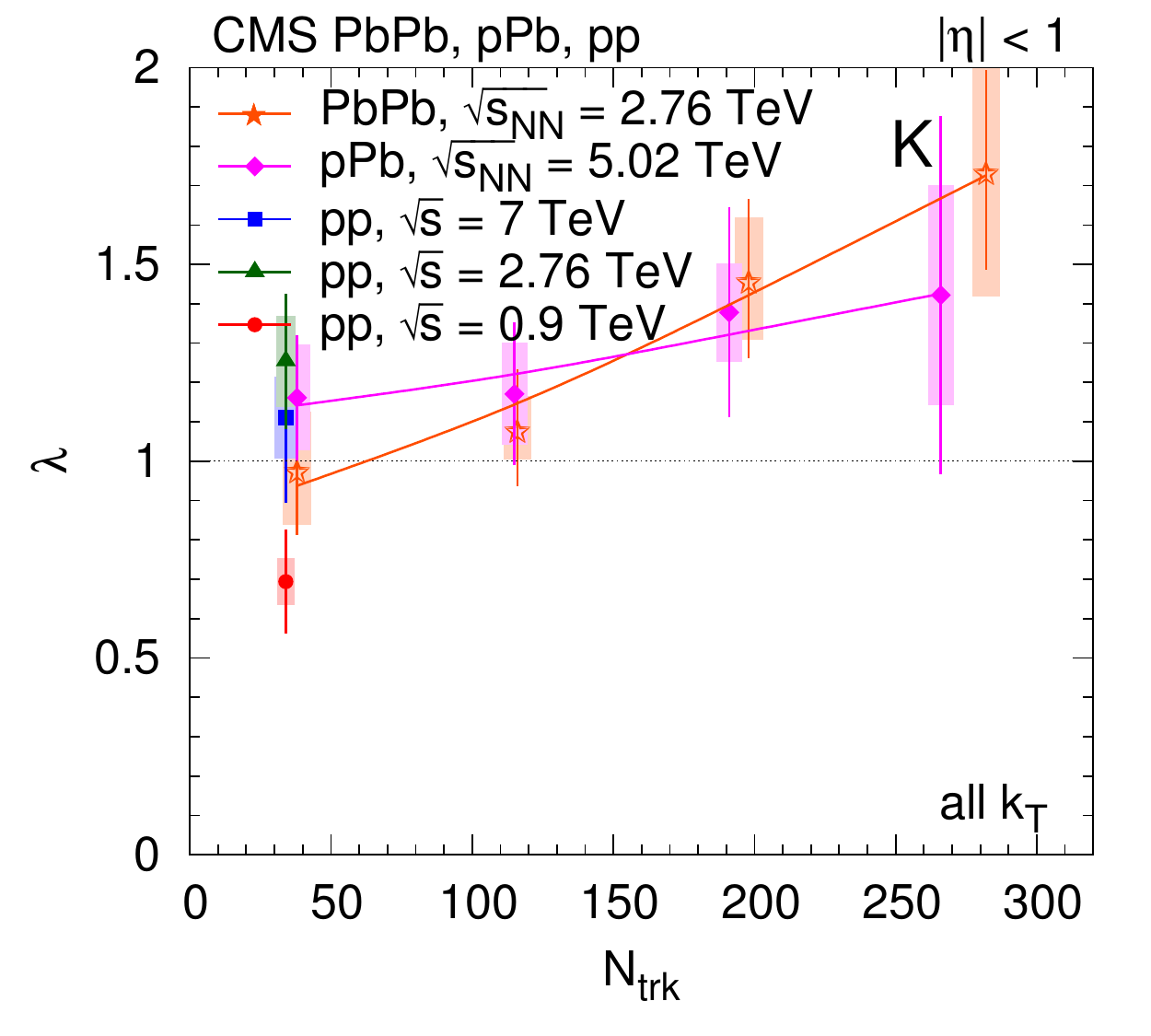}
  \caption{Track-multiplicity ($\Nt$) dependence of the 1D kaon radius parameter
$\Ri$ (\cmsLeft) and intercept parameter $\lambda$ (\cmsRight), obtained from 1D fits
(integrated over all $\kt$) for all collision systems studied.
For better visibility, only every second measurement is plotted.  Lines are
drawn to guide the eye.}
  \label{fig:kaon_1D}
\end{figure}

The $\Nt$ dependence of the 1D radius $\Ri$ and intercept $\lambda$ parameters
for pions (Eq.~\eqref{eq:pid_c2_1d}), averaged over several \kt bins, for all
studied systems are shown in Fig.~\ref{fig:pi_radius_qinv}.
The radii found for pions from $\Pp\Pp$ collisions span a similar range to those
found previously~\cite{cms-hbt-1st,cms-hbt-2nd}, as well as those obtained for
charged hadrons via the double ratio technique, as reported in Section
\ref{subsec:legacy-results-1d}. Although illustrated in
Fig.~\ref{fig:pi_radius_qinv} for the \kt-integrated case only, the dependence
on the number of tracks of the 1D pion radius parameter is remarkably similar
for $\Pp\Pp$ and $\Pp$Pb collisions in all \kt bins. This also applies to PbPb collisions
when $\kt > 0.4\GeV$.
The values of the pion intercept parameter $\lambda$ are usually below unity,
in the range 0.7--1, and seem to remain approximately constant as a function of
$\Nt$ for all systems and $\kt$ bins. Similar results obtained for kaons in the
1D analysis are displayed in Fig.~\ref{fig:kaon_1D}. The 1D kaon radius
parameter increases with $\Nt$, but with a smaller slope than seen for pions.

\begin{figure*}[!ht]
 \centering
  \includegraphics[width=0.49\textwidth]{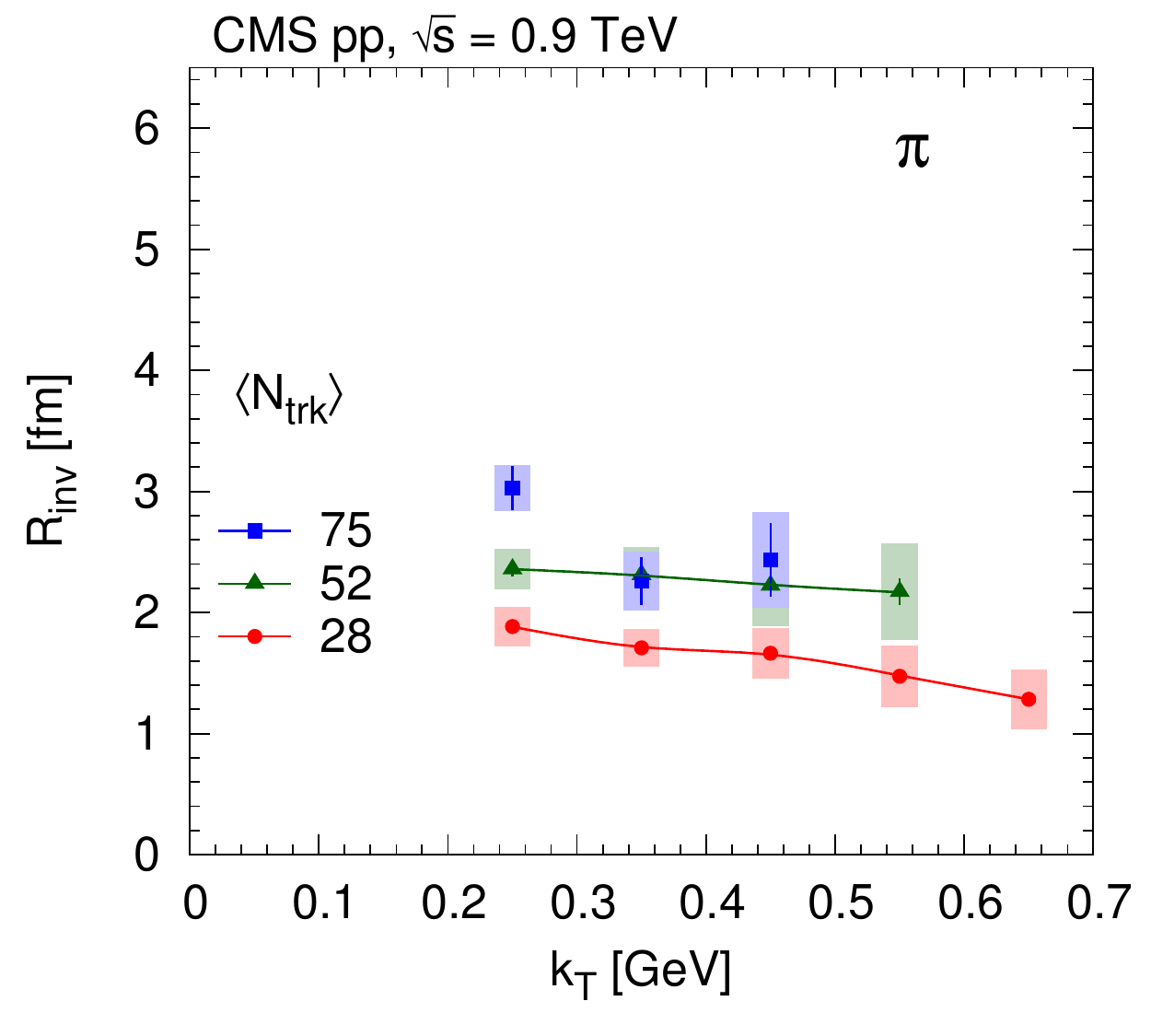}
  \includegraphics[width=0.49\textwidth]{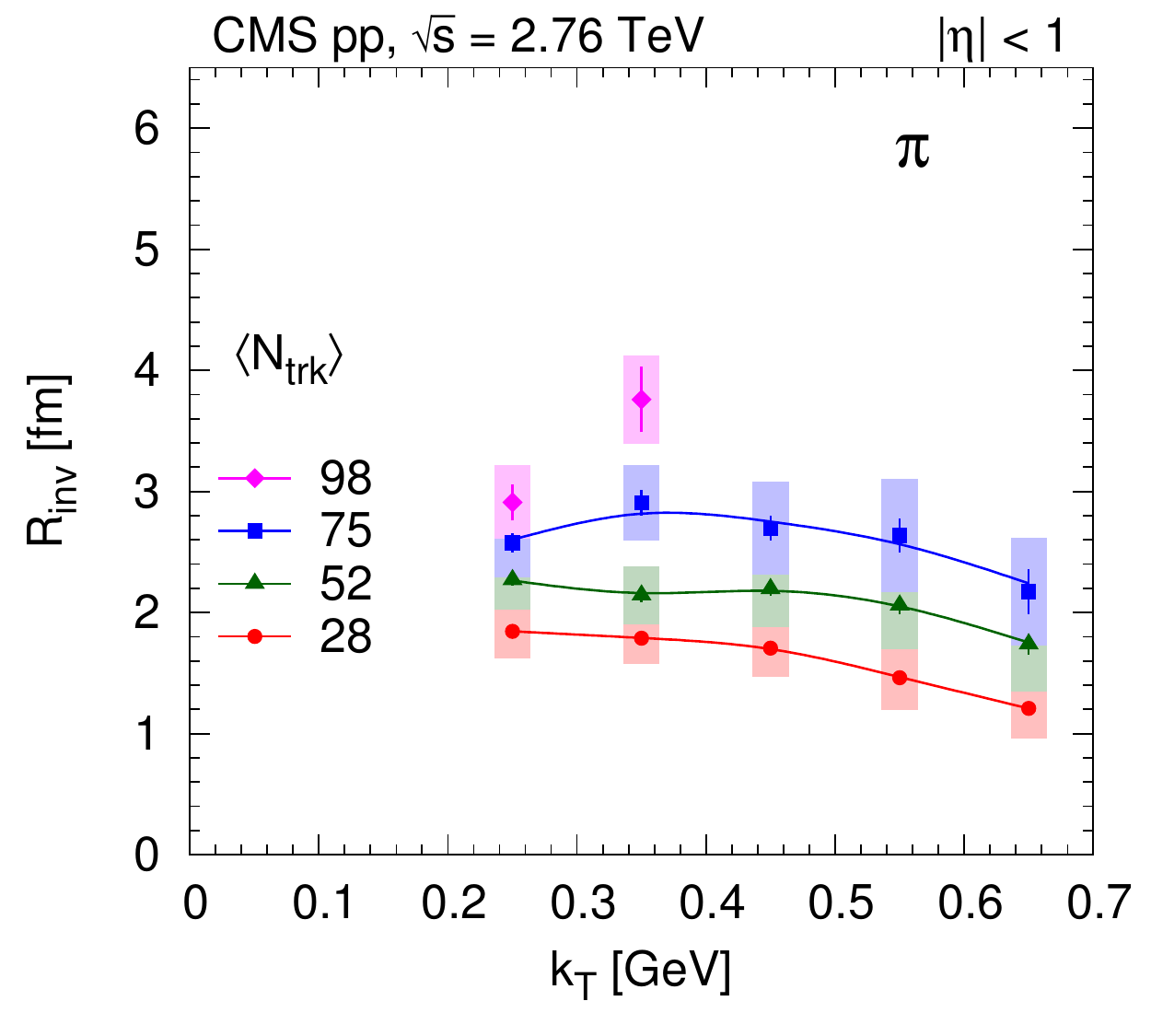}

  \includegraphics[width=0.49\textwidth]{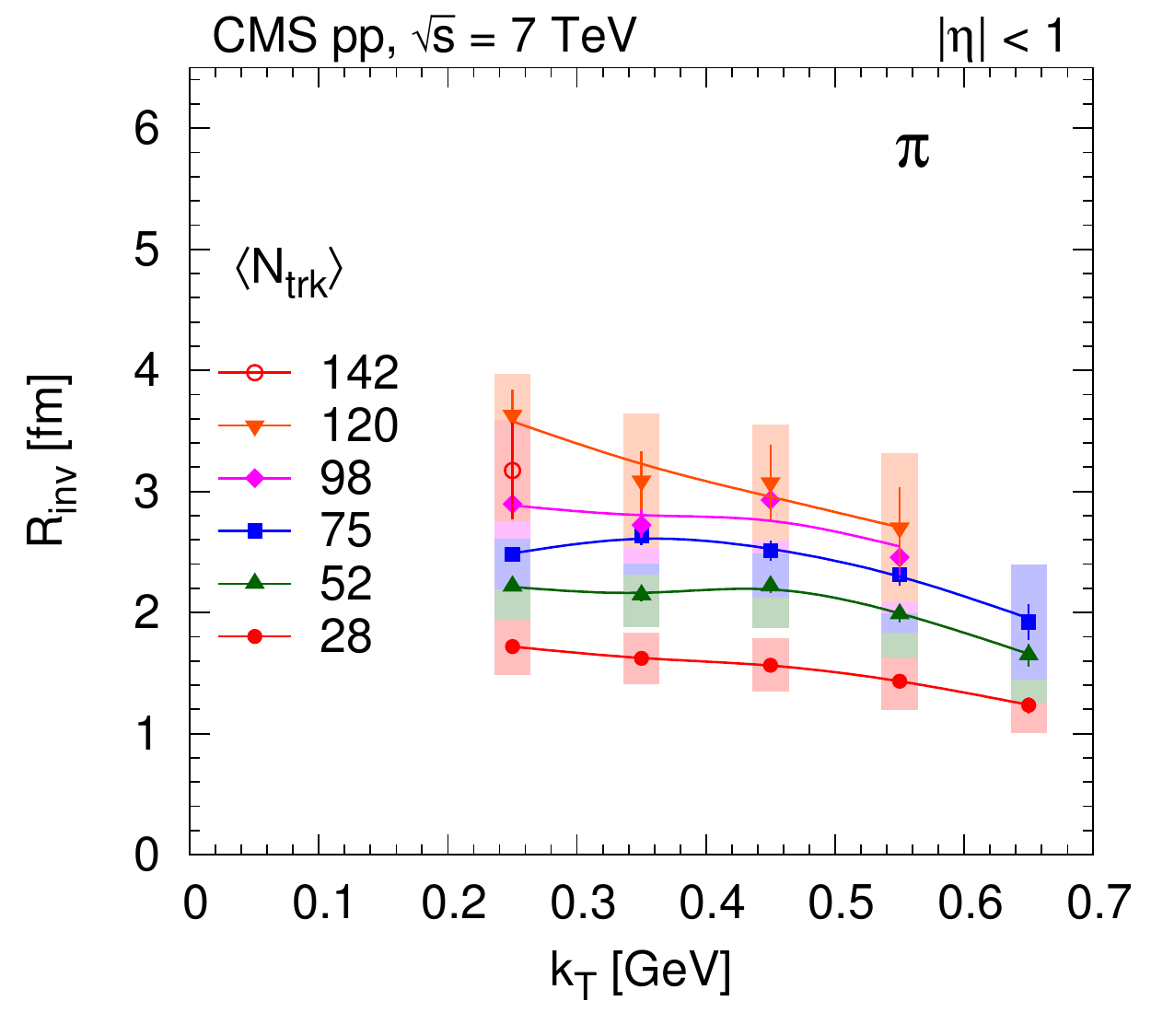}
  \includegraphics[width=0.49\textwidth]{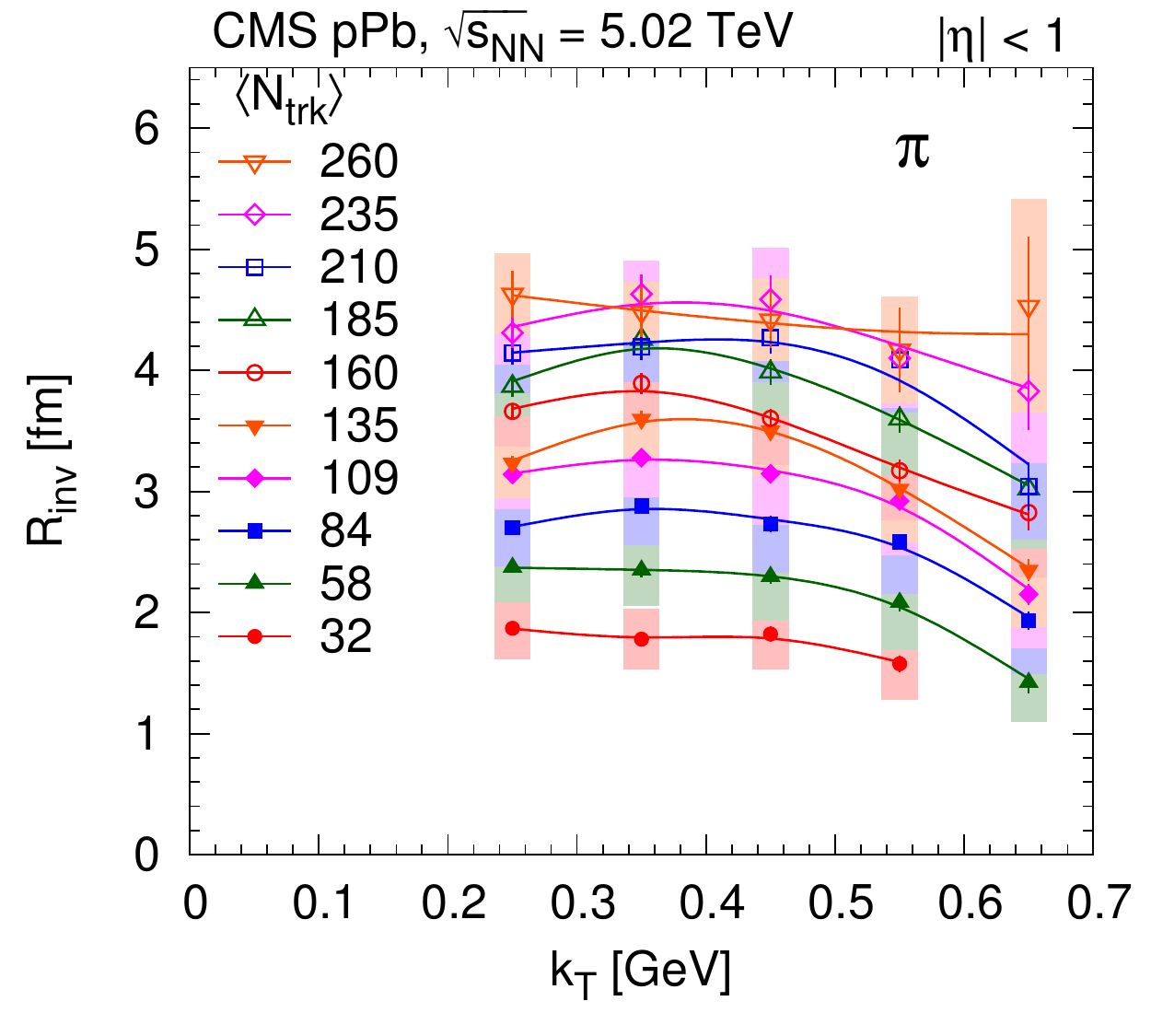}
 \caption{$\kt$-dependence of the 1D pion radius $\Ri$, for several $\Nt$ bins,
extracted for the $\Pp\Pp$ and $\Pp$Pb systems. Lines are drawn to guide the eye.}
 \label{fig:radius12_vs_kt}
\end{figure*}

The \kt dependencies of the 1D radius parameters for pions, in several $\Nt$
bins for the $\Pp\Pp$ and $\Pp$Pb systems are plotted in Fig.~\ref{fig:radius12_vs_kt}.
They show a decrease with increasing $\kt$ for $\kt > 0.4\GeV$.

\begin{figure*}[!htbp]
 \centering
  \includegraphics[width=0.49\textwidth]{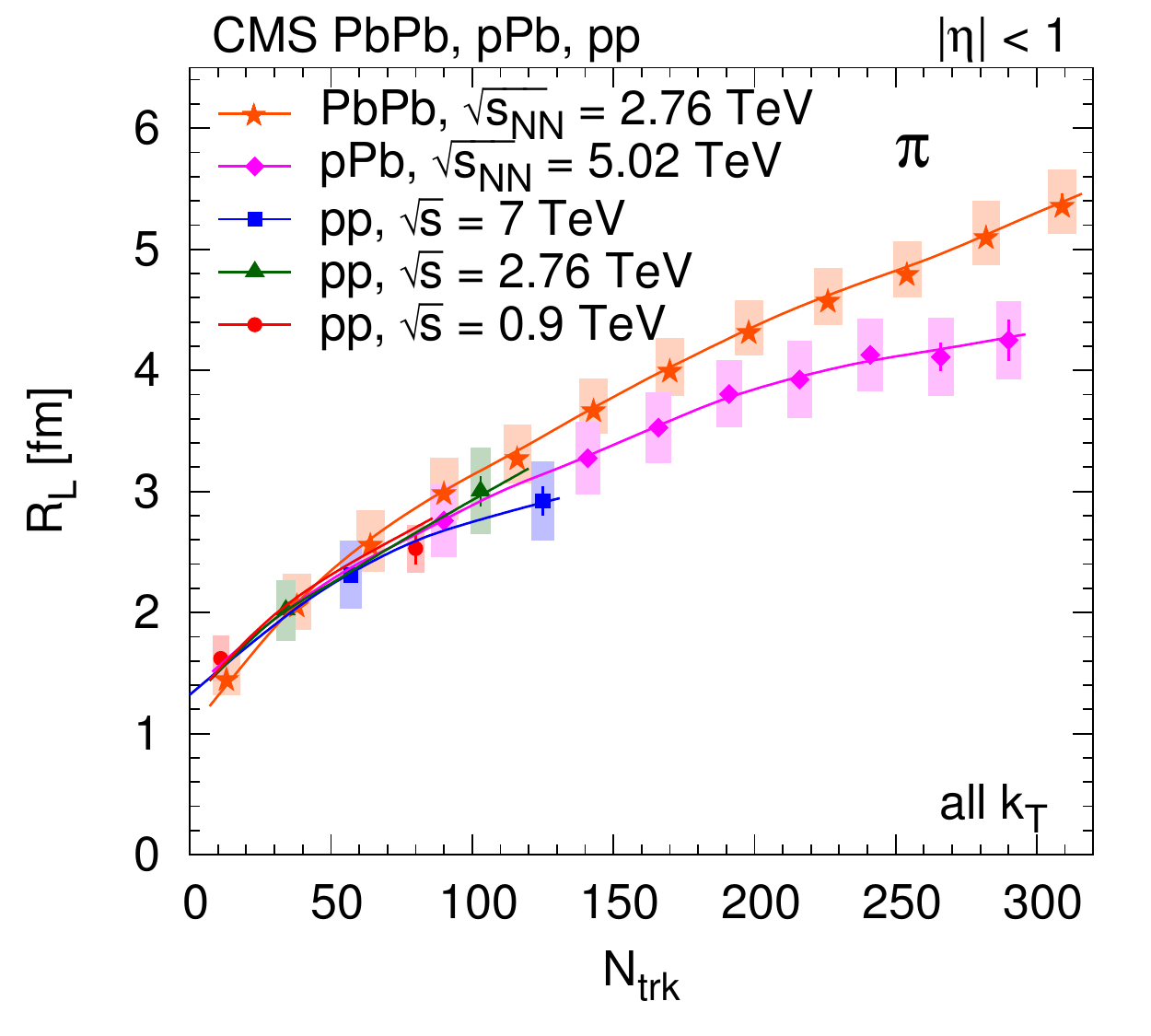}
  \includegraphics[width=0.49\textwidth]{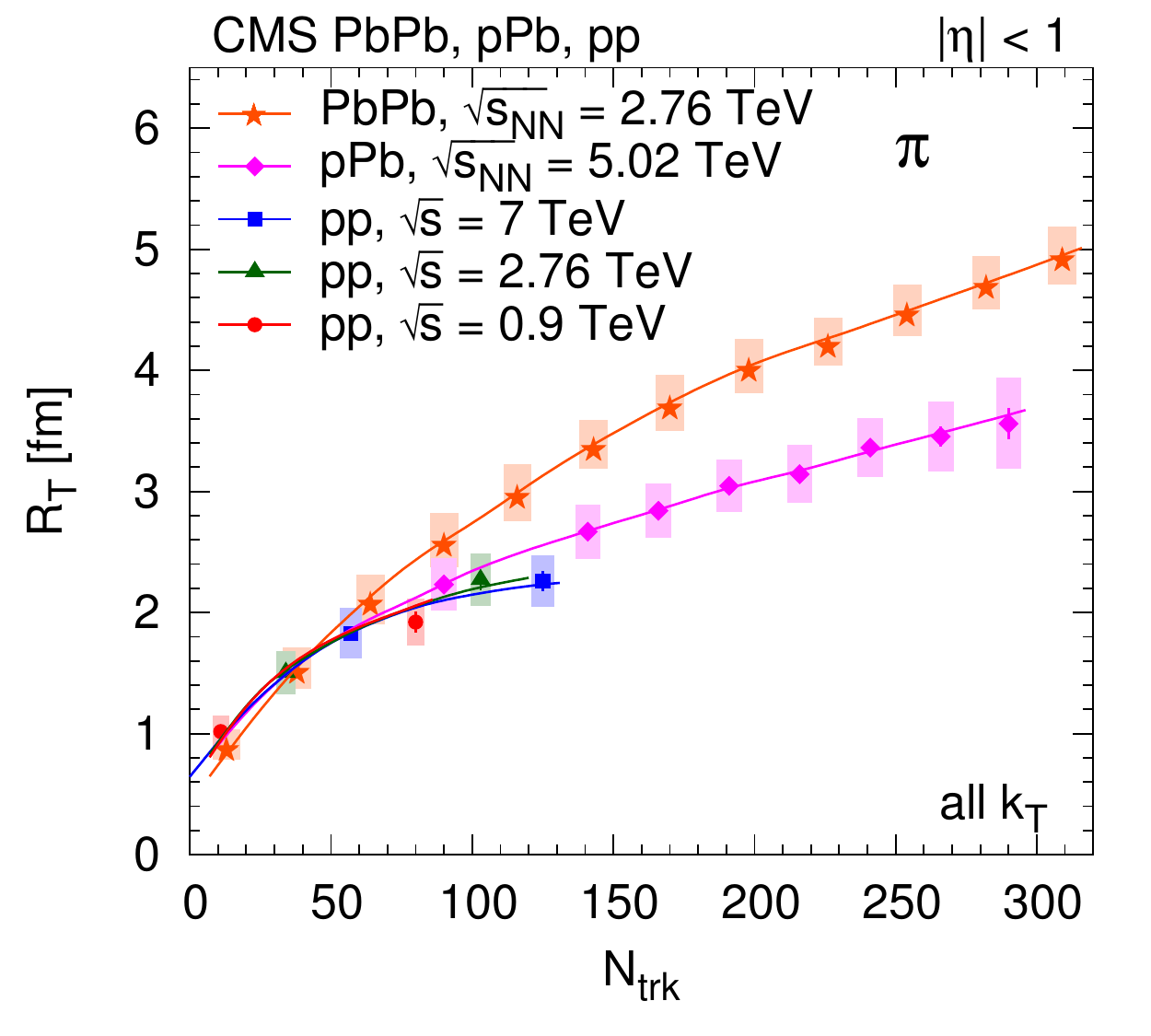}
  \includegraphics[width=0.49\textwidth]{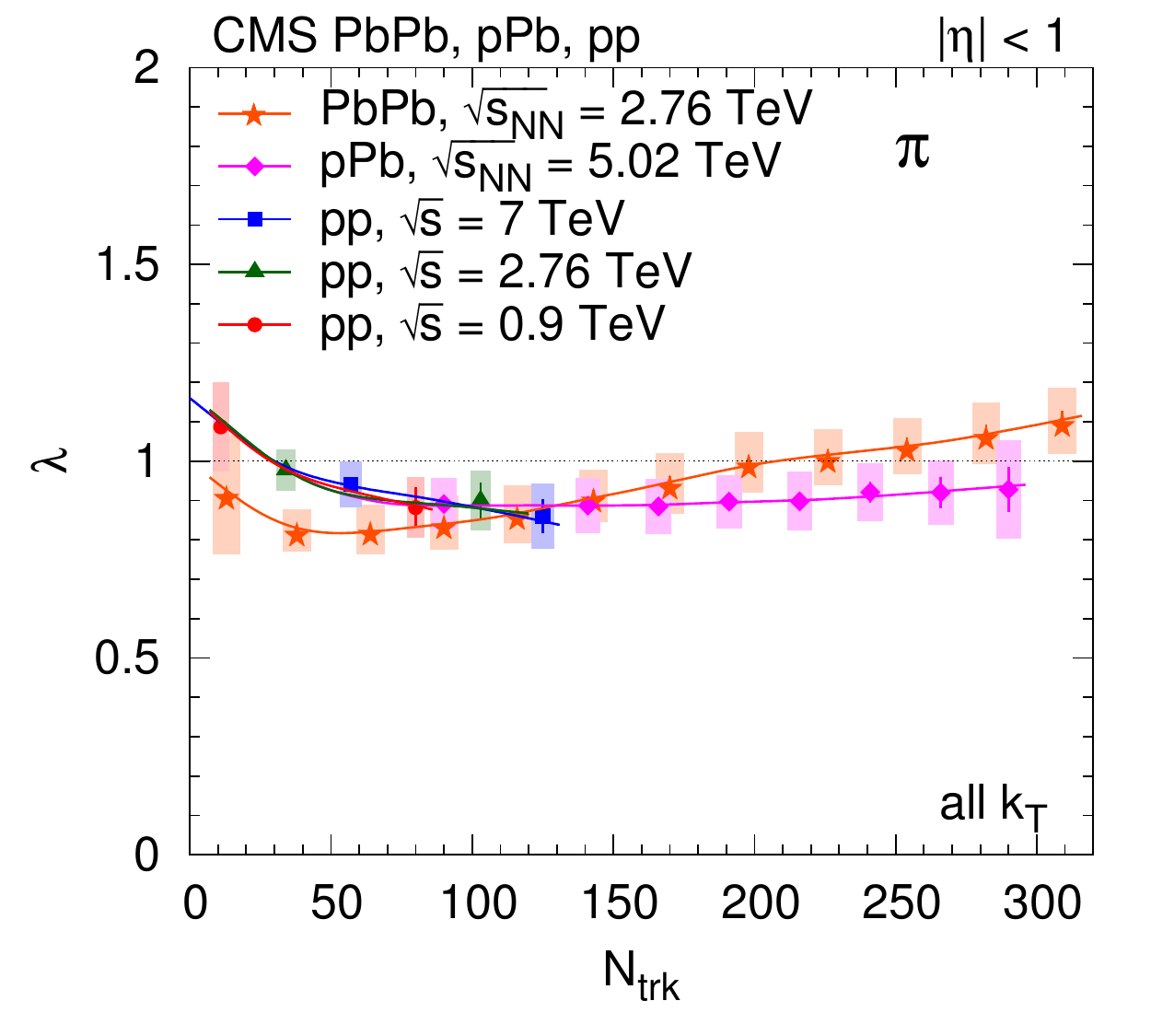}
 \caption{Track-multiplicity ($\Nt$) dependence of the 2D pion radius parameters
$\Rl$ and $\Rt$ (top), and intercept parameter $\lambda$ (bottom), obtained from
fits (integrated over all $\kt$) for all collision systems studied. For
better visibility, only every second measurement is plotted. Lines are drawn to
guide the eye.}
 \label{fig:pi_radius_qlqt}
\end{figure*}

\begin{figure*}[!ht]
 \centering
  \includegraphics[width=0.49\textwidth]{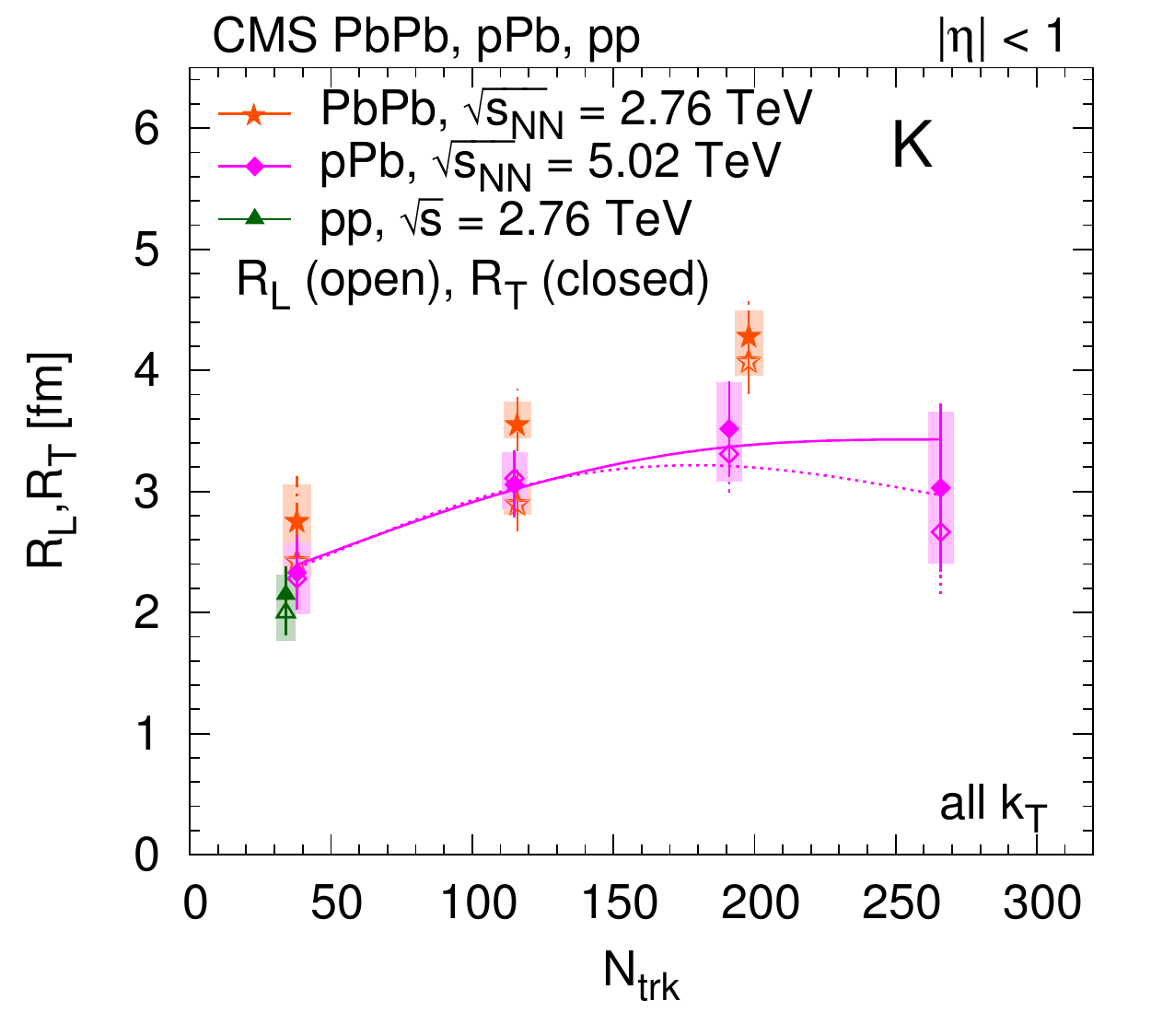}
  \includegraphics[width=0.49\textwidth]{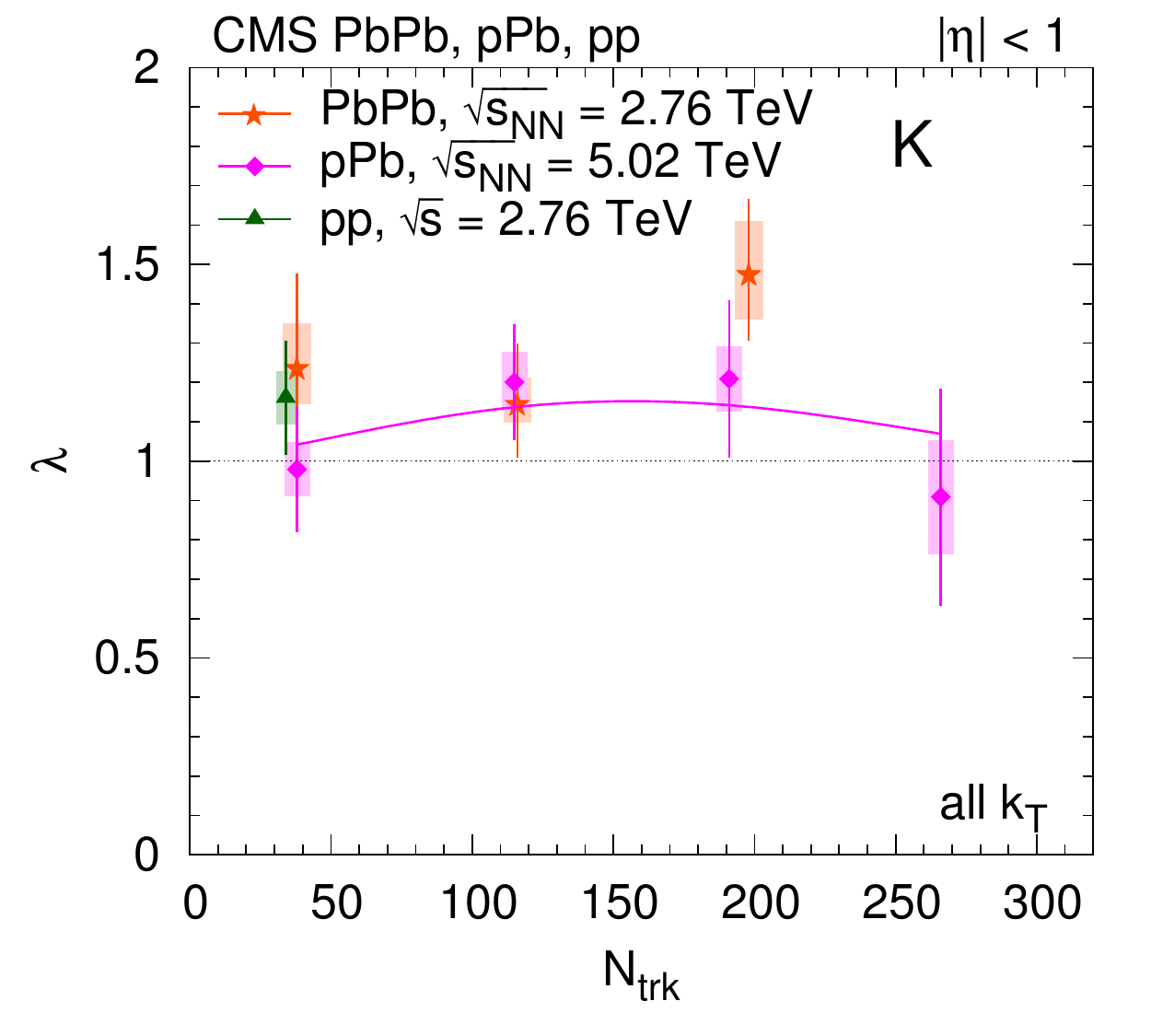}
 \caption{Track-multiplicity ($\Nt$) dependence of the 2D kaon radius
parameters (left; $\Rl$ -- open symbols, $\Rt$ -- closed symbols) and
intercept parameter $\lambda$ (right), obtained from fits (integrated over all
$\kt$) for all collision systems studied.  For better visibility, only every
second measurement is plotted. Lines are drawn to guide the eye.}
 \label{fig:kaon_2D}
\end{figure*}

The $\Nt$ dependence of the 2D radii and the corresponding intercept parameters
for pions (Eq.~\eqref{eq:pid_c2_23d}), in several \kt bins, for all studied
systems are shown in Fig.~\ref{fig:pi_radius_qlqt}.
The $\Nt$ dependence of 2D pion radius parameters, $\Rl$ and $\Rt$, is similar
for $\Pp\Pp$ and $\Pp$Pb collisions in all \kt bins. In general $\Rl > \Rt$, indicating
that the source is elongated in the beam direction, a behavior also
seen in $\Pp\Pp$ collisions with the double ratio technique, as discussed in
Section~\ref{subsec:legacy-results-1d}. In the case of PbPb collisions $\Rl
\approx \Rt$, indicating that the source is approximately symmetric.
In the case of kaons, the results corresponding to the 2D case are shown in
Fig.~\ref{fig:kaon_2D}.

\begin{figure*}[!ht]
 \centering
  \includegraphics[width=0.49\textwidth]{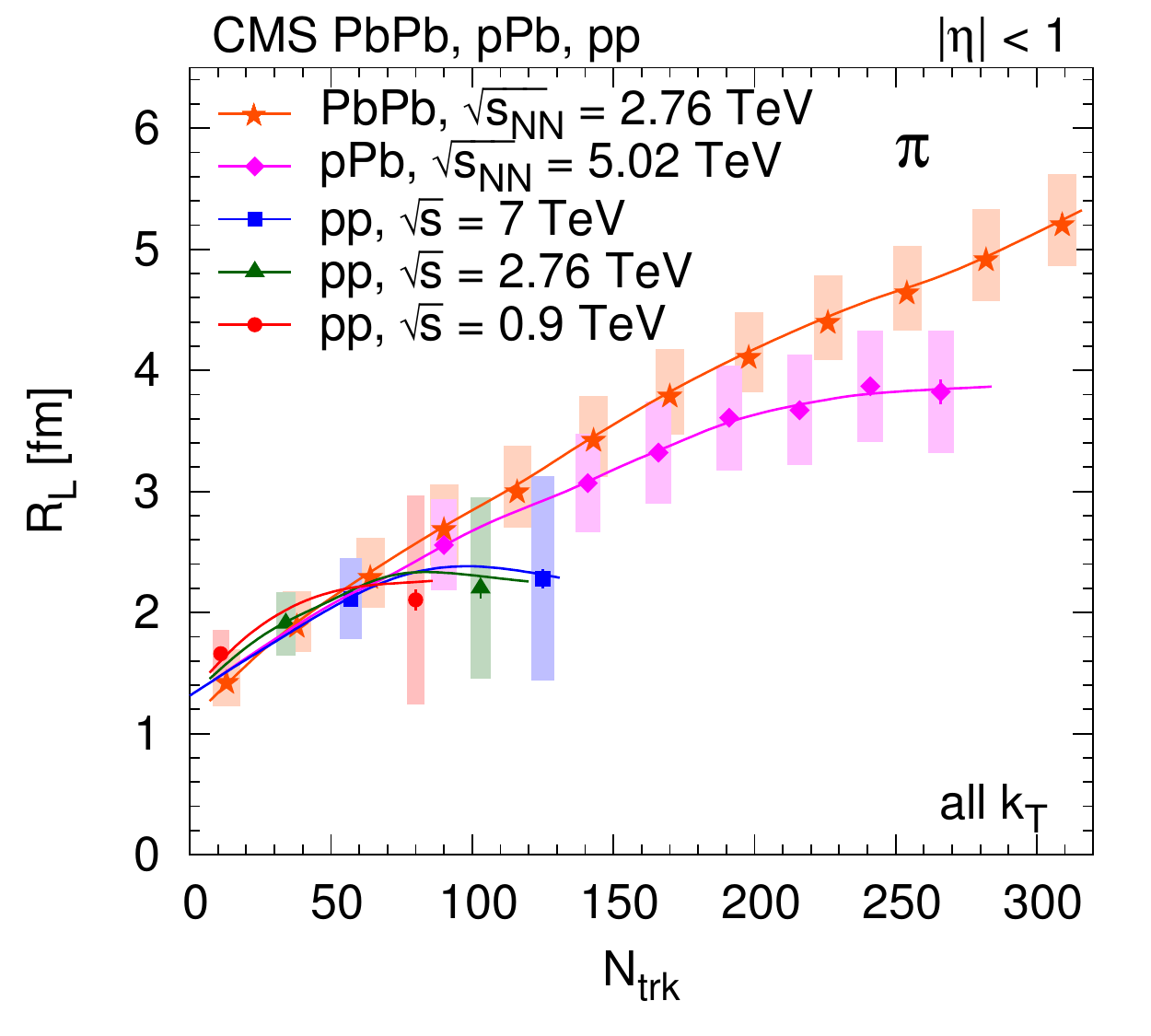}
  \includegraphics[width=0.49\textwidth]{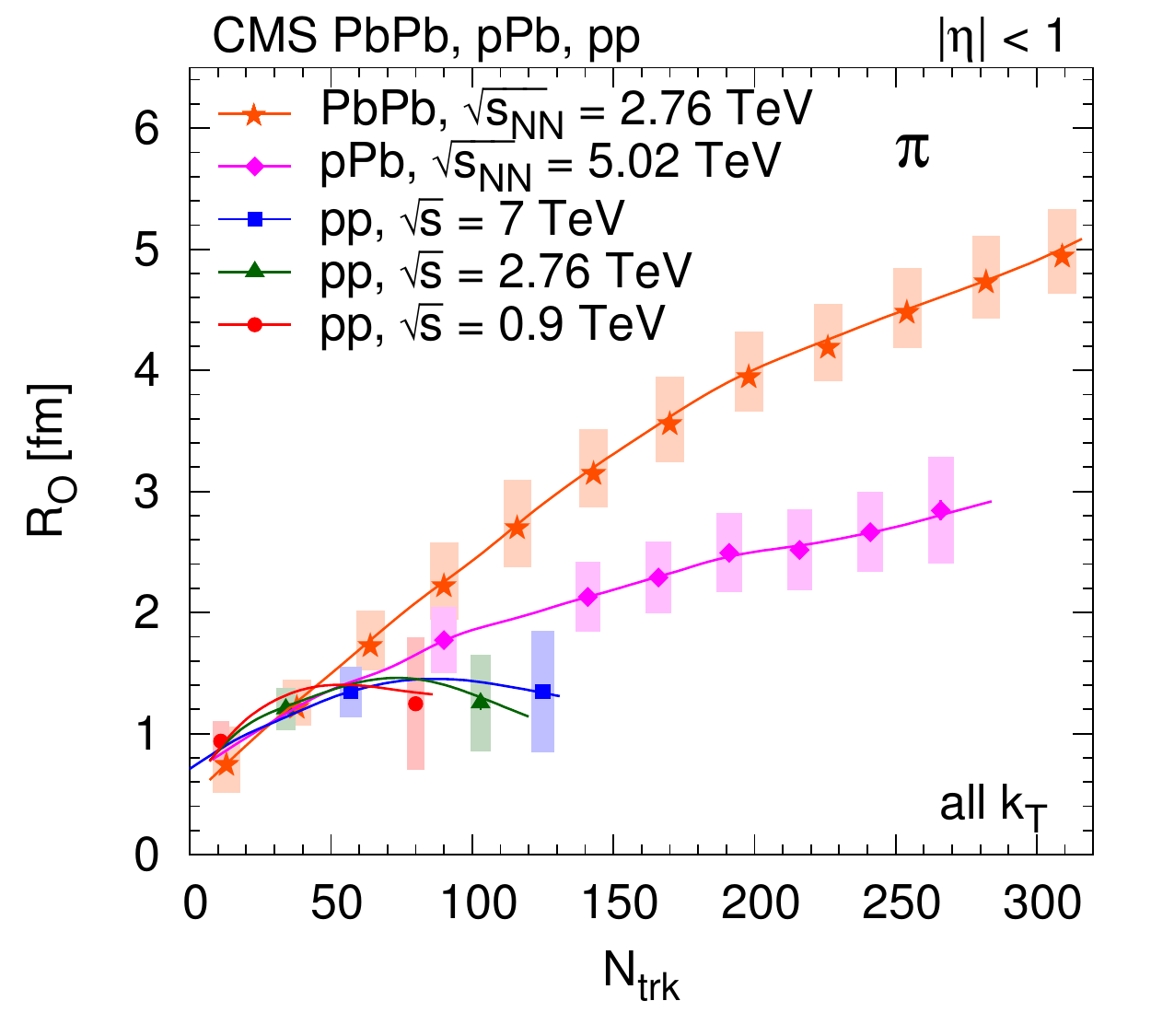}
  \includegraphics[width=0.49\textwidth]{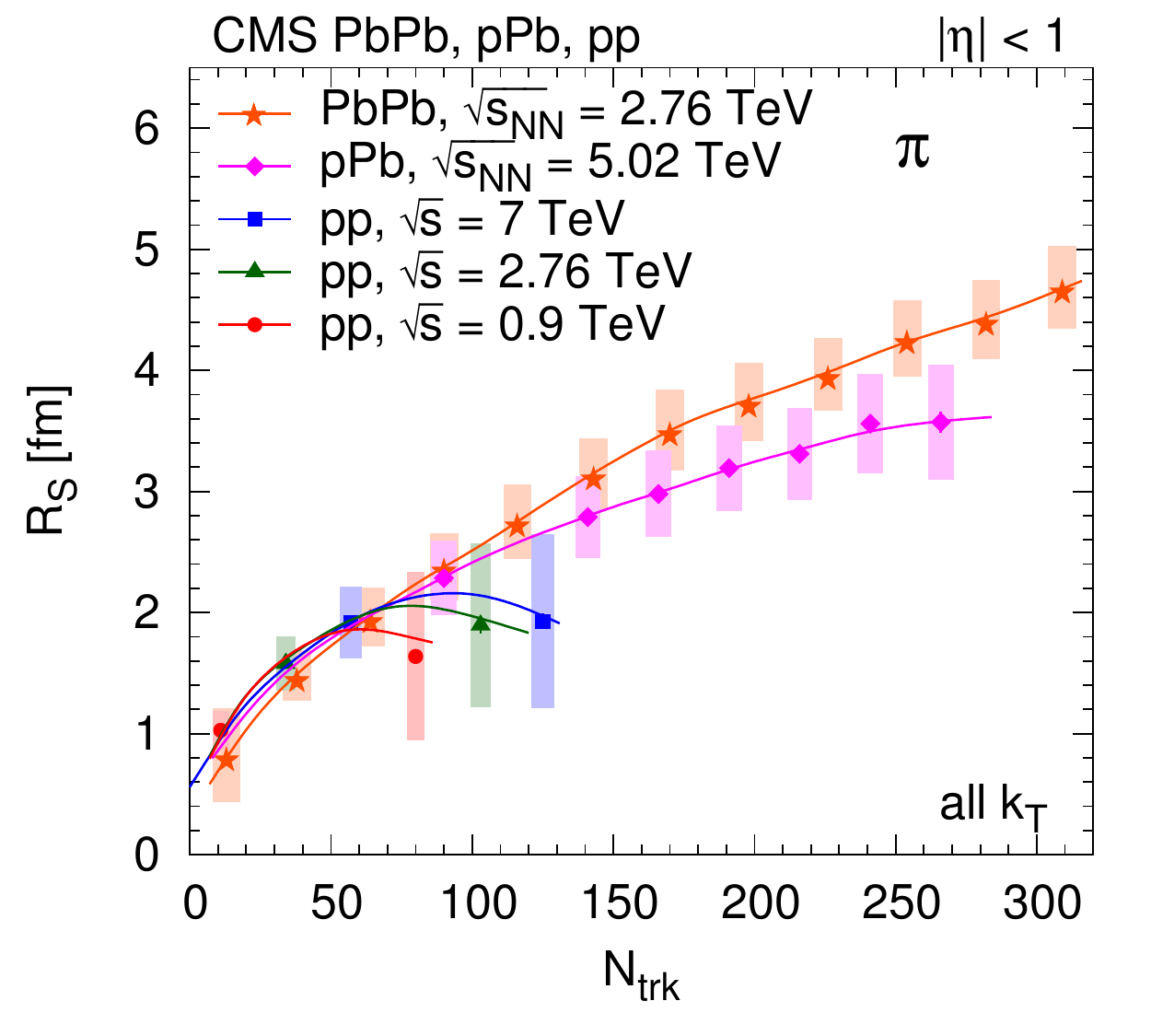}
  \includegraphics[width=0.49\textwidth]{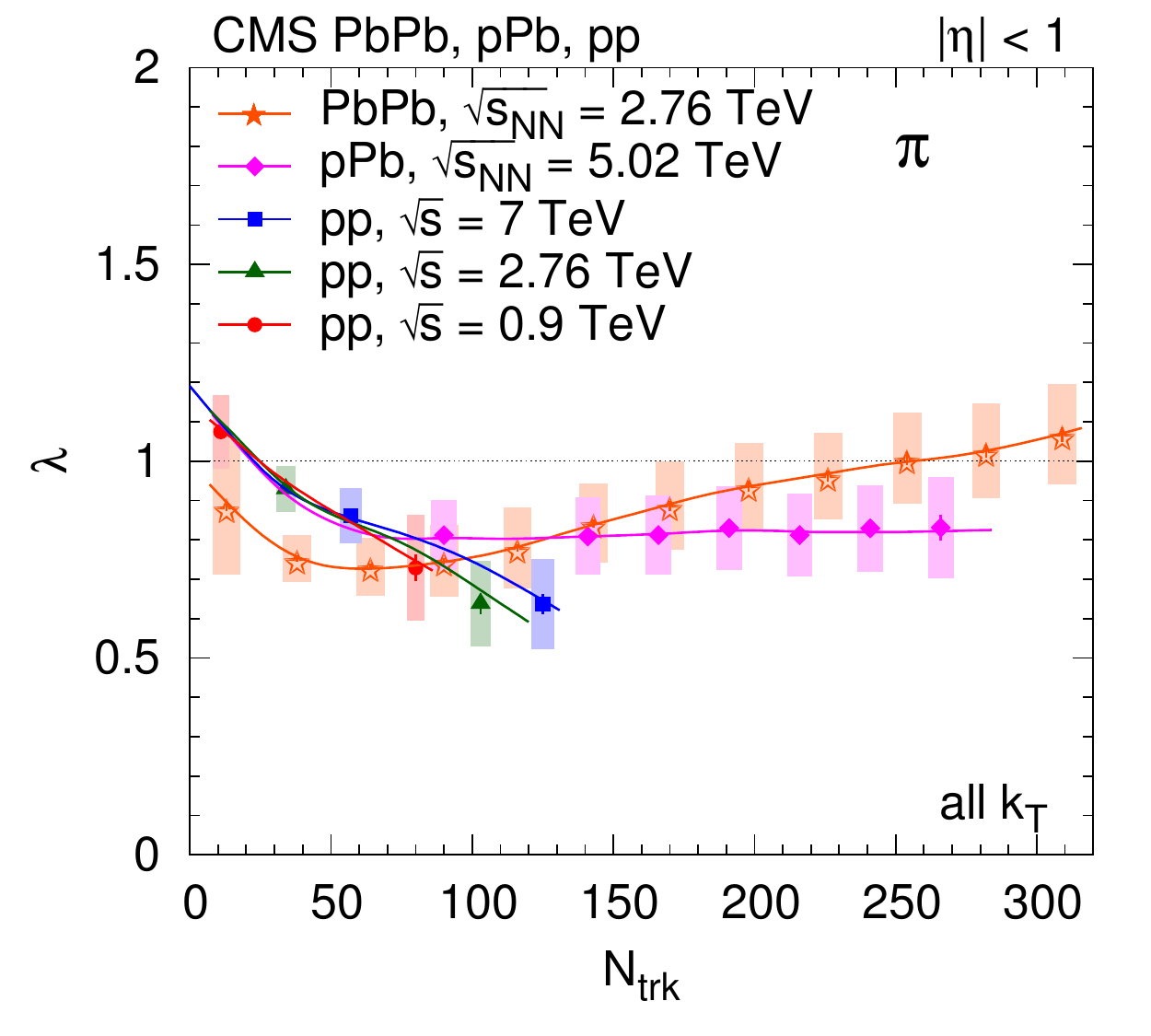}
 \caption{Track-multiplicity ($\Nt$) dependence of the 3D pion radius parameters
$\Rl$ (top left), $\Ro$ (top right), $\Rs$ (bottom left), and the intercept
parameter $\lambda$ (bottom right), obtained from fits (integrated over all
$\kt$) for all collision systems studied. For better visibility, only every
second measurement is plotted. Lines are drawn to guide the eye.}
 \label{fig:pi_radius_qlos_ql}
\end{figure*}

The $\Nt$ dependence of the 3D radii and the corresponding intercept parameters
for pions (Eq.~\eqref{eq:pid_c2_23d}), in several \kt bins for all studied
systems are shown in Fig.~\ref{fig:pi_radius_qlos_ql}.
The 3D pion radius parameters also show a similar pattern. The values of $\Rl$,
$\Ro$, and $\Rt$ are similar for $\Pp\Pp$ and $\Pp$Pb collisions in all \kt bins.
In general $\Rl > \Rs > \Ro$, indicating once again that the source is
elongated along the beam, which once more coincides with the behavior
seen in $\Pp\Pp$ collisions with the double ratio technique in
Section~\ref{subsec:legacy-results-1d}. In the case of PbPb collisions, $\Rl
\approx \Ro \approx \Rs$, indicating that the source is approximately
symmetric. In addition, PbPb data show a slightly different $\Nt$ dependence.
The $\Rl$ radii extracted from the two- and three-dimensional fits differ
slightly. While they are the same within statistical uncertainties for $\Pp\Pp$ collisions at $\sqrt{s} = 0.9$\TeV, the 3D radii are on average smaller by
0.15\unit{fm} for $\Pp\Pp$ collisions at $\sqrt{s} =  2.76$\TeV, by 0.3\unit{fm} for $\Pp\Pp$ collisions at $\sqrt{s} =  7$\TeV and $\Pp$Pb collisions at $\ssNN = 5.02$\TeV, and
by 0.5\unit{fm} for PbPb collisions at $\ssNN = 2.76$\TeV.
The most visible distinction between $\Pp\Pp$, $\Pp$Pb and PbPb data is observed in
$\Ro$, which could point to a different lifetime, $\tau$, of the created
systems in those collisions, since the outward radius is related to the
emitting source lifetime by $\Ro^2 = (R_O^{^G})^2 + \tau^2 \beta_T^2$, as
discussed in Section~\ref{sub:fitting_functions}.

\subsection{Scaling for pions}

The extracted radius parameters for pions are in the 1--5\unit{fm} range, reaching
their largest values for very high multiplicity $\Pp$Pb collisions, as well as for
similar multiplicity PbPb collisions, and decrease with increasing $\kt$. By
fitting the radii with a product of two independent functions of $\Nt$ and
$\kt$, the dependencies on multiplicity and pair momentum can be factorized.
For that purpose, we have used the following functional form:
\begin{equation}
 \label{eq:frakR}
 {\Rp}(\Nt,\kt) =
   \bigl[a^2 + (b \Nt^\beta)^2\bigr]^{1/2} \,
    \left(0.2\GeV/\kt\right)^\gamma,
\end{equation}
where the minimal radius $a$ and the exponent $\gamma$ of $\kt$ are
kept fixed for a given radius component, for all collision types.
This choice of parametrization is based on previous results~\cite{Lisa:2005js},
namely the minimal radius can be connected to the size of the proton, while the
power law dependence on $\Nt$ is often attributed to the freeze-out density of
hadrons.

\begin{figure*}
 \centering
   \includegraphics[width=0.45\textwidth]{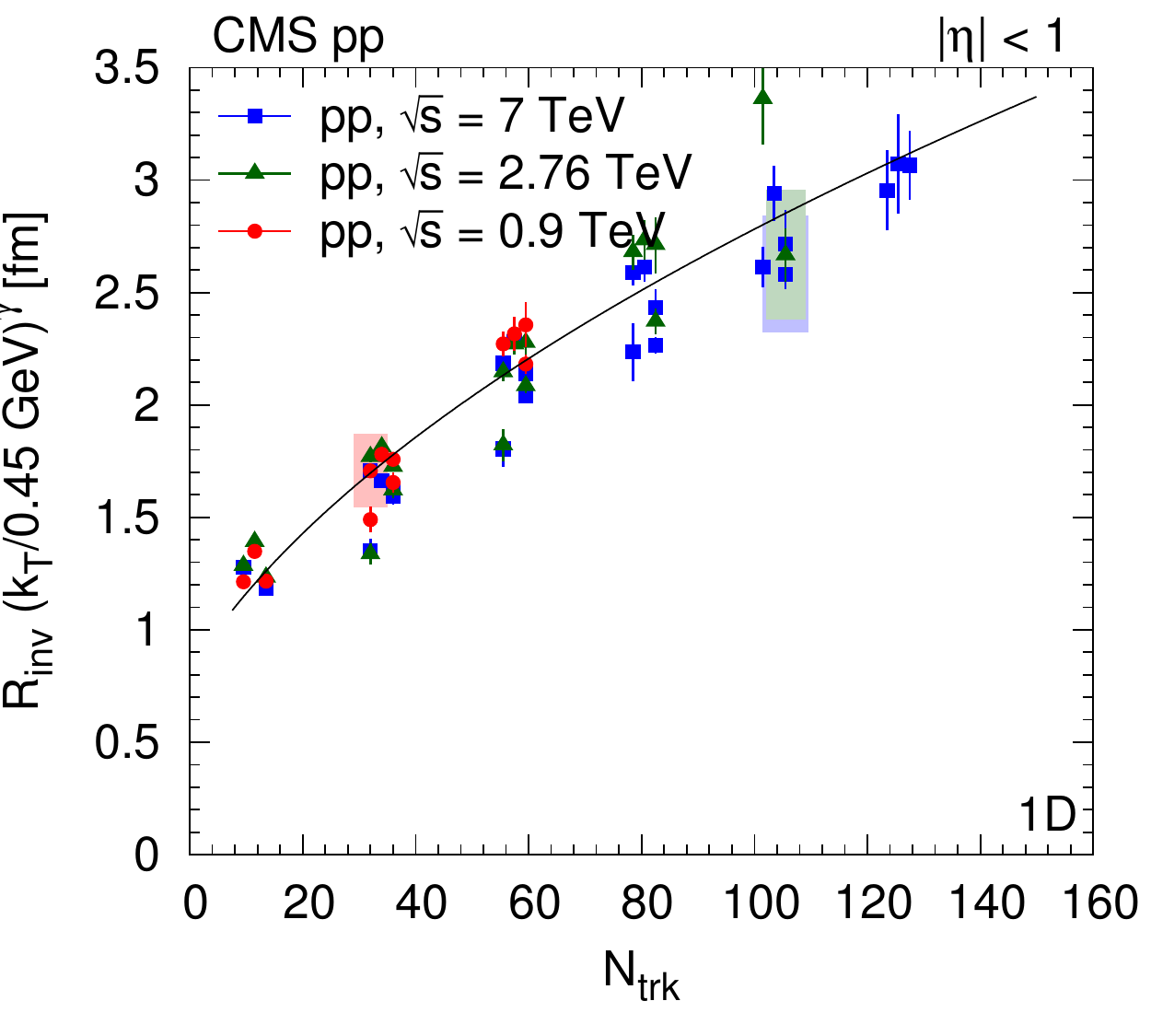}
  \includegraphics[width=0.45\textwidth]{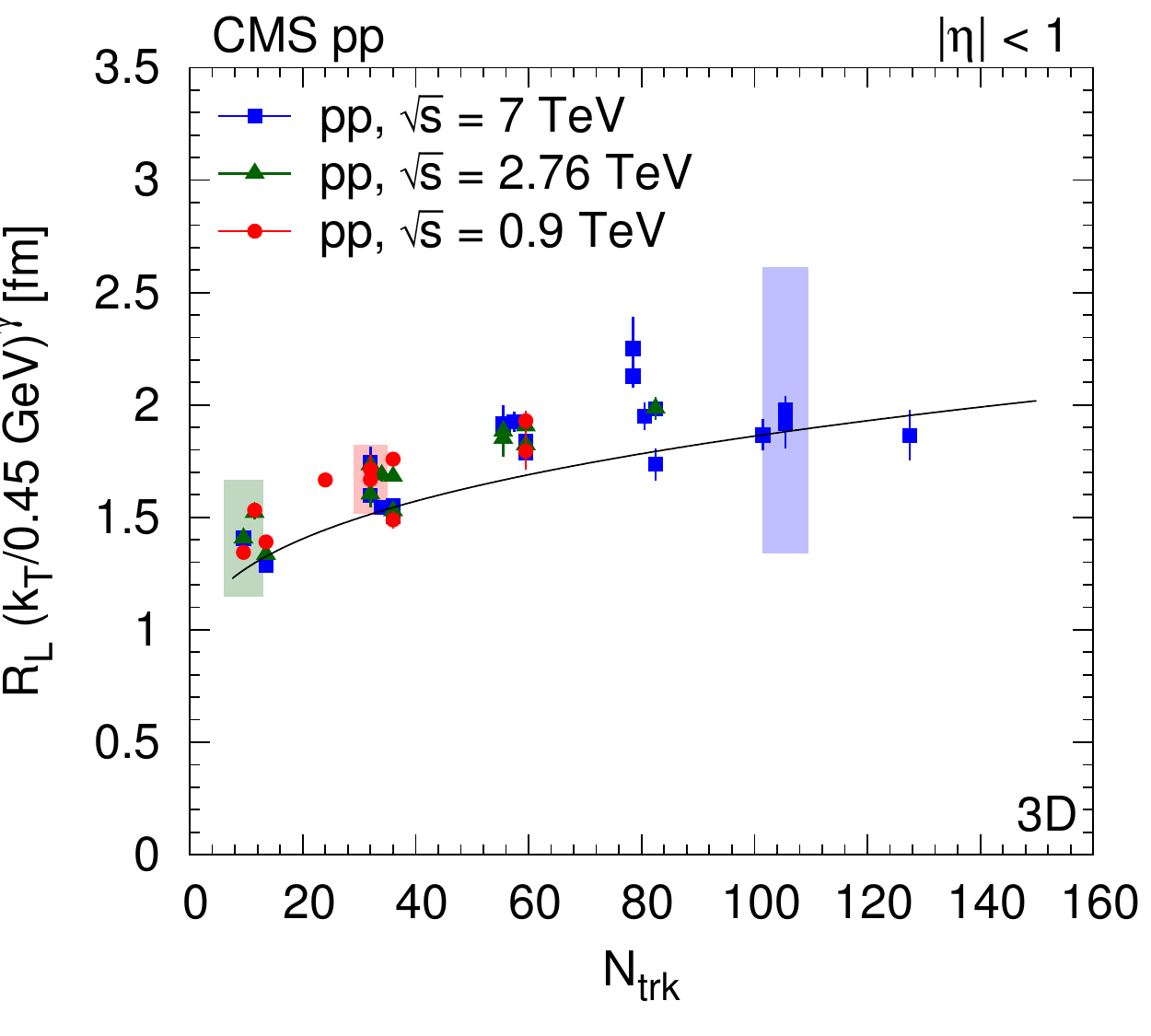}
  \includegraphics[width=0.45\textwidth]{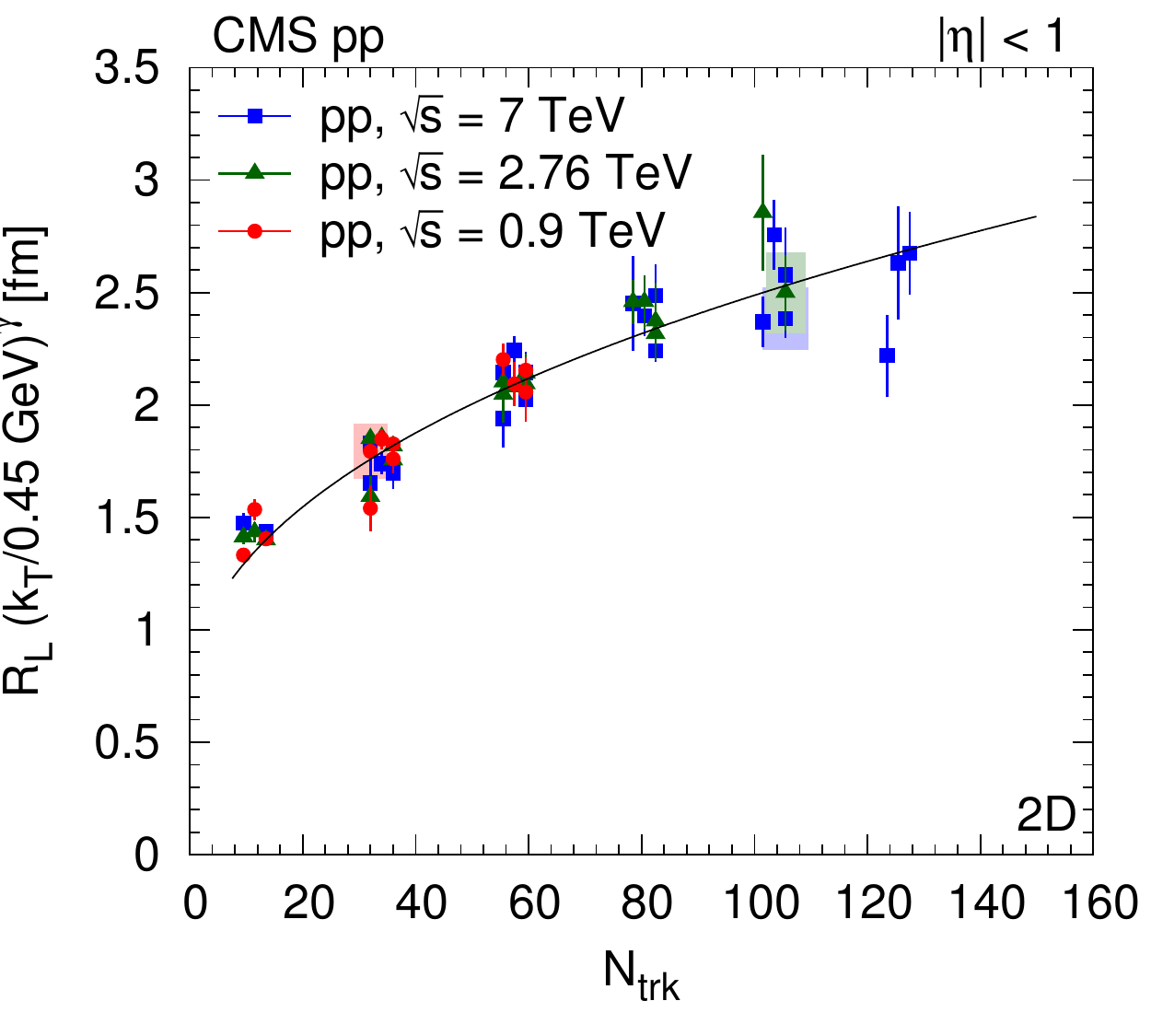}
  \includegraphics[width=0.45\textwidth]{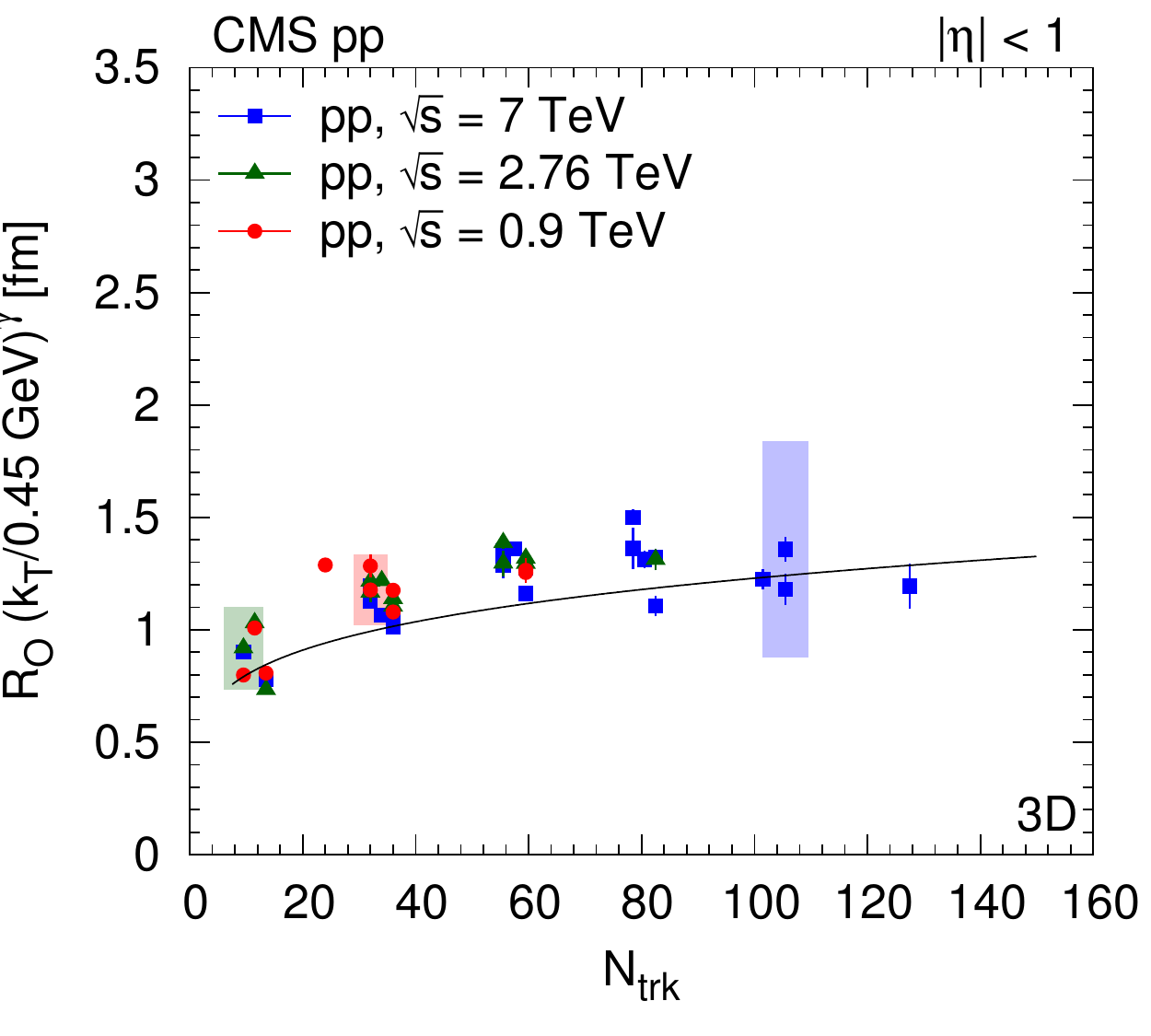}
  \includegraphics[width=0.45\textwidth]{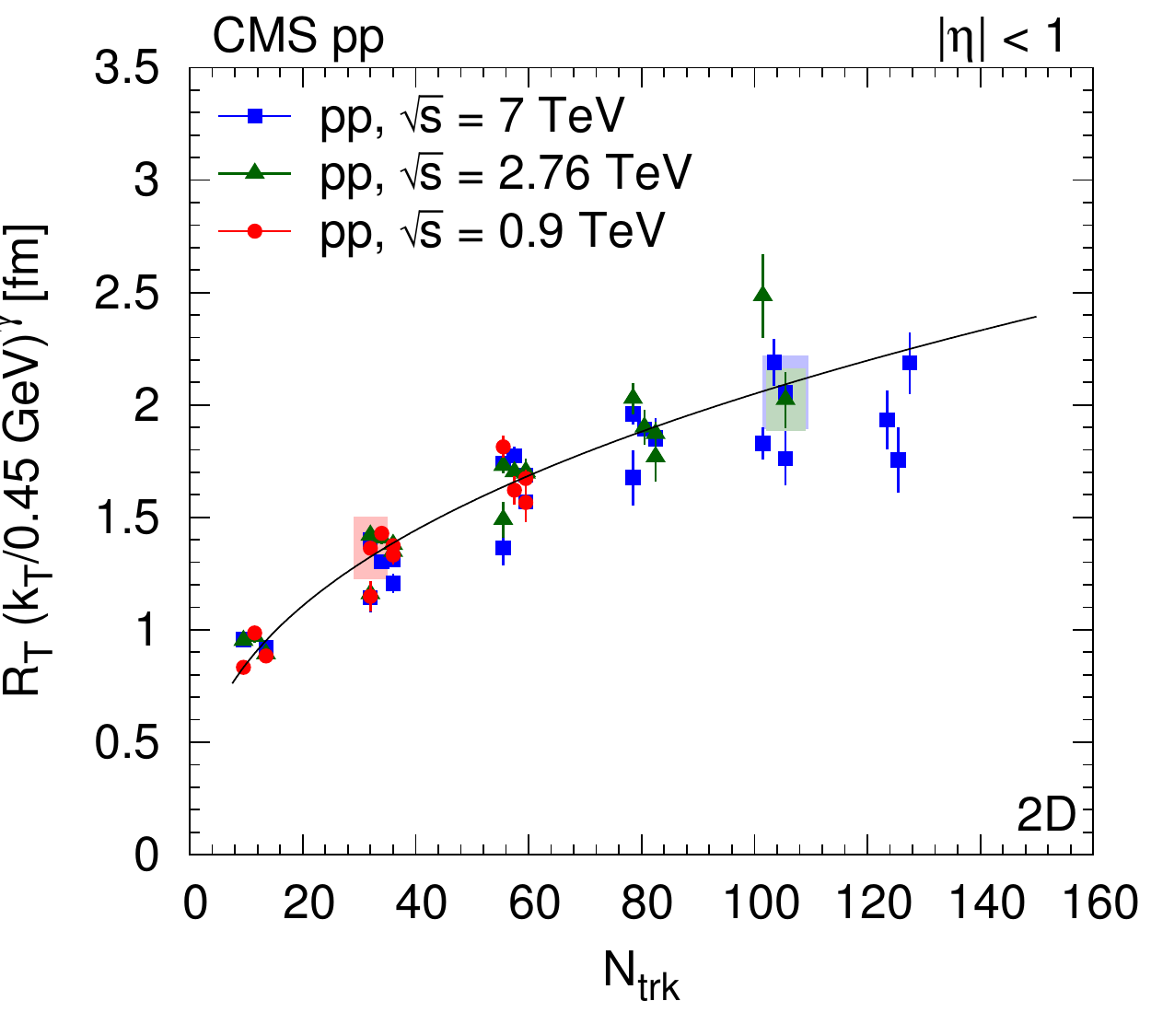}
  \includegraphics[width=0.45\textwidth]{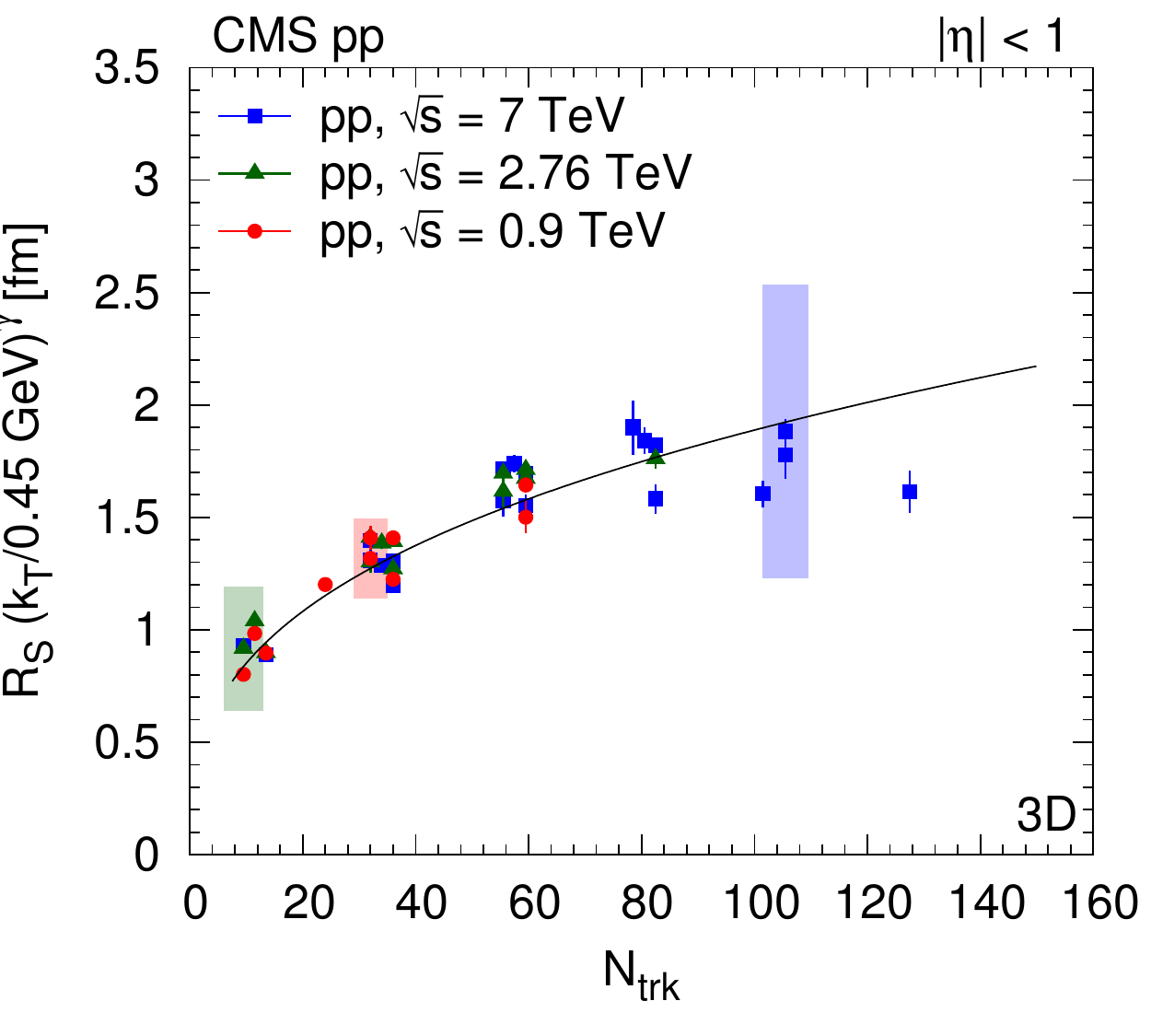}
 \caption{Radius parameters for pions as a function of $\Nt$ scaled to $\kt =
0.45\GeV$ with help of the parametrization, as $R (\kt/0.45\GeV)^\gamma$
(Eq.~\eqref{eq:frakR}) for $\Pp\Pp$ collisions. (For better visibility, systematic
uncertainties are indicated with shaded boxes, only for a fraction of the
points.) Left column, from upper to lower: $\Ri$ from the 1D ($\qi$) analysis,
$\Rl$ and $\Rt$ from the 2D $(\ql,\qt)$ analysis. Right column, from upper to
lower: $\Rl$, $\Ro$, and $\Rs$ from the 3D $(\ql,\qo,\qs)$ analysis. Fit
results are indicated in the figures, see text for details.}
 \label{fig:radius123_scale_pp_only}
\end{figure*}

\begin{figure*}[htbp]
 \centering
  \includegraphics[width=0.45\textwidth]{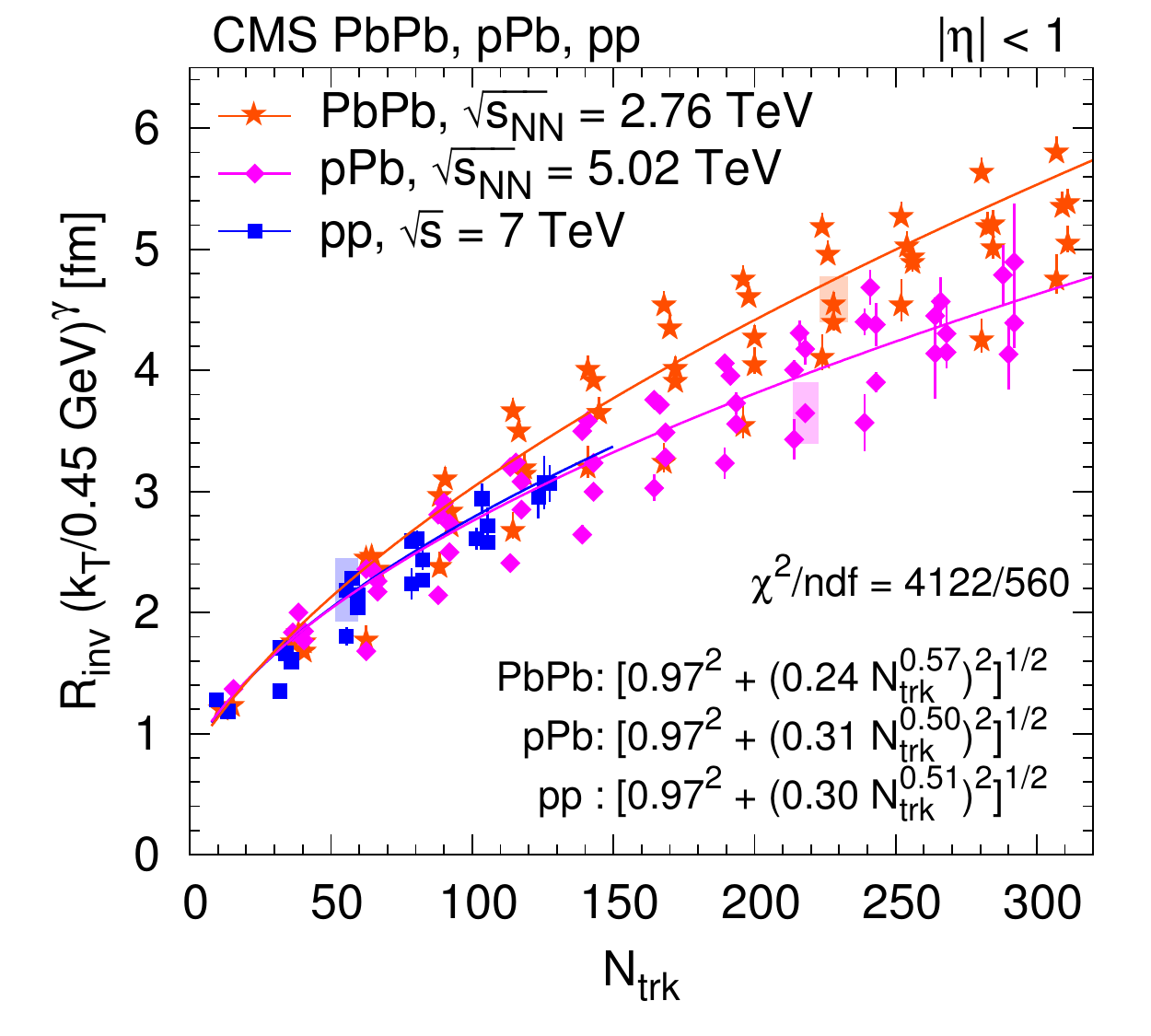}
  \includegraphics[width=0.45\textwidth]{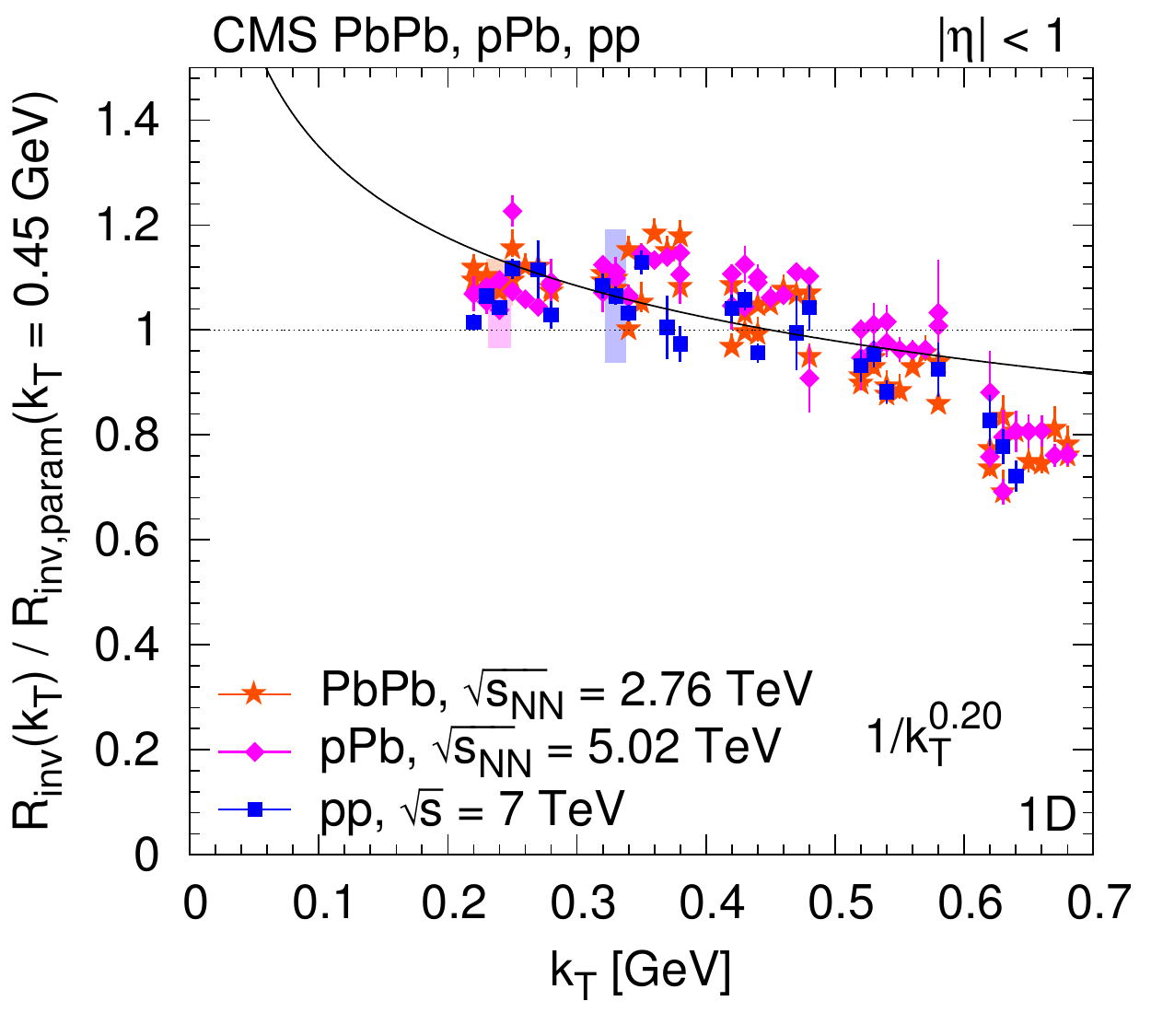}
  \includegraphics[width=0.45\textwidth]{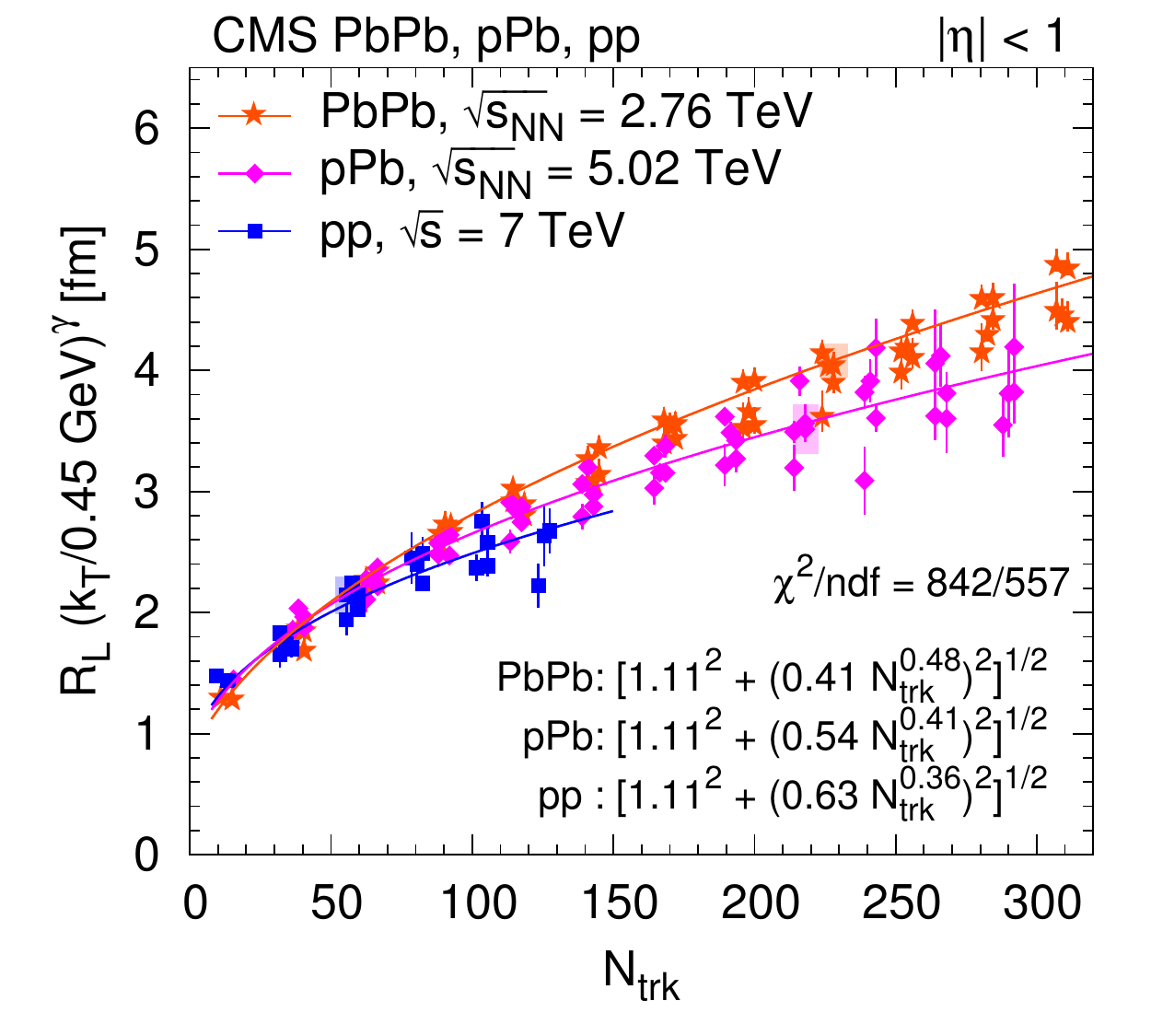}
  \includegraphics[width=0.45\textwidth]{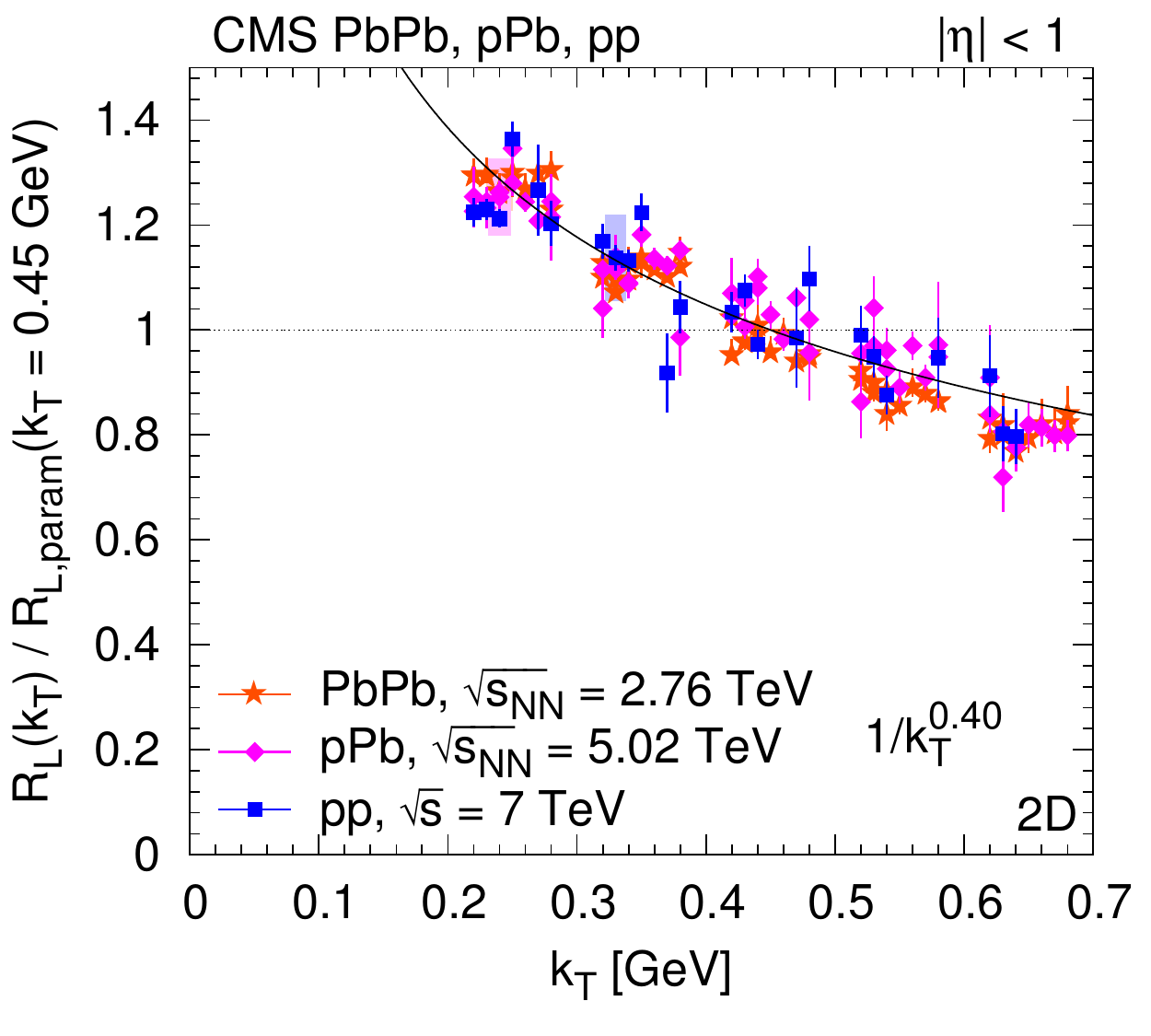}
  \includegraphics[width=0.45\textwidth]{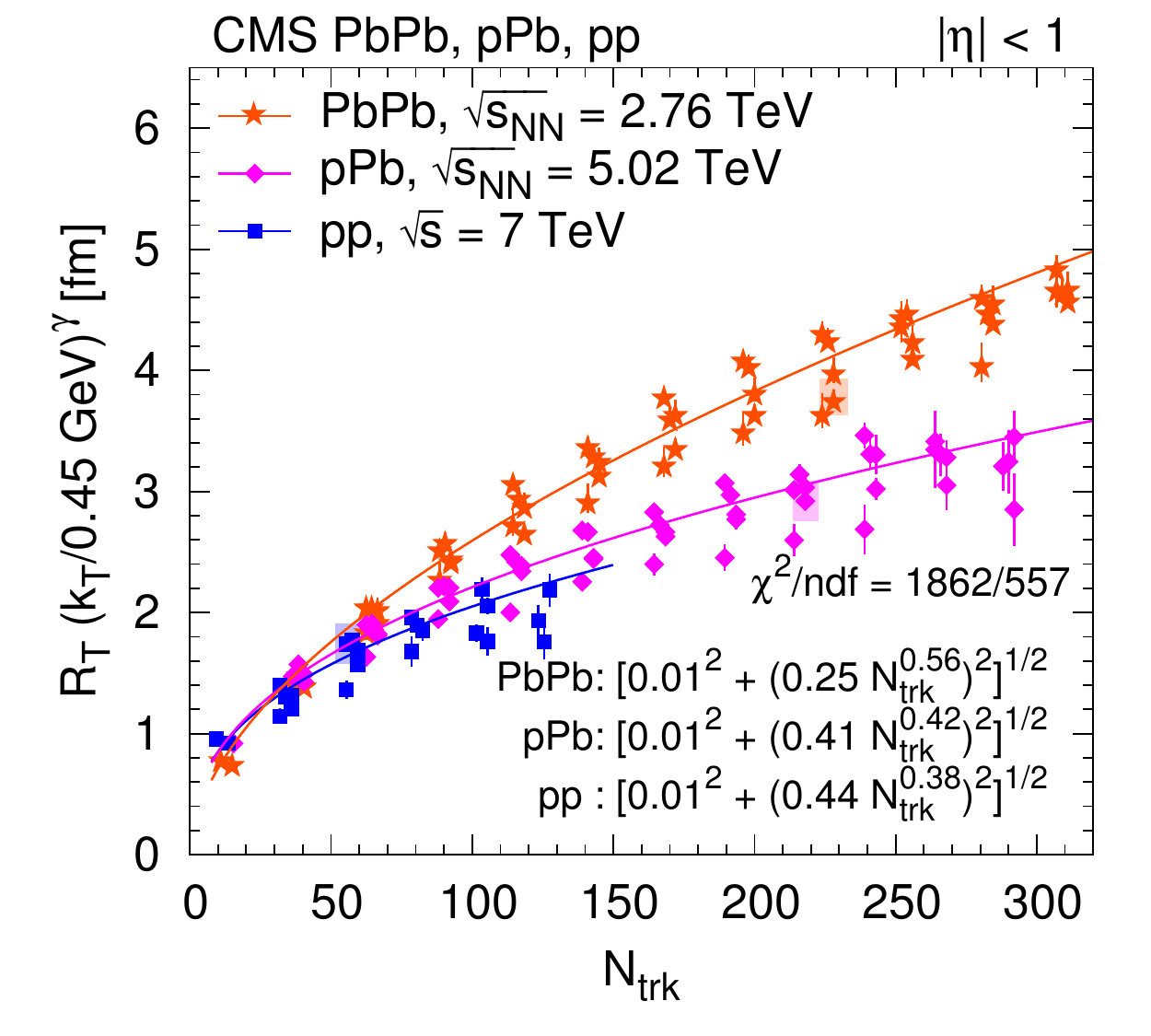}
  \includegraphics[width=0.45\textwidth]{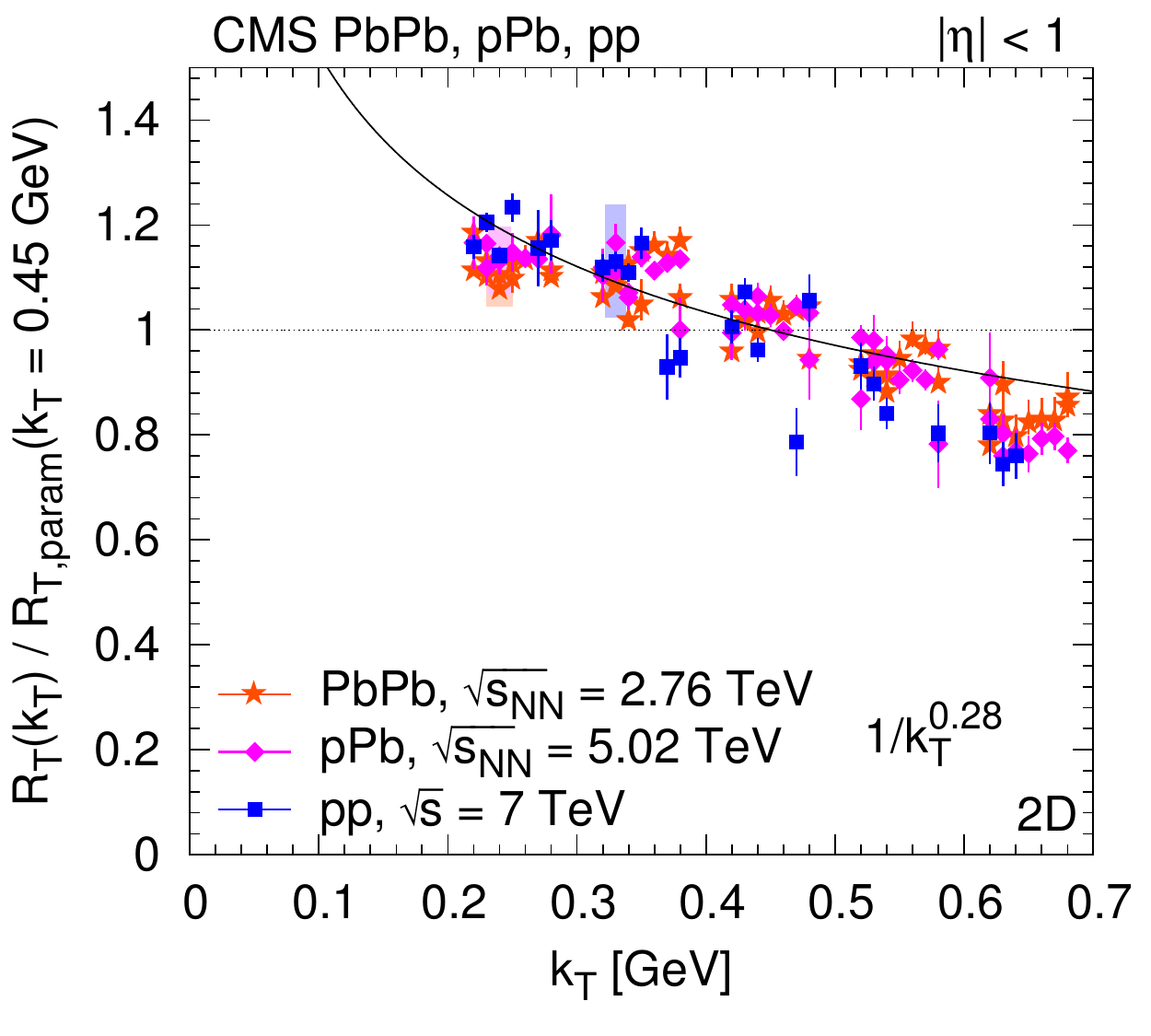}
 \caption{Left: Radius parameters for pions as a function of $\Nt$ scaled to
$\kt = 0.45\GeV$ with help of the parametrization, as $R (\kt/0.45\GeV)^\gamma$
(Eq.~\eqref{eq:frakR}). Right: Ratio of the radius parameter and the value of
the parametrization $\Rp$ (Eq.~\eqref{eq:frakR}) at $\kt = 0.45\GeV$ as a
function $\kt$. (For better visibility, points are shifted to left and to right
with respect to the center of the $\kt$ bin. Systematic uncertainties are
indicated with shaded boxes, only for a fraction of the points.) Upper row:
$\Ri$ from the 1D ($\qi$) analysis. Middle row: $\Rl$ from the 2D $(\ql,\qt)$
analysis. Lower row: $\Rt$ from the 2D $(\ql,\qt)$ analysis. Fit results are
indicated in the figures, see text for details.}
 \label{fig:radius12_scale}
\end{figure*}

We demonstrate the performance of the functional form of Eq.~\eqref{eq:frakR} by
scaling one of the two variables ($\Nt$ or $\kt$) to a common value and showing
the dependence on the other.
Radius parameters scaled to $\kt = 0.45\GeV$, that is $R
(\kt/0.45\GeV)^\gamma$, as a function of $\Nt$ are shown in the left panels
of Figs.~\ref{fig:radius123_scale_pp_only} and
~\ref{fig:radius12_scale}.
The ratio of the radius parameter over the value of the above parametrization at
$\kt = 0.45\GeV$, that is $R/\Rp(\Nt,\kt = 0.45\GeV)$, as a function $\kt$ is
shown in the right panels of Figs.~\ref{fig:radius123_scale_pp_only} and
~\ref{fig:radius12_scale}.
The obtained parameter values are listed in Table~\ref{tab:scaling}. The
estimated uncertainty on the parameters is about 10\%, based on the
joint goodness-of-fits with the above functional form. Overall the proposed
factorization works reasonably well.

\begin{table*}[!hbt]
 \topcaption{Values of the 1D, 2D, and 3D fit parameters
for pions using Eq.~\eqref{eq:frakR} in $\Pp\Pp$, $\Pp$Pb and PbPb collisions.
The estimated uncertainty in the parameters is about 10\%.}
 \label{tab:scaling}
 \centering
 \begin{scotch}{ccccc{c}@{\hspace*{5pt}}cc{c}@{\hspace*{5pt}}ccc}
  &
  \multirow{2}{*}{Radius} & $a$
                          & \multicolumn{2}{c}{pp}
                          &&\multicolumn{2}{c}{$\Pp$Pb}
                          && \multicolumn{2}{c}{PbPb}
                          & \multirow{2}{*}{$\gamma$} \\ \cline{4-5} \cline{7-8} \cline{10-11}
  & & [fm] & $b$ [fm] & $\beta$ && $b$ [fm] & $\beta$ && $b$ [fm] & $\beta$ & \\
  \hline
 1D
 & $\Ri$ & 1.02 & 0.27 & 0.53 && 0.32 & 0.49 && 0.24 & 0.57 & 0.18 \\[\cmsTabSkip]
 \multirow{2}{*}{2D}
 & $\Rl$ & 1.18 & 0.60 & 0.37 && 0.54 & 0.40 && 0.39 & 0.49 & 0.39 \\
 & $\Rt$ & 0.01 & 0.46 & 0.38 && 0.43 & 0.41 && 0.24 & 0.56 & 0.29 \\[\cmsTabSkip]
 \multirow{3}{*}{3D}
 & $\Rl$ & 1.39 & 0.69 & 0.26 && 0.42 & 0.44 && 0.21 & 0.59 & 0.47 \\
 & $\Ro$ & 0.03 & 0.93 & 0.18 && 0.56 & 0.37 && 0.24 & 0.58 & 0.71 \\
 & $\Rs$ & 0.03 & 0.47 & 0.34 && 0.33 & 0.44 && 0.18 & 0.58 & 0.20 \\
 \end{scotch}
\end{table*}

\clearpage

\section{Summary and conclusions}
\label{sec:summ_conclusions}

Detailed studies have been presented on femtoscopic correlations between pairs
of same-sign hadrons produced in $\Pp\Pp$ collisions at $\sqrt{s} = 0.9$, 2.76, and
7\TeV, as well as in $\Pp$Pb collisions at $\ssNN = 5.02$\TeV and peripheral PbPb
collisions at $\ssNN = 2.76$\TeV. The characteristics of two-particle
Bose--Einstein correlations are investigated in one ($\Ri$), two ($\Rt$ and
$\Rl$), and three ($\Rl$, $\Rs$, and $\Ro$) dimensions, as functions of charged
multiplicity $\Nt$ and pair average transverse momentum $\kt$.

Two different analysis techniques are employed. The first ``double ratio''
technique is used to study Bose--Einstein correlations of pairs of charged
hadrons emitted in $\Pp\Pp$ collisions at $\sqrt{s} = 2.76$ and 7\TeV, both in the
collision center-of-mass frame and in the local co-moving system.
The second ``particle identification and cluster subtraction''
method is used to study the characteristics of two-particle correlation
functions of identical charged pions and kaons, identified via their energy
loss in the CMS silicon tracker, in collisions of $\Pp\Pp$, $\Pp$Pb, and peripheral PbPb
collisions at various center-of-mass energies. The similarities and differences
between the three colliding systems have been investigated.

The quantum correlations are well described by an exponential parametrization
as a function of the relative momentum of the particle pair, both in one and in
multiple dimensions, consistent with a Cauchy-Lorentz spatial source
distribution.
The fitted radius parameters of the emitting source, obtained for inclusive
charged hadrons as well as for identified pions, increase along with
charged-particle multiplicity for all colliding systems and center-of-mass
energies, for one, two, and three dimensions alike. The radii are in
the range 1--5\unit{fm}, reaching the largest values for very high multiplicity $\Pp$Pb
collisions, close to those observed in peripheral PbPb collisions with similar
multiplicity.
In the one-dimensional case, $\Ri$ in $\Pp\Pp$ collisions steadily increases with the
charged multiplicity following a $\Nt^{1/3}$ dependence, as
well as showing an approximate independence of the center-of-mass energy.  This
behavior is also observed for the longitudinal radius parameter ($\Rl^*$ in the
center-of-mass frame, and $\Rl$ in the local co-moving system) in the
two-dimensional case.

The multiplicity dependence of $\Rl$ and $\Rt$ is similar for $\Pp\Pp$ and $\Pp$Pb
collisions in all $\kt$ bins, and that similarity also applies to peripheral
PbPb collisions for $\kt >$ 0.4\GeV. In general, the observed orderings, $\Rl >
\Rt$ and $\Rl \gtrsim \Rs > \Ro$, indicate that the $\Pp\Pp$ and $\Pp$Pb sources are
elongated in the beam direction. For peripheral PbPb collisions, the source is
quite symmetric, and shows a slightly different $\Nt$ dependence, with the
largest differences between the systems for $\Rt$ and $\Ro$, while
$\Rl$ and $\Rs$ are approximately equal. The most visible differences between
pp, $\Pp$Pb, and PbPb collisions are seen in $\Ro$, which could point to a
different lifetime of the emitting systems in the three cases.
The kaon radius parameters show a smaller increase with $\Nt$ than observed for
pions.
The radius parameters of charged hadrons and pions decrease with increasing
$\kt$, an observation compatible with expanding emitting sources. The
$\kt$- and $\Nt$-dependences of the radius parameters factorize,
and are largely independent of beam energy and colliding system for lower
multiplicities, although some system dependence appears at higher values of
$\Nt$.

A shallow anticorrelation (a region where the correlation function falls below
one) observed using the double ratio technique in a previous analysis of $\Pp\Pp$ collisions, is also seen in the present data. The depth of this dip in the
one-dimensional correlation functions decreases with increasing particle
multiplicity and also decreases slightly with increasing $\kt$.

Finally, the similar multiplicity dependencies of the radii extracted for $\Pp\Pp$, $\Pp$Pb, and peripheral PbPb collisions suggest a common freeze-out density of
hadrons for all collision systems, since the correlation techniques measure the
characteristic size of the system near the time of the last interactions.
In the case of $\Pp\Pp$ collisions, extending the investigation of these similarities
to higher multiplicities than those accessible with the current data can
provide new information on final state collective effects in such events.

\begin{acknowledgments}
\hyphenation{Bundes-ministerium Forschungs-gemeinschaft Forschungs-zentren Rachada-pisek} We congratulate our colleagues in the CERN accelerator departments for the excellent performance of the LHC and thank the technical and administrative staffs at CERN and at other CMS institutes for their contributions to the success of the CMS effort. In addition, we gratefully acknowledge the computing centers and personnel of the Worldwide LHC Computing Grid for delivering so effectively the computing infrastructure essential to our analyses. Finally, we acknowledge the enduring support for the construction and operation of the LHC and the CMS detector provided by the following funding agencies: the Austrian Federal Ministry of Science, Research and Economy and the Austrian Science Fund; the Belgian Fonds de la Recherche Scientifique, and Fonds voor Wetenschappelijk Onderzoek; the Brazilian Funding Agencies (CNPq, CAPES, FAPERJ, and FAPESP); the Bulgarian Ministry of Education and Science; CERN; the Chinese Academy of Sciences, Ministry of Science and Technology, and National Natural Science Foundation of China; the Colombian Funding Agency (COLCIENCIAS); the Croatian Ministry of Science, Education and Sport, and the Croatian Science Foundation; the Research Promotion Foundation, Cyprus; the Secretariat for Higher Education, Science, Technology and Innovation, Ecuador; the Ministry of Education and Research, Estonian Research Council via IUT23-4 and IUT23-6 and European Regional Development Fund, Estonia; the Academy of Finland, Finnish Ministry of Education and Culture, and Helsinki Institute of Physics; the Institut National de Physique Nucl\'eaire et de Physique des Particules~/~CNRS, and Commissariat \`a l'\'Energie Atomique et aux \'Energies Alternatives~/~CEA, France; the Bundesministerium f\"ur Bildung und Forschung, Deutsche Forschungsgemeinschaft, and Helmholtz-Gemeinschaft Deutscher Forschungszentren, Germany; the General Secretariat for Research and Technology, Greece; the National Scientific Research Foundation, and National Innovation Office, Hungary; the Department of Atomic Energy and the Department of Science and Technology, India; the Institute for Studies in Theoretical Physics and Mathematics, Iran; the Science Foundation, Ireland; the Istituto Nazionale di Fisica Nucleare, Italy; the Ministry of Science, ICT and Future Planning, and National Research Foundation (NRF), Republic of Korea; the Lithuanian Academy of Sciences; the Ministry of Education, and University of Malaya (Malaysia); the Mexican Funding Agencies (BUAP, CINVESTAV, CONACYT, LNS, SEP, and UASLP-FAI); the Ministry of Business, Innovation and Employment, New Zealand; the Pakistan Atomic Energy Commission; the Ministry of Science and Higher Education and the National Science Centre, Poland; the Funda\c{c}\~ao para a Ci\^encia e a Tecnologia, Portugal; JINR, Dubna; the Ministry of Education and Science of the Russian Federation, the Federal Agency of Atomic Energy of the Russian Federation, Russian Academy of Sciences, the Russian Foundation for Basic Research and the Russian Competitiveness Program of NRNU ``MEPhI"; the Ministry of Education, Science and Technological Development of Serbia; the Secretar\'{\i}a de Estado de Investigaci\'on, Desarrollo e Innovaci\'on, Programa Consolider-Ingenio 2010, Plan de Ciencia, Tecnolog\'{i}a e Innovaci\'on 2013-2017 del Principado de Asturias and Fondo Europeo de Desarrollo Regional, Spain; the Swiss Funding Agencies (ETH Board, ETH Zurich, PSI, SNF, UniZH, Canton Zurich, and SER); the Ministry of Science and Technology, Taipei; the Thailand Center of Excellence in Physics, the Institute for the Promotion of Teaching Science and Technology of Thailand, Special Task Force for Activating Research and the National Science and Technology Development Agency of Thailand; the Scientific and Technical Research Council of Turkey, and Turkish Atomic Energy Authority; the National Academy of Sciences of Ukraine, and State Fund for Fundamental Researches, Ukraine; the Science and Technology Facilities Council, UK; the US Department of Energy, and the US National Science Foundation.

Individuals have received support from the Marie-Curie program and the European Research Council and Horizon 2020 Grant, contract No. 675440 (European Union); the Leventis Foundation; the A. P. Sloan Foundation; the Alexander von Humboldt Foundation; the Belgian Federal Science Policy Office; the Fonds pour la Formation \`a la Recherche dans l'Industrie et dans l'Agriculture (FRIA-Belgium); the Agentschap voor Innovatie door Wetenschap en Technologie (IWT-Belgium); the Ministry of Education, Youth and Sports (MEYS) of the Czech Republic; the Council of Scientific and Industrial Research, India; the HOMING PLUS program of the Foundation for Polish Science, cofinanced from European Union, Regional Development Fund, the Mobility Plus program of the Ministry of Science and Higher Education, the National Science Center (Poland), contracts Harmonia 2014/14/M/ST2/00428, Opus 2014/13/B/ST2/02543, 2014/15/B/ST2/03998, and 2015/19/B/ST2/02861, Sonata-bis 2012/07/E/ST2/01406; the National Priorities Research Program by Qatar National Research Fund; the Programa Severo Ochoa del Principado de Asturias; the Thalis and Aristeia programs cofinanced by EU-ESF and the Greek NSRF; the Rachadapisek Sompot Fund for Postdoctoral Fellowship, Chulalongkorn University and the Chulalongkorn Academic into Its 2nd Century Project Advancement Project (Thailand); the Welch Foundation, contract C-1845; and the Weston Havens Foundation (USA).
\end{acknowledgments}

\bibliography{auto_generated}
\cleardoublepage \appendix\section{The CMS Collaboration \label{app:collab}}\begin{sloppypar}\hyphenpenalty=5000\widowpenalty=500\clubpenalty=5000\input{FSQ-14-002-authorlist.tex}\end{sloppypar}
\end{document}

%% file: FSQ-14-002-authorlist.tex
\vskip\cmsinstskip
\textbf{Yerevan~Physics~Institute,~Yerevan,~Armenia}\\*[0pt]
A.M.~Sirunyan, A.~Tumasyan
\vskip\cmsinstskip
\textbf{Institut~f\"{u}r~Hochenergiephysik,~Wien,~Austria}\\*[0pt]
W.~Adam, F.~Ambrogi, E.~Asilar, T.~Bergauer, J.~Brandstetter, E.~Brondolin, M.~Dragicevic, J.~Er\"{o}, M.~Flechl, M.~Friedl, R.~Fr\"{u}hwirth\cmsAuthorMark{1}, V.M.~Ghete, J.~Grossmann, J.~Hrubec, M.~Jeitler\cmsAuthorMark{1}, A.~K\"{o}nig, N.~Krammer, I.~Kr\"{a}tschmer, D.~Liko, T.~Madlener, I.~Mikulec, E.~Pree, D.~Rabady, N.~Rad, H.~Rohringer, J.~Schieck\cmsAuthorMark{1}, R.~Sch\"{o}fbeck, M.~Spanring, D.~Spitzbart, J.~Strauss, W.~Waltenberger, J.~Wittmann, C.-E.~Wulz\cmsAuthorMark{1}, M.~Zarucki
% ...
\vskip\cmsinstskip
% ...
1:~Also at~Vienna~University~of~Technology,~Vienna,~Austria\\
% ...
